\documentclass[12pt]{article}  
\usepackage{cite}
\usepackage{epsfig}
\usepackage{graphicx}
\usepackage{amsmath}
\usepackage{amssymb}
\usepackage{ulem}
\usepackage{mdwlist}          
\usepackage{color}              

\usepackage{a41}
\usepackage{color}
\usepackage[rflt]{floatflt}
\usepackage{float}
\usepackage{slashed}

\usepackage{array}

\usepackage[english]{babel}
\usepackage[latin1]{inputenc}
\usepackage[T1]{fontenc}
\usepackage{ae}

\usepackage{url}

\usepackage{amsmath, amsthm, amssymb}
\newtheorem{thm}{Theorem}[section]

\newtheorem{definition}[thm]{Definition}

\newcommand{\gsim}{\raisebox{-0.07cm   }
{$\, \stackrel{>}{{\scriptstyle\sim}}\, $}}
 \newcommand{\GeV}{\mathrm{GeV}}

 \newcommand{\MOM}{\sf MOM}
 \newcommand{\MS}{\overline{\sf MS}}
 \newcommand{\adag}{/\!\!\! }
 
 \newcommand{\Atiltil}{\tilde{\hspace*{0mm}\tilde{A}}}
 
 \newcommand{\Ahathat}{\hat{\hspace*{0mm}\hat{A}}}
 \newcommand{\Athathat}{\hat{\hspace*{0mm}\hat{\tilde{A}}}}
 
 \newcommand{\Ctil}{\tilde{C}}
 \newcommand{\Brack}[2]{\genfrac{[}{]}{0pt}{}{#1}{#2}}

\newcommand{\NN}{\nonumber}

\newcommand{\N}{\mathbb N}

\newcommand{\Li}{{\rm Li}}
\newcommand{\HA}{{\rm H}}

\newcommand{\Mvec}{{\rm\bf M}}

\newcommand{\ep}{\varepsilon}

\usepackage{rotating}

\usepackage{graphicx}

\newcounter{mmacnt}
\def\restartmma{\setcounter{mmacnt}{0}}
\restartmma \catcode`|=\active
\def|#1|{\mathrm{#1}}
\catcode`|=12
\newenvironment{mma}{
 \par\smallskip
 \catcode`|=\active
 \parskip=0pt\parindent=0pt 
 \small
 \def\In##1\\{%
   \def\linebreak{\hfill\break\null\qquad}%
   \refstepcounter{mmacnt}
   \hangindent=2.5em\hangafter=0
   \leavevmode
   \llap{\tiny\sffamily In[\arabic{mmacnt}]:=\kern.5em}%
   \mathversion{bold}\footnotesize$\displaystyle##1$\normalsize
   \mathversion{normal}\par
 }%
 \def\Print##1\\{%
   \def\linebreak{\hfill\break}%
   \hangindent=2.5em\hangafter=0
   \leavevmode ##1\par}%
 \def\Out##1\\{%
   \def\linebreak{$\hfill\break\null\hfill$}%
   \kern\abovedisplayskip\par
   \hangindent=2.5em\hangafter=0
   \leavevmode
   \llap{\tiny\sffamily Out[\arabic{mmacnt}]=\kern.5em}
   \footnotesize$\displaystyle##1$\normalsize\hfill\null\par
   \kern\belowdisplayskip
 }%
 \def\Warning##1##2\\{%
   \def\linebreak{\hfill\break}%
   \hangindent=2.5em\hangafter=0
   \leavevmode
   {\scriptsize##1 : ##2}\par}%
}{%
 \par\smallskip
}

\usepackage{color}

\newenvironment{fshaded}{%
\MakeFramed {\FrameRestore}
}%
{\endMakeFramed}

\allowdisplaybreaks[4]

\begin{document}
\setlength{\baselineskip}{0.515cm}
\sloppy
\thispagestyle{empty}
\begin{flushleft}
DESY 14--019
\\
DO--TH 15/08\\
May  2017\\
\end{flushleft}

\mbox{}
\vspace*{\fill}
\begin{center}

{\LARGE\bf Three Loop Massive Operator Matrix}

\vspace*{3mm} 
{\LARGE\bf Elements and Asymptotic Wilson Coefficients}

\vspace*{3mm} 
{\LARGE\bf  with Two Different Masses}

\vspace{3cm}
\large
J.~Ablinger$^a$, 
J.~Bl\"umlein$^b$, 
A.~De Freitas$^b$, 
A.~Hasselhuhn$^a$,\footnote{Present address:~Institut f\"ur Theoretische Teilchenphysik
Campus S\"ud, Karlsruher Institut f\"ur Technologie (KIT), D-76128 Karlsruhe, Germany.} 
C.~Schneider$^a$, \\  and  F.~Wi\ss{}brock$^{a,b,c}$ 

\vspace{1.cm}
\normalsize
{\it $^a$~Research Institute for Symbolic Computation (RISC),\\
                          Johannes Kepler University, Altenbergerstra\ss{}e 69,
                          A--4040, Linz, Austria}\\

\vspace*{3mm}
{\it  $^b$ Deutsches Elektronen--Synchrotron, DESY,}\\
{\it  Platanenallee 6, D-15738 Zeuthen, Germany}
\\

\vspace*{3mm}
{\it  $^c$ IHES, 35 Route de Chartres, F-91440 Bures-sur-Yvette, France.} 
\\

\end{center}
\normalsize
\vspace{\fill}
\begin{abstract}
\noindent
Starting at 3-loop order, the massive Wilson coefficients for deep-inelastic scattering and the massive operator
matrix elements describing the variable flavor number scheme receive contributions of Feynman diagrams carrying 
quark lines with two different masses. In the case of the charm and bottom quarks, the usual decoupling of one heavy 
mass at a time no longer holds, since the ratio of the respective masses, $\eta = m_c^2/m_b^2 \sim 1/10$, is not 
small enough. Therefore, the usual variable flavor number scheme (VFNS) has to be generalized. The renormalization 
procedure in the two--mass case is different from the single mass case derived in \cite{Bierenbaum:2009mv}. We present 
the moments $N=2,4$ and $6$ for all contributing operator matrix elements, expanding in the ratio $\eta$. We calculate 
the analytic results for general values of the Mellin variable $N$ in the flavor non-singlet case, as well as for 
transversity and the matrix element $A_{gq}^{(3)}$. We also calculate the two-mass scalar integrals of all topologies 
contributing to the gluonic operator matrix element $A_{gg}$. As it turns out, the expansion in $\eta$ is usually 
inapplicable for general values of $N$. We therefore derive the result for general values of the mass ratio. From the 
single pole terms we derive, now in a two-mass calculation, the corresponding contributions to the 3-loop anomalous 
dimensions. We introduce a new general class of iterated integrals and study their relations and present special values. 
The corresponding functions are implemented in computer-algebraic form.
\end{abstract}

\vspace*{\fill}
\noindent
\numberwithin{equation}{section}

\newpage
\section{Introduction}
\label{sec:1}

\vspace*{1mm}
\noindent
The heavy flavor corrections to deep-inelastic scattering for pure photon exchange are known to leading \cite{HLO} and
next-to-leading order (NLO) \cite{HNLO}\footnote{For a precise implementation of the Wilson coefficients in Mellin space see
\cite{Alekhin:2003ev}.}. The present accuracy of the deep-inelastic world data requires next-to-next-to leading order (NNLO)
QCD analyses in order to determine the strong coupling constant $\alpha_s(M_Z^2)$ 
\cite{Bethke:2011tr,Blumlein:2006be,Alekhin:2017kpj} to $\sim 1\%$ accuracy at NNLO, to obtain highly 
accurate  values for the charm and bottom quark masses $m_c$ and $m_b$, and to make precise determinations 
of the parton distribution functions. 
All of this is in turn needed to describe precision measurements at the LHC \cite{LHC} and at facilities
planned for the future \cite{Boer:2011fh,AbelleiraFernandez:2012cc}. 

In the region of large scales $Q^2 \gg m^2$, analytic 
expressions for the heavy flavor Wilson coefficients have been obtained at NLO \cite{Buza:1995ie,Bierenbaum:2007qe}. A 
factorization relation valid in this asymptotic region was given in Refs.~\cite{Buza:1995ie,Buza:1996wv}. For the structure function 
$F_2(x,Q^2)$, the asymptotic corrections are sufficient at scales $Q^2/m^2 \gsim 10$,~cf.~\cite{Buza:1995ie}. The 
massless corrections at NNLO to the deep-inelastic structure functions are available 
\cite{WIL2,Moch:2004pa,Vogt:2004mw}, while for the corresponding massive corrections in the asymptotic limit, a series of moments 
has been calculated in the single heavy mass case \cite{Bierenbaum:2009mv} for all contributing terms in neutral current 
deep-inelastic scattering. The calculation of the general expressions for the Wilson coefficients is still underway. The 
asymptotic Wilson coefficients for the structure function $F_L(x,Q^2)$ have been 
completed \cite{Blumlein:2006mh,Behring:2014eya}. Here the first genuine two-mass contributions 
emerge at fourth order in the coupling constant.
In the case of the structure function $F_2(x,Q^2)$, all corrections to the color factors $O(N_F T_F^2 C_{A,F})$ have been obtained in 
\cite{Ablinger:2010ty,Blumlein:2012vq}, which provides the complete results for two out of five contributing Wilson coefficients, cf. also 
\cite{Behring:2014eya}. The flavor non-singlet corrections have been calculated in Ref.~\cite{Ablinger:2014vwa} and the flavor 
pure singlet terms in Ref.~\cite{Ablinger:2014nga}. 
The massive operator matrix elements (OMEs) calculated in \cite{Ablinger:2010ty,Behring:2014eya,Ablinger:2014vwa,Ablinger:2014nga} are also needed to describe
the variable flavor number scheme (VFNS) in the case of a single heavy quark transition \cite{Buza:1996wv}, for which also
the gluonic contributions $A_{gq,Q}$ and $A_{gg,Q}$ are required and have been calculated at 3-loop order in \cite{Ablinger:2014lka} and in 
\cite{Blumlein:2012vq,Ablinger:2014uka,AGG}, respectively.\footnote{For a recent survey on these calculations see \cite{Blumlein:2014zxa}.}
Technical aspects of these calculations have been described in 
\cite{Ablinger:2012qm,Ablinger:2014yaa,Ablinger:2015tua}. Heavy quark corrections to charged current deep-inelastic processes 
have been dealt with in Refs.~\cite{CC}. 

In the calculations mentioned above, besides internal massless fermion lines, only a single heavy mass is attached
to massive fermion lines. However, starting at 3-loop order, there are also diagrams with two different masses attached to the massive lines.
In the present paper, we consider corrections of this type. As before in the single heavy mass case \cite{Bierenbaum:2009mv},
a series of finite moments for all massive OMEs and the Wilson coefficients in the asymptotic
region $Q^2 \gg m_{c,b}^2$ is calculated. In some cases, we also compute the results at general values of the 
Mellin 
variable $N$ and the momentum fraction $z$. Furthermore, we present the scalar two-mass integrals contributing 
to the OME $A_{gg}$ both in $z$- and $N$-space, in extension to the single mass case in Ref.~\cite{Ablinger:2014uka}.
In the present paper, we concentrate on the calculation of the two-mass effects in the case of massive OMEs, playing a 
central role in the variable flavor number scheme, and leave phenomenological studies of the contributions to various 
deep-inelastic structure functions for a separate publication.

The paper is organized as follows. In Section~2, the general formalism is outlined, describing the Wilson coefficients in the 
asymptotic region in the case of two massive quarks and the representation of the deep-inelastic structure functions. We also 
present the transition relations between a representation of three and five massless quarks to 3-loop order, which is governed 
by the massive OMEs and describes the matching conditions in the VFNS. In Section~3, the renormalization of the massive OMEs is 
described in the case of two massive flavors. Here we also derive the structure of the massive OMEs, which now receives logarithmic 
contributions depending on two masses. The fixed moments for $N = 2, 4$ and $6$ are calculated for all massive OMEs 
in Section~4,
for which we also present numerical illustrations. We have reported on a few results already briefly in 
\cite{Ablinger:2011pb,Ablinger:2012qj,Wissbrock:2015faa}.
In the flavor non-singlet and $gq$-cases, we have calculated the massive 
OMEs for general values of the Mellin variable $N$. These are presented in Section~5 and are numerically illustrated. 
In Section~6, we turn to the more involved case of the genuine two-mass contributions to the massive 
OME $A_{gg,Q}^{(3)}$, and outline the calculation strategy, which is significantly different from those of the easier cases 
being dealt with in Section~5. In the present paper, we limit the consideration to the calculation of all scalar\footnote{ 
That is, not including in the numerator any term other than the one coming from the operator insertion.}
3-loop diagrams contributing to $A_{gg,Q}^{(3)}$, both in $N$- and $z$-space, leading to new functional structures. Unlike the case for the moments,
cf.~Section~4, where we can expand in the mass ratio of the heavy quarks, this is in general not possible in the case of the 
diagrams contributing to $A_{gg,Q}^{(3)}$ for general values of $N$. Therefore, as in Section~5, we derive the analytic 
solution for general values of the mass ratio. Section~7 contains the conclusions. The $z$-space results of a series of OMEs 
are given in the Appendix~\ref{APP1}, and a collection of new root-valued iterated integrals is presented in 
Appendix~\ref{APP2}. 

\section{Massive OMEs and Wilson Coefficients with two masses
{\label{Sec-OMEs2m}}}

\vspace*{1mm}
\noindent
Starting at 3--loop order, Feynman diagrams carrying internal fermion lines of different mass contribute to the 
OMEs. The relevant masses are those of the charm and bottom quark, $m_c$ and $m_b$. In the following, we will work 
in the on-shell scheme. Here the masses are given by \cite{Alekhin:2012vu,PDG}
\begin{eqnarray}
\label{eq:mc}
m_c &=& 1.59~\GeV
\\
\label{eq:mb}
m_b &=& 4.78~\GeV~.
\end{eqnarray}
The ratio 
\begin{eqnarray}
\eta =  \frac{m_2^2}{m_1^2}
\end{eqnarray}
with $m_2 = m_c, m_1 = m_b$, amounts to $\eta \sim 1/10$. Later we will also use the symbol 
$\eta_1 = \sqrt{\eta}$. The two masses do not form a strong hierarchy and charm cannot be assumed to be 
massless at $\mu^2 = m_b^2$. 
The asymptotic decoupling thus rather proceeds under the condition
\begin{eqnarray}
Q^2, \mu^2 \gg m_c^2, m_b^2~,
\label{eq:cond}
\end{eqnarray}
with $Q^2$ the virtuality of the exchanged gauge boson in the deep-inelastic process and $\mu^2$ the factorization scale,
which we will set equal to the renormalization scale in the following. The transition relation to the
$\overline{\rm MS}$-scheme for the mass renormalization will be given in Section~\ref{MRS}.
We refer to the on-shell scheme in the following for computational reasons, rather than
giving preference to this scheme. In any data analysis, the mass effects shall be expressed in the $\overline{\rm 
MS}$-scheme, which provides perturbative stability.

In view of this, the associated variable flavor number scheme (VFNS) differs from the one in which only a single heavy quark
is decoupled at the time \cite{Buza:1996wv,Bierenbaum:2009mv}, which also works up to 2--loop order since there no diagrams 
containing fermion lines 
of different mass contribute. 

In the following, we will mainly work in Mellin space to take advantage of the simplicity of the emerging convolution
formulae, which are given by ordinary products. The Mellin transform of a function $f(x)$ is defined by
\begin{eqnarray}
\Mvec[f(x)](N) = \int_0^1 dx x^{N-1} f(x)~.
\label{eq:MT}
\end{eqnarray}
The convolution formula of two functions reads
\begin{eqnarray}
\left[f\otimes g\right](z) = \int_0^1 dx_1 \int_0^1 dx_2 \delta(z  - x_1 x_2) f(x_1) g(x_2). 
\end{eqnarray}
Its Mellin transform factors into the Mellin transforms of both functions
\begin{eqnarray}
\Mvec[f(z) \otimes g(z)](N) = 
\Mvec[f(z)](N) \cdot \Mvec[g(z)](N).
\end{eqnarray}
In what follows, we will use the Mellin transform to map between the $z$- and the Mellin $N$-spaces.

Let us now derive the massive Wilson coefficients for deep-inelastic scattering in the kinematic range of large virtualities $Q^2$, 
cf.~(\ref{eq:cond}). We generalize the considerations in the case of a single heavy quark mass in 
Refs.~\cite{Buza:1996wv,Bierenbaum:2009mv} and obtain the following factorization relation in the non--singlet case:
\begin{eqnarray}
   C_{q,(2,L)}^{\sf NS}\Bigl(N,N_F,\frac{Q^2}{\mu^2}\Bigr)
     + L_{q,(2,L)}^{\sf NS}
          \Bigl(N,N_F+2,\frac{Q^2}{\mu^2},\frac{m_1^2}{\mu^2},\frac{m_2^2}{\mu^2}\Bigr)
     &=&
\NN\\ &&
       \hspace{-50mm}
        A_{qq,Q}^{\sf NS}\Bigl(N,N_F+2,\frac{m_1^2}{\mu^2},\frac{m_2^2}{\mu^2}\Bigr)
        C_{q,(2,L)}^{\sf NS}\Bigl(N,N_F+2,\frac{Q^2}{\mu^2}\Bigr)~.
        \NN\\&&
        \label{LNSFAC2M}
\end{eqnarray}
Here $N_F$ denotes the number of massless flavors (with $N_F = 3$ in QCD). $C_i^j$ and $A_{kl}$ are the massless 
Wilson coefficients, cf. \cite{WIL1,WIL2,Vermaseren:2005qc} and massive operator matrix elements (OMEs), 
respectively.

For the pure singlet and singlet contributions the corresponding relations read
\begin{eqnarray}
     C_{q,(2,L)}^{\sf PS}(N_F)
       +L_{q,(2,L)}^{\sf PS}
            (N_F+2)
     &=&
        \Bigl[
               A_{qq,Q}^{\sf NS}(N_F+2)
              +A_{qq,Q}^{\sf PS}(N_F+2)
              +A_{Qq}^{\sf PS}(N_F+2)
         \Bigr]
\NN\\ &&
         \times
	 N_F {\tilde{C}}_{q,(2,L)}^{\sf PS}(N_F+2)
         \NN\\&&
        +A_{qq,Q}^{\sf PS}(N_F+2)
         C_{q,(2,L)}^{\sf NS}(N_F+2)
\NN\\ &&        
+A_{gq,Q}(N_F+2)
N_F {\tilde{C}}_{g,(2,L)}(N_F+2)~, \NN\\ 
                 \label{LPSFAC2M} \\
      C_{g,(2,L)}(N_F)
     +L_{g,(2,L)}(N_F+2)
    &=&
           A_{gg,Q}(N_F+2)
	   N_F {\tilde{C}}_{g,(2,L)}(N_F+2)
\NN\\ &&
         + A_{qg,Q}(N_F+2)
           C_{q,(2,L)}^{\sf NS}(N_F+2)
\NN\\ &&         
+\Bigl[
                A_{qg,Q}(N_F+2)
               +A_{Qg}(N_F+2)
          \Bigr]
	  N_F{\tilde{C}}_{q,(2,L)}^{\sf PS}(N_F+2)~,\NN\\
        \label{LgFAC2M}
\\
     \tilde{\tilde{H}}_{q,(2,L)}^{\sf PS}(N_F+2)
     &=&  
        A_{Qq}^{\sf PS}(N_F+2)
           \Bigl[ 
                 C_{q,(2,L)}^{\sf NS}(N_F+2)
		 +{\tilde C_{q,(2,L)}}^{\sf PS}
                         (N_F+2)
          \Bigr]
\NN\\ &&
          +\Bigl[ 
                A_{qq,Q}^{\sf NS}(N_F+2)
               +A_{qq,Q}^{\sf PS}(N_F+2)
         \Bigr]
	 {\tilde{C}}_{q,(2,L)}^{\sf PS}(N_F+2)
\NN\\ &&
       +A_{gq,Q}(N_F+2)
       {\tilde{C}}_{g,(2,L)}(N_F+2)~,         \label{HPSFAC2M} \\
     \tilde{\tilde{H}}_{g,(2,L)}(N_F+2)
      &=&
         A_{gg,Q}(N_F+2)
	 {\tilde{C}}_{g,(2,L)}(N_F+2)
\nonumber\\ &&
        +A_{qg,Q}(N_F+2)
	{\tilde{C}}_{q,(2,L)}^{\sf PS}(N_F+2)
\NN\\ &&
        + A_{Qg}(N_F+2)
          \Bigl[ C_{q,(2,L)}^{\sf NS}(N_F+2)
		  +{\tilde{C}}_{q,(2,L)}^{\sf PS}(N_F+2)
             \Bigr]~.         \label{HgFAC2M}
\nonumber\\
\end{eqnarray}
Due to the heavy quark charge, Eqs.~(\ref{HPSFAC2M}) and (\ref{HgFAC2M}) are still generic and its specification is 
given later
in Eqs.~(\ref{eq:WILPS2M}) and (\ref{eq:WILS2M}). In the following, the 
mass-, $Q^2$-, and 
$\mu^2$-dependence of the Wilson coefficients and operator matrix elements have been suppressed for brevity. Here Wilson 
coefficients are denoted by $L$ if the exchanged gauge boson couples to a massless quark line and by $\tilde{\tilde{H}}$ if it couples to a 
massive quark line. Only in the case of $L_{q,(2,L)}^{\sf NS}, \tilde{\tilde{H}}_{q,(2,L)}^{\sf PS}$  and $\tilde{\tilde{H}}_{g,(2,L)}$ genuine 
two-mass terms 
contribute at 3-loop order. For the other Wilson coefficients \cite{Ablinger:2010ty,Behring:2014eya} contributions of this 
type emerge with 4-loop order for the first time.

Above and in what follows we use the notation 
\begin{eqnarray}
	\tilde{f}(x)&=&\frac{f(x)}{x}~, \label{tildenotation}
	\\
	\hat{f}(x)&=&f(x+2)-f(x)~. \label{hatnotation}
\end{eqnarray}
The double tilde in $\tilde{\tilde{H}}_{q,(2,L)}^{\sf PS}$ and $\tilde{\tilde{H}}_{g,(2,L)}$
should not be interpreted as applying Eq.~(\ref{tildenotation}) twice. Instead, it is used to differentiate these 
Wilson coefficients 
from those of the single mass case, indicating now the required sum over charges as made explicit later in 
Eqs.~(\ref{eq:WILPS2M}) and (\ref{eq:WILS2M}).

The massive operator matrix elements are the expectation values 
\begin{eqnarray}
A_{ij} = \langle j|O_i| j\rangle,~~~~~~j = q,g
\end{eqnarray}
of the local twist $\tau =2$ operators $O_j$, obtained in the light cone expansion \cite{LCE} of the
products of electromagnetic currents, 
\begin{eqnarray}
\label{COMP1}
O^{\sf NS}_{q,r;\mu_1, \ldots, \mu_N} &=& i^{N-1} {\bf S} [\overline{\psi}\gamma_{\mu_1} D_{\mu_2}
\ldots D_{\mu_N} \frac{\lambda_r}{2}\psi] - {\rm trace~terms}~,
\\
\label{COMP2}
O^{\sf S}_{q;\mu_1, \ldots, \mu_N} &=& i^{N-1} {\bf S} [\overline{\psi}\gamma_{\mu_1} D_{\mu_2} \ldots
D_{\mu_N} \psi] - {\rm trace~terms}~,
\\
\label{COMP3} O^{\sf S}_{g;\mu_1, \ldots, \mu_N} &=& 2 i^{N-2}
{\bf S} {\rm \bf Sp}[F_{\mu_1 \alpha}^a D_{\mu_2} \ldots D_{\mu_{N-1}}
F_{\mu_N}^{\alpha,a}] - {\rm trace~terms}~.
\end{eqnarray}
The partonic states $|i(p)\rangle$, with $i=q \, {\rm (quark)}$ or $i=g \, {\rm (gluon)}$, are on-shell with $p^2 = 0$.
In Eqs.~(\ref{COMP1}--\ref{COMP3}) $\rm \bf Sp$ is the color-trace, and $\bf S$ denotes the 
symmetrization operator\footnote{
The sum in Eq.~(\ref{SymmOp}) is over the words $w$ given by the different orderings of the Lorentz indices. For 
example, 
for $M=3$ one obtains, 
${\bf S} f_{\mu_1,\mu_2,\mu_3} = \frac{1}{6} \left(
f_{\mu_1,\mu_2,\mu_3}+f_{\mu_1,\mu_3,\mu_2}+f_{\mu_2,\mu_1,\mu_3}+f_{\mu_2,\mu_3,\mu_1}+f_{\mu_3,\mu_1,\mu_2}+f_{\mu_3,\mu_2,\mu_1}
\right)$.} 
    \begin{eqnarray}
    {\bf S} f_{\mu_1, \ldots,\mu_M}&=& \frac 1 {M !} \sum_{w} f_w~, {\label{SymmOp}}
    \end{eqnarray}
of the Lorentz indices $\mu_1, \ldots, \mu_N$. 
$D_{\mu}$ is the covariant derivative, $\psi$ and $\overline{\psi}$ are
the quark and anti--quark fields, and $F_{\mu \nu}^a$ the gluonic field strength tensor, with $a$ the color index in the adjoint
representation. Furthermore,  $\lambda_r$ is the flavor matrix of $SU(N_F)$. The labels $q,g$ on the left-hand side of 
Eqs.~(\ref{COMP1}--\ref{COMP3}) 
distinguish quarkonic and gluonic operators.

For convenience we will express the strong coupling constant by $a_s = \alpha_s/(4\pi) \equiv g_s^2/(4\pi)^2$ in the following.
Expanding the expressions (\ref{LNSFAC2M}--\ref{HgFAC2M}) up to $O\left(a_s^3\right)$ we obtain: 
\begin{eqnarray}
     \label{eqWIL12M}
     L_{q,(2,L)}^{\sf NS}(N_F+2) &=& 
     a_s^2 \Bigl[A_{qq,Q}^{(2), {\sf NS}}(N_F+2)~\delta_2 +
     \hat{C}^{(2), {\sf NS}}_{q,(2,L)}(N_F)\Bigr]
     \NN\\
     &+&
     a_s^3 \Bigl[A_{qq,Q}^{(3), {\sf NS}}(N_F+2)~\delta_2
     +  A_{qq,Q}^{(2), {\sf NS}}(N_F+2) C_{q,(2,L)}^{(1), {\sf NS}}(N_F+2)
       \NN \\
     && \hspace*{5mm}
     + \hat{C}^{(3), {\sf NS}}_{q,(2,L)}(N_F)\Bigr]~,  \\
      \label{eqWIL22M}
      L_{q,(2,L)}^{\sf PS}(N_F+2) &=& 
     a_s^3 \Bigl[~A_{qq,Q}^{(3), {\sf PS}}(N_F+2)~\delta_2
     +  A_{gq,Q}^{(2)}(N_F+2) N_F\Ctil_{g,(2,L)}^{(1)}(N_F+2) \NN \\
     && \hspace*{5mm}
     + N_F \hat{\Ctil}^{(3), {\sf PS}}_{q,(2,L)}(N_F)\Bigr]~,
     \\
     \label{eqWIL32M}
      L_{g,(2,L)}^{\sf S}(N_F+2) &=& 
     a_s^2 A_{gg,Q}^{(1)}(N_F+2)N_F \Ctil_{g,(2,L)}^{(1)}(N_F+2)
     \NN\\ &+&
      a_s^3 \Bigl[~A_{qg,Q}^{(3)}(N_F+2)~\delta_2 
     +  A_{gg,Q}^{(1)}(N_F+2) N_F\Ctil_{g,(2,L)}^{(2)}(N_F+2)
     \NN\\ && \hspace*{5mm}
     +  A_{gg,Q}^{(2)}(N_F+2) N_F\Ctil_{g,(2,L)}^{(1)}(N_F+2)
     \NN\\ && \hspace*{5mm}
     +  ~A^{(1)}_{Qg}(N_F+2) N_F\Ctil_{q,(2,L)}^{(2), {\sf PS}}(N_F+2)
     + N_F \hat{\Ctil}^{(3)}_{g,(2,L)}(N_F)\Bigr]~,
 \\ \NN \\
\label{eq:WILPS2M}
     \tilde{\tilde{H}}_{q,(2,L)}^{\sf PS}(N_F+2)
     &=& \sum_{i=1}^2 e_{Q_i}^2 a_s^2 \Bigl[~A_{Qq}^{(2), {\sf PS}}(N_F+2,m_i^2)~\delta_2
     +~\Ctil_{q,(2,L)}^{(2), {\sf PS}}(N_F+2)\Bigr]
     \\
     &+& a_s^3 \Bigl[~\tilde{\tilde{A}}_{Qq}^{(3), {\sf PS}}(N_F+2)~\delta_2
     + \sum_{i=1}^2 e_{Q_i}^2 \Bigl[~\Ctil_{q,(2,L)}^{(3), {\sf PS}}(N_F+2) \NN\\
 && 
     + A_{gq,Q}^{(2)}(N_F+2)~\Ctil_{g,(2,L)}^{(1)}(N_F+2) 
     \NN\\&&
     + A_{Qq}^{(2), {\sf PS}}(N_F+2)~C_{q,(2,L)}^{(1), {\sf NS}}(N_F+2) 
        \Bigr] \Bigr]~,       \label{eqWIL42M}
         \NN\\ 
\label{eq:WILS2M}
     \tilde{\tilde{H}}_{g,(2,L)}^{\sf S}(N_F+2) &=&   \sum_{i=1}^2 e_{Q_i}^2\Bigl[
a_s \Bigl[~A_{Qg}^{(1)}(N_F+2)~\delta_2
+~\Ctil^{(1)}_{g,(2,L)}(N_F+2) \Bigr] \NN\\
     &+& a_s^2 \Bigl[~A_{Qg}^{(2)}(N_F+2)~\delta_2
     +~A_{Qg}^{(1)}(N_F+2)~C^{(1), {\sf NS}}_{q,(2,L)}(N_F+2)\NN\\ && 
     \hspace*{5mm}
     +~A_{gg,Q}^{(1)}(N_F+2)~\Ctil^{(1)}_{g,(2,L)}(N_F+2) 
     +~\Ctil^{(2)}_{g,(2,L)}(N_F+2) \Bigr]\Bigr]
     \NN\\ &+&
     a_s^3 \Bigl[~\tilde{\tilde{A}}_{Qg}^{(3)}(N_F+2)~\delta_2
     + \sum_{i=1}^2 e_{Q_i}^2 \Bigl[A_{Qg}^{(2)}(N_F+2)~C^{(1), {\sf NS}}_{q,(2,L)}(N_F+2)
     \NN\\ &&
     \hspace*{5mm}
     +~A_{gg,Q}^{(2)}(N_F+2)~\Ctil^{(1)}_{g,(2,L)}(N_F+2)
     \NN\\ && \hspace*{5mm}
     +~A_{Qg}^{(1)}(N_F+2)\Bigl\{
     C^{(2), {\sf NS}}_{q,(2,L)}(N_F+2)
     +~\Ctil^{(2), {\sf PS}}_{q,(2,L)}(N_F+2)\Bigr\}
     \NN\\ && \hspace*{5mm}
     +~A_{gg,Q}^{(1)}(N_F+2)~\Ctil^{(2)}_{g,(2,L)}(N_F       +2)
     +~\Ctil^{(3)}_{g,(2,L)}(N_F+2) \Bigr]\Bigr]~.
\label{eqWIL52M}
\end{eqnarray}
Here the symbol $\delta_2$ takes the values
\begin{eqnarray}
\delta_2 = \left\{ \begin{array}{lll} 1 & \text{for} & F_2 \\
                                      0 & \text{for} & F_L. 
\end{array} \right.
\end{eqnarray}
Because of the coupling of the exchanged gauge boson to the heavy quark line in the case of the Wilson coefficients 
denoted by $\tilde{\tilde{H}}$, we have still to present the detailed structure of the 3-loop OMEs $A_{ij}^{(3)}$ 
in this case. They consist of the two equal mass terms $A_{ij}^{{\sf Eq.},(3)}$ and the unequal mass term 
$A_{ij}^{{\sf nEq.},(3)}$,
\begin{eqnarray}
A_{ij}^{{\sf nEq.},(3)}(m_1,m_2) = 
  \bar{A}_{ij}^{{\sf nEq.},(3)}(m_1,m_2)
+ \bar{A}_{ij}^{{\sf nEq.},(3)}(m_2,m_1)
\label{splitAs}
\end{eqnarray}
which is symmetric in $m_1$ and $m_2$. The representation given in Eq. (\ref{splitAs}) is only relevant in 
the case of $A_{Qg}^{(3)}$ and $A_{Qq}^{(3), \sf PS}$. 
Here $\bar{A}_{ij}^{{\sf nEq.},(3)}(m_1,m_2)$ denotes the part for which 
the current couples to the fermion-loop of the heavy quark of mass $m_1$. 
This line is carrying the respective local operator. In general, the following representation holds
\begin{eqnarray}
A_{ij}^{(3)} = A_{ij}^{{\sf Eq.},(3)}(m_1) + A_{ij}^{{\sf Eq.},(3)}(m_2) + A_{ij}^{{\sf nEq.},(3)}(m_1,m_2)~.
\end{eqnarray}
The charge-weighted OME is thus given by 
\begin{eqnarray}
\tilde{\tilde{A}}_{ij}^{(3)} = e_{Q_1}^2 A_{ij}^{{\sf Eq.},(3)}(m_1) + e_{Q_2}^2 A_{ij}^{{\sf Eq.},(3)}(m_2) 
+   e_{Q_1}^2 \bar{A}_{ij}^{{\sf nEq.},(3)}(m_1,m_2)
+   e_{Q_2}^2 \bar{A}_{ij}^{{\sf nEq.},(3)}(m_2,m_1)~.
\end{eqnarray}
In the case of the structure function $F_L(x,Q^2)$, the asymptotic massive 3-loop corrections are obtained
by the massive OMEs up to 2-loop order only and therefore do not contain genuine two-mass contributions, 
cf.~\cite{Blumlein:2006mh,Behring:2014eya}.

The {\it inclusive} deep inelastic structure functions $F_{i}(x,Q^2),~~i = 2,L$ can be represented in the fixed flavor number 
scheme in terms of their purely massless contributions and the remaining terms consisting of the real and virtual heavy quark 
contributions,
\begin{eqnarray}
F_i(x,Q^2) = F_i^{\rm massless}(x,Q^2) + F_i^{\rm heavy}(x,Q^2)~.
\end{eqnarray}
Since the parton distribution functions are related to massless partons only, $F_i^{\rm massless}(x,Q^2)$
is obtained in a completely massless calculation. One finds
\begin{eqnarray}
 \frac{1}{x} F_i^{\rm massless}(x,Q^2)&=& \sum_{q} e_q^2 \Biggl\{\frac{1}{N_F} \Biggl[
 \Sigma(x,\mu^2) \otimes C_{i,Q}^{\rm S}\left(x,\frac{Q^2}{\mu^2}\right)
 +G\left(x,\mu^2\right) \otimes C_{i,g}\left(x,\frac{Q^2}{\mu^2}\right)\Biggr]
 \NN\\&&
 + \Delta_{q}(x,\mu^2) \otimes C_{i,q}^{\rm NS}\left(x,\frac{Q^2}{\mu^2}\right) \Biggr\}~,
 \quad i=2,L~,
\label{eq:F2L_Factorization}
\end{eqnarray}
with $\Sigma$ and $\Delta_k$ the flavor singlet and non-singlet distributions given by 
\begin{eqnarray}
\Sigma   &=& \sum_{k=1}^{N_F} (f_k+f_{\overline{k}})~,
\\
\Delta_k &=& f_k + f_{\overline{k}} - \frac{1}{N_F} \Sigma~,
\end{eqnarray} 
and $G$ denoting the gluon density. The heavy quark part is given by
\begin{eqnarray}
\label{eq:F2}
    \frac{1}{x}   F_{(2,L)}^{\rm  heavy}(x,N_F\!\!\!&+&\!\!\!2,Q^2,m_1^2,m_2^2) =\NN\\
       &&\sum_{k=1}^{N_F}e_k^2\Biggl\{
                   L_{q,(2,L)}^{\sf NS}\left(x,N_F+2,\frac{Q^2}{\mu^2}
                                                ,\frac{m^2_1}{\mu^2}
                                                ,\frac{m^2_2}{\mu^2}
\right)
                \otimes
                   \Bigl[f_k(x,\mu^2,N_F)+f_{\overline{k}}(x,\mu^2,N_F)\Bigr]
\NN\\ &&\hspace{14mm}
               +\frac{1}{N_F}L_{q,(2,L)}^{\sf PS}\left(x,N_F+2,\frac{Q^2}{\mu^2}
                                                ,\frac{m^2_1}{\mu^2}
                                                ,\frac{m^2_2}{\mu^2}
\right)
                \otimes
                   \Sigma(x,\mu^2,N_F)
\NN\\ &&\hspace{14mm}
               +\frac{1}{N_F}L_{g,(2,L)}^{\sf S}\left(x,N_F+2,\frac{Q^2}{\mu^2}
                                                 ,\frac{m^2_1}{\mu^2}
                                                 ,\frac{m^2_2}{\mu^2}
\right)
                \otimes
                   G(x,\mu^2,N_F)
                             \Biggr\}
\NN\\ && \hspace{7mm}
                   +\tilde{\tilde{H}}_{q,(2,L)}^{\sf PS}\left(x,N_F+2,\frac{Q^2}{\mu^2}
                                        ,\frac{m^2_1}{\mu^2}
                                        ,\frac{m^2_2}{\mu^2}
\right)
                \otimes
                   \Sigma(x,\mu^2,N_F)
\NN\\ &&\hspace{7mm}
                  +\tilde{\tilde{H}}_{g,(2,L)}^{\sf S}\left(x,N_F+2,\frac{Q^2}{\mu^2}
                                           ,\frac{m^2_1}{\mu^2}
                                           ,\frac{m^2_2}{\mu^2}
\right)
                \otimes
                   G(x,\mu^2,N_F)~.
\end{eqnarray}
The presence of diagrams with $c$- and $b$-quarks at $3$--loop order also yields
power corrections in $\eta$ to the massive operator matrix
elements\footnote{They may emerge as non-logarithmic contributions in terms of higher transcendental functions.}. 
One obtains the following transition relations decoupling both
the charm and bottom contributions at high scales $\mu^2$~: 
\begin{eqnarray}
\label{eq:VFNS1}
&&
\hspace*{-2.4cm}
      f_k(N_F+2,N,\mu^2,m_1^2,m_2^2) + f_{\overline{k}}(N_F+2,N,\mu^2,m_1^2,m_2^2)= 
\NN\\ &&  \hspace{20mm}
A_{qq,Q}^{\sf NS}\left(N,N_F+2,\frac{m_1^2}{\mu^2},\frac{m_2^2}{\mu^2}\right)
          \cdot\bigl[f_k(N_F,N,\mu^2)+f_{\overline{k}}(N_F,N,\mu^2)\bigr]
\NN\\ &&  \hspace{20mm}
       +\frac{1}{N_F}A_{qq,Q}^{\sf PS}\left(N,N_F+2,\frac{m_1^2}{\mu^2},\frac{m_2^2}{\mu^2}\right)
          \cdot\Sigma(N_F,N,\mu^2)
\NN\\ &&  \hspace{20mm}
       +\frac{1}{N_F}A_{qg,Q}\left(N,N_F+2,\frac{m_1^2}{\mu^2},\frac{m_2^2}{\mu^2}\right)
          \cdot G(N_F,N,\mu^2), \label{HPDF12M} \\
     && 
\hspace*{-2.4cm}\label{fQQB2M}
        f_Q(N_F+2,N,\mu^2,m^2_1,m_2^2) + f_{\overline{Q}}(N_F+2,N,\mu^2,m^2_1,m_2^2)=
\NN\\ &&  \hspace{20mm}
        A_{Qq}^{\sf PS}\left(N,N_F+2,\frac{m_1^2}{\mu^2},
\frac{m_2^2}{\mu^2},
\right)
          \cdot \Sigma(N_F,N,\mu^2)
\NN\\ &&  \hspace{20mm}
       +A_{Qg}\left(N,N_F+2,\frac{m_1^2}{\mu^2},\frac{m_2^2}{\mu^2}\right)
          \cdot G(N_F,N,\mu^2)~.   
    \end{eqnarray}
The flavor singlet, non--singlet and gluon densities for $(N_F+2)$ flavors are given by
    \begin{eqnarray}
     \Sigma(N_F+2,N,\mu^2,m^2_1,m_2^2) 
      &=& \Biggl[
             A_{qq,Q}^{\sf NS}\left(N,N_F+2,\frac{m_1^2}{\mu^2},\frac{m_2^2}{\mu^2}\right)
            +A_{qq,Q}^{\sf PS}\left(N,N_F+2,\frac{m_1^2}{\mu^2},\frac{m_2^2}{\mu^2}\right)
\NN \\ && \hspace*{-20mm}
            +A_{Qq}^{\sf PS}\left(N,N_F+2,\frac{m_1^2}{\mu^2},\frac{m_2^2}{\mu^2}\right)
          \Biggr]  \cdot \Sigma(N_F,N,\mu^2)
\NN \\ && \hspace*{-23mm} 
          +\left[
                A_{qg,Q}\left(N,N_F+2,\frac{m_1^2}{\mu^2},\frac{m_2^2}{\mu^2}\right)
               +A_{Qg}\left(N,N_F+2,\frac{m_1^2}{\mu^2},\frac{m_2^2}{\mu^2}\right)
          \right]   \cdot G(N_F,N,\mu^2)~,
\NN\\ \\
     \Delta_k(N_F+2,N,\mu^2,m_1^2,m_2^2)
      &=& f_k(N_F+2,N,\mu^2,m_1^2,m_2^2)+f_{\overline{k}}(N_F+2,N,\mu^2,m_1^2,m_2^2)
\NN\\ &&
         -\frac{1}{N_F+2}\Sigma(N_F+2,N,\mu^2,m_1^2,m_2^2)~, \\
     \label{HPDF22M}
     G(N_F+2,N,\mu^2,m^2_1, m^2_2) 
      &=& A_{gq,Q}\left(N,N_F+2,\frac{m_1^2}{\mu^2},\frac{m_2^2}{\mu^2}\right) 
                    \cdot \Sigma(N_F,N,\mu^2)
\NN\\ && 
         +A_{gg,Q}\left(N,N_F+2,\frac{m_1^2}{\mu^2},\frac{m_2^2}{\mu^2}\right) 
                    \cdot G(N_F,N,\mu^2)~.
\end{eqnarray}
Here $f_{k(\overline{k})}(N_F), \Sigma(N_F)$ and $G(N_F)$ denote the massless quarkonic parton 
densities. 
Note that the above process independent leading twist OMEs $A_{i,j}$ for fixed moments 
$N$ contain besides logarithmic corrections in $\eta$ also power corrections. For general values of $N$ the 
$\eta$-dependence is more involved and requests at least generalized harmonic sums \cite{Moch:2001zr,Ablinger:2013cf} 
and binomially weighted generalized harmonic sums \cite{Ablinger:2014bra} as will be shown below.\footnote{For 
recent surveys on these function spaces see Refs.~\cite{Ablinger:2013eba,Ablinger:2013jta}.}
We would like to mention, that although $\Delta_k$ is the genuine flavor non-singlet distribution, sometimes the combination
$f_k + f_{\overline{k}}$ may be considered to take its role, \cite{Buza:1995ie,Ablinger:2014vwa}.

The presence of 2-mass terms in Eqs.~(\ref{eq:VFNS1}--\ref{HPDF22M}) only allows to define the new parton 
densities at $(N_F+2)$ out of those at $N_F$ at sufficiently high decoupling scales $\mu^2 \gg m_1^2, m_2^2$ at 
3-loop order, while up to 2-loop order, flavors can technically be decoupled one by one, if $m_2^2 \gg m_1^2$ 
(which is not the case, however for $b$- and $c$-quarks). The picture of an individual charm and bottom quark 
density does therfore not hold from 3-loop order onwards. The quantities $f_k + f_{\bar{k}}$, $\Sigma$, 
$\Delta_k$ and $G$ are not affected, as they depend on all heavy quark masses in a symmetric way. The two-mass
generalization (\ref{fQQB2M}) of the single mass case \cite{Bierenbaum:2009mv,Buza:1996wv}, is a formal relation 
as 
it stands. It can be rewritten expressing the charm and bottom quark densities in the variable flavor scheme,
still requesting
\begin{eqnarray}
Q^2 \gg m_c^2~~~~~~\text{and}~~~~~~Q^2 \gg m_b^2
\end{eqnarray}
by
\begin{eqnarray}
\label{eq:FC}
&&        f_c(N_F+2,N,\mu^2,m^2_1,m_2^2) + f_{\overline{c}}(N_F+2,N,\mu^2,m^2_1,m_2^2)=
\NN\\ &&  \hspace{60mm}
        \bar{\bar{A}}_{Qq}^{{\sf PS},c(b)}\left(N,N_F+2,\frac{m_1^2}{\mu^2},
\frac{m_2^2}{\mu^2},
\right)
          \cdot \Sigma(N_F,N,\mu^2)
\NN\\ &&  \hspace{60mm}
       +\bar{\bar{A}}_{Qg}^{c(b)}\left(N,N_F+2,\frac{m_1^2}{\mu^2},\frac{m_2^2}{\mu^2}\right)
          \cdot G(N_F,N,\mu^2)
\\
\label{eq:FB}
&&        f_b(N_F+2,N,\mu^2,m^2_1,m_2^2) + f_{\overline{b}}(N_F+2,N,\mu^2,m^2_1,m_2^2)=
\NN\\ &&  \hspace{60mm}
        \bar{\bar{A}}_{Qq}^{{\sf PS},b(c)}\left(N,N_F+2,\frac{m_1^2}{\mu^2},
\frac{m_2^2}{\mu^2},
\right)
          \cdot \Sigma(N_F,N,\mu^2)
\NN\\ &&  \hspace{60mm}
       +\bar{\bar{A}}_{Qg}^{b(c)}\left(N,N_F+2,\frac{m_1^2}{\mu^2},\frac{m_2^2}{\mu^2}\right)
          \cdot G(N_F,N,\mu^2)~,
\end{eqnarray}
where
\begin{eqnarray}
\bar{\bar{A}}_{ij}^{c(b)} =  A_{ij}^{{\sf Eq.},(3)}(m_c) + \bar{A}_{ij}^{{\sf nEq.},(3)}(m_c,m_b),
\end{eqnarray}
and $\bar{\bar{A}}_{ij}^{b(c)}$ is obtained by $c \leftrightarrow b$. Eq.~(\ref{fQQB2M}) is the sum of
Eqs.~(\ref{eq:FC}) and (\ref{eq:FB}).

We turn now to the calculation of the massive two-mass OMEs and discuss first their renormalization in the case of two heavy quark 
masses.
\section{Renormalization of the Massive Operator Matrix Elements}
\label{sec:ren}

\vspace*{1mm}
\noindent
The Feynman integrals contributing to the various operator matrix elements contain mass, coupling, ultraviolet operator 
singularities, and collinear divergences, due to massless sub-graphs. They are regularized by applying dimensional 
regularization \cite{'tHooft:1972fi} in $D=4+\ep$ dimensions. The singularities appear as poles in the Laurent series in 
$\ep$, with the highest pole corresponding to the loop order. At one and two loop order the two--mass massive operator 
matrix elements $A_{ij}$ are given in terms of the known single mass contributions since they do not contain more than one 
internal massive fermion 
line~\cite{Buza:1995ie,Buza:1996wv,Bierenbaum:2007qe,Bierenbaum:2007dm,Bierenbaum:2008yu,Bierenbaum:2009zt,Blumlein:2006mh,Behring:2014eya}.

The first single particle irreducible diagrams with two masses emerge at $O(\alpha_s^3)$. In the 
following, 
we consider the 
renormalization of the two mass contributions in individual terms together with the genuine two-mass contributions. The latter
terms will then be obtained subtracting the former ones, cf.~Ref.~\cite{Bierenbaum:2009mv}. The unrenormalized
OMEs are given by
\begin{eqnarray}
    \Ahathat_{ij}^{(l)}\Bigl(\frac{m_1^2}{\mu^2},\frac{m_2^2}{\mu^2}\Bigr)=
    \left[\left(\frac{m_1^2}{\mu^2}\right)^{l/2 \ep} +
\left(\frac{m_2^2}{\mu^2}\right)^{l/2 \ep}\right] 
 \Ahathat_{ij}^{(l,{\rm sing})}
 +\Athathat_{ij}^{(l)}
\Bigl(\frac{m_1^2}{\mu^2},\frac{m_2^2}{\mu^2}\Bigr),
\label{AhathatDecomp}
\end{eqnarray}
where $\Ahathat_{ij}^{(l), \rm sing}$ are the single-mass OMEs \cite{Bierenbaum:2009mv}
and $\Athathat_{ij}^{(l)}$ are the 
new two-mass contributions. The last term in Eq.~(\ref{AhathatDecomp}) for $l = 3$ contains a factor 
$(m_1 m_2 /\mu^2)^{3/(2\ep)}$. 
Furthermore, a change in the renormalization scheme as in Eqs.~(\ref{asmoma}, \ref{asmsa}) 
generally introduces a mixing between the different components of Eq.~(\ref{AhathatDecomp}). 
 
In the main steps we follow the renormalization procedure outlined in Ref.~\cite{Bierenbaum:2009mv}, 
incorporating the necessary modifications for the two-mass case. We consider the case of $N_F$ massless and two massive 
quark flavors as this covers the physical case of contributions e.g. due to the charm and bottom quarks.

We first consider mass and coupling constant renormalization, followed by the renormalization of the 
ultraviolet singularity of the local operators, and the factorization of the collinear singularities.

\begin{subsection}{\bf\boldmath Mass Renormalization}
\label{SubSec-RENMa}

\vspace*{1mm}
\noindent
The schemes most frequently used for the mass renormalization are the $\MS$-- and the on--mass shell scheme (OMS). 
In the following, we renormalize the mass in the OMS and provide the finite renormalization to switch to the $\MS$-mass at 
a later stage, cf. Eq.~(\ref{eq:zm}). We perform the mass renormalization first, i.e. the respective expressions
are still containing the bare coupling $\hat{a}_s = \hat{g}_s^2/(4 \pi)^2$.\footnote{Note that our notation 
therefore agrees with \cite{Gray:1990yh}, but e.g. differs form the notation 
in~\cite{Melnikov:2000zc,Bekavac:2007tk,Marquard:2016dcn},
where also the charge renormalization has been carried out.}

The bare masses $\hat{m_i},~ i \in \{1,2\}$ are expressed by the renormalized on--shell masses $m_i$ via
\begin{align}
\hat{m_i}=Z_{m,i}(m_1,m_2)~ m_i 
           =& m_i \Bigl[ 1 
                       + \hat{a}_s
                       \Bigl(\frac{m_i^2}{\mu^2}\Bigr)^{\ep/2}
                                   \delta m_1 
                       + \hat{a}_s^2
                       \Bigl(\frac{m_i^2}{\mu^2}\Bigr)^{\ep} \delta m_{2,i}\left(m_1,m_2\right)                       
                 \Bigr] 
                 + O(\hat{a}_s^3)~,
            \label{mren1}
\end{align}
and  
\begin{eqnarray}
\delta m_{2,i}\left(m_1,m_2\right)=
                                     \delta m_2^0 +\tilde{\delta}{m_2}^{i}(m_1,m_2)~.
{\label{dm2twomass}}
\end{eqnarray}                                    
Here $\delta m_2^0$ is the single mass-contribution, whereas $\tilde{\delta}{m_2}^i$ denotes the additional contribution 
emerging in the case of two massive flavors.  Note that from order $O(\hat{a}_s^2)$ onward the $Z$-factor 
renormalizing 
$\hat{m}_1$ depends on $m_2$ and vice versa. For the massive operator matrix elements this can be observed at $3$--loop 
order for the first time. The coefficients $\delta m_1$ and $\delta m_2$ have been derived in
\cite{Tarrach:1980up,Nachtmann:1981zg} up to $O(\ep^0)$ and $O(\ep^{-1})$, respectively. The constant part of $\delta m_2$ 
was given in \cite{Gray:1990yh,Broadhurst:1991fy,Fleischer:1998dw} and the $O(\ep)$-term of $\delta m_1$ in
\cite{Bierenbaum:2009mv}. One obtains
\begin{eqnarray}
    \delta m_1 &=&C_F
                  \left[\frac{6}{\ep}-4+\left(4+\frac{3}{4}\zeta_2\right)\ep
                  \right] \label{delm1}  \\
               &\equiv&  \frac{\delta m_1^{(-1)}}{\ep}
                        +\delta m_1^{(0)}
                        +\delta m_1^{(1)}\ep~, \label{delm1exp} \\
    \delta m_2^0 &=& C_F
                   \Biggl[\frac{1}{\ep^2}\left(18 C_F-22 C_A+8T_F(N_F+1)
                    \right)
                  +\frac{1}{\ep}\Biggl(-\frac{45}{2}C_F+\frac{91}{2}C_A
\NN\\&&
                  -14T_F
                   (N_F+1)\Biggr)
                  +C_F\left(\frac{199}{8}-\frac{51}{2}\zeta_2+48\ln(2)\zeta_2
                   -12\zeta_3\right)
                   +C_A\Biggl(-\frac{605}{8}
                  \NN\\&&
                  +\frac{5}{2}\zeta_2-24\ln(2)\zeta_2+6\zeta_3
                  \Biggr)
                  +T_F\left[N_F\left(\frac{45}{2}+10\zeta_2\right)+
                  \frac{69}{2}-14\zeta_2\right]\Biggr]
                  \label{delm2}  \\
               &\equiv&  \frac{\delta m_2^{0,(-2)}}{\ep^2}
                        +\frac{\delta m_2^{0,(-1)}}{\ep}
                        +\delta m_2^{0,(0)}~, \label{delm2exp}
\\
\tilde{\delta}{m_2}^{i}(m_1,m_2) &=& C_F T_F \Biggl\{
  \frac{8}{\ep^2}
 -\frac{14}{\ep}
 +8 r_i^4 \HA_{0}^2(r_i)
 -8 (r_i+1)^2 \left(r_i^2-r_i+1\right) \HA_{-1,0}(r_i)
 \NN\\&&
 +8 (r_i-1)^2 \left(r_i^2+r_i+1\right) \HA_{1,0}(r_i)
 +8 r_i^2 \HA_0(r_i)
 +\frac{3}{2} \left(8 r_i^2+15\right)
 \NN\\&&
+2 \Bigl[4 r_i^4-12 r_i^3-12 r_i+5\Bigr] \zeta_2 \Biggr\}
 \\
               &\equiv&  \frac{\tilde{\delta}{m_2}^{(-2)}}{\ep^2}
                        +\frac{\tilde{\delta}{m_2}^{(-1)}}{\ep}
                        +\tilde{\delta}{m_2}^{i,(0)}~, \label{delm2mixexp}
\label{dm2mixfull}
\end{eqnarray}
cf.~\cite{Gray:1990yh}, with $C_F = (N_c^2-1)/(2 N_c), C_A = N_c, T_F = 1/2$, $N_c = 3$ in the case of 
QCD, $i\in \{1,2\}$ and  
\begin{eqnarray}
r_1 = \sqrt{\eta}~~~~\text{and}~~~~r_2=\frac{1}{\sqrt{\eta}}.
\end{eqnarray}
Here $\zeta_k = \sum_{l=1}^\infty (1/l^k), k \in \mathbb{N}, k \geq 2$ denotes the Riemann's $\zeta$-function at integer 
arguments\footnote{In Feynman graph calculations at higher orders also multiple zeta values contribute, cf.~\cite{Blumlein:2009cf}.}. 
The superscript $i$ for the coefficients 
$\tilde{\delta} m_2^{(-2)}$ and $\tilde{\delta} m_2^{(-2)}$ has been dropped as they are independent of the renormalized 
mass $m_i$. Furthermore, $\HA_{\vec{a}}(\zeta)$ are the harmonic polylogarithms (HPLs) \cite{Remiddi:1999ew}
\begin{eqnarray}
\HA_0(\zeta)      &=& \ln(\zeta)
\\
\HA_{-1,0}(\zeta) &=& \Li_2(-\zeta)+\ln(\zeta) \ln(1+\zeta)
\\
\HA_{1,0}(\zeta)  &=& \Li_2(1-\zeta)-\zeta_2~.
\end{eqnarray}
Eq.~(\ref{dm2mixfull}) states the complete analytic form of the contribution of the respective other massive flavor to 
the renormalization of the bare masses. In the present analysis we will focus on $m_1$, $m_2$ being the masses of the
bottom and charm quarks, respectively. Due to the size of the ratio 
\begin{eqnarray}
\eta\sim 0.1~,
\end{eqnarray}
it is enough to do the expansion up to $O\left(\eta^4 \ln(\eta) \right)$, as we will do in general for
the fixed Mellin moments of the OMEs. The mixed-mass terms are given by 
\begin{eqnarray}
\tilde{\delta} {m_2}^{1,(0)}(m_1,m_2) &=& C_F T_F
\Biggl[
\frac{45}{2}
+10 \zeta_2
-24 \zeta_2 \eta^{1/2}
+24 \eta
-24 \zeta_2 \eta^{3/2}
\NN\\ &&
+\left(2 \ln^2(\eta)-\frac{26}{3} \ln(\eta)+8 \zeta_2+\frac{151}{9}\right) \eta^2
\NN\\&&
+\left(\frac{16}{15} \ln(\eta)-\frac{152}{75} \right) \eta^3\Biggr] 
+O\left(\eta^4 \ln(\eta)\right)~,
\\
\tilde{\delta} {m_2}^{2,(0)}(m_1,m_2) &=& C_F T_F
\Biggl[
-2 \ln^2(\eta)
+\frac {26}{3} \ln(\eta)
+2 \zeta_2
+\frac{103}{18}
\NN\\&&
+\left(-\frac{16}{15} \ln(\eta)+\frac{152}{75}\right) \eta
+\left(-\frac{9}{35} \ln(\eta)+\frac{1389}{4900}\right) \eta^2
\NN\\&&
+\left(-\frac{32}{315} \ln(\eta)+\frac{7976}{99225} \right) \eta^3 
\Biggr]
+O\left(\eta^4 \ln(\eta)\right)~.
\end{eqnarray}
Applying Eq.~(\ref{mren1}) we obtain the mass renormalized operator matrix elements by
\begin{eqnarray}
    \Ahathat_{ij}\Bigl(\frac{m_1^2}{\mu^2},\frac{m_2^2}{\mu^2},\ep,N\Bigr) 
                 &=& \delta_{ij}
                 +\hat{a}_s~ 
                   \Ahathat_{ij}^{(1)}\Bigl(\frac{m_1^2}{\mu^2},\frac{m_2^2}{\mu^2},\ep,N\Bigr) 
                         + \hat{a}_s^2 \Biggl\{
                                        \Ahathat^{(2)}_{ij}
                                        \Bigl(\frac{m_1^2}{\mu^2},\frac{m_2^2}{\mu^2},\ep,N\Bigr) 
\NN\\&&
                                      + {\delta m_1} 
                                        \Bigl[\Bigl(\frac{m_1^2}{\mu^2}\Bigr)^{\ep/2}
                                        m_1 \frac{d}{d m_1}
                                      +\Bigl(\frac{m_2^2}{\mu^2}\Bigr)^{\ep/2}
                                        m_2 \frac{d}{d m_2}
                                      \Bigr]
                                                   \Ahathat_{ij}^{(1)}
                                           \Bigl(\frac{m_1^2}{\mu^2},\frac{m_2^2}{\mu^2},\ep,N\Bigr)                        
                                \Biggr\}
\NN\\ &&
                         + \hat{a}_s^3 \Biggl\{ 
                                         \Ahathat^{(3)}_{ij}
                                           \Bigl(\frac{m_1^2}{\mu^2},\frac{m_2^2}{\mu^2},\ep,N\Bigr) 
\NN\\ &&
                                        +{\delta m_1} 
                                         \left[
                                           \Bigl(\frac{m_1^2}{\mu^2}\Bigr)^{\ep/2}
                                         m_1\frac{d}{d m_1} 
                                         +
                                           \Bigl(\frac{m_2^2}{\mu^2}\Bigr)^{\ep/2}
                                         m_2 \frac{d}{d m_2} 
                                         \right]
                                                    \Ahathat_{ij}^{(2)}
                                           \Bigl(\frac{m_1^2}{\mu^2},\frac{m_2^2}{\mu^2},\ep,N\Bigr)
\NN\\ &&
                                        +
                                        \delta m_{2,1}(m_1,m_2) 
                                    \Bigl(\frac{m_1^2}{\mu^2}\Bigr)^{\ep} m_1    \frac{d}{d m_1}
                                                    \Ahathat_{ij}^{(1)}\Bigl(\frac{m_1^2}{\mu^2},\frac{m_2^2}{\mu^2},\ep,N\Bigr)
\NN\\ &&
                                          +\delta m_{2,2}(m_1,m_2)
                                    \Bigl(\frac{m_2^2}{\mu^2}\Bigr)^{\ep} m_2    \frac{d}{d m_2} 
                                                    \Ahathat_{ij}^{(1)}\Bigl(\frac{m_1^2}{\mu^2},\frac{m_2^2}{\mu^2},\ep,N\Bigr) 
\NN\\&&
                                        +
                                        \frac{(\delta m_1)^2}{2}
                                        \left[
                                          \Bigl(\frac{m_1^2}{\mu^2}\Bigr)^{\ep}         
                                              m_1^2     \frac{d^2}{{d m_1}^2}
                                                   +
                                            \Bigl(\frac{m_2^2}{\mu^2}\Bigr)^{\ep}
                                               m_2^2    \frac{d^2}{{d m_2}^2}
                                               \right]
                                                     \Ahathat_{ij}^{(1)}
                                           \Bigl(\frac{m_1^2}{\mu^2},\frac{m_2^2}{\mu^2},\ep,N\Bigr) 
\NN\\&&
                                         + (\delta m_1)^2 
                                        \Bigl(\frac{m_1^2}{\mu^2}\Bigr)^{\ep/2}
                                          \Bigl(\frac{m_2^2}{\mu^2}\Bigr)^{\ep/2}
                                       m_1  \frac{d}{d m_1}
                                       m_2  \frac{d}{d m_2}
                                                   \Ahathat_{ij}^{(1)}
                                           \Bigl(\frac{m_1^2}{\mu^2},\frac{m_2^2}{\mu^2},\ep,N\Bigr)
    \Biggr\}~,
\NN\\&&
\label{maren}
\end{eqnarray}
which generalizes Eq.~(3.10) of Ref.~\cite{Bierenbaum:2009mv}.
The OMEs are symmetric under the interchange of the masses $m_1$ and $m_2$.
\end{subsection}
\begin{subsection}{\bf\boldmath Renormalization of the Coupling}
\label{SubSec-RENCo}
When renormalizing the coupling constant, it is important to note that the factorization relation (\ref{LNSFAC2M}--\ref{HgFAC2M}) 
strictly requires the external massless partonic legs of the operator matrix elements to be on--shell, i.e. 
\begin{eqnarray}
	p^2=0~{\label{ONSHELL}}, 
\end{eqnarray} 
with $p$ the external momentum of the OME. This condition would be violated by naively applying
massive loop corrections to the gluon propagator.  We follow \cite{Bierenbaum:2009mv} and absorb these 
corrections uniquely into the coupling constant by using the background field method
\cite{Abbott:1980hw,Rebhan:1985yf,Jegerlehner:1998zg} to maintain the Slavnov--Taylor identities of QCD.  In this 
way, one first obtains the coupling constant in a {\sf{MOM}}--scheme. A finite renormalization to transform to
the {$\MS$}--scheme is applied subsequently.

The light flavor contributions to the unrenormalized coupling constant in terms of the renormalized coupling constant
in the {$\MS$}--scheme read 
\begin{eqnarray}
   \hat{a}_s             &=& {Z_g^{\MS}}^2(\ep,N_F) 
                             a^{\MS}_s(\mu^2) \NN\\
                         &=& a^{\MS}_s(\mu^2)\left[
                                   1 
                                 + \delta a^{\MS}_{s, 1}(N_F) 
                                      a^{\MS}_s(\mu^2)
                                 + \delta a^{\MS}_{s, 2}(N_F) 
                                      {a^{\MS}_s}^2(\mu)    
                                     \right] + O({a^{\MS}_s}^3)~. 
                            \label{asrenMSb}
\end{eqnarray}
Here the coefficients $\delta a^{\MS}_{s, i}(N_F)$ are given by
\begin{eqnarray}
    \delta a^{\MS}_{s, 1}(N_F) &=& \frac{2}{\ep} \beta_0(N_F)~,
                             \label{deltasMSb1} \\
    \delta a^{\MS}_{s, 2}(N_F) &=& \frac{4}{\ep^2} \beta_0^2(N_F)
                           + \frac{1}{\ep} \beta_1(N_F),
                             \label{deltasMSb2}
\end{eqnarray}
with $\beta_k(N_F)$ the expansion coefficients of the QCD $\beta$-function
\cite{Gross:1973id,Politzer:1973fx,tHooft:unpub,Khriplovich:1969aa,Caswell:1974gg,Jones:1974mm}
\begin{eqnarray}
   \beta_0(N_F)
                 &=& \frac{11}{3} C_A - \frac{4}{3} T_F N_F \label{beta0}~, \\
   \beta_1(N_F)
                 &=& \frac{34}{3} C_A^2 
               - 4 \left(\frac{5}{3} C_A + C_F\right) T_F N_F \label{beta1}~.
\end{eqnarray}

We split the renormalized gluon self--energy $\Pi$ into the purely light and the heavy flavor contributions, $\Pi_L$ and 
$\Pi_H$,
\begin{equation}
 {\Pi}\left(p^2,m_1^2,m_2^2\right)={\Pi}_{L}\left(p^2\right)+{\Pi}_{H}\left(p^2,m_1^2,m_2^2\right)~.
\end{equation}
The heavy quarks are required to decouple from the running coupling constant and the renormalized OMEs for $\mu^2<m_1^2,m_2^2$ which 
implies{\cite{Buza:1995ie}} 
\begin{equation}
{\Pi}_{H}(0,m_1^2,m_2^2)=0~ \label{DecouplingCond}. 
\end{equation}
We apply the background field method, which has the advantage of producing gauge-invariant results also for off--shell Green's functions, 
to compute the heavy flavor contributions to the unrenormalized gluon polarization function {\cite{DeWitt:1967ub,Abbott:1980hw}}.
Applying the respective Feynman rules\cite{YND} one obtains  
  \begin{eqnarray}
   \hat{\Pi}^{\mu\nu}_{H,ab,\mbox{\tiny{BF}}}(p^2,m_1^2,m_2^2,\mu^2,\ep,\hat{a}_s)&=&
                                i (-p^2g^{\mu\nu}+p^{\mu}p^{\nu})\delta_{ab}
\hat{\Pi}_{H,\mbox{\tiny{BF}}}(p^2,m_1^2,m_2^2,\mu^2,\ep,\hat{a}_s)~, 
\\
   \hat{\Pi}_{H,\mbox{\tiny{BF}}}(0,m_1^2,m_2^2,\mu^2,\ep,\hat{a}_s)&=&
                    \hat{a}_s   \frac{2\beta_{0,Q}}{\ep}
                         \left[\Bigl(\frac{m_1^2}{\mu^2}\Bigr)^{\ep/2}
                           +\Bigl(\frac{m_2^2}{\mu^2}\Bigr)^{\ep/2} \right]
                          \exp \Bigl(\sum_{i=2}^{\infty}\frac{\zeta_i}{i}
                          \Bigl(\frac{\ep}{2}\Bigr)^{i}\Bigr)
\NN\\ &&
                   +\hat{a}_s^2 \left[\Bigl(\frac{m_1^2}{\mu^2}\Bigr)^{\ep}
                   +\Bigl(\frac{m_2^2}{\mu^2}\Bigr)^{\ep}\right]
                        \Biggl[
                       \frac{1}{\ep}\Bigl(
                                          -\frac{20}{3}T_FC_A
                                          -4T_FC_F
                                    \Bigr)
\NN\\&&
                      -\frac{32}{9}T_FC_A
                      +15T_FC_F
\NN\\&&
                     +\ep            \Bigl(
                                          -\frac{86}{27}T_FC_A
                                          -\frac{31}{4}T_FC_F
                                          -\frac{5}{3}\zeta_2T_FC_A
                                          -\zeta_2T_FC_F
                                   \Bigr)
\NN\\ &&
                        +2 \left(\frac{2\beta_{0,Q}}{\ep}\right)^2
                         \Bigl(\frac{m_1^2}{\mu^2}\Bigr)^{\ep/2}
                           \Bigl(\frac{m_2^2}{\mu^2}\Bigr)^{\ep/2} 
                          \exp \Bigl(2 \sum_{i=2}^{\infty}\frac{\zeta_i}{i}
                          \Bigl(\frac{\ep}{2}\Bigr)^{i}\Bigr)
                         \Biggl]
\NN\\&&                         
                         + O(\hat{a}_s^3)~, \label{GluSelfBack}
\end{eqnarray}
where the masses $m_1$ and $m_2$ have been renormalized in the on--shell scheme (\ref{mren1}). In order to write (\ref{GluSelfBack}) 
in a more compact form we use the notation
\begin{eqnarray}
   f(\ep)&\equiv&
   \left[
                 \Bigl(\frac{m_1^2}{\mu^2}\Bigr)^{\ep/2}
                 +\Bigl(\frac{m_2^2}{\mu^2}\Bigr)^{\ep/2}\right]
    \exp \left[\sum_{i=2}^{\infty}\frac{\zeta_i}{i}
                       \Bigl(\frac{\ep}{2}\Bigr)^{i}\right]~, \label{fep}
\end{eqnarray}
and keep this factor unexpanded in the dimensional regularization parameter
$\ep$ for the moment.
Furthermore, we denote the contributions to the 
QCD $\beta$-function coefficients by $\beta_{i,Q}^{(j)}$ 
\cite{Gross:1973id,Politzer:1973fx,tHooft:unpub,Khriplovich:1969aa,Caswell:1974gg,Jones:1974mm,Buza:1995ie,Bierenbaum:2009mv}
  \begin{eqnarray}
   \beta_{0,Q} &=&-\frac{4}{3}T_F~, \label{b0Q} \\
   \beta_{1,Q} &=&- 4 \left(\frac{5}{3} C_A + C_F \right) T_F~, \label{b1Q} \\
   \beta_{1,Q}^{(1)}&=&
                           -\frac{32}{9}T_FC_A
                           +15T_FC_F~, \label{b1Q1} \\
   \beta_{1,Q}^{(2)}&=&
                               -\frac{86}{27}T_FC_A
                               -\frac{31}{4}T_FC_F
                               -\zeta_2\left(\frac{5}{3}T_FC_A
                                        +T_FC_F\right)~. \label{b1Q2}
  \end{eqnarray}
Eq.~(\ref{GluSelfBack}) differs from the sum of the two individual single--mass 
contributions\cite{Bierenbaum:2009mv} by the last term 
only, which is due to additional reducible Feynman diagrams in the cases of two heavy quark flavors of different mass.

The background field is renormalized using the $Z$-factor $Z_A$ which is split into light and heavy quark contributions, $Z_{A,L}$ 
and $Z_{A,H}$. It is related to the $Z$-factor renormalizing the coupling constant $g$ via
  \begin{eqnarray} 
   Z_g=Z_A^{-\frac 1 2}=\frac{1}{\left(Z_{A,L}+Z_{A,H}\right)^{1/2}}~. \label{ZAZg}
  \end{eqnarray}
Concerning the light flavors, we require the renormalization to correspond to the $\MS$--scheme with $N_F$ light flavors
\begin{eqnarray}
Z_{A,l}(N_F)&=&{Z_g^{\MS}}^{1/2}~ \label{ZAl}.
\end{eqnarray}
The heavy flavor contributions are fixed by condition (\ref{DecouplingCond}) which implies
\begin{eqnarray}
   \Pi_{H,\mbox{\tiny{BF}}}(0,\mu^2, a_s, m_1^2,m_2^2)+Z_{A,H}\equiv 0~. 
	\label{ZAcond}
\end{eqnarray}
The $Z$-factor in the $\MOM$--scheme  is read off by combining Eqs.~(\ref{ZAZg}), (\ref{DecouplingCond}), 
(\ref{GluSelfBack}) and (\ref{ZAcond})
\begin{eqnarray}
   Z^{\MOM}_g(\ep,N_F+2,\mu,m_1^2,m_2^2)
         \equiv \frac{1}{(Z_{A,l}+Z_{A,H})^{1/2}}~. \label{Zgnfp1}
\end{eqnarray}
Up to $O({a^{\MOM}_s}^3)$ one obtains the renormalization constant
\begin{eqnarray}
   {Z_g^{\MOM}}^2(\ep,N_F+2,\mu,m_1^2,m_2^2)&=&
                  1+a^{\MOM}_s(\mu^2) \Bigl[
                              \frac{2}{\ep} (\beta_0(N_F)+\beta_{0,Q}f(\ep))
                        \Bigr]
\NN\\ &&
                  +{a^{\MOM}_s}^2(\mu^2) \Bigl[
                                \frac{\beta_1(N_F)}{\ep}
                         +\frac{4}{\ep^2} (\beta_0(N_F)+\beta_{0,Q}f(\ep))^2
\NN\\ &&
                          +\frac{1}{\ep}\left(\Bigl(\frac{m_1^2}{\mu^2}\Bigr)^{\ep}
                            +\Bigl(\frac{m_2^2}{\mu^2}\Bigr)^{\ep}\right)
                           \Bigl(\beta_{1,Q}+\ep\beta_{1,Q}^{(1)}
                                            +\ep^2\beta_{1,Q}^{(2)}
                           \Bigr)
                          \Bigr]
\NN\\ &&
+O({a^{\MOM}_s}^3)~. \label{Zgheavy2}
\end{eqnarray}.
  
  We define the coefficients of the $\MOM$--scheme $Z$-factor, $\delta
  a_{s,1}^{\MOM}$ and $\delta a_{s,2}^{\MOM}$, analogously to those of the
  $\MS$--coefficients in (\ref{asrenMSb})
  \begin{eqnarray}
   \delta a_{s,1}^{\MOM}&=&\frac{2\beta_0(N_F)}{\ep}
                           +\frac{2\beta_{0,Q}}{\ep}f(\ep)
                            ~,\label{dela1} \\
   \delta a_{s,2}^{\MOM}&=&\frac{\beta_1(N_F)}{\ep}+
                            \left(\frac{2\beta_0(N_F)}{\ep}
                              +\frac{2\beta_{0,Q}}{\ep}f(\ep)\right)^2
\NN\\&&
                          +\frac{1}{\ep}\left(\Bigl(\frac{m_1^2}{\mu^2}\Bigr)^{\ep}+
                            \Bigl(\frac{m_2^2}{\mu^2}\Bigr)^{\ep}\right)
                           \Bigl(\beta_{1,Q}+\ep\beta_{1,Q}^{(1)}
                                            +\ep^2\beta_{1,Q}^{(2)}
                           \Bigr) + O(\ep^2)~.\label{dela2}
\nonumber\\
  \end{eqnarray}
Finally, we express our results in the $\MS$--scheme. For this transition we
assume the decoupling of the heavy quark flavors. 

The transformation to the ${\MS}$ scheme is then implied by
  \begin{eqnarray}
      {Z_g^{\MS}}^2(\ep,N_F+2) a^{\MS}_s(\mu^2) = 
      {Z_g^{\MOM}}^2(\ep,N_F+2,\mu,m_1^2,m_2^2) a^{\MOM}_s(\mu^2) \label{condas1}~.
  \end{eqnarray}
Solving (\ref{condas1}) perturbatively one obtains
  \begin{eqnarray}
   a_s^{\MOM}&=& a_s^{\MS}
                -\beta_{0,Q} \left(\ln \Bigl(\frac{m_1^2}{\mu^2}\Bigr)+\ln \Bigl(\frac{m_2^2}{\mu^2}\Bigr)\right) {a_s^{\MS}}^2
                +\Biggl[ \beta^2_{0,Q} \left(\ln
                  \Bigl(\frac{m_1^2}{\mu^2}\Bigr)+\ln
                  \Bigl(\frac{m_2^2}{\mu^2}\Bigr)\right)^2 
    \NN\\&&
                        -\beta_{1,Q} \left(\ln \Bigl(\frac{m_1^2}{\mu^2}\Bigr) +\ln \Bigl(\frac{m_2^2}{\mu^2}\Bigr)\right)
                        -2 \beta_{1,Q}^{(1)}
                 \Biggr] {a_s^{\MS}}^3+O\left({a_s^{\MS}}^4\right)~, \label{asmoma}
  \end{eqnarray}
  or, 
  \begin{eqnarray}
   a_s^{\MS}&=&
               a_s^{\MOM}
              +{a_s^{\MOM}}^2\Biggl(
                          \delta a^{\MOM}_{s, 1}
                         -\delta a^{\MS}_{s, 1}(N_F+2)
                             \Biggr)
              +{a_s^{\MOM}}^{3}\Biggl(
                          \delta a^{\MOM}_{s, 2}
                         -\delta a^{\MS}_{s, 2}(N_F+2)
     \NN\\ &&
                        -2\delta a^{\MS}_{s, 1}(N_F+2)\Bigl[
                             \delta a^{\MOM}_{s, 1}
                            -\delta a^{\MS}_{s, 1}(N_F+2)
                                                      \Bigr]
                             \Biggr)+O({a_s^{\MOM}}^4)~. \label{asmsa}
  \end{eqnarray}
Note that, unlike in Eq.~(\ref{asrenMSb}), in Eq.~(\ref{asmoma}) and (\ref{asmsa}) $a_s^{\MS} 
\equiv a_s^{\MS}\left(N_F+2\right)$. Applying the coupling renormalization (\ref{Zgheavy2}) to 
(\ref{maren}) we obtain the OME after mass and coupling renormalization
\begin{eqnarray}
   {\hat{A}}_{ij}  &=&  
\delta_{ij}
                 +{a}_s^{\MOM}~ 
                   \Ahathat_{ij}^{(1)} 
                         + {{a}_s^{\MOM}}^2 \Biggl[
                                        \Ahathat^{(2)}_{ij}                                        
                                        +\delta a^{\MOM}_{s, 1}   \Ahathat_{ij}^{(1)}                                           
\NN\\&&
                                      + {\delta m_1} 
                                        \left(\Bigl(\frac{m_1^2}{\mu^2}\Bigr)^{\ep/2}
                                        m_1 \frac{d}{dm_1}
                                      +\Bigl(\frac{m_2^2}{\mu^2}\Bigr)^{\ep/2}
                                        m_2 \frac{d}{dm_2}
                                      \right)
                                                   \Ahathat_{ij}^{(1)}                                                                   
                                \Biggr]
\NN\\ &&
                         + {{a}_s^{\MOM}}^3 \Biggl[ 
                                         \Ahathat^{(3)}_{ij}                                           
                                    +\delta a^{\MOM}_{s, 2}
                                    \Ahathat_{ij}^{(1)}
                                    + 2 \delta a^{\MOM}_{s,1} 
                                    \Biggl[
                                        \Ahathat^{(2)}_{ij}                                        
\NN\\&&
                                      + {\delta m_1} 
                                        \left(\Bigl(\frac{m_1^2}{\mu^2}\Bigr)^{\ep/2}
                                        m_1 \frac{d}{dm_1}
                                      +\Bigl(\frac{m_2^2}{\mu^2}\Bigr)^{\ep/2}
                                        m_2 \frac{d}{dm_2}
                                      \right)
                                                   \Ahathat_{ij}^{(1)}                                                                   
\Biggr]
\NN\\ &&
                                        +{\delta m_1} 
                                         \left(
                                           \Bigl(\frac{m_1^2}{\mu^2}\Bigr)^{\ep/2}
                                         m_1 \frac{d}{dm_1} 
                                         +
                                           \Bigl(\frac{m_2^2}{\mu^2}\Bigr)^{\ep/2}
                                         m_2 \frac{d}{dm_2} 
                                         \right)
                                                    \Ahathat_{ij}^{(2)}                                           
\NN\\ &&
                                        + \left(\delta m_{2,1}(m_1,m_2) \Bigl(\frac{m_1^2}{\mu^2}\Bigr)^{\ep}
                                          m_1 \frac{d}{dm_1}
                                               +\delta m_{2,2}(m_1,m_2) \Bigl(\frac{m_2^2}{\mu^2}\Bigr)^{\ep}
                                          m_2 \frac{d}{dm_2} \right) 
                                                    \Ahathat_{ij}^{(1)}                                            
\NN\\&&
                                        +
                                        \frac{(\delta m_1)^2}{2}
                                        \left(
                                          \Bigl(\frac{m_1^2}{\mu^2}\Bigr)^{\ep}
                                                   m_1^2
                                                   \frac{d^2}{{dm_1}^2}
                                                   +
                                            \Bigl(\frac{m_2^2}{\mu^2}\Bigr)^{\ep}
                                                   m_2^2  \frac{d^2}{{dm_2}^2}
                                               \right)
                                                     \Ahathat_{ij}^{(1)}                                            
\NN\\&&
                                         + (\delta m_1)^2 
                                        \Bigl(\frac{m_1^2}{\mu^2}\Bigr)^{\ep/2}
                                          \Bigl(\frac{m_2^2}{\mu^2}\Bigr)^{\ep/2}
                                        m_1 \frac{d}{dm_1}
                                        m_2 \frac{d}{dm_2}
                                                   \Ahathat_{ij}^{(1)}                                           
    \Biggr]~,
    \label{macoren}
  \end{eqnarray}
  where we have suppressed the dependence on the masses, $\ep$ and $N$ in the arguments of the OMEs.
\end{subsection}
\begin{subsection}{\bf\boldmath Operator Renormalization}
\label{SubSec-RENOp}

\vspace*{1mm}
\noindent
Next we remove the ultraviolet divergence of the different local operators defined in Eqs.~(\ref{COMP1}--\ref{COMP3})
by introducing the respective  $Z$-factors
\begin{eqnarray}
        O^{\sf NS}_{q,r;\mu_1,...,\mu_N}&=&
                    Z^{\sf NS}(\mu^2)\hat{O}^{\sf NS}_{q,r;\mu_1,...,\mu_N}~,
                    \label{ZNSdef}\\
        O^{\sf S}_{i;\mu_1,...,\mu_N}&=&
                  Z^{\sf S}_{ij}(\mu^2)
                  \hat{O}^{\sf S}_{j;\mu_1,...,\mu_N}~,\quad~i=q,g~.
                  \label{ZSijdef}
\end{eqnarray}
In the singlet case, the operator renormalization introduces a mixing between the different operators as they carry the same quantum 
numbers. Analogously to the OMEs, here the $Z$-factors are  split into the flavor pure singlet 
($\sf{PS}$) and non-singlet ($\sf{NS}$) contributions
   \begin{eqnarray}
    Z_{qq}^{-1}&=&Z_{qq}^{-1, {\sf PS}}+Z_{qq}^{-1, {\sf NS}} 
\label{ZPSNS1}~.
   \end{eqnarray}
Each $Z$-factor is associated with an anomalous dimension $\gamma_{ij}$ via 
   \begin{eqnarray}
    \gamma_{qq}^{\sf NS}(a_s^{\MS},N_F,N)&=&
                           \mu \frac{d}{d\mu} \ln Z_{qq}^{\sf 
NS}(a_s^{\MS},N_F,\ep,N)~,
                                               \label{gammazetNS}\\
    \gamma_{ij}(a_s^{\MS},N_F,N)&=&
                           \mu \frac{d}{d\mu} 
Z_{ij}(a_s^{\MS},N_F,\ep,N)~.
                                               \label{gammazetS}
   \end{eqnarray}
Here both the anomalous dimensions and the operator $Z$-factors obey perturbative series expansions in the coupling constant
   \begin{eqnarray}
    \gamma_{ij}^{{\sf S,~PS,~NS}}(a_s^{\MS},N_F,N)
        &=&\sum_{l=1}^{\infty}{a^{\MS}_s}^l 
        \gamma_{ij}^{(l-1), {\sf S,~PS,~NS}}(N_F,N)
            \label{pertgamma}
            \\
            Z_{ij} &=& \delta_{ij} + \sum_{k=1}^\infty a_s^k Z_{ij}^{(k)}
            \\
            Z_{ij}^{-1} &=& \delta_{ij} + \sum_{k=1}^\infty a_s^k Z_{ij}^{-1, (k)}~.
   \end{eqnarray}
In order to renormalize the respective operators, we first consider operator matrix elements with off-shell external legs as a sum of 
massive and massless contributions:
\begin{eqnarray}
 \hat{A}_{ij}\left(p^2,m_1^2,m_2^2,\mu^2,a_s^{\MOM},N_F+2\right)
 &=&\hat{A}_{ij}\left(\frac{-p^2}{\mu^2},a_s^{\MS},N_F\right)
 \NN
 \\&&
 +\hat{A}_{ij}^Q\left(p^2,m_1^2,m_2^2,\mu^2,a_s^{\MOM},N_F+2\right)~.
 \label{AijSplit}
\end{eqnarray}
Here the massless contribution depends on $a_s^{\MS}$ since the $\MOM$--scheme, cf. Section \ref{SubSec-RENCo}, has been constructed in 
such a way that it corresponds to the $\MS$--scheme concerning the renormalization of the light quark flavor and gluon contributions.
$\hat{A}_{ij}^Q$ denotes any massive OME we consider. 
The term $\delta_{ij}$, which appears in the expansion of the OMEs (see Eqs.~(\ref{maren}) and (\ref{macoren})), 
does not have any mass-dependence and is considered a part of the light flavor part 
$\hat{A}_{ij}\left(\frac{-p^2}{\mu^2},a_s^{\MS},N_F\right)$.

We first consider the renormalization of the purely massless contribution in the $\MS$--scheme \cite{Matiounine:1998ky}
   \begin{eqnarray}
    A_{qq}^{\sf NS}\Bigl(\frac{-p^2}{\mu^2},a_s^{\MS},N_F,N\Bigr)
           &=&Z^{-1,{\sf NS}}_{qq}(a_s^{\MS},N_F,\ep,N)
              \hat{A}_{qq}^{\sf NS}\Bigl(\frac{-p^2}{\mu^2},a_s^{\MS},N_F,\ep,N\Bigr)
              \label{renAqqnf}
\\
    A_{ij}\Bigl(\frac{-p^2}{\mu^2},a_s^{\MS},N_F,N\Bigr)
           &=&Z^{-1}_{il}(a_s^{\MS},N_F,\ep,N)
              \hat{A}_{lj}\Bigl(\frac{-p^2}{\mu^2},a_s^{\MS},N_F,\ep,N\Bigr)
                   ~,~i,j,l=q,g~.
              \label{renAijnf}
\nonumber\\
   \end{eqnarray}
Solving (\ref{gammazetNS}--\ref{gammazetS}) yields the $Z$-factors in the singlet case 
\begin{eqnarray}
   Z_{ij}(a^{\MS}_s,N_F) &=&
                            \delta_{ij}
                           +a^{\MS}_s \frac{\gamma_{ij}^{(0)}}{\ep}
                           +{a^{\MS}_s}^2 \Biggl\{
                                 \frac{1}{\ep^2} \Bigl(
                                     \frac{1}{2} \gamma_{il}^{(0)}
                                                 \gamma_{lj}^{(0)}
                                   + \beta_0 \gamma_{ij}^{(0)}
                                                 \Bigr)
                               + \frac{1}{2 \ep} \gamma_{ij}^{(1)}
                                   \Biggr\}
\nonumber 
\\ 
&&
                           + {a^{\MS}_s}^3 \Biggl\{
                                 \frac{1}{\ep^3} \Bigl(
                                     \frac{1}{6}\gamma_{il}^{(0)}
                                                \gamma_{lk}^{(0)}
                                                \gamma_{kj}^{(0)}
                                   + \beta_0 \gamma_{il}^{(0)} 
                                             \gamma_{lj}^{(0)}
                                   + \frac{4}{3} \beta_0^2 \gamma_{ij}^{(0)}
                                                  \Bigr)
\NN\\ &&
                               + \frac{1}{\ep^2}  \Bigl(
                                     \frac{1}{6} \gamma_{il}^{(1)} 
                                                 \gamma_{lj}^{(0)}
                                   + \frac{1}{3} \gamma_{il}^{(0)} 
                                                 \gamma_{lj}^{(1)}
                                   + \frac{2}{3} \beta_0 \gamma_{ij}^{(1)} 
                                   + \frac{2}{3} \beta_1 \gamma_{ij}^{(0)}
                                                  \Bigr)
                              + \frac{\gamma_{ij}^{(2)}}{3 \ep}
                                   \Biggr\} \label{Zijnf}.
\end{eqnarray}
In the non-singlet and pure singlet cases one has
\begin{eqnarray}
   Z_{qq}^{\sf NS}(a^{\MS}_s,N_F) &=& 
                             1 
                           +a^{\MS}_s \frac{\gamma_{qq}^{(0),{\sf NS}}}{\ep}
                           +{a^{\MS}_s}^2 \Biggl\{
                                 \frac{1}{\ep^2} \Bigl(
                                     \frac{1}{2}{\gamma_{qq}^{(0),{\sf NS}}}^2 
                                   + \beta_0 \gamma_{qq}^{(0),{\sf NS}}
                                                 \Bigr)
                              + \frac{1}{2 \ep} \gamma_{qq}^{(1),{\sf NS}} 
                                        \Biggr\}
\NN\\ &&
                           +{a^{\MS}_s}^3 \Biggl\{
                                 \frac{1}{\ep^3} \Bigl(
                                     \frac{1}{6} {\gamma_{qq}^{(0),{\sf NS}}}^3
                                   + \beta_0 {\gamma_{qq}^{(0),{\sf NS}}}^2 
                                   + \frac{4}{3} \beta_0^2 
                                                 \gamma_{qq}^{(0),{\sf NS}}
                                                 \Bigr)
\NN\\ &&
+\frac{1}{\ep^2} \Bigl(
\frac{1}{2} \gamma_{qq}^{(0),{\sf NS}} \gamma_{qq}^{(1),{\sf NS}}
+\frac{2}{3} \beta_0 \gamma_{qq}^{(1),{\sf NS}}
+\frac{2}{3} \beta_1 \gamma_{qq}^{(0),{\sf NS}} \Bigr)
+\frac{1}{3 \ep} \gamma_{qq}^{(2),{\sf NS}}
\Biggl\}
\\
   Z_{qq}^{\sf PS}(a^{\MS}_s,N_F) &=&
                            {a^{\MS}_s}^2 \Biggl\{
                                 \frac{1}{2\ep^2} \gamma_{qg}^{(0)}
                                                  \gamma_{gq}^{(0)}   
                               + \frac{1}{2\ep}   \gamma_{qq}^{(1), {\sf PS}}
                                        \Biggr\}
                           +{a^{\MS}_s}^3 \Biggl\{
                                 \frac{1}{\ep^3} \Bigl(
                                     \frac{1}{3}\gamma_{qq}^{(0)} 
                                      \gamma_{qg}^{(0)}
                                      \gamma_{gq}^{(0)}
\NN\\ &&
                                    +\frac{1}{6}\gamma_{qg}^{(0)}
                                     \gamma_{gg}^{(0)} \gamma_{gq}^{(0)}
                                    +\beta_0 \gamma_{qg}^{(0)}
                                             \gamma_{gq}^{(0)}
                                                 \Bigr)
                               + \frac{1}{\ep^2} \Bigl(
                                     \frac{1}{3}\gamma_{qg}^{(0)} 
                                                \gamma_{gq}^{(1)}
\NN\\ &&
                                    +\frac{1}{6}\gamma_{qg}^{(1)} 
                                                \gamma_{gq}^{(0)}
                                    +\frac{1}{2} \gamma_{qq}^{(0)}
                                                 \gamma_{qq}^{(1), {\sf PS}}
                                    +\frac{2}{3} \beta_0
                                                 \gamma_{qq}^{(1), {\sf PS}}
                                                 \Bigr)
                                    +\frac{\gamma_{qq}^{(2), {\sf PS}}}{3\ep} 
                                        \Biggr\}~, \label{ZqqPSnf}
\end{eqnarray}
respectively. The $Z$-factors describing the ultraviolet renormalization of the complete operator matrix elements 
$\hat{A}_{ij}\left(p^2,m_1^2,m_2^2,\mu^2,a_s^{\MOM},N_F+2\right)$  are obtained by inverting (\ref{Zijnf}--\ref{ZqqPSnf}) and 
replacing $N_F\rightarrow N_F+2$. Finally, the transformation (\ref{asmsa}) is applied. 
The resulting operator $Z$-factors read:
  \begin{eqnarray}
   Z_{ij}^{-1}(a_s^{\MOM},N_F+2,\mu)&=&
      \delta_{ij}
     -a_s^{\MOM}\frac{\gamma_{ij}^{(0)}}{\ep}
     +{a^{\MOM}_s}^2\Biggl[
          \frac{1}{\ep}\Bigl(
                       -\frac{1}{2}\gamma_{ij}^{(1)}
                       -\delta a^{\MOM}_{s,1}\gamma_{ij}^{(0)}
                       \Bigr)
                       \NN\\&&
         +\frac{1}{\ep^2}\Bigl(
                        \frac{1}{2}\gamma_{il}^{(0)}\gamma_{lj}^{(0)}
                       +\beta_0\gamma_{ij}^{(0)}
                        \Bigr)
         \Biggr]
     +{a^{\MOM}_s}^3\Biggl[
          \frac{1}{\ep}\Bigl(
                       -\frac{1}{3}\gamma_{ij}^{(2)}
                       -\delta a^{\MOM}_{s,1}\gamma_{ij}^{(1)}
                       \NN\\&&
                       -\delta a^{\MOM}_{s,2}\gamma_{ij}^{(0)}
                       \Bigr)
         +\frac{1}{\ep^2}\Bigl(
                        \frac{4}{3}\beta_0\gamma_{ij}^{(1)}
                       +2\delta a^{\MOM}_{s,1}\beta_0\gamma_{ij}^{(0)}
                       +\frac{1}{3}\beta_1\gamma_{ij}^{(0)}
\NN\\ &&
                       +\delta a^{\MOM}_{s,1}\gamma_{il}^{(0)}\gamma_{lj}^{(0)}
                       +\frac{1}{3}\gamma_{il}^{(1)}\gamma_{lj}^{(0)}
                       +\frac{1}{6}\gamma_{il}^{(0)}\gamma_{lj}^{(1)}
                        \Bigr)
         +\frac{1}{\ep^3}\Bigl(
                       -\frac{4}{3}\beta_0^{2}\gamma_{ij}^{(0)}
\NN\\&&
                       -\beta_0\gamma_{il}^{(0)}\gamma_{lj}^{(0)}
                       -\frac{1}{6}\gamma_{il}^{(0)}\gamma_{lk}^{(0)}
                                    \gamma_{kj}^{(0)} 
                     \Bigr)
       \Biggr]~, \label{ZijInfp1}
\end{eqnarray}
  {\small
  \begin{eqnarray}
   Z_{qq}^{-1,{\sf NS}}(a_s^{\MOM},N_F+2)&=&
      1
     -a^{\MOM}_s\frac{\gamma_{qq}^{(0),{\sf NS}}}{\ep}
     +{a^{\MOM}_s}^2\Biggl[
          \frac{1}{\ep}\Bigl(
                       -\frac{1}{2}\gamma_{qq}^{(1),{\sf NS}}
                       -\delta a^{\MOM}_{s,1}\gamma_{qq}^{(0),{\sf NS}}
                       \Bigr)
\nonumber\\&&
         +\frac{1}{\ep^2}\Bigl(
                        \beta_0\gamma_{qq}^{(0),{\sf NS}}
                       +\frac{1}{2}{\gamma_{qq}^{(0),{\sf NS}}}^{2}
                        \Bigr)
         \Biggr]
     +{a^{\MOM}_s}^3\Biggl[
          \frac{1}{\ep}\Bigl(
                       -\frac{1}{3}\gamma_{qq}^{(2),{\sf NS}}
                       -\delta a^{\MOM}_{s,1}\gamma_{qq}^{(1),{\sf NS}}
\NN\\  && 
                       -\delta a^{\MOM}_{s,2}\gamma_{qq}^{(0),{\sf NS}}
                       \Bigr)
         +\frac{1}{\ep^2}\Bigl(
                       \frac{4}{3}\beta_0\gamma_{qq}^{(1),{\sf NS}}
                      +2\delta a^{\MOM}_{s,1}\beta_0\gamma_{qq}^{(0),{\sf NS}}
                       +\frac{1}{3}\beta_1\gamma_{qq}^{(0),{\sf NS}}
\NN\\  && 
                     +\frac{1}{2}\gamma_{qq}^{(0),{\sf NS}}
                                   \gamma_{qq}^{(1),{\sf NS}}
                       +\delta a^{\MOM}_{s,1}{\gamma_{qq}^{(0),{\sf NS}}}^{2}
                        \Bigr)
         +\frac{1}{\ep^3}\Bigl(
                       -\frac{4}{3}\beta_0^{2}\gamma_{qq}^{(0),{\sf NS}}
                       -\beta_0{\gamma_{qq}^{(0),{\sf NS}}}^{2} \NN\\ &&
                       -\frac{1}{6}{\gamma_{qq}^{(0),{\sf NS}}}^{3}
                     \Bigr)
       \Biggr]~, \label{ZNSInfp1} %
\\
   Z_{qq}^{-1,{\sf PS}}(a_s^{\MOM},N_F+2)&=&
      {a^{\MOM}_s}^2\Biggl[
          \frac{1}{\ep}\Bigl(
                       -\frac{1}{2}\gamma_{qq}^{(1), {\sf PS}}
                       \Bigr)
         +\frac{1}{\ep^2}\Bigl(
                        \frac{1}{2}\gamma_{qg}^{(0)}\gamma_{gq}^{(0)}
                        \Bigr)
         \Biggr]
     +{a^{\MOM}_s}^3\Biggl[
          \frac{1}{\ep}\Bigl(
                       -\frac{1}{3}\gamma_{qq}^{(2), {\sf PS}}
 \NN\\  &&
                       -\delta a^{\MOM}_{s,1}\gamma_{qq}^{(1), {\sf PS}}
                       \Bigr)
         +\frac{1}{\ep^2}\Bigl(
                        \frac{1}{6}\gamma_{qg}^{(0)}\gamma_{gq}^{(1)}
                       +\frac{1}{3}\gamma_{gq}^{(0)}\gamma_{qg}^{(1)}
                       +\frac{1}{2}\gamma_{qq}^{(0)}\gamma_{qq}^{(1), {\sf PS}}
 \NN\\  &&
                       +\frac{4}{3}\beta_0\gamma_{qq}^{(1), {\sf PS}}
                       +\delta a^{\MOM}_{s,1}\gamma_{qg}^{(0)}\gamma_{gq}^{(0)}
                        \Bigr)
         +\frac{1}{\ep^3}\Bigl(
                       -\frac{1}{3}\gamma_{qg}^{(0)}\gamma_{gq}^{(0)}
                                   \gamma_{qq}^{(0)}
                       -\frac{1}{6}\gamma_{gq}^{(0)}\gamma_{qg}^{(0)}
                                   \gamma_{gg}^{(0)}
  \NN\\  &&
                       -\beta_0\gamma_{qg}^{(0)}\gamma_{gq}^{(0)}
                     \Bigr)
       \Biggr]~. \label{ZPSInfp1}
  \end{eqnarray}}

\noindent
Here and in the Eqs.~(\ref{Zijnf}--\ref{ZqqPSnf}) we have dropped the $N_F$-dependence of the anomalous dimensions 
$\gamma_{ij}$ and 
$\beta_i$ for brevity. The inverse $Z$-factors for the purely light-parton case correspond to (\ref{ZijInfp1}--\ref{ZPSInfp1}) 
after 
substituting $N_F+2\rightarrow N_F$ and $\delta a^{\MOM}_{s,i}\rightarrow \delta a^{\MS}_{s,i}$.

We are only interested in performing the ultraviolet renormalization for the massive contributions to the operator matrix element in 
(\ref{AijSplit}) and thus subtract the contributions stemming from purely light parts again
\begin{eqnarray}
   \Atiltil^Q_{ij}(p^2,m_1^2,m_2^2,\mu^2,a_s^{\MOM},N_F+2)&=&
               Z^{-1}_{il}(a_s^{\MOM},N_F+2,\mu) 
                  \hat{A}^Q_{ij}(p^2,m_1^2,m_2^2,\mu^2,a_s^{\MOM},N_F+2)
\NN\\ &&
              +Z^{-1}_{il}(a_s^{\MOM},N_F+2,\mu) 
                   \hat{A}_{ij}\Bigl(\frac{-p^2}{\mu^2},a_s^{\MS},N_F\Bigr)
\NN\\ &&
              -Z^{-1}_{il}(a_s^{\MS},N_F,\mu)
                   \hat{A}_{ij}\Bigl(\frac{-p^2}{\mu^2},a_s^{\MS},N_F\Bigr)~.
\label{eqXX}
\end{eqnarray}

\noindent
Finally, the limit $p^2\rightarrow 0$ is performed. Since scale-less diagrams vanish if computed in dimensional regularization, only 
the Born piece of the massless OME contributes
\begin{eqnarray}
    \hat{A}_{ij}\left(0,\alpha_s^{\MS},N_F\right)=\delta_{ij}~.
\end{eqnarray}
One obtains the UV--renormalization prescription
\begin{eqnarray}
\Atiltil^Q_{ij}\Bigl(\frac{m_1^2}{\mu^2},\frac{m_2^2}{\mu^2},a_s^{\MOM},N_F+2\Bigr) &=& 
            a_s^{\MOM}\Biggl( \hat{A}_{ij}^{(1),Q}
                              \Bigl(\frac{m_1^2}{\mu^2},\frac{m_2^2}{\mu^2}\Bigr)
                     +Z^{-1,(1)}_{ij}(N_F+2,\mu)
                     -Z^{-1,(1)}_{ij}(N_F)
               \Biggr)\nonumber
\\ 
&& + {a_s^{\MOM}}^2\Biggl( \hat{A}_{ij}^{(2),Q}
                                    \Bigl(\frac{m_1^2}{\mu^2},\frac{m_2^2}{\mu^2}\Bigr)
                       +Z^{-1,(2)}_{ij}(N_F+2,\mu)
                       -Z^{-1,(2)}_{ij}(N_F)
\nonumber\\ &&                  
     +Z^{-1,(1)}_{ik}(N_F+2,\mu)
                        \hat{A}_{kj}^{(1),Q}\Bigl(\frac{m_1^2}{\mu^2},\frac{m_2^2}{\mu^2}\Bigr)
               \Biggr)
\NN\\ &&
           +{a_s^{\MOM}}^3\Biggl( \hat{A}_{ij}^{(3),Q}
                                   \Bigl(\frac{m_1^2}{\mu^2},\frac{m_2^2}{\mu^2}\Bigr)
                       +Z^{-1,(3)}_{ij}(N_F+2,\mu)
                       \NN\\&&
                       -Z^{-1,(3)}_{ij}(N_F)
                       +Z^{-1,(1)}_{ik}(N_F+2,\mu)
                        \hat{A}_{kj}^{(2),Q}\Bigl(\frac{m_1^2}{\mu^2},\frac{m_2^2}{\mu^2}\Bigr)
  \NN\\ &&\phantom{{a_s^{\MOM}}^3\Biggl(}
                       +Z^{-1,(2)}_{ik}(N_F+2,\mu)
                        \hat{A}_{kj}^{(1),Q}\Bigl(\frac{m_1^2}{\mu^2},\frac{m_2^2}{\mu^2}\Bigr)
                        \Biggr)~. \label{GenRen1}
\end{eqnarray}
Here $Z$-factors at $N_F+2$ flavors describe the massive case (\ref{ZijInfp1}--\ref{ZPSInfp1}) while those with argument $N_F$ 
denote the $Z$-factors for the massless case.
\end{subsection}
\begin{subsection}{\bf\boldmath Collinear Factorization}
\label{SubSec-RENOp1}
At this point only collinear singularities remain. They arise from massless subgraphs only and are therefore independent of the 
additional heavy quark flavor considered in these analyses. We thus follow {\cite{Bierenbaum:2009mv}} directly and 
remove the 
collinear singularities via mass factorization
  \begin{eqnarray}
   A_{ij}\Bigl(\frac{m_1^2}{\mu^2},\frac{m_2^2}{\mu^2},a_s^{\MOM},N_F+2\Bigr)&=&
     \Atiltil^Q_{il}\Bigl(\frac{m_1^2}{\mu^2},\frac{m_2^2}{\mu^2},a_s^{\MOM},N_F+2\Bigr)
     \Gamma_{lj}^{-1}~. \label{genren1}
  \end{eqnarray}
Note that in a fully massless scenario the transition functions $\Gamma_{ij}$ would be related to the light flavor renormalization 
constant via 
  \begin{eqnarray}
      \Gamma_{ij} \left(N_F\right)= Z^{-1}_{ij}\left(N_F\right)~, \label{GammaZ}
  \end{eqnarray}
cf.~\cite{Buza:1995ie}. However, in the presence of one or more heavy quark flavors the transition functions
stem from the corresponding massless subgraphs only. Due to this and the subtraction of the $\delta_{ij}$--term 
in the OMEs after ultraviolet renormalization $\Atiltil^Q_{ij}$ the transition functions contribute up to $O(\alpha_s^2)$ only. 

The renormalized OME is then obtained by
  \begin{eqnarray}
   && A_{ij}\Bigl(\frac{m_1^2}{\mu^2},\frac{m_2^2}{\mu^2},a_s^{\MOM},N_F+2\Bigr)=
\NN\\&&\phantom{+}
                a^{\MOM}_s~\Biggl(
                      \hat{A}_{ij}^{(1),Q}\Bigl(\frac{m_1^2}{\mu^2},\frac{m_2^2}{\mu^2}\Bigr)
                     +Z^{-1,(1)}_{ij}(N_F+2)
                     -Z^{-1,(1)}_{ij}(N_F)
                           \Biggr)
\NN\\&&
           +{a^{\MOM}_s}^2\Biggl( 
                        \hat{A}_{ij}^{(2),Q}\Bigl(\frac{m_1^2}{\mu^2},\frac{m_2^2}{\mu^2}\Bigr)
                       +Z^{-1,(2)}_{ij}(N_F+2)
                       -Z^{-1,(2)}_{ij}(N_F)               
\NN\\
&&\phantom{+{a^{\MOM}_s}^2\Biggl(}
 +Z^{-1,(1)}_{ik}(N_F+2)\hat{A}_{kj}^{(1),Q}
                                              \Bigl(\frac{m_1^2}{\mu^2},\frac{m_2^2}{\mu^2}\Bigr)
                       +\Bigl[ \hat{A}_{il}^{(1),Q}
                               \Bigl(\frac{m_1^2}{\mu^2},\frac{m_2^2}{\mu^2}\Bigr)
                              +Z^{-1,(1)}_{il}(N_F+2)
\NN\\
&&\phantom{+{a^{\MOM}_s}^2\Biggl(}
                              -Z^{-1,(1)}_{il}(N_F)
                        \Bigr] 
                             \Gamma^{-1,(1)}_{lj}(N_F)
                         \Biggr)
\NN\\ 
&&
          +{a^{\MOM}_s}^3\Biggl( 
                        \hat{A}_{ij}^{(3),Q}\Bigl(\frac{m_1^2}{\mu^2},\frac{m_2^2}{\mu^2}\Bigr)
                       +Z^{-1,(3)}_{ij}(N_F+2)
                       -Z^{-1,(3)}_{ij}(N_F)
\NN\\ 
&&\phantom{+{a^{\MOM}_s}^3\Biggl(}
                       +Z^{-1,(1)}_{ik}(N_F+2)\hat{A}_{kj}^{(2),Q}
                                              \Bigl(\frac{m_1^2}{\mu^2},\frac{m_2^2}{\mu^2}\Bigr)
                       +Z^{-1,(2)}_{ik}(N_F+2)\hat{A}_{kj}^{(1),Q}
                                              \Bigl(\frac{m_1^2}{\mu^2},\frac{m_2^2}{\mu^2}\Bigr)
 \NN\\ &&\phantom{+{a^{\MOM}_s}^3\Biggl(}
                        +\Bigl[ 
                               \hat{A}_{il}^{(1),Q}
                                 \Bigl(\frac{m_1^2}{\mu^2},\frac{m_2^2}{\mu^2}\Bigr)
                              +Z^{-1,(1)}_{il}(N_F+2)
                              -Z^{-1,(1)}_{il}(N_F)
                        \Bigr]
                              \Gamma^{-1,(2)}_{lj}(N_F)
 \NN\\ &&\phantom{+{a^{\MOM}_s}^3\Biggl(}                              
                       +\Bigl[ 
                               \hat{A}_{il}^{(2),Q}
                                 \Bigl(\frac{m_1^2}{\mu^2},\frac{m_2^2}{\mu^2}\Bigr)
                              +Z^{-1,(2)}_{il}(N_F+2)
                              -Z^{-1,(2)}_{il}(N_F)
 \NN\\ 
&&\phantom{+{a^{\MOM}_s}^3\Biggl(}
                              +Z^{-1,(1)}_{ik}(N_F+2)\hat{A}_{kl}^{(1),Q}
                                              \Bigl(\frac{m_1^2}{\mu^2},\frac{m_2^2}{\mu^2}\Bigr)
                        \Bigr]
                              \Gamma^{-1,(1)}_{lj}(N_F)
                        \Biggr) 
+O\left({a^{\MOM}_s}^4\right)
\label{GenRen3}.
\end{eqnarray}
Eq.~(\ref{GenRen3}) differs from the corresponding renormalization and factorization prescription for one heavy quark flavor 
{\cite{Bierenbaum:2009mv}}
only by the definition of the renormalization constants $Z^{-1,(k)}_{ij}(N_F+2)$. Now the term $\delta_{ij}$ is added back to the 
massive OME. In a final step, the coupling constant is transformed to that in the $\MS$--scheme via Eq.~(\ref{asmoma}).
\end{subsection}

\begin{subsection}{One--particle reducible contributions}
We will perform the renormalization of the massive operator matrix elements starting from the set of 
Feynman diagrams which also include the one-particle reducible contributions. These terms contribute 
from $O(\alpha_s^2)$ onward and are obtained by quark and gluon self--energy contributions to the 
external legs of lower order one-particle irreducible 
diagrams. From 3-loop order onward the reducible contributions to the OMEs $A_{Qg}$ and $A_{gg,Q}$ may contain three
different heavy flavors, while this is not the case for the irreducible contributions. Note that the inclusion of 
the top quark in a loop of the irreducible terms for $A_{ij}^{(3)}$ would demand to consider the 
energy range $Q^2 \gg m_t^2$. At a scale $\mu^2 \simeq m_t^2$, both charm and bottom can be dealt with as effectively massless. The emergence
of massive top loops in the reducible contributions is accounted for by renormalization. In the following we will 
strictly consider the case of two heavy flavors only.

\begin{subsubsection}{Self--energy contributions}
\label{Sec-elf}
The scalar self--energies are obtained by projecting out the Lorentz--structure
\begin{eqnarray}
       \hat{\Pi}_{\mu\nu}^{ab}(p^2,\hat{m}_1^2,\hat{m}_2^2,\mu^2,\hat{a}_s) &=& i\delta^{ab}
                            \left[-g_{\mu\nu}p^2 +p_\mu p_\nu\right] 
                            \hat{\Pi}(p^2,\hat{m}_1^2,\hat{m}_2^2,\mu^2,\hat{a}_s)~,  \\
                            \hat{\Pi}(p^2,\hat{m}_1^2,\hat{m}_2^2,\mu^2,\hat{a}_s)&=&
                            \sum_{k=1}^{\infty}\hat{a}_s^k
                            \hat{\Pi}^{(k)}(p^2,\hat{m}_1^2,\hat{m}_2^2,\mu^2)
                            ~,\\
                            \hat{\Sigma}_{ij}(p^2,\hat{m}_1^2,\hat{m}_2^2,\mu^2,\hat{a}_s)&=&
        \label{pertPiGlu}
                      i \, \delta_{ij} \, \adag p \,
                      \hat{\Sigma}(p^2,\hat{m}_1^2,\hat{m}_2^2,\mu^2,\hat{a_s})~, \\
   \hat{\Sigma}(p^2,\hat{m}_1^2,\hat{m}_2^2,\mu^2,\hat{a}_s)&=&
   \sum_{k=2}^{\infty}\hat{a}_s^k\hat{\Sigma}^{(k)}(p^2,\hat{m}_1^2,\hat{m}_2^2,\mu^2)~.
        \label{pertSiQu}
\end{eqnarray}
We decompose the irreducible two--mass self--energies into contributions which depend on one mass 
only and an additional part stemming from diagrams containing both heavy quark flavors
\begin{eqnarray}
\hat{\Pi}^{(k)}\left(p^2,\hat{m}_1^2,\hat{m}_2^2,\mu^2\right)&=&
\hat{\Pi}^{(k)}\Bigl(p^2,\frac{\hat{m}_1^2}{\mu^2}\Bigr)+
\hat{\Pi}^{(k)}\Bigl(p^2,\frac{\hat{m}_2^2}{\mu^2}\Bigr)
+\hat{\tilde{\Pi}}^{(k)}\left(p^2,\hat{m}_1^2,\hat{m}_2^2,\mu^2\right)~,
\label{gSelf2m}
\end{eqnarray}
\begin{eqnarray}
\hat{\Sigma}^{(j)}\left(p^2,\hat{m}_1^2,\hat{m}_2^2,\mu^2\right)&=&
\hat{\Sigma}^{(j)}\Bigl(p^2,\frac{\hat{m}_1^2}{\mu^2}\Bigr)+
\hat{\Sigma}^{(j)}\Bigl(p^2,\frac{\hat{m}_2^2}{\mu^2}\Bigr)
+\hat{\tilde{\Sigma}}^{(j)}\left(p^2,\hat{m}_1^2,\hat{m}_2^2,\mu^2\right)
~.\label{qSelf2m}
\end{eqnarray}
Up to two--loop order no diagrams with two heavy flavors contribute

\begin{eqnarray}
\hat{\tilde{\Pi}}^{(k)}(p^2,\hat{m}_1^2,\hat{m}_2^2,\mu^2)&=&0~\text{for}~k\in\{1,2\}~,
\\
\hat{\tilde{\Sigma}}^{(2)}(p^2,\hat{m}_1^2,\hat{m}_2^2,\mu^2)&=&0~.
\end{eqnarray}
The single-mass contributions for the gluon are known from
\cite{CS1,Chetyrkin:2008jk,Bierenbaum:2009mv}
  \begin{eqnarray}
  \label{eqPI1}
   \hat{\Pi}^{(1)}\Bigl(0,\frac{\hat{m}^2}{\mu^2}\Bigr)&=&
            T_F\Bigl(\frac{\hat{m}^2}{\mu^2}\Bigr)^{\ep/2}
                        \left[
             -\frac{8}{3\ep}
              \exp \Bigl(\sum_{i=2}^{\infty}\frac{\zeta_i}{i}
                       \Bigl(\frac{\ep}{2}\Bigr)^{i}\Bigr)
             \right]~,
               ~\label{GluSelf1} 
               \\
   \hat{\Pi}^{(2)}\Bigl(0,\frac{\hat{m}^2}{\mu^2}\Bigr)&=&
      T_F\Bigl(\frac{\hat{m}^2}{\mu^2}\Bigr)^{\ep}\Biggl\{
      -\frac{4}{\ep^2} C_A + \frac{1}{\ep} \left(5 C_A-12 C_F\right) 
      + C_A \Bigl(\frac{13}{12} -\zeta_2\Bigr)
      - \frac{13}{3} C_F   
\NN\\ &&
      + \ep \left[C_A \Bigl(\frac{169}{144} + \frac{5}{4} \zeta_2 - 
      \frac{\zeta_3}{3} \Bigr) 
     - C_F \Bigl(\frac{35}{12}+3 \zeta_2\Bigr) 
     \right]\Biggr\}
       + 
            O(\ep^2)~,  \label{GluSelf2}
\\
             \hat{\Pi}^{(3)}\Bigl(0,\frac{\hat{m}^2}{\mu^2}\Bigr)&=&
       T_F\Bigl(\frac{\hat{m}^2}{\mu^2}\Bigr)^{3\ep/2}\Biggl\{
                        \frac{1}{\ep^3}\left[
                                 -\frac{32}{9}T_F C_A \left(2 N_F+1\right)
                                 + \frac{164}{9} C_A^2      
                                       \right]
\NN\\ &&
                       +\frac{1}{\ep^2}\left[
                               \frac{80}{27} (C_A-6 C_F) N_FT_F
                              +\frac{8}{27} (35 C_A-48 C_F) T_F
                              -\frac{781}{27} C_A^2 \right.                                                 
\NN\\ &&
                       \left.       +\frac{712}{9}C_AC_F
                                       \right]
                       +\frac{1}{\ep}\biggl[ 
                                \frac{4}{27}\big(
                                                       C_A(-101-18\zeta_2)
                                                      -62C_F
                                                \big)N_FT_F 
\NN\\ &&
                              -\frac{2}{27}   \big(
                                                       C_A(37+18 \zeta_2)
                                                       +80 C_F
                                                \big) T_F
                              +C_A^2            \Bigl(
                                                  -12\zeta_3
                                                  +\frac{41}{6}\zeta_2
                                                  +\frac{3181}{108}
                                                \Bigr)
\NN\\ &&
                              +C_A C_F           \Bigl(
                                                   16\zeta_3
                                                  -\frac{1570}{27}
                                                \Bigr)
                              +\frac{272}{3}C_F^2
                                       \biggr]
\NN\\ &&
                       +N_FT_F    \biggl[
                                       C_A\Bigl(
                                             \frac{56}{9}\zeta_3
                                            +\frac{10}{9}\zeta_2
                                            -\frac{3203}{243}
                                          \Bigr)
                                      -C_F\Bigl(
                                            \frac{20}{3}\zeta_2
                                            +\frac{1942}{81}
                                          \Bigr)
                                       \biggr]
\NN\\ &&
                       +T_F      \biggl[
                                       C_A\Bigl(
                                            -\frac{295}{18}\zeta_3
                                            +\frac{35}{9}\zeta_2
                                            +\frac{6361}{486}
                                          \Bigr)
                                      -C_F\Bigl(
                                            7\zeta_3
                                            +\frac{16}{3}\zeta_2
                                            +\frac{218}{81}
                                          \Bigr)
                                   \biggr]
\NN\\ &&
                       +C_A^2      \biggl(
                                       4{\sf B_4}
                                      -27\zeta_4
                                      +\frac{1969}{72}\zeta_3
                                      -\frac{781}{72}\zeta_2
                                      +\frac{42799}{3888}
                                   \biggr)
\NN\\ &&
                       +C_A C_F      \biggl(
                                      -8{\sf B_4}
                                      +36\zeta_4
                                      -\frac{1957}{12}\zeta_3
                                                                            +\frac{89}{3}\zeta_2
                                      +\frac{10633}{81}
                                   \biggr)
\NN\\ &&
                       +C_F^2      \biggl(
                                      \frac{95}{3}\zeta_3
                                      +\frac{274}{9}
                                   \biggr)
                                                   \Biggr\} + O(\ep)~, 
                                          \label{GluSelf3}
\end{eqnarray}
and for the quark self--energy,
\begin{eqnarray}
    \hat{\Sigma}^{(2)}\Bigl(0,\frac{\hat{m}^2}{\mu^2}\Bigr) &=&
        T_F C_F \Bigl(\frac{\hat{m}^2}{\mu^2}\Bigr)^{\ep} 
        \left[\frac{2}{\ep} 
        +\frac{5}{6} + \left(\frac{89}{72} + \frac{\zeta_2}{2}\right) 
             \ep \right] + O(\ep^2)~. \label{QuSelf2} \\
               \hat{\Sigma}^{(3)}\Bigl(0,\frac{\hat{m}^2}{\mu^2}\Bigr) &=&
        T_F C_F \Bigl(\frac{\hat{m}^2}{\mu^2}\Bigr)^{3\ep/2}
        \Biggl\{
                        \frac{8}{3\ep^3} C_A
                       +\frac{1}{\ep^2} \biggl[
                                    \frac{32}{9}T_F(N_F+2)
                                   -\frac{40}{9} C_A
                                   -\frac{8}{3}C_F
                                        \biggl]
\NN\\&&
                       +\frac{1}{\ep} \biggl[
                                       \frac{40}{27}T_F(N_F+2)
                                      +C_A\Bigl(
                                             \zeta_2
                                            +\frac{454}{27}
                                          \Bigr)
                                       -26C_F
                                        \biggl]
\NN\\&&
                       +N_FT_F\Bigl(
                                \frac{4}{3}\zeta_2
                               +\frac{674}{81}
                              \Bigr)
                       +  T_F\Bigl(
                                \frac{8}{3}\zeta_2
                               +\frac{604}{81}
                              \Bigr)
                       +  C_A\Bigl(
                                \frac{17}{3}\zeta_3
                               -\frac{5}{3}\zeta_2
                               +\frac{1879}{162}\Bigr)
\NN\\&&
                       -  C_F\Bigl(
                                8\zeta_3
                               +\zeta_2
                               +\frac{335}{18}
                              \Bigr)
        \Biggr\} 
+ O(\ep)~.  \label{QuSelf3}
\end{eqnarray}
Similarly to other massive processes
   \cite{B4cite,Bierenbaum:2009mv}
   the constant
   \begin{eqnarray}
          {\sf B_4}&=&-4\zeta_2\ln^2(2) +\frac{2}{3}\ln^4(2) 
           -\frac{13}{2}\zeta_4
                  +16 {\rm Li}_4\Bigl(\frac{1}{2}\Bigr)
                 ~\approx~  -1.762800093...~  \label{B4}
   \end{eqnarray}
emerges in Eq.~(\ref{GluSelf3}).
At $O(\alpha_s^3)$ irreducible diagrams with two different masses contribute
for the first time. For the gluonic case, we compute the respective diagrams up
to $O(\eta^3)$ using the codes {\tt Q2E}/{\tt
Exp}\cite{Harlander:1997zb,Seidensticker:1999bb},
\begin{eqnarray}
\hat{\tilde{\Pi}}^{(3)}\left(0,\hat{m}_1^2,\hat{m}_2^2,\mu^2\right)&=&
T_F^2 C_F \Biggl\{
-\frac{256}{9 \ep^2}
-\frac{1}{\ep} \left[
\frac{320}{27}
+\frac{64}{3} \ln\left(\frac{\hat{m}_1^2}{\mu^2}\right)
+\frac{64}{3} \ln\left(\frac{\hat{m}_2^2}{\mu^2}\right)
\right]
\NN\\&&
-\left(
\frac{40}{3}
+\frac{32}{35} \eta^2
+\frac{128}{315} \eta^3
\right) \ln(\eta)^2
-32 \ln\left(\frac{\hat{m}_1^2}{\mu^2}\right) \ln\left(\frac{\hat{m}_2^2}{\mu^2}\right)
\NN\\&&
+\left(
-\frac{64}{9}
+\frac{64}{15} \eta
+\frac{10208}{3675} \eta^2
+\frac{39616}{99225} \eta^3
\right) \ln(\eta)
-\frac{160}{9} \ln\left(\frac{\hat{m}_1^2}{\mu^2}\right)
\NN\\&&
-\frac{1504}{81}
-\frac{32}{3} \zeta_2
+\frac{416}{225} \eta
-\frac{1987136}{385875} \eta^2
-\frac{7026016}{31255875} \eta^3
+O(\eta^4)
\Biggr\}
\NN\\&&
+T_F^2 C_A \Biggl\{
-\frac{64}{9 \ep^3}
+\frac{1}{\ep^2} \left[
\frac{560}{27}
-\frac{16}{3} \ln\left(\frac{\hat{m}_1^2}{\mu^2}\right)
-\frac{16}{3} \ln\left(\frac{\hat{m}_2^2}{\mu^2}\right)
\right]
\NN\\&&
+\frac{1}{\ep} \left[
-\frac{148}{27}
-\frac{8}{3} \zeta_2
-4 \ln^2\left(\frac{\hat{m}_1^2}{\mu^2}\right)
-4 \ln^2\left(\frac{\hat{m}_2^2}{\mu^2}\right) 
+\frac{140}{9} \ln\left(\frac{\hat{m}_1^2}{\mu^2}\right)
\right. \NN\\&& \left.
+\frac{140}{9} \ln\left(\frac{\hat{m}_2^2}{\mu^2}\right)
\right]
+\frac{4}{9} \ln^3(\eta)
-2 \ln^3\left(\frac{\hat{m}_1^2}{\mu^2}\right)
-2 \ln^3\left(\frac{\hat{m}_2^2}{\mu^2}\right)
\NN\\&&
+\left(
\frac{65}{9}
-\frac{2}{15} \eta
-\frac{16}{21} \eta^2
-\frac{50}{189} \eta^3
\right) \ln^2(\eta)
+\frac{70}{3} \ln\left(\frac{\hat{m}_1^2}{\mu^2}\right) \ln\left(\frac{\hat{m}_2^2}{\mu^2}\right)
\NN\\&&
+\left(
\frac{167}{27}
-2 \zeta_2
+\frac{1924}{225} \eta
+\frac{6392}{2205} \eta^2
+\frac{20284}{59535} \eta^3
\right) \ln(\eta)
\NN\\&&
-\left(
\frac{74}{9}
+4 \zeta_2
\right) \ln\left(\frac{\hat{m}_1^2}{\mu^2}\right) 
-\frac{1139}{243}
+\frac{70}{9} \zeta_2
+\frac{56}{9} \zeta_3
-\frac{34144}{3375} \eta
\NN\\&&
-\frac{1292594}{231525} \eta^2
-\frac{4231264}{18753525} \eta^3
+O(\eta^4)
\Biggr\}
+O(\ep)
\end{eqnarray}
The quarkonic self--energy contributions have been computed analytically in $\eta$,
\begin{eqnarray}
 \hat{\tilde{\Sigma}}^{(3)}(0,\hat{m}_1^2,\hat{m}_2^2,\mu^2) &=&
 T_F^2 C_F \Bigl(\frac{\hat{m}_1 \hat{m}_2}{\mu^2}\Bigr)^{\frac{3}{2} \ep} \left[
 \frac{128}{9 \ep^2}
 +\frac{160}{27 \ep}
 +\frac{4}{3} \ln^2(\eta)
 +\frac{16}{3} \zeta_2
\right. \NN\\&& \left. 
 +\frac{1208}{81}
 +O(\ep)\right]~.
\end{eqnarray}
\end{subsubsection}

\begin{subsubsection}{The reducible operator matrix elements}

\vspace*{1mm}
\noindent
As in Eqs.~(\ref{gSelf2m}--\ref{qSelf2m}) we define the two--mass OMEs at
one--loop order and the irreducible OMEs at $O(\alpha_s^2)$ by
    \begin{eqnarray}
        \Ahathat^{(1)}_{ij}\Bigl(\frac{\hat{m}_1^2}{\mu^2},\frac{\hat{m}_2^2}{\mu^2}\Bigl)&=&
        \Ahathat^{(1)}_{ij}\left(\frac{\hat{m}_1^2}{\mu^2}\right)
        +\Ahathat^{(1)}_{ij}\left(\frac{\hat{m}_2^2}{\mu^2}\right)~,\label{Aij12m}
        \\
        \Ahathat^{(2),\rm{irr}}_{ij}\Bigl(\frac{\hat{m}_1^2}{\mu^2},\frac{\hat{m}_2^2}{\mu^2}\Bigl)&=&
        \Ahathat^{(2),\rm{irr}}_{ij}\left(\frac{\hat{m}_1^2}{\mu^2}\right)
        +\Ahathat^{(2),\rm{irr}}_{ij}\left(\frac{\hat{m}_2^2}{\mu^2}\right)~,
        \label{Aij2irr2m}
    \end{eqnarray}
    where the $A_{ij}$'s with one argument denote the usual single--mass OMEs.
    Using the definitions (\ref{gSelf2m}--\ref{qSelf2m}) and (\ref{Aij12m}--\ref{Aij2irr2m}) we
    compose the reducible massive operator matrix elements at $O(\alpha_s^2)$ by
    \begin{eqnarray}
     \Ahathat_{qq}^{(2),\rm{NS}}\Bigl(\frac{\hat{m}_1^2}{\mu^2},\frac{\hat{m}_2^2}{\mu^2}\Bigl)&=&
     \Ahathat_{qq}^{(2),\rm{NS},\rm{irr}}\Bigl(\frac{\hat{m}_1^2}{\mu^2},\frac{\hat{m}_2^2}{\mu^2}\Bigl)
     -\hat{\Sigma}^{(2)}\left(0,\hat{m}_1^2,\hat{m}_2^2,\mu^2\right)~, {\label{Aqq2NSred}}
\\
\Ahathat_{Qg}^{(2)}\Bigl(\frac{\hat{m}_1^2}{\mu^2},\frac{\hat{m}_2^2}{\mu^2}\Bigl)&=&
\Ahathat_{Qg}^{(2),\rm{irr}}\Bigl(\frac{\hat{m}_1^2}{\mu^2},\frac{\hat{m}_2^2}{\mu^2}\Bigl)
-\Ahathat_{Qg}^{(1)}\Bigl(\frac{\hat{m}_1^2}{\mu^2},\frac{\hat{m}_2^2}{\mu^2}\Bigl)
\hat{\Pi}^{(1)}\left(0,\hat{m}_1^2,\hat{m}_2^2,\mu^2\right)~, {\label{AQg2red}}
\\
 \Ahathat_{gg}^{(2)}\Bigl(\frac{\hat{m}_1^2}{\mu^2},\frac{\hat{m}_2^2}{\mu^2}\Bigl)&=&
 \Ahathat_{gg}^{(2),\rm{irr}}\Bigl(\frac{\hat{m}_1^2}{\mu^2},\frac{\hat{m}_2^2}{\mu^2}\Bigl)
 -\hat{\Pi}^{(2)}\left(0,\hat{m}_1^2,\hat{m}_2^2,\mu^2\right)
\NN\\&&
 -\Ahathat_{gg}^{(1)}\Bigl(\frac{\hat{m}_1^2}{\mu^2},\frac{\hat{m}_2^2}{\mu^2}\Bigl)
  \hat{\Pi}^{(1)}\left(0,\hat{m}_1^2,\hat{m}_2^2,\mu^2\right)~, {\label{Agg2red}}
     \end{eqnarray}
and at $O(\alpha_s^3)$ by
     \begin{eqnarray}
     \Ahathat_{qq}^{(3),\rm{NS}}\Bigl(\frac{\hat{m}_1^2}{\mu^2},\frac{\hat{m}_2^2}{\mu^2}\Bigl)&=&
     \Ahathat_{qq}^{(3),\rm{NS},\rm{irr}}\Bigl(\frac{\hat{m}_1^2}{\mu^2},\frac{\hat{m}_2^2}{\mu^2}\Bigl)
     -\hat{\Sigma}^{(3)}\left(0,\hat{m}_1^2,\hat{m}_2^2,\mu^2\right)
\\
 \Ahathat_{Qg}^{(3)}\Bigl(\frac{\hat{m}_1^2}{\mu^2},\frac{\hat{m}_2^2}{\mu^2}\Bigl)&=&
 \Ahathat_{Qg}^{(3),\rm{irr}}\Bigl(\frac{\hat{m}_1^2}{\mu^2},\frac{\hat{m}_2^2}{\mu^2}\Bigl)
 -
 \Ahathat_{Qg}^{(2)}\Bigl(\frac{\hat{m}_1^2}{\mu^2},\frac{\hat{m}_2^2}{\mu^2}\Bigl)
 \hat{\Pi}^{(1)}\left(0,\hat{m}_1^2,\hat{m}_2^2,\mu^2\right)
 \NN\\&&
  -
 \Ahathat_{Qg}^{(1)}\Bigl(\frac{\hat{m}_1^2}{\mu^2},\frac{\hat{m}_2^2}{\mu^2}\Bigl)
 \hat{\Pi}^{(2)}\left(0,\hat{m}_1^2,\hat{m}_2^2,\mu^2\right)
 \\
 \Ahathat_{gg}^{(3)}\Bigl(\frac{\hat{m}_1^2}{\mu^2},\frac{\hat{m}_2^2}{\mu^2}\Bigl)&=&
 \Ahathat_{gg}^{(3),\rm{irr}}\Bigl(\frac{\hat{m}_1^2}{\mu^2},\frac{\hat{m}_2^2}{\mu^2}\Bigl)
 -\Ahathat_{gg}^{(2)}\Bigl(\frac{\hat{m}_1^2}{\mu^2},\frac{\hat{m}_2^2}{\mu^2}\Bigl) 
  \hat{\Pi}^{(1)}\left(0,\hat{m}_1^2,\hat{m}_2^2,\mu^2\right)
 \NN\\&&
 -\Ahathat_{gg}^{(1)}\Bigl(\frac{\hat{m}_1^2}{\mu^2},\frac{\hat{m}_2^2}{\mu^2}\Bigl) 
  \hat{\Pi}^{(2)}\left(0,\hat{m}_1^2,\hat{m}_2^2,\mu^2\right)
 -\hat{\Pi}^{(3)}\left(0,\hat{m}_1^2,\hat{m}_2^2,\mu^2\right)
\,.
\end{eqnarray}
We can subtract the single--mass contributions to these equations using Eq. (\ref{AhathatDecomp}), keeping
only the genuine two--mass contributions. At three loops we obtain

\begin{eqnarray}
\Athathat_{qq}^{(3),\rm{NS}}\Bigl(\frac{\hat{m}_1^2}{\mu^2},\frac{\hat{m}_2^2}{\mu^2}\Bigl)&=&
\Athathat_{qq}^{(3),\rm{NS},\rm{irr}}\Bigl(\frac{\hat{m}_1^2}{\mu^2},\frac{\hat{m}_2^2}{\mu^2}\Bigl)
-\hat{\tilde{\Sigma}}^{(3)}\left(0,\hat{m}_1^2,\hat{m}_2^2,\mu^2\right)
\\
\Athathat_{Qg}^{(3)}\Bigl(\frac{\hat{m}_1^2}{\mu^2},\frac{\hat{m}_2^2}{\mu^2}\Bigl)&=&
\Athathat_{Qg}^{(3),\rm{irr}}\Bigl(\frac{\hat{m}_1^2}{\mu^2},\frac{\hat{m}_2^2}{\mu^2}\Bigl)
 \NN\\&&
+\Ahathat_{Qg}^{(1)}\left(\frac{\hat{m}_1^2}{\mu^2}\right)
 \left[2 \hat{\Pi}^{(1)}\Bigl(0,\frac{\hat{m}_1^2}{\mu^2}\Bigr)+\hat{\Pi}^{(1)}\Bigl(0,\frac{\hat{m}_2^2}{\mu^2}\Bigr)\right]
 \hat{\Pi}^{(1)}\Bigl(0,\frac{\hat{m}_2^2}{\mu^2}\Bigr)
 \NN\\&&
+\Ahathat_{Qg}^{(1)}\left(\frac{\hat{m}_2^2}{\mu^2}\right)
 \left[2 \hat{\Pi}^{(1)}\Bigl(0,\frac{\hat{m}_2^2}{\mu^2}\Bigr)+\hat{\Pi}^{(1)}\Bigl(0,\frac{\hat{m}_1^2}{\mu^2}\Bigr)\right]
 \hat{\Pi}^{(1)}\Bigl(0,\frac{\hat{m}_1^2}{\mu^2}\Bigr)
 \NN\\&&
-\Ahathat_{Qg}^{(2)}\left(\frac{\hat{m}_1^2}{\mu^2}\right)
 \hat{\Pi}^{(1)}\Bigl(0,\frac{\hat{m}_2^2}{\mu^2}\Bigr)
-\Ahathat_{Qg}^{(2)}\left(\frac{\hat{m}_2^2}{\mu^2}\right)
 \hat{\Pi}^{(1)}\Bigl(0,\frac{\hat{m}_1^2}{\mu^2}\Bigr)
 \NN\\&&
-\Ahathat_{Qg}^{(1)}\left(\frac{\hat{m}_1^2}{\mu^2}\right)
 \hat{\Pi}^{(2)}\Bigl(0,\frac{\hat{m}_2^2}{\mu^2}\Bigr)
-\Ahathat_{Qg}^{(1)}\left(\frac{\hat{m}_2^2}{\mu^2}\right)
 \hat{\Pi}^{(2)}\Bigl(0,\frac{\hat{m}_1^2}{\mu^2}\Bigr)
 \\
\Athathat_{gg}^{(3)}\Bigl(\frac{\hat{m}_1^2}{\mu^2},\frac{\hat{m}_2^2}{\mu^2}\Bigl)&=&
\Athathat_{gg}^{(3),\rm{irr}}\Bigl(\frac{\hat{m}_1^2}{\mu^2},\frac{\hat{m}_2^2}{\mu^2}\Bigl)
-\hat{\tilde{\Pi}}^{(3)}\left(0,\hat{m}_1^2,\hat{m}_2^2,\mu^2\right)
 \NN\\&&
 -\Ahathat_{gg}^{(2),\rm{irr}}\left(\frac{\hat{m}_1^2}{\mu^2}\right) 
  \hat{\Pi}^{(1)}\Bigl(0,\frac{\hat{m}_2^2}{\mu^2}\Bigr)
 -\Ahathat_{gg}^{(2),\rm{irr}}\left(\frac{\hat{m}_2^2}{\mu^2}\right) 
  \hat{\Pi}^{(1)}\Bigl(0,\frac{\hat{m}_1^2}{\mu^2}\Bigr)
 \NN\\&&
 -2 \Ahathat_{gg}^{(1)}\left(\frac{\hat{m}_1^2}{\mu^2}\right) 
  \hat{\Pi}^{(2)}\Bigl(0,\frac{\hat{m}_2^2}{\mu^2}\Bigr)
 -2 \Ahathat_{gg}^{(1)}\left(\frac{\hat{m}_2^2}{\mu^2}\right) 
  \hat{\Pi}^{(2)}\Bigl(0,\frac{\hat{m}_1^2}{\mu^2}\Bigr)
 \NN\\&&
 +\Ahathat_{gg}^{(1)}\left(\frac{\hat{m}_1^2}{\mu^2}\right) 
  \left[2 \hat{\Pi}^{(1)}\Bigl(0,\frac{\hat{m}_1^2}{\mu^2}\Bigr)+ \hat{\Pi}^{(1)}\Bigl(0,\frac{\hat{m}_2^2}{\mu^2}\Bigr)\right]
  \hat{\Pi}^{(1)}\Bigl(0,\frac{\hat{m}_2^2}{\mu^2}\Bigr)
 \NN\\&&
 +\Ahathat_{gg}^{(1)}\left(\frac{\hat{m}_2^2}{\mu^2}\right) 
  \left[2 \hat{\Pi}^{(1)}\Bigl(0,\frac{\hat{m}_2^2}{\mu^2}\Bigr)+\hat{\Pi}^{(1)}\Bigl(0,\frac{\hat{m}_1^2}{\mu^2}\Bigr)\right]
  \hat{\Pi}^{(1)}\Bigl(0,\frac{\hat{m}_1^2}{\mu^2}\Bigr)
\,.
\end{eqnarray}
\end{subsubsection}
\end{subsection}
\begin{subsection}{\bf\boldmath The General Structure of the Massive Operator Matrix Elements}
\label{SubSec-RENPred}

\vspace*{1mm}
\noindent
In the following, we present the structure of the different unrenormalized and renormalized OMEs for the genuine two-mass 
contributions. 

In the case of only one heavy quark flavor with mass $m$ \cite{Bierenbaum:2009mv}, the mass dependence of the 
unrenormalized massive operator matrix element at order $\alpha_s^l$ is given by
\begin{eqnarray}
\Ahathat_{ij}^{(l)}\Bigl(\frac{\hat{m}^2}{\mu^2},\ep,N\Bigr)&=&
\left(\frac{\hat{m}^2}{\mu^2}\right)^{\frac{l \ep}{2}} \Ahathat_{ij}^{(l)}\left(\ep,N\right)~.
\label{1MassFactor}
\end{eqnarray}
Here the OME $\Ahathat_{ij}^{(l)}\Bigl(\ep,N\Bigr)$ does not depend on the mass explicitely anymore. It exhibits poles
in the dimensional parameter $\ep$ up to $\ep^{-l}$
\begin{eqnarray}
       \Ahathat_{ij}^{(l)}\left(\ep,N\right) &=&
        \sum_{k=0}^{\infty}
        \frac{a^{(l,k)}_{ij}}{\ep^{l-k}}
        ~. \label{GenStructure}
\end{eqnarray}
We adopt the notation of Ref. \cite{Bierenbaum:2009mv} and denote
\begin{eqnarray}
       a^{(l,l)}\equiv a^{(l)}~
       , \quad a^{(l,l+1)}\equiv \overline{a}^{(l)}.
       \label{deflittlea}
\end{eqnarray}
The unrenormalized operator matrix elements with two massive fermion flavors
with masses $m_1 \neq m_2$ are split into the respective single-mass
contributions (\ref{1MassFactor}, \ref{GenStructure}) and a part
$\Athathat_{ij}^{(l)}\left(\frac{\hat{m}_1^2}{\mu^2},\frac{\hat{m}_2^2}{\mu^2},\ep,N\right)$
depending on both masses
\begin{eqnarray}
&& \hspace*{-10mm}
 \Ahathat_{ij}^{(l)}\Bigl(\frac{\hat{m}_1^2}{\mu^2},\frac{\hat{m}_2^2}{\mu^2},\ep,N\Bigr) =
 \left[\left(\frac{\hat{m}_1^2}{\mu^2}\right)^{\frac{l \ep}{2}}
 +\left(\frac{\hat{m}_2^2}{\mu^2}\right)^{\frac{l \ep}{2}} \right] \Ahathat_{ij}^{(l)}\Bigl(\ep,N\Bigr)
 +\Athathat_{ij}^{(l)}\Bigl(\frac{\hat{m}_1^2}{\mu^2},\frac{\hat{m}_2^2}{\mu^2},\ep,N\Bigr)~.
\end{eqnarray}

\noindent
The two--flavor contributions 
$\Athathat_{ij}^{(l)}\Bigl(\frac{\hat{m}_1^2}{\mu^2},\frac{\hat{m}_2^2}{\mu^2},\ep,N\Bigr)$,
   $m_1 \neq m_2$, to the massive OMEs do not obey a factorization relation 
 as (\ref{1MassFactor}) and the mass 
   dependence is pulled into the coefficients of the Laurent expansion
   
   \begin{eqnarray}
       \Athathat_{ij}^{(l)}\Bigl(\frac{\hat{m}_1^2}{\mu^2},\frac{\hat{m}_2^2}{\mu^2},\ep,N\Bigr) &=&
        \sum_{k=0}^{\infty}
        \frac{\tilde{a}^{(l,k)}_{ij}\Bigl(\frac{\hat{m}_1^2}{\mu^2},\frac{\hat{m}_2^2}{\mu^2}\Bigr)}{\ep^{l-k}}
        ~. \label{GenStructure2m}
   \end{eqnarray}
   Analogously to (\ref{deflittlea}) we define
   
      \begin{eqnarray}
       \tilde{a}^{(l,l)}\Bigl(\frac{\hat{m}_1^2}{\mu^2},\frac{\hat{m}_2^2}{\mu^2}\Bigr)\equiv 
       \tilde{a}^{(l)}\Bigl(\frac{\hat{m}_1^2}{\mu^2},\frac{\hat{m}_2^2}{\mu^2}\Bigr).
       \label{deflittlea2m}
      \end{eqnarray}
In the following, $a^{(l,k)},~a^{(l)},~\overline{a}^{(l)}$ without argument will
denote the single mass--quantities corresponding to the definitions in
(\ref{GenStructure}, \ref{deflittlea}), while
$\tilde{a}^{(l,l)}\Bigl(\frac{\hat{m}_1^2}{\mu^2},\frac{\hat{m}_2^2}{\mu^2}\Bigr)$ refers
to the two-mass contribution.  From Eq.~(\ref{GenRen3}) it is obvious that the
renormalization of the $3$--loop OMEs requires the knowledge of the one--loop
OMEs $A_{ij}^{(1)}(m_1,m_2)$ up to $O(\ep^2)$ and the two--loop OMEs
$A_{ij}^{(2)}(m_1,m_2)$ up to $O(\ep)$. Up to $O(\alpha_s^2)$, these two mass
quantities can be traced back to the corresponding single--mass quantities by
Eqs.~(\ref{Aij12m}--\ref{Aij2irr2m}) and (\ref{Aqq2NSred}--\ref{Agg2red}). 

It is technically advantageous to perform the renormalization on the complete
two--flavor OMEs
$\Ahathat_{ij}^{(l)}\Bigl(\frac{\hat{m}_1^2}{\mu^2},\frac{\hat{m}_2^2}{\mu^2},\ep,N\Bigr)$.
For brevity we will present the renormalization formulas for the two-mass
contribution
$\Athathat_{ij}^{(l)}\Bigl(\frac{\hat{m}_1^2}{\mu^2},\frac{\hat{m}_2^2}{\mu^2},\ep,N\Bigr)$
only, which is obtained after subtracting the respective single-mass
contributions \cite{Bierenbaum:2009mv,Klein:2009ig}.

The analytic expressions for the respective single mass contributions and renormalization constants
to two-loop order, which appear in subsequent relations, have been given in 
Refs.~\cite{Bierenbaum:2007qe,Bierenbaum:2008yu,Bierenbaum:2009mv,Moch:2004pa,Vogt:2004mw}
and references therein.
%
%
%
\subsubsection{\bf\boldmath $A_{qq,Q}^{\sf NS}$}
  \label{Sec-NS}
The lowest non--trivial flavor non-singlet (NS) contribution is of $O(a_s^2)$,
\begin{eqnarray}
   A_{qq,Q}^{\sf NS}&=&1
                       +a_s^2 A_{qq,Q}^{(2), {\sf NS}}
                       +a_s^3 A_{qq,Q}^{(3), {\sf NS}}
                       +O(a_s^4)~. \label{NSpert}
\end{eqnarray}
Starting from $O(a_s^3)$ it exhibits a non-trivial two--mass contribution
\begin{eqnarray}
   \tilde{A}_{qq,Q}^{\sf NS}&=&1
                       +a_s^3 \tilde{A}_{qq,Q}^{(3), {\sf NS}}
                       +O(a_s^4)~. \label{NSpert2m}
\end{eqnarray}
The renormalized two-mass OME in the ${\sf MOM}$--scheme is obtained from the bare quantities combining 
Eqs.~(\ref{macoren},~\ref{GenRen3}). It is given by
\begin{eqnarray}
A_{qq,Q}^{(3), \sf NS, \MOM}\left(N_F +2\right)&=&
                 \hat{A}_{qq,Q}^{(3), {\sf NS},\MOM}
                    +Z^{-1,(3), {\sf NS}}_{qq}(N_F+N_H)
                    -Z^{-1,(3), {\sf NS}}_{qq}(N_F)
\NN\\ &&
                    +Z^{-1,(1), {\sf NS}}_{qq}(N_F+N_H)
                     \hat{A}_{qq,Q}^{(2), {\sf NS},\MOM}
                    +\Bigl[ \hat{A}_{qq,Q}^{(2), {\sf NS},\MOM}
\NN\\ &&
                           +Z^{-1,(2), {\sf NS}}_{qq}(N_F+N_H)
                           -Z^{-1,(2), {\sf NS}}_{qq}(N_F)
                     \Bigr]\Gamma^{-1,(1)}_{qq}(N_F)
                ~. \label{3LNSRen1}
\end{eqnarray}
After a finite renormalization to the $\MS$--scheme and the subtraction of the 
single-mass 
contributions one obtains the pole-structure of the two--flavor piece by 
\begin{eqnarray}
 \Athathat_{qq,Q}^{(3), \rm{NS}} &=&
 -\frac{16}{3 \ep^3} \gamma_{qq}^{(0)} \beta_{0,Q}^2
 +\frac{1}{\ep^2}
\Biggl[
-\frac{8}{3} \beta_{0,Q} \hat{\gamma}_{qq}^{\rm{NS},(1)}
-4 \gamma_{qq}^{(0)} \beta_{0,Q}^2 \left(L_1+L_2\right)
\Biggr] 
\NN\\&&
+\frac{1}{\ep} \Biggl[
  -2 \beta_{0,Q} \hat{\gamma}_{qq}^{\rm{NS},(1)} \left(L_2+L_1\right)
  -2 \gamma_{qq}^{(0)} \beta_{0,Q}^2 \left(L_1^2+ L_2 L_1+L_2^2\right)
  \NN\\&&
  -8 a_{qq}^{\rm{NS},(2)} \beta_{0,Q}
  +\frac{2}{3} \hat{\tilde{\gamma}}_{qq}^{(2),\rm{NS}}
\Biggr]
+\tilde{a}_{qq,Q}^{(3),\rm{NS}}\left(m_1^2,m_2^2,\mu^2\right)~,
{\label{AthhNS3}}
   \end{eqnarray}
with 
\begin{equation}
	L_1=\ln\left(\frac{m_1^2}{\mu^2}\right)~,\,\,\,L_2=\ln\left(\frac{m_2^2}{\mu^2}\right)~.
\end{equation}
The renormalized expression in the $\MS$--scheme is given by
\begin{eqnarray}
\tilde{A}_{qq,Q}^{(3), \MS, \rm{NS}} &=&
 \gamma_{qq}^{(0)} \beta_{0,Q}^2 \left(\frac{2}{3} L_1^3+\frac{2}{3} L_2^3+\frac{1}{2} L_2^2 L_1+ \frac{1}{2} L_1^2 L_2 \right)
+\beta_{0,Q} \hat{\gamma}_{qq}^{\rm{NS},(1)} \left(L_1^2+L_2^2\right)
\NN\\&&
+\Biggl\{
4 a_{qq}^{\rm{NS},(2)} \beta_{0,Q}
+\frac{1}{2} \beta_{0,Q}^2 \gamma_{qq}^{(0)} \zeta_2
\Biggr\} \left(L_1+L_2\right)
+8 \overline{a}_{qq}^{\rm{NS},(2)} \beta_{0,Q}
\NN\\&&
+\tilde{a}_{qq,Q}^{(3),\rm{NS}}\left(m_1^2,m_2^2,\mu^2\right)~.
\end{eqnarray}
For $N=1$ the OME vanishes due to fermion number conservation; this applies both for the anomalous 
dimensions $\gamma_{qq}^{(l)}$ and the expansion coefficients of the OMEs $a_{qq}^{\rm NS,(2)}, 
\overline{a}_{qq}^{\rm NS,(2)}$ and $\tilde{a}_{qq,Q}^{\rm (3),NS}$. 
%
%
%
 \subsubsection{\bf\boldmath $A_{Qq}^{\sf PS}$}
  \label{SubSec-PS}
Depending on whether the operator couples to a heavy or a light fermion, there
are two pure--singlet contributions\cite{Bierenbaum:2009mv}
\begin{eqnarray}
   A_{Qq}^{\sf PS}&=&
                       a_s^2A_{Qq}^{(2), {\sf PS}}
                       +a_s^3A_{Qq}^{(3), {\sf PS}}
                       +O(a_s^4)~, \label{PSQqpert}\\
   A_{qq,Q}^{\sf PS}&=&
                        a_s^3A_{qq,Q}^{(3), {\sf PS}}
                       +O(a_s^4)~. \label{PSqqQpert}
\end{eqnarray}
Up to $O(a_s^3)$ only the OME $A_{Qq}$ contains a generic two--mass contribution, since $A_{qq,Q}^{\sf PS}$ emerges only at 
$O(a_s^3)$ and contains one internal massless fermion line. One has
\begin{eqnarray}
   \tilde{A}_{Qq}^{\sf PS}&=&
                       a_s^3 \tilde{A}_{Qq}^{(3), {\sf PS}}
                       +O(a_s^4)~. \label{PSQqpert2m}
\end{eqnarray}
The combined renormalization relation at third order is given by
\begin{eqnarray}
   &&A_{Qq}^{(3), \sf PS, \MOM}+
     A_{qq,Q}^{(3), \sf PS, \MOM}=
                     \hat{A}_{Qq}^{(3), {\sf PS}, \MOM} 
                    +\hat{A}_{qq,Q}^{(3), {\sf PS}, \MOM}
                    +Z^{-1,(3), {\sf PS}}_{qq}(N_F+N_H)
\NN\\ && \phantom{abc}
                    -Z^{-1,(3), {\sf PS}}_{qq}(N_F)
                    +Z^{-1,(1)}_{qq}(N_F+N_H)\hat{A}_{Qq}^{(2), {\sf PS}, \MOM}
                    +Z^{-1,(1)}_{qg}(N_F+N_H)\hat{A}_{gq,Q}^{(2), \MOM}
\NN\\ && \phantom{abc}
                    +\Bigl[
                            \hat{A}_{Qg}^{(1), \MOM}
                           +Z^{-1,(1)}_{qg}(N_F+N_H)
                           -Z^{-1,(1)}_{qg}(N_F)
                     \Bigr]\Gamma^{-1,(2)}_{gq}(N_F)
                    +\Bigl[ \hat{A}_{Qq}^{(2), {\sf PS}, \MOM}
\NN\\ && \phantom{abc}
                           +Z^{-1,(2), {\sf PS}}_{qq}(N_F+N_H)
                           -Z^{-1,(2), {\sf PS}}_{qq}(N_F)
                     \Bigr]\Gamma^{-1,(1)}_{qq}(N_F)
                    +\Bigl[ \hat{A}_{Qg}^{(2), \MOM}
                           +Z^{-1,(2)}_{qg}(N_F+N_H)
\NN\\ && \phantom{abc}
                           -Z^{-1,(2)}_{qg}(N_F)
                           +Z^{-1,(1)}_{qq}(N_F+N_H)A_{Qg}^{(1), \MOM}
                           +Z^{-1,(1)}_{qg}(N_F+N_H)A_{gg,Q}^{(1), \MOM}
                     \Bigr]\Gamma^{-1,(1)}_{gq}(N_F)~.\NN\\ 
                  \label{AQqq3PSRen}
\end{eqnarray}
This yields the generic pole structure for the {\sf PS} two--mass contribution
\begin{eqnarray}
\Athathat_{Qq}^{(3),\rm{PS}} &=&
\frac{16}{3 \ep^3} \gamma_{gq}^{(0)} \hat{\gamma}_{qg}^{(0)} \beta_{0,Q}
+\frac{1}{\ep^2}\Biggl[
4 \gamma_{gq}^{(0)} \hat{\gamma}_{qg}^{(0)} \beta_{0,Q} \left(L_1+L_2\right)
+\frac{2}{3} \hat{\gamma}_{qg}^{(0)} \hat{\gamma}_{gq}^{(1)}
-\frac{8}{3} \beta_{0,Q} \hat{\gamma}_{qq}^{\rm{PS},(1)}
\Biggr] 
\NN\\&&
+\frac{1}{\ep}
\Biggl[
2 \gamma_{gq}^{(0)} \hat{\gamma}_{qg}^{(0)} \beta_{0,Q} \left(L_1^2+L_1 L_2+L_2^2\right)
+\Biggl\{\frac{1}{2} \hat{\gamma}_{qg}^{(0)} \hat{\gamma}_{gq}^{(1)}
-2 \beta_{0,Q} \hat{\gamma}_{qq}^{\rm{PS},(1)}
\Biggr\} \left(L_2+L_1\right)
\NN\\&&
+\frac{2}{3} \hat{\tilde{\gamma}}_{qq}^{(2),\rm{PS}}
-8 a_{Qq}^{(2),\rm{PS}} \beta_{0,Q}
+2 \hat{\gamma}_{qg}^{(0)} a_{gq}^{(2)}
\Biggr] 
+\tilde{a}_{Qq}^{(3),\rm{PS}}\left(m_1^2,m_2^2,\mu^2\right)
                   \label{Ahhhqq3PSQ}~.
   \end{eqnarray}
In the ${\MS}$--scheme one obtains the renormalized expression by
\begin{eqnarray}
\tilde{A}_{Qq}^{(3), \MS, \rm{PS}} &=&
-\gamma_{gq}^{(0)} \hat{\gamma}_{qg}^{(0)} \beta_{0,Q} 
\left(\frac{1}{2} L_2^2 L_1+ \frac{1}{2} L_1^2 L_2+ \frac{2}{3} L_1^3+\frac{2}{3} L_2^3\right)
\NN\\&&
+\Biggl\{
-\frac{1}{4} \hat{\gamma}_{qg}^{(0)} \hat{\gamma}_{gq}^{(1)}
+\beta_{0,Q} \hat{\gamma}_{qq}^{\rm{PS},(1)}
\Biggr\}
\left(L_2^2+L_1^2\right)
\NN\\&&
+\Biggl\{
4 a_{Qq}^{(2), \rm{PS}} \beta_{0,Q}
-\hat{\gamma}_{qg}^{(0)} a_{gq}^{(2)}
-\frac{1}{2} \beta_{0,Q} \zeta_2 \gamma_{gq}^{(0)} \hat{\gamma}_{qg}^{(0)}
\Biggr\} \left(L_1+L_2\right)
\NN\\&&
+8 \overline{a}_{Qq}^{(2), \rm{PS}} \beta_{0,Q}
-2 \hat{\gamma}_{qg}^{(0)} \overline{a}_{gq}^{(2)}
+\tilde{a}_{Qq}^{(3), \rm{PS}}\left(m_1^2,m_2^2,\mu^2\right)~.
\label{Aqq3PSQMSren}
\end{eqnarray}
%
%
%
\subsubsection{\bf\boldmath $A_{Qg}$}
\label{SubSec-AQqg}
Like in the  ${\sf PS}$ case, there are two different contributions to the OME
$A_{Qg}$
  \begin{eqnarray}
   A_{Qg}&=&
                        a_s  A_{Qg}^{(1)}
                       +a_s^2A_{Qg}^{(2)}
                       +a_s^3A_{Qg}^{(3)}
                       +O(a_s^4)~. \label{AQgpert}\\
   A_{qg,Q}&=&
                        a_s^3A_{qg,Q}^{(3)}
                       +O(a_s^4)~. \label{AqgQpert}
  \end{eqnarray}
  Of these OMEs only $A_{Qg}$ contains two--flavor contributions starting from $O(a_s^2)$
  \begin{eqnarray}
   \tilde{A}_{Qg}&=&                      
                       a_s^2 \tilde{A}_{Qg}^{(2)}
                       +a_s^3 \tilde{A}_{Qg}^{(3)}
                       +O(a_s^4)~. \label{AQgpert2m}
\end{eqnarray}

\noindent
In Eq.~(\ref{AQgpert2m}) the $O(a_s^2)$ contribution consists of one--particle reducible diagrams only, see Eq.~(\ref{AQg2red}). As 
a 
consequence the flavor dependence factorizes in the $O(a_s^2)$ terms.
  
The renormalized {\sf MOM}--scheme two--loop contribution is obtained by
\begin{eqnarray} 
    A_{Qg}^{(2), \MOM}&=&
                    \hat{A}_{Qg}^{(2), \MOM}
                   +Z^{-1,(2)}_{qg}(N_F+N_H)
                   -Z^{-1,(2)}_{qg}(N_F)
                   +Z^{-1,(1)}_{qg}(N_F+N_H)\hat{A}_{gg,Q}^{(1), \MOM}
 \NN\\ &&
                   +Z^{-1,(1)}_{qq}(N_F+N_H)\hat{A}_{Qg}^{(1), \MOM}
                   +\Bigl[ \hat{A}_{Qg}^{(1), \MOM}
                          +Z_{qg}^{-1,(1)}(N_F+N_H)
\NN\\ &&
                          -Z_{qg}^{-1,(1)}(N_F)
                    \Bigr]\Gamma^{-1,(1)}_{gg}(N_F)~. \label{RenAQg2MOM}
\end{eqnarray}
The unrenormalized terms are given by 
\begin{eqnarray}
   \Athathat_{Qg}^{(2)}&=& - \frac{4}{\ep^2} \beta_{0,Q} \hat{\gamma}_{qg}^{(0)}
			    -\frac{2} {\ep} \beta_{0,Q} \hat{\gamma}_{qg}^{(0)}
			     \left(
L_1+L_2
\right)
			     +\tilde{a}_{Qg}^{(2)}\left(\frac{m_1^2}{\mu^2},\frac{m_2^2}{\mu^2}\right)
			     \NN\\&&
			     +\ep \overline{\tilde{a}}_{Qg}^{(2)}\left(\frac{m_1^2}{\mu^2},\frac{m_2^2}{\mu^2}\right)
                         ~.\label{AhhhQg2}
  \end{eqnarray}
The coefficients $\tilde{a}_{Qg}^{(2)}\left(\frac{m_1^2}{\mu^2},\frac{m_2^2}{\mu^2}\right)$ and $\overline{\tilde{a}}_{Qg}^{(2)}\left(\frac{m_1^2}{\mu^2},\frac{m_2^2}{\mu^2}\right)$ 
are read off from Eq.~(\ref{AQg2red}) 
\begin{eqnarray}
\label{eq:aqg2a}
a_{Qg}^{(2)} &=&
-\beta_{0,Q} \hat{\gamma}_{qg}^{(0)} 
\Biggl\{
\frac{1}{2} \left(L_1+L_2\right)^2
+\zeta_2
\Biggr\}~,
\\
\label{eq:aqg2b}
\overline{a}_{Qg}^{(2)} &=&
\beta_{0,Q} \hat{\gamma}_{qg}^{(0)}
\Biggl\{
-\frac{1}{12} \left(L_1+L_2\right)^3
-\frac{1}{2}  \zeta_2 \left(L_1+L_2\right)
-\frac{1}{3} \zeta_3 
\Biggr\}~.
\end{eqnarray}
The renormalized expression at 2 loops then reads
\begin{eqnarray}
    \tilde{A}_{Qg}^{(2), \MS}&=& \frac{1}{2} \beta_{0,Q} \hat{\gamma}_{qg}^{(0)} 
    \left(L_1^2+L_2^2\right)
    +\zeta_2 \beta_{0,Q} \hat{\gamma}_{qg}^{(0)} +\tilde{a}_{Qg}^{(2)}~.
\end{eqnarray}
The renormalized 3--loop OMEs in the $\MOM$--scheme are obtained from the charge-- and mass--renormalized OMEs by
\begin{eqnarray}
    && A_{Qg}^{(3), \MOM}+A_{qg,Q}^{(3), \MOM}
            =
                     \hat{A}_{Qg}^{(3), \MOM}
                    +\hat{A}_{qg,Q}^{(3), \MOM}
                    +Z^{-1,(3)}_{qg}(N_F+N_H)
                    -Z^{-1,(3)}_{qg}(N_F)
\NN\\ && \phantom{abc} 
                    +Z^{-1,(2)}_{qg}(N_F+N_H)\hat{A}_{gg,Q}^{(1), \MOM}
                    +Z^{-1,(1)}_{qg}(N_F+N_H)\hat{A}_{gg,Q}^{(2), \MOM}
                    +Z^{-1,(2)}_{qq}(N_F+N_H)\hat{A}_{Qg}^{(1), \MOM}
\NN\\ && \phantom{abc} 
                    +Z^{-1,(1)}_{qq}(N_F+N_H)\hat{A}_{Qg}^{(2), \MOM}
                    +\Bigl[
                            \hat{A}_{Qg}^{(1), \MOM}
                           +Z^{-1,(1)}_{qg}(N_F+N_H)
\NN\\ && \phantom{abc} 
                           -Z^{-1,(1)}_{qg}(N_F)
                     \Bigr]\Gamma^{-1,(2)}_{gg}(N_F)
                    +\Bigl[ \hat{A}_{Qg}^{(2), \MOM}
                           +Z^{-1,(2)}_{qg}(N_F+N_H) 
                           -Z^{-1,(2)}_{qg}(N_F)
\NN\\ && \phantom{abc} 
                           +Z^{-1,(1)}_{qq}(N_F+N_H)A_{Qg}^{(1), \MOM}
                           +Z^{-1,(1)}_{qg}(N_F+N_H)A_{gg,Q}^{(1), \MOM}
                     \Bigr]\Gamma^{-1,(1)}_{gg}(N_F) 
\NN\\ && \phantom{abc} 
                    +\Bigl[ \hat{A}_{Qq}^{(2), {\sf PS}, \MOM}
                           +Z^{-1,(2), {\sf PS}}_{qq}(N_F+N_H)
                           -Z^{-1,(2), {\sf PS}}_{qq}(N_F)
                     \Bigr]\Gamma^{-1,(1)}_{qg}(N_F)
\NN\\ && \phantom{abc} 
                    +\Bigl[ \hat{A}_{qq,Q}^{(2), {\sf NS}, \MOM}
                           +Z^{-1,(2), {\sf NS}}_{qq}(N_F+N_H)
                           -Z^{-1,(2), {\sf NS}}_{qq}(N_F)
                     \Bigr]\Gamma^{-1,(1)}_{qg}(N_F)~.
\end{eqnarray}
The structure of the unrenormalized OME is more complex than in the ${\sf NS}$-- or ${\sf PS}$ case. It is given by
\begin{eqnarray}
   \Athathat_{Qg}^{(3)}&=&
\frac{1}{\ep^3} \Biggl[
\frac{28}{3} \beta_{0} \beta_{0,Q} \hat{\gamma}_{qg}^{(0)}
-\frac{8}{3} \hat{\gamma}_{qg}^{(0)} \gamma_{qq}^{(0)} \beta_{0,Q}
+\frac{14}{3} \beta_{0,Q} \hat{\gamma}_{qg}^{(0)} \gamma_{gg}^{(0)}
+24 \beta_{0,Q}^2 \hat{\gamma}_{qg}^{(0)}
+\frac{1}{3} \gamma_{gq}^{(0)} \left(\hat{\gamma}_{qg}^{(0)}\right)^2\Biggr]
\NN\\&&
+\frac{1}{\ep^2} \Biggl[
\Biggl\{
\frac{1}{4} \gamma_{gq}^{(0)} \left(\hat{\gamma}_{qg}^{(0)}\right)^2
+18 \beta_{0,Q}^2 \hat{\gamma}_{qg}^{(0)}
+7 \beta_{0} \beta_{0,Q} \hat{\gamma}_{qg}^{(0)}
-2 \hat{\gamma}_{qg}^{(0)} \gamma_{qq}^{(0)} \beta_{0,Q}
+\frac{7}{2} \beta_{0,Q} \hat{\gamma}_{qg}^{(0)} \gamma_{gg}^{(0)}
\Biggr\} 
\NN\\&&
\times \left(L_1+L_2\right)
+\frac{1}{3} \hat{\gamma}_{qg}^{(0)} \hat{\gamma}_{qq}^{\rm{PS},(1)}
+\frac{1}{3} \hat{\gamma}_{qg}^{(0)} \hat{\gamma}_{qq}^{\rm{NS},(1)}
-\frac{10}{3} \beta_{0,Q} \hat{\gamma}_{qg}^{(1)}
+\frac{2}{3} \hat{\gamma}_{qg}^{(0)} \hat{\gamma}_{gg}^{(1)}
-\frac{2}{3} \hat{\gamma}_{qg}^{(0)} \beta_{1,Q}
\NN\\&&
+10 \hat{\gamma}_{qg}^{(0)} \beta_{0,Q} \delta m_1^{(-1)}\Biggr] 
+\frac{1}{\ep} 
\Biggl[
\Biggl\{
\frac{1}{4} \hat{\gamma}_{qg}^{(0)} \hat{\gamma}_{qq}^{\rm{NS},(1)}
+\frac{15}{2} \hat{\gamma}_{qg}^{(0)} \beta_{0,Q} \delta m_1^{(-1)}
+\frac{1}{4} \hat{\gamma}_{qg}^{(0)} \hat{\gamma}_{qq}^{\rm{PS},(1)}
-\frac{5}{2} \beta_{0,Q} \hat{\gamma}_{qg}^{(1)}
\NN\\&&
+\frac{1}{2} \hat{\gamma}_{qg}^{(0)} \hat{\gamma}_{gg}^{(1)}
-\frac{1}{2} \hat{\gamma}_{qg}^{(0)} \beta_{1,Q}
\Biggr\} \left(L_1+L_2\right)
+\Biggl\{
\frac{13}{4} \beta_{0} \beta_{0,Q} \hat{\gamma}_{qg}^{(0)}
+\frac{13}{8} \beta_{0,Q} \hat{\gamma}_{qg}^{(0)} \gamma_{gg}^{(0)}
+\frac{15}{2} \beta_{0,Q}^2 \hat{\gamma}_{qg}^{(0)}
\NN\\&&
+\frac{3}{16} \gamma_{gq}^{(0)} \left(\hat{\gamma}_{qg}^{(0)}\right)^2
-\hat{\gamma}_{qg}^{(0)} \gamma_{qq}^{(0)} \beta_{0,Q}
\Biggr\} \left(L_1^2+L_2^2\right)
+ \Biggl\{
-\hat{\gamma}_{qg}^{(0)} \gamma_{qq}^{(0)} \beta_{0,Q}
+4 \beta_{0} \beta_{0,Q} \hat{\gamma}_{qg}^{(0)}
+12 \beta_{0,Q}^2 \hat{\gamma}_{qg}^{(0)}
\NN\\&&
+2 \beta_{0,Q} \hat{\gamma}_{qg}^{(0)} \gamma_{gg}^{(0)}
\Biggr\} L_1 L_2
+\frac{2}{3} \hat{\tilde{\gamma}}_{qg}^{(2)}
-8 \beta_{0,Q} a_{Qg}^{(2)}
-\frac{1}{8} \left(\hat{\gamma}_{qg}^{(0)}\right)^2 \zeta_2 \gamma_{gq}^{(0)}
+2 \hat{\gamma}_{qg}^{(0)} a_{gg,Q}^{(2)}
-2 \hat{\gamma}_{qg}^{(0)} \tilde{\delta}m_2^{(-1)}
\NN\\&&
+9 \hat{\gamma}_{qg}^{(0)} \zeta_2 \beta_{0,Q}^2
+\frac{1}{4} \beta_{0,Q} \zeta_2 \hat{\gamma}_{qg}^{(0)} \gamma_{gg}^{(0)}
+\frac{1}{2} \hat{\gamma}_{qg}^{(0)} \zeta_2 \beta_{0,Q} \beta_{0}
+8 \delta m_1^{(0)} \beta_{0,Q} \hat{\gamma}_{qg}^{(0)}
\Biggr]
\NN\\&&
+\tilde{a}_{Qg}^{(3)}\left(m_1^2,m_2^2,\mu^2\right)~.
\label{AhhhQg3} 
\end{eqnarray}

For the renormalized operator matrix element in the $\MS$ scheme we finally obtain,
\begin{eqnarray}
   \tilde{A}_{Qg}^{(3), \MS}&=&
   \Biggl\{
-\frac{9}{4} \beta_{0,Q}^2 \hat{\gamma}_{qg}^{(0)}
-\frac{7}{96} \gamma_{gq}^{(0)} \left(\hat{\gamma}_{qg}^{(0)}\right)^2
+\frac{1}{3} \hat{\gamma}_{qg}^{(0)} \gamma_{qq}^{(0)} \beta_{0,Q}
-\frac{25}{48} \beta_{0,Q} \hat{\gamma}_{qg}^{(0)} \gamma_{gg}^{(0)}
-\frac{25}{24} \beta_{0} \beta_{0,Q} \hat{\gamma}_{qg}^{(0)}
\Biggr\}
\NN\\&&
\times
\left(L_1^3+L_2^3\right)
+\Biggl\{
\frac{1}{4} \hat{\gamma}_{qg}^{(0)} \gamma_{qq}^{(0)} \beta_{0,Q}
-\beta_{0} \beta_{0,Q} \hat{\gamma}_{qg}^{(0)}
-\frac{1}{2} \beta_{0,Q} \hat{\gamma}_{qg}^{(0)} \gamma_{gg}^{(0)}
-3 \beta_{0,Q}^2 \hat{\gamma}_{qg}^{(0)}
\Biggr\} 
\NN\\&&
\times
\left(L_1^2 L_2+L_2^2 L_1\right)
+\Biggl\{
-\frac{1}{16} \hat{\gamma}_{qg}^{(0)} \hat{\gamma}_{qq}^{\rm{PS},(1)}
-\frac{1}{16} \hat{\gamma}_{qg}^{(0)} \hat{\gamma}_{qq}^{\rm{NS},(1)}
+\frac{9}{8} \beta_{0,Q} \hat{\gamma}_{qg}^{(1)}
-\frac{1}{4} \hat{\gamma}_{qg}^{(0)} \hat{\gamma}_{gg}^{(1)}
\NN\\&&
-\frac{29}{8} \hat{\gamma}_{qg}^{(0)} \beta_{0,Q} \delta m_1^{(-1)}
+\frac{1}{8} \hat{\gamma}_{qg}^{(0)} \beta_{1,Q}
\Biggr\} \left(L_1^2+L_2^2\right)
-4 L_1 L_2 \hat{\gamma}_{qg}^{(0)} \beta_{0,Q} \delta m_1^{(-1)}
  +\Biggl\{
  \frac{3}{2} \hat{\gamma}_{qg}^{(0)} \tilde{\delta}m_2^{(-1)}
\NN\\&&
+\frac{1}{32} \left(\hat{\gamma}_{qg}^{(0)}\right)^2 \zeta_2 \gamma_{gq}^{(0)}
+\frac{1}{4} \hat{\gamma}_{qg}^{(0)} \zeta_2 \beta_{0,Q} \gamma_{qq}^{(0)}
-6 \delta m_1^{(0)} \beta_{0,Q} \hat{\gamma}_{qg}^{(0)}
-\hat{\gamma}_{qg}^{(0)} a_{gg,Q}^{(2)}
-\frac{9}{16} \beta_{0,Q} \zeta_2 \hat{\gamma}_{qg}^{(0)} \gamma_{gg}^{(0)}
  \NN\\&&
-\frac{27}{4} \hat{\gamma}_{qg}^{(0)} \zeta_2 \beta_{0,Q}^2
-\frac{9}{8} \hat{\gamma}_{qg}^{(0)} \zeta_2 \beta_{0,Q} \beta_{0}
+4 \beta_{0,Q} a_{Qg}^{(2)}
\Biggr\} \left(L_1+L_2\right)
+8 \overline{a}_{Qg}^{(2)} \beta_{0,Q}
-\frac{1}{8} \hat{\gamma}_{qg}^{(0)} \zeta_2 \hat{\gamma}_{qq}^{\rm{PS},(1)}
  \NN\\&&
-\frac{1}{8} \hat{\gamma}_{qg}^{(0)} \zeta_2 \hat{\gamma}_{qq}^{\rm{NS},(1)}
+\frac{1}{24} \left(\hat{\gamma}_{qg}^{(0)}\right)^2 \zeta_3 \gamma_{gq}^{(0)}
-3 \hat{\gamma}_{qg}^{(0)} \beta_{0,Q}^2 \zeta_3
+\frac{1}{4} \hat{\gamma}_{qg}^{(1)} \beta_{0,Q} \zeta_2
-8 \delta m_1^{(1)} \beta_{0,Q} \hat{\gamma}_{qg}^{(0)}
  \NN\\&&
+\frac{1}{4} \hat{\gamma}_{qg}^{(0)} \zeta_2 \beta_{1,Q}
-2 \hat{\gamma}_{qg}^{(0)} \overline{a}_{gg,Q}^{(2)}
-\frac{1}{6} \hat{\gamma}_{qg}^{(0)} \beta_{0} \beta_{0,Q} \zeta_3
-\frac{1}{12} \beta_{0,Q} \zeta_3 \hat{\gamma}_{qg}^{(0)} \gamma_{gg}^{(0)}
+\hat{\gamma}_{qg}^{(0)} \Bigl(
\tilde{\delta}{m_2}^{1,(0)}
  \NN\\&&
+\tilde{\delta}{m_2}^{2,(0)}
\Bigr)
-\frac{9}{4} \hat{\gamma}_{qg}^{(0)} \zeta_2 \beta_{0,Q} \delta m_1^{(-1)}
+\tilde{a}_{Qg}^{(3)}\left(m_1^2,m_2^2,\mu^2\right)~.
\end{eqnarray}
%
%
%
\subsubsection{\bf\boldmath$A_{gq,Q}$}
\label{SubSec-AgqQ}
The matrix element $A_{gq,Q}$ contains contributions starting at $O(a_s^2)$,
\begin{eqnarray}
   A_{gq,Q}&=&
                       a_s^2A_{gq,Q}^{(2)}
                       +a_s^3A_{gq,Q}^{(3)}
                       +O(a_s^4)~. \label{AgqQpert}
  \end{eqnarray}
  Diagrams with two different masses, however, contribute only from $O(a_s^3)$
  \begin{eqnarray}
      \tilde{A}_{gq,Q}&=&
      a_s^3 \tilde{A}_{gq,Q}^{(3)}
                       +O(a_s^4)~. \label{AgqQpert2m}
\end{eqnarray}

The renormalization in the {\sf MOM}--scheme is performed using
\begin{eqnarray}
    A_{gq,Q}^{(2),\MOM}&=&
                     \hat{A}_{gq,Q}^{(2),\MOM}
                    +Z_{gq}^{-1,(2)}(N_F+N_H)
                    -Z_{gq}^{-1,(2)}(N_F)
\NN\\ 
&&
                    +\Bigl(
                           \hat{A}_{gg,Q}^{(1),\MOM}
                          +Z_{gg}^{-1,(1)}(N_F+N_H)
                          -Z_{gg}^{-1,(1)}(N_F)
                      \Bigr)\Gamma_{gq}^{-1,(1)}~, \\
    A_{gq,Q}^{(3),\MOM}&=& 
                        \hat{A}_{gq,Q}^{(3),\MOM}
                       +Z^{-1,(3)}_{gq}(N_F+N_H)
                       -Z^{-1,(3)}_{gq}(N_F)
                       +Z^{-1,(1)}_{gg}(N_F+N_H)\hat{A}_{gq,Q}^{(2),\MOM}
\NN\\ &&
                       +Z^{-1,(1)}_{gq}(N_F+N_H)\hat{A}_{qq}^{(2),\MOM}
                       +\Bigl[ \hat{A}_{gg,Q}^{(1),\MOM}
                              +Z^{-1,(1)}_{gg}(N_F+N_H)
\NN\\ &&
                              -Z^{-1,(1)}_{gg}(N_F)
                        \Bigr]
                              \Gamma^{-1,(2)}_{gq}(N_F)
                       +\Bigl[ \hat{A}_{gq,Q}^{(2),\MOM}
                              +Z^{-1,(2)}_{gq}(N_F+N_H)
\NN\\ &&
                              -Z^{-1,(2)}_{gq}(N_F)
                        \Bigr]
                              \Gamma^{-1,(1)}_{qq}(N_F)
                       +\Bigl[ \hat{A}_{gg,Q}^{(2),\MOM}
                              +Z^{-1,(2)}_{gg}(N_F+N_H)
\NN\\ &&
                              -Z^{-1,(2)}_{gg}(N_F)
                              +Z^{-1,(1)}_{gg}(N_F+N_H)\hat{A}_{gg,Q}^{(1),\MOM}
\NN\\ &&
                              +Z^{-1,(1)}_{gq}(N_F+N_H)\hat{A}_{Qg}^{(1),\MOM}
                        \Bigr]
                              \Gamma^{-1,(1)}_{gq}(N_F)
                      \label{AgqQRen1}~.
\end{eqnarray}
Applying Eq.~(\ref{AgqQRen1}) yields the unrenormalized expression
\begin{eqnarray}
\Athathat_{gq,Q}^{(3)}&=&
-\frac{16}{\ep^3} \gamma_{gq}^{(0)} \beta_{0,Q}^2
+\frac{1}{\ep^2} \Bigl[
-12 \gamma_{gq}^{(0)} \beta_{0,Q}^2 \left(L_2+L_1\right)
-4 \beta_{0,Q} \hat{\gamma}_{gq}^{(1)}
\Bigr]
\NN\\&&
+\frac{1}{\ep} \Biggl[
-6 \gamma_{gq}^{(0)} \beta_{0,Q}^2 
\left(L_2^2+L_1 L_2+L_1^2\right)
-3 \beta_{0,Q} \hat{\gamma}_{gq}^{(1)} \left(L_2+L_1\right)
  \NN\\&&
+\frac{2}{3} \hat{\tilde{\gamma}}_{gq}^{(2)}
-12 a_{gq}^{(2)} \beta_{0,Q}
\Biggr] 
+\tilde{a}_{gq,Q}^{(3)}\left(m_1^2,m_2^2,\mu^2\right)
        ~, \label{AhhhgqQ3}
   \end{eqnarray}
   and the renormalized operator matrix element reads
   \begin{eqnarray}
    \tilde{A}_{gq,Q}^{(3), \MS}&=&
\gamma_{gq}^{(0)} \beta_{0,Q}^2 \left(2 L_2^3+2 L_1^3 + \frac{3}{2} L_2^2 L_1+\frac{3}{2} L_1^2 L_2\right)    
+\frac{3}{2} \beta_{0,Q} \hat{\gamma}_{gq}^{(1)} \left(L_2^2+L_1^2\right)
\NN\\&&
+\Biggl\{
6 a_{gq}^{(2)} \beta_{0,Q}
+\frac{3}{2} \gamma_{gq}^{(0)} \beta_{0,Q}^2 \zeta_2
\Biggr\} \left(L_2+L_1\right)
+12 \overline{a}_{gq}^{(2)} \beta_{0,Q}
+\tilde{a}_{gq,Q}^{(3)}\left(m_1^2,m_2^2,\mu^2\right)
                     ~. \label{Agq3QMSren}
\nonumber\\
\end{eqnarray}
%
%
%
 \subsubsection{\bf\boldmath$A_{gg,Q}$}
  \label{SubSec-AggQ}
Finally, the matrix element $A_{gg,Q}$ obeys the expansion
\begin{eqnarray}
A_{gg,Q}&=&1+
                        a_sA_{gg,Q}^{(1)}
                       +a_s^2A_{gg,Q}^{(2)}
                       +a_s^3A_{gg,Q}^{(3)}
                       +O(a_s^4)~, \label{AggQpert}
\end{eqnarray}
with two--mass contributions starting at $O(a_s^2)$,
\begin{eqnarray}
   \tilde{A}_{gg,Q}&=&
                        a_s^2\tilde{A}_{gg,Q}^{(2)}
                       +a_s^3\tilde{A}_{gg,Q}^{(3)}
                       +O(a_s^4)~. \label{AggQpert2m}
  \end{eqnarray}
  
  The renormalization formulae in the {\sf MOM}--scheme read
  \begin{eqnarray}
    A_{gg,Q}^{(2), \MOM}&=&
                    \hat{A}_{gg,Q}^{(2), \MOM}
                   +Z^{-1,(2)}_{gg}(N_F+N_H)
                   -Z^{-1,(2)}_{gg}(N_F)
 \NN\\ &&
                   +Z^{-1,(1)}_{gg}(N_F+N_H)\hat{A}_{gg,Q}^{(1), \MOM}
                   +Z^{-1,(1)}_{gq}(N_F+N_H)\hat{A}_{Qg}^{(1), \MOM}
 \NN\\ &&
                   +\Bigl[ \hat{A}_{gg,Q}^{(1), \MOM}
                          +Z_{gg}^{-1,(1)}(N_F+N_H)
                          -Z_{gg}^{-1,(1)}(N_F)
                    \Bigr]\Gamma^{-1,(1)}_{gg}(N_F)
    ~, \label{AggQ1ren2} \\
    A_{gg,Q}^{(3), \MOM}&=&
                     \hat{A}_{gg,Q}^{(3), \MOM}
                    +Z^{-1,(3)}_{gg}(N_F+N_H)
                    -Z^{-1,(3)}_{gg}(N_F)
                    +Z^{-1,(2)}_{gg}(N_F+N_H)\hat{A}_{gg,Q}^{(1), \MOM}
\NN\\ &&
                    +Z^{-1,(1)}_{gg}(N_F+N_H)\hat{A}_{gg,Q}^{(2), \MOM}
                    +Z^{-1,(2)}_{gq}(N_F+N_H)\hat{A}_{Qg}^{(1), \MOM}
\NN\\ &&
                    +Z^{-1,(1)}_{gq}(N_F+N_H)\hat{A}_{Qg}^{(2), \MOM}
                    +\Bigl[
                            \hat{A}_{gg,Q}^{(1), \MOM}
                           +Z^{-1,(1)}_{gg}(N_F+N_H)
  \NN\\ &&
                           -Z^{-1,(1)}_{gg}(N_F)
                     \Bigr]\Gamma^{-1,(2)}_{gg}(N_F)
                    +\Bigl[ \hat{A}_{gg,Q}^{(2), \MOM}
                           +Z^{-1,(2)}_{gg}(N_F+N_H)
\NN\\ &&
                           -Z^{-1,(2)}_{gg}(N_F)
                           +Z^{-1,(1)}_{gq}(N_F+N_H)A_{Qg}^{(1), \MOM}
\NN\\ &&
                           +Z^{-1,(1)}_{gg}(N_F+N_H)A_{gg,Q}^{(1), \MOM}
                     \Bigr]\Gamma^{-1,(1)}_{gg}(N_F)
\NN\\ &&
                    +\Bigl[ \hat{A}_{gq,Q}^{(2), \MOM}
                           +Z^{-1,(2)}_{gq}(N_F+N_H)
                           -Z^{-1,(2)}_{gq}(N_F)
                     \Bigr]\Gamma^{-1,(1)}_{qg}(N_F)
         ~.\label{AggQ1ren3}
\end{eqnarray}

After subtracting all single--mass contributions we obtain the unrenormalized
two--flavor contribution at 2 loops
\begin{eqnarray}
   \Athathat_{gg,Q}^{(2)}&=&\frac{8 \beta_{0,Q}}{\ep^2}+\frac{4 \beta_{0,Q}^2}{\ep} \left(L_1 + L_2\right)
   +\tilde{a}_{gg,Q}\left(m_1^2,m_2^2,\mu^2\right)+\ep \overline{\tilde{a}}_{gg,Q}
\left(m_1^2,m_2^2,\mu^2\right)
\end{eqnarray}
and the renormalized expression 
\begin{eqnarray}
\tilde{A}_{gg,Q}^{(2), \MS}&=& 
- \beta_{0,Q}^2 \left(
\ln^2\left(\frac{m_b^2}{\mu^2}\right)
+\ln^2\left(\frac{m_b^2}{\mu^2}\right)\right)
-2 \beta_{0,Q}^2 \zeta_2
+\tilde{a}_{gg,Q}\left(m_1^2,m_2^2,\mu^2\right)
~.   \label{AggQ2MSren}
\end{eqnarray}
The $O(a_s^2)$ contribution consists of one particle reducible contributions only and the coefficients 
follow from Eq.~(\ref{Agg2red})
\begin{eqnarray}
\label{eq:aggQ2a}
 a_{gg,Q}^{(2)} &=&
 \beta_{0,Q}^2 \left(L_2+L_1\right)^2+2 \beta_{0,Q}^2 \zeta_2~,
 \\
\label{eq:aggQ2b}
 \overline{a}_{gg,Q}^{(2)} &=&
 \frac{1}{6} \beta_{0,Q}^2 \left(L_1+L_2\right)^3
+\beta_{0,Q}^2 \zeta_2 \left(L_2+L_1\right)
+\frac{2}{3} \beta_{0,Q}^2 \zeta_3~.
\end{eqnarray}

The unrenormalized 3-loop contribution from two masses reads 
\begin{eqnarray}
    \Ahathat_{gg,Q}^{(3)}&=&
    \frac{1}{\ep^3} \Biggl[
-\frac{10}{3} \hat{\gamma}_{qg}^{(0)} \beta_{0,Q} \gamma_{gq}^{(0)}
-\frac{56}{3} \beta_{0} \beta_{0,Q}^2
-\frac{28}{3} \beta_{0,Q}^2 \gamma_{gg}^{(0)}
-48 \beta_{0,Q}^3
\Biggr] 
+\frac{1}{\ep^2} \Biggl[
\Biggl\{
-7 \beta_{0,Q}^2 \gamma_{gg}^{(0)}
\NN\\&&
-14 \beta_{0} \beta_{0,Q}^2
-\frac{5}{2} \hat{\gamma}_{qg}^{(0)} \beta_{0,Q} \gamma_{gq}^{(0)}
-36 \beta_{0,Q}^3
\Biggr\} \left(L_1+L_2\right)
+\frac{1}{3} \hat{\gamma}_{qg}^{(0)} \hat{\gamma}_{gq}^{(1)}
-\frac{14}{3} \beta_{0,Q} \hat{\gamma}_{gg}^{(1)}
\NN\\&&
+\frac{4}{3} \beta_{1,Q} \beta_{0,Q}
-20 \delta m_1^{(-1)} \beta_{0,Q}^2
\Biggr]
+\frac{1}{\ep} \Biggl[
\Biggl\{
\frac{1}{4} \hat{\gamma}_{qg}^{(0)} \hat{\gamma}_{gq}^{(1)}
-15 \delta m_1^{(-1)} \beta_{0,Q}^2
-\frac{7}{2} \beta_{0,Q} \hat{\gamma}_{gg}^{(1)}
\NN\\&&
+\beta_{1,Q} \beta_{0,Q}
\Biggr\} \left(L_1+L_2\right)
+\Biggl\{
-15 \beta_{0,Q}^3
-\frac{11}{8} \hat{\gamma}_{qg}^{(0)} \beta_{0,Q} \gamma_{gq}^{(0)}
-\frac{13}{2} \beta_{0} \beta_{0,Q}^2
-\frac{13}{4} \beta_{0,Q}^2 \gamma_{gg}^{(0)}
\Biggr\}
\NN\\&&
\times
\left(L_1^2+L_2^2\right)
+\Biggl\{
-4 \beta_{0,Q}^2 \gamma_{gg}^{(0)}
-24 \beta_{0,Q}^3
-8 \beta_{0} \beta_{0,Q}^2
-\hat{\gamma}_{qg}^{(0)} \beta_{0,Q} \gamma_{gq}^{(0)}
\Biggr\} L_1 L_2
-\frac{1}{2} \beta_{0,Q}^2 \zeta_2 \gamma_{gg}^{(0)}
\NN\\&&
+\frac{2}{3} \hat{\tilde{\gamma}}_{gg}^{(2)}
-12 \beta_{0,Q} a_{gg,Q}^{(2)}
-18 \beta_{0,Q}^3 \zeta_2
+\frac{1}{4} \beta_{0,Q} \zeta_2 \gamma_{gq}^{(0)} \hat{\gamma}_{qg}^{(0)}
-\beta_{0} \beta_{0,Q}^2 \zeta_2
-16 \delta m_1^{(0)} \beta_{0,Q}^2
\NN\\&&
+4 \beta_{0,Q} \tilde{\delta} m_2^{(-1)}
\Biggr]
+\tilde{a}_{gg,Q}^{(3)}\left(m_1^2,m_2^2,\mu^2\right)~.
\end{eqnarray}

The renormalized result in the $\MS$--scheme is given by
\begin{eqnarray}
\tilde{A}_{gg,Q}^{(3), \MS}
&=&
\Bigg\{
\frac{25}{24} \beta_{0,Q}^2 \gamma_{gg}^{(0)}
+\frac{25}{12} \beta_{0} \beta_{0,Q}^2
+\frac{9}{2} \beta_{0,Q}^3
+\frac{23}{48} \hat{\gamma}_{qg}^{(0)} \beta_{0,Q} \gamma_{gq}^{(0)}
\Biggr\} \left(L_1^3+L_2^3\right)
+\Biggl\{
\frac{1}{4} \hat{\gamma}_{qg}^{(0)} \beta_{0,Q} \gamma_{gq}^{(0)}
\NN\\&&
+\beta_{0,Q}^2 \gamma_{gg}^{(0)}
+2 \beta_{0} \beta_{0,Q}^2
+6 \beta_{0,Q}^3
\Biggr\} \left(L_1^2 L_2+L_2^2 L_1\right)
+\Biggl\{
-\frac{1}{4} \beta_{1,Q} \beta_{0,Q}
+\frac{13}{8} \beta_{0,Q} \hat{\gamma}_{gg}^{(1)}
\NN\\&&
+\frac{29}{4} \delta m_1^{(-1)} \beta_{0,Q}^2
-\frac{1}{16} \hat{\gamma}_{qg}^{(0)} \hat{\gamma}_{gq}^{(1)}
\Biggr\} \left(L_1^2+L_2^2\right)
+8 L_2 L_1 \delta m_1^{(-1)} \beta_{0,Q}^2
+\Biggl\{
 \frac{9}{4} \beta_{0} \beta_{0,Q}^2 \zeta_2
\NN\\&&
+\frac{27}{2} \beta_{0,Q}^3 \zeta_2
-3 \beta_{0,Q} \tilde{\delta}m_2^{(-1)}
+\frac{9}{8} \zeta_2 \beta_{0,Q}^2 \gamma_{gg}^{(0)}
+12 \delta m_1^{(0)} \beta_{0,Q}^2
+\frac{3}{16} \beta_{0,Q} \zeta_2 \gamma_{gq}^{(0)} \hat{\gamma}_{qg}^{(0)}
\NN\\&&
+6 \beta_{0,Q} a_{gg,Q}^{(2)}
\Biggr\} \left(L_2+L_1\right)
-\frac{1}{8} \hat{\gamma}_{qg}^{(0)} \zeta_2 \hat{\gamma}_{gq}^{(1)}
+\frac{1}{4} \beta_{0,Q} \zeta_2 \hat{\gamma}_{gg}^{(1)}
+\frac{1}{3} \beta_{0} \beta_{0,Q}^2 \zeta_3
+12 \beta_{0,Q} \overline{a}_{gg,Q}^{(2)}
\NN\\&&
+6 \beta_{0,Q}^3 \zeta_3+16 \delta m_1^{(1)} \beta_{0,Q}^2
+\frac{1}{6} \beta_{0,Q}^2 \zeta_3 \gamma_{gg}^{(0)}
-2 \beta_{0,Q} \left(
\tilde{\delta} {m_2}^{1,(0)}
+\tilde{\delta} {m_2}^{2,(0)}
\right)
\NN\\&&
+\frac{9}{2} \delta m_1^{(-1)} \beta_{0,Q}^2 \zeta_2
-\frac{1}{12} \zeta_3 \beta_{0,Q} \gamma_{gq}^{(0)} \hat{\gamma}_{qg}^{(0)}
-\frac{1}{2} \zeta_2 \beta_{0,Q} \beta_{1,Q}
+\tilde{a}_{gg,Q}^{(3)}\left(m_1^2,m_2^2,\mu^2\right)~. \label{Agg3QMSren}
   \end{eqnarray}
 \end{subsection}
\begin{subsection}{Mass renormalization schemes \label{MRS}}

\vspace*{1mm}
\noindent
The heavy quark masses in the ${\MS}$ and on-shell renormalization schemes are related by
\begin{eqnarray}
\hat{m}=Z_m^{\MS} \overline{m} = Z_m m,
\end{eqnarray}
where $\overline{m}$ denotes the mass in the ${\MS}$ scheme and $m$ in the OMS scheme. The ratio of these two masses for
a quark of mass $m_2$ in the presence of a second heavy quark of mass $m_1$ is given by 
\begin{eqnarray}
\label{eq:zm}
z_m = \frac{\overline{m}}{m} = 1 + \sum_{k=1}^\infty \left(\frac{\alpha_s^{\overline{\sf MS}}}{\pi}\right)^k 
z_m^{(k)}
\end{eqnarray}
with \cite{Gray:1990yh,Melnikov:2000zc,Bekavac:2007tk,Marquard:2016dcn}\footnote{We thank P. Marquard for providing
this relation.}, 

\vspace{1mm}\noindent
\begin{eqnarray}
z_m^{(1)} &=& - C_F\Biggl[1 - \frac{3}{4} L_\mu\Biggr] 
\\
z_m^{(2)} &=& 
C_F T_F
\Biggl[
  \frac{143}{96}
- \frac{\pi^2}{6}
+ \frac{13}{24} L_\mu
+ \frac{1}{8} L_\mu^2
\Biggr]
+ C_F N_F T_F
\Biggl[
\frac{71}{96}
+ \frac{\pi^2}{12}
+ \frac{13}{24} L_\mu
+ \frac{1}{8} L_\mu^2
\Biggr]
\nonumber\\ &&
+ C_F T_F
\Biggl[
\frac{71}{96}
+ \frac{\pi^2}{12}
- \frac{x}{4} \pi^2
+ \frac{3}{4} x^2
- \frac{\pi^2}{4}  x^3
+ \frac{\pi^2}{12} x^4
+ \frac{1}{2} x^2 H_0(x)
\nonumber\\ &&
- \frac{1}{2} \left(1
+ x
+ x^3
+ x^4 \right) H_{-1,0}(x)
+ \frac{1}{2} x^4 H_0^2(x)
+ \frac{1}{2} \left[1 -x -x^3 +x^4\right] H_{1,0}(x)
\nonumber\\ &&
+ \frac{13}{24} L_\mu
+ \frac{1}{8} L_\mu^2
\Biggr]
+ C_F^2 \Biggl[
\frac{7}{128} - \frac{5 \pi^2}{16}
+ \frac{1}{2} \pi^2 \ln(2) + \frac{21}{32} L_\mu
+ \frac{9}{32} L_\mu^2
- \frac{3}{4} \zeta_3
\Biggr]
\nonumber\\ &&
+C_A C_F \Biggl[
-\frac{1111}{384} + \frac{\pi^2}{12} - \frac{1}{4} \pi^2 \ln(2)
          - \frac{185}{96} L_\mu
          - \frac{11}{32}  L_\mu^2
          + \frac{3}{8} \zeta_3 \Biggr],
\end{eqnarray}
with $L_\mu = \ln(\mu^2/m^2)$ and $x = m_1/m_2$.

In data analyses one usually fits the $\overline{\sf MS}$-mass $\overline{m}$, which is free of infrared 
renormalon ambiguities, unlike the on-shell mass, which grows significantly order-by-order in 
perturbation theory \cite{Marquard:2016dcn}. 
\end{subsection}

\section{Fixed Moments of the Massive Operator Matrix Elements}
\label{sec:x1}

\vspace*{1mm}
\noindent
In Ref. {\cite{Bierenbaum:2009mv}} a series of fixed Mellin moments of all massive operator matrix elements at 3-loop
order have been calculated in the single mass case by projecting the corresponding integrals onto massive tadpoles and 
evaluating them using the code {\tt MATAD}\cite{Steinhauser:2000ry}. These moments serve as important reference points
for the general $N$ solution. In the following, we will calculate the Mellin moments  $N = 2,4,6$ in the case of unequal masses.
The number of moments is less than in the equal mass case, where values of $N = 10 ... 14$ could be reached, which is due 
to the presence of the second variable $\eta$ and the performance of the codes {\tt Q2e}/{\tt Exp}~\cite{Harlander:1997zb, 
Seidensticker:1999bb}, which we are going to use. The full calculation took about one CPU year. We still obtain very useful
reference points by this.

The Feynman diagrams are generated using the code {\tt QGRAF} \cite{Nogueira:1991ex}. In order to take into account the 
local operator insertions, we introduce new additional propagators which either carry an operator insertion or which
generate an operator on an attached vertex. In the case of operator insertions on a gauge boson line, this method leads to a 
double counting of some vertex diagrams which has to be removed. For the calculation of the color algebra of the expressions we 
used the code  {\tt Color} \cite{vanRitbergen:1998pn}. 
   
After inserting the Feynman rules, cf. Section 8.1 \cite{Bierenbaum:2009mv}, and the projection operators, the momentum 
integrals take the form 
\begin{eqnarray} 
	I^{(l)}(p,m_1,m_2,n_1\ldots n_j) &\equiv& \int
	\frac{d^Dk_1}{(2\pi)^D}\ldots \int \frac{d^Dk_l}{(2\pi)^D}
	(\Delta.q_1)^{n_1}\ldots (\Delta.q_j)^{n_j} f(k_1\ldots
	k_l,p,m_1,m_2)~. \NN \\ \label{ExInt1} 
\end{eqnarray} 
Here $p$ denotes the external momentum, $p^2=0$, $\Delta$ is an arbitrary light--like vector $\Delta^2=0$ and $q_i$ are 
linear combinations of the loop momenta $k_j$ and the external momentum $p$. The exponents $n_i$ are integer-valued and obey 
$\sum n_i =N$, while the function $f(k_1\ldots k_l,p,m_1,m_2)$ contains the remaining numerator structure and denominators. 
In Eq. (\ref{ExInt1}), we have omitted possible summations over indices on which the exponents $n_i$ might depend.

We may represent (\ref{ExInt1}) as
\begin{eqnarray}
    I^{(l)}\left(p,m_1,m_2,n_1\ldots n_j\right)=\prod_{j=1}^N \Delta^{\mu_j} 
    \tilde{I}^{(l)}_{\mu_1,\ldots,\mu_N}\left(p,m_1,m_2,n_1\ldots n_j\right)~.
\end{eqnarray}
Since $\prod_{j=1}^N \Delta^{\mu_j}$ constitutes a completely symmetric tensor only the purely symmetric part of  
$\tilde{I}^{(l)}_{\mu_1,\ldots,\mu_N}$ contributes. We thus symmetrize by shuffling the indices, \cite{Blumlein:2003gb}, 
and normalize it by dividing by the number of terms. For the general integral (\ref{ExInt1}) the symmetrized tensor is given 
by
   \begin{eqnarray}
    I^{(l)}_{\mu_1,\ldots,\mu_M}\left(p,m_1,m_2,n_1\ldots n_j\right)
    &=&{\bf S}~~\tilde{I}^{(l)}_{\mu_1,\ldots,\mu_M}\left(p,m_1,m_2,n_1\ldots n_j\right)~,
    \end{eqnarray}
where ${\bf S}$ is
the symmetrization operator given in Eq.~(\ref{SymmOp}).
The result of the original integral (\ref{ExInt1}) may then be obtained again by applying the projection operator 
\cite{Bierenbaum:2009mv}
  \begin{eqnarray}
    \Pi_{\mu_1 \ldots \mu_N}&=&F(N)
                              \sum_{i=1}^{[N/2]+1}C(i,N)
                              \left(\prod_{l=1}^{[N/2]-i+1}
                                   \frac{g_{\mu_{2l-1}\mu_{2l}}}{p^2} 
                              \right)
                              \left(\prod_{k=2[N/2]-2i+3}^N
                                   \frac{p_{\mu_k}}{p^2}
                              \right)~. \label{Proj1}
   \end{eqnarray}
The pre-factors $F(N)$ and the combinatorial factors $C(i,N)$ for odd values of $N$ are given by
   \begin{eqnarray}
    C^{odd}(k,N)&=&(-1)^{N/2+k+1/2}
                 \frac{2^{2k-N/2-3/2}\Gamma(N+1)\Gamma(D/2+N/2+k-3/2)}
                      {\Gamma(N/2-k+3/2)\Gamma(2k)\Gamma(D/2+N/2-1/2)}~,\\
    F^{odd}(N)  &=&\frac{2^{3/2-N/2}\Gamma(D/2+1/2)}
                        {(D-1)\Gamma(N/2+D/2-1)}~,
                  \end{eqnarray}
and read
    \begin{eqnarray}
    C^{even}(k,N)&=&(-1)^{N/2+k+1}
                    \frac{2^{2k-N/2-2}\Gamma(N+1)\Gamma(D/2+N/2-2+k)}
                         {\Gamma(N/2-k+2)\Gamma(2k-1)\Gamma(D/2+N/2-1)}~, \\
    F^{even}(N)  &=&\frac{2^{1-N/2}\Gamma(D/2+1/2)}
                         {(D-1)\Gamma(N/2+D/2-1/2)}~,
   \end{eqnarray}
for even values of $N$. The pre-factors $F^{odd}(N), F^{even}(N)$ are chosen such that the projector (\ref{Proj1})
is normalized
   \begin{eqnarray}
    \Pi_{\mu_1\ldots \mu_N}p^{\mu_1}\ldots p^{\mu_N}=1~. 
   \end{eqnarray}
The integrals with a local operator insertion for fixed values of $N$ are thus represented in terms of tadpole diagrams with a 
modified numerator structure. The projection operators (\ref{Proj1}) become sizeable for large values of $N$, which leads to 
an exponential increase in the computation time. 

In the calculation, the projected Feynman integrals are first expanded in the mass ratio $\eta$ by an expansion in subgraphs
\cite{Chetyrkin:1988zz,Chetyrkin:1988cu,Tkachov:1997gz,Gorishnii:1986gn} using the codes 
{\tt Q2e}/{\tt Exp}~\cite{Harlander:1997zb, Seidensticker:1999bb}, which also rely on {\tt MATAD} to evaluate the single-mass 
tadpole diagrams, using {\tt Form} and {\tt TForm} \cite{FORM}.
   
The pole structure of the unrenormalized OMEs corresponds to the one which was deduced from the renormalization prescription 
given in Section~\ref{sec:ren}. As a by-product of the present calculation, also the terms in these 3--loop anomalous 
dimensions for the moments $N = 2, 4, 6$,~~$\propto T_F$ are obtained, cf.~\cite{ANDI1}, here in a two--mass calculation.

The moments of the OMEs calculated in the following depend on the logarithms
\begin{align}
	L_1=\ln \left(\frac{m_1^2}{\mu^2}\right),\,
	L_2=\ln \left(\frac{m_2^2}{\mu^2}\right),\,
	L_{\eta}=\ln \left(\eta\right)\equiv \ln \left(\frac{m_2^2}{m_1^2}\right), \eta < 1. 
\end{align}
We expand up to remaining terms of 
\begin{align}
O(|\eta^4 L^3_\eta|) \simeq 0.15 \%.
\end{align}
The pole terms in the dimensional parameter $\ep$ do not contain any power corrections in $\eta$.

In the following, we present the moments $N = 2, 4$ and $6$ for the two--flavor contributions to the constant
parts of the various operator matrix elements as defined in Eq.~(\ref{AhathatDecomp}).\footnote{We have presented 
a few of these results before in \cite{Ablinger:2011pb,Ablinger:2012qj}.} 

The flavor non-singlet contribution two-mass contribution is obtained by
\begin{eqnarray}
 \tilde{a}_{qq,Q}^{\sf NS, (3)}(N=2)&=&
 C_F T_F^2 
\Biggl\{
	\Biggl(
-\frac{1024}{8505} L_\eta^2
-\frac{190500608}{843908625}
-\frac{747008}{2679075} L_\eta
\Biggr) \eta^3
\NN\\&&
+\Biggl(
-\frac{7176352}{1157625}
-\frac{64}{105} L_\eta^2
-\frac{33856}{11025} L_\eta
\Biggr) \eta^2
+\Biggl(
-\frac{12032}{675}
-\frac{512}{45} L_\eta
\Biggr) \eta
\NN\\&&
+\Biggl(
-\frac{1024}{81}
-\frac{128}{9} \left(L_2+L_1\right)
\Biggr) \zeta_2
-\frac{153856}{2187}
+\frac{512}{81} \zeta_3
-\frac{14080}{243} L_1
-\frac{2048}{81} L_2
\NN\\&&
-\frac{1024}{81} L_2 L_1
-\frac{256}{27} L_1 L_2^2
-\frac{512}{81} L_2^3
-\frac{1024}{81} \left(L_2^2+L_1^2\right)
-\frac{640}{81} L_1^3
-\frac{128}{27} L_1^2 L_2\Biggr\}
\NN\\&&
+O(\eta^4 L_\eta^3)~,
\\
\tilde{a}_{qq,Q}^{\sf NS, (3)}(N=4) &=&
 C_F T_F^2 
\Biggl\{
	 \Biggl(
-\frac{10048}{42525} L_\eta^2
-\frac{1869287216}{4219543125}
-\frac{7330016}{13395375} L_\eta
\Biggr) \eta^3
\NN\\&&
+\Biggl(
-\frac{70417954}{5788125}
-\frac{628}{525} L_\eta^2
-\frac{332212}{55125} L_\eta
\Biggr) \eta^2
+\Biggl(
-\frac{118064}{3375}
-\frac{5024}{225} L_\eta\Biggr) \eta
\NN\\&&
+\Biggl(
-\frac{53084}{2025}
-\frac{1256}{45} \left(L_2+L_1\right)
\Biggr) \zeta_2
-\frac{388370299}{2733750}
+\frac{5024}{405} \zeta_3
-\frac{3509323}{30375} L_1
\NN\\&&
-\frac{520841}{10125} L_2
-\frac{53084}{2025} L_2 L_1
-\frac{2512}{135} L_1 L_2^2
-\frac{5024}{405} L_2^3
-\frac{53084}{2025} \left(L_2^2+L_1^2\right)
\NN\\&&
-\frac{1256}{81} L_1^3
-\frac{1256}{135} L_1^2 L_2
\Biggr\} +O\left(\eta^4 L_\eta^3\right)~,
\\
\tilde{a}_{qq,Q}^{\sf NS, (3)}(N=6) &=&
 C_F T_F^2 
\Biggl\{
	\Biggl(
-\frac{90752}{297675} L_\eta^2
-\frac{16883116384}{29536801875}
-\frac{66203584}{93767625} L_\eta
\Biggr) \eta^3
\NN\\&&
+\Biggl(
-\frac{636004196}{40516875}
-\frac{5672}{3675} L_\eta^2
-\frac{3000488}{385875} L_\eta
\Biggr) \eta^2
+\Biggl(
-\frac{1066336}{23625}
-\frac{45376}{1575} L_\eta\Biggr) \eta
\NN\\&&
+\Biggl(
-\frac{3424952}{99225}
-\frac{11344}{315} \left(L_2+L_1\right)
\Biggr) \zeta_2
-\frac{202733427313}{1093955625}
+\frac{45376}{2835} \zeta_3
\NN\\&&
-\frac{520819486}{3472875} L_1
-\frac{700881658}{10418625} L_2
-\frac{3424952}{99225} L_2 L_1
-\frac{22688}{945} L_1 L_2^2
\NN\\&&
-\frac{45376}{2835} L_2^3
-\frac{3424952}{99225} \left(L_2^2+L_1^2\right)
-\frac{11344}{567} L_1^3
-\frac{11344}{945} L_1^2 L_2
\Biggr\}
\NN\\&&
+O\left(\eta^4 L_\eta^3\right).
\end{eqnarray}

The constant two-mass contribution to the OME $A_{Qq}^{\sf PS, (3)}$ is given by
\begin{eqnarray}
 \tilde{a}_{Qq}^{\sf PS, (3)}(N=2) &=&
 C_F T_F^2\Biggl\{
 \Biggl(
-\frac{381001216}{843908625}
-\frac{1494016}{2679075} L_\eta
-\frac{2048}{8505} L_\eta^2
\Biggr) \eta^3
+\Biggl(
-\frac{14352704}{1157625}
\NN\\&&
-\frac{128}{105} L_\eta^2
-\frac{67712}{11025} L_\eta
\Biggr) \eta^2
+\Biggl(
-\frac{24064}{675}
-\frac{1024}{45} L_\eta
\Biggr) \eta
+\Biggl(
-\frac{1472}{81}
\NN\\&&
-\frac{256}{9} \left(L_2+L_1\right)
\Biggr)\zeta_2
-\frac{1280}{81} L_1^3
-\frac{256}{27} L_1^2 L_2
+\frac{1024}{81} \zeta_3
-\frac{26720}{243} L_1
-\frac{3616}{81} L_2
\NN\\&&
-\frac{1472}{81} L_2 L_1
-\frac{512}{27} L_1 L_2^2
-\frac{266528}{2187}
-\frac{1024}{81} L_2^3
-\frac{1472}{81} \left(L_2^2+L_1^2\right)
\Biggr\}
\NN\\&&
+O\left(\eta^4 L_\eta^3\right)~,
\\
 \tilde{a}_{Qq}^{\sf PS, (3)}(N=4) &=&
 C_F T_F^2\Biggl\{
 \Biggl(\frac{190292193776}{1123242379875}
+\frac{8509216}{324168075} L_\eta
-\frac{1472}{93555} L_\eta^2\Biggr) \eta^3
+\Biggl(
-\frac{71844302}{31255875}
\NN\\&&
-\frac{76}{315} L_\eta^2
-\frac{89252}{99225} L_\eta\Biggr) \eta^2
+\Biggl(
-\frac{5008}{945}
-\frac{32}{9} L_\eta
\Biggr) \eta
+\Biggl(
-\frac{2236}{2025}
-\frac{968}{225} \Bigl(L_2
\NN\\&&
+L_1\Bigr)
\Biggr)\zeta_2
-\frac{968}{405} L_1^3
-\frac{968}{675} L_1^2 L_2
+\frac{3872}{2025} \zeta_3
-\frac{2406319}{151875} L_1
-\frac{297941}{50625} L_2
\NN\\&&
-\frac{2236}{2025} L_2 L_1
-\frac{1936}{675} L_1 L_2^2
-\frac{195482623}{13668750}
-\frac{3872}{2025} L_2^3
-\frac{2236}{2025} \left(L_2^2+L_1^2\right)
\Biggr\}
\NN\\&&
+O\left(\eta^4 L_\eta^3\right)~,
\\
\tilde{a}_{Qq}^{\sf PS, (3)}(N=6) &=&
C_F T_F^2 \Biggl\{
\Biggl(\frac{19353315711436064}{86371722800488125}
+\frac{112677158848}{1917454163625} L_\eta
+\frac{385408}{42567525} L_\eta^2
\Biggr) \eta^3
\NN\\&&
+\Biggl(
-\frac{5015464079432}{4368164810625}
-\frac{45616}{363825} L_\eta^2
-\frac{432844912}{1260653625} L_\eta
\Biggr) \eta^2
+\Biggl(
-\frac{2455328}{1157625}
\NN\\&&
-\frac{1984}{1323} L_\eta\Biggr) \eta+(
-\frac{15184}{99225}
-\frac{3872}{2205} \left(L_2+L_1\right)) \zeta_2
-\frac{3872}{3969} L_1^3
-\frac{3872}{6615} L_1^2 L_2
\NN\\&&
+\frac{15488}{19845} \zeta_3
-\frac{52387796}{8103375} L_1
-\frac{172633556}{72930375} L_2
-\frac{15184}{99225} L_2 L_1
-\frac{7744}{6615} L_1 L_2^2
\NN\\&&
-\frac{7819198418}{1531537875}
-\frac{15488}{19845} L_2^3
-\frac{15184}{99225} \left(L_2^2+L_1^2\right)
\Biggr\}+O\left(\eta^4 L_\eta^3\right).
\end{eqnarray}

For $\tilde{a}_{Qg}^{(3)}$ one obtains
\begin{eqnarray}
\tilde{a}_{Qg}^{(3)}(N=2) &=&
 C_A T_F^2 \Biggl\{
 \Biggl(
 \frac{56086736}{843908625}
-\frac{164464}{2679075} L_\eta
-\frac{2552}{8505} L_\eta^2\Biggr) \eta^3
+\Biggl(\frac{6008}{4725} L_\eta
+\frac{1565036}{496125}
\NN\\&&
-\frac{8}{45} L_\eta^2
\Biggr) \eta^2
+\Biggl(
\frac{256304}{10125}
+\frac{7184}{675} L_\eta
-\frac{8}{45} L_\eta^2
\Biggr) \eta
+\Biggl(
-\frac{74}{81}
+\frac{140}{9} \left(L_2+L_1\right)\Biggr) \zeta_2
\NN\\&&
-\frac{5}{3} L_2
+\frac{772}{81} L_1^3
-\frac{848}{81} \zeta_3
+\frac{9355}{243} L_1
+\frac{280}{27} L_1 L_2^2
-\frac{35}{81} \left(L_2^2+L_1^2\right)
+\frac{104}{27} L_1^2 L_2
\NN\\&&
-\frac{152}{81} L_2 L_1
+\frac{596}{81} L_2^3
+\frac{78229}{2187}
\Biggr\}
\NN\\&&
+C_F T_F^2\Biggl\{
\Biggl(
\frac{826805984}{843908625}
+\frac{5893184}{2679075} L_\eta
+\frac{23872}{8505} L_\eta^2
\Biggr) \eta^3
+\Biggl(\frac{1028192}{99225} L_\eta
\NN\\&&
+\frac{169892864}{10418625}
+\frac{4768}{945} L_\eta^2
\Biggr) \eta^2
+\Biggl(
-\frac{758944}{30375}
+\frac{22976}{2025} L_\eta
+\frac{448}{135} L_\eta^2
\Biggr) \eta
\NN\\&&
+\Biggl(\frac{320}{27}
-\frac{64}{9} \left(L_2+L_1\right)
\Biggr)\zeta_2
+\frac{6752}{243} L_2
-\frac{704}{81} L_1^3
+\frac{1792}{81} \zeta_3
+\frac{128}{81} L_1
-\frac{128}{27} L_1 L_2^2
\NN\\&&
+\frac{968}{81} \left(L_2^2+L_1^2\right)
+\frac{128}{27} L_1^2 L_2
+\frac{944}{81} L_2 L_1
-\frac{448}{81} L_2^3
+\frac{64}{243}
\Biggr\}
+O\left(\eta^4 L_\eta^3\right)~,
\\
 \tilde{a}_{Qg}^{(3)}(N=4) &=&
 C_A T_F^2 \Biggl\{
 \Biggl(\frac{250077164867}{11232423798750}
-\frac{156082853}{3241680750} L_\eta
-\frac{744283}{1871100} L_\eta^2
\Biggr) \eta^3
\NN\\&&
+\Biggl(
\frac{1634774}{1488375} L_\eta
+\frac{1255194149}{468838125}
-\frac{142}{525} L_\eta^2
\Biggr) \eta^2
+\Biggl(
\frac{496855133}{14883750}
+\frac{1877399}{141750} L_\eta
\NN\\&&
+\frac{707}{2700} L_\eta^2
\Biggr) \eta
+\Biggl(
\frac{5807}{360}
+\frac{17963}{900} \left(L_2+L_1\right)
\Biggr) \zeta_2
+\frac{47956573}{1620000} L_2
+\frac{3817}{324} L_1^3
\NN\\&&
-\frac{23573}{2025} \zeta_3
+\frac{384762007}{4860000} L_1
+\frac{17963}{1350} L_1 L_2^2
+\frac{532373}{32400} \left(L_2^2+L_1^2\right)
+\frac{7579}{1350} L_1^2 L_2
\NN\\&&
+\frac{62893}{4050} L_2 L_1
+\frac{74657}{8100} L_2^3
+\frac{4887988511}{48600000}
\Biggr\}
\NN\\&&
+C_F T_F^2 \Biggl\{
\Biggl(\frac{23024568781}{44929695195}
+\frac{285046646}{324168075} L_\eta
+\frac{879808}{467775} L_\eta^2
\Biggr) \eta^3
+\Biggl(
\frac{2876423}{595350} L_\eta
\NN\\&&
+\frac{582667691}{75014100}
+\frac{27101}{9450} L_\eta^2
\Biggr) \eta^2
+\Biggl(
-\frac{59657237}{4134375}
+\frac{184214}{39375} L_\eta
+\frac{2228}{1125} L_\eta^2
\Biggr) \eta
\NN\\&&
+\Biggl(
\frac{1473641}{405000}
-\frac{18601}{4500} \left(L_2+L_1\right)
\Biggr)\zeta_2
+\frac{76621423}{8100000} L_2
-\frac{204611}{40500} L_1^3
+\frac{130207}{10125} \zeta_3
\NN\\&&
-\frac{37307959}{4860000} L_1
-\frac{18601}{6750} L_1 L_2^2
+\frac{530371}{162000} \left(L_2^2+L_1^2\right)
+\frac{18601}{6750} L_1^2 L_2
+\frac{442267}{101250} L_2 L_1
\NN\\&&
-\frac{130207}{40500} L_2^3
-\frac{33406758667}{2187000000}
\Biggr\}
+O\left(\eta^4 L_\eta^3\right)~,
\\
 \tilde{a}_{Qg}^{(3)}(N=6) &=&
 C_A T_F^2 \Biggl\{
 \Biggl(
-\frac{84840004938801319}{1381947564807810000}
-\frac{2287164970759}{15339633309000} L_\eta
-\frac{31340489}{68108040} L_\eta^2
\Biggr) \eta^3
\NN\\&&
+\Biggl(\frac{105157957}{360186750} L_\eta
+\frac{755537213056}{624023544375}
-\frac{49373}{103950} L_\eta^2
\Biggr) \eta^2
+\Biggl(\frac{832369820129}{29172150000}
\NN\\&&
+\frac{1406143531}{138915000} L_\eta
+\frac{112669}{1323000} L_\eta^2
\Biggr) \eta
+\Biggl(\frac{1316809}{79380}
+\frac{39248}{2205} \left(L_2+L_1\right)
\Biggr) \zeta_2
\NN\\&&
+\frac{11771644229}{388962000} L_2
+\frac{206404}{19845} L_1^3
-\frac{197648}{19845} \zeta_3
+\frac{83755534727}{1166886000} L_1
+\frac{78496}{6615} L_1 L_2^2
\NN\\&&
+\frac{2668087}{158760} \left(L_2^2+L_1^2\right)
+\frac{34166}{6615} L_1^2 L_2
+\frac{64117}{3969} L_2 L_1
+\frac{162074}{19845} L_2^3
\NN\\&&
+\frac{69882273800453}{735138180000}
\Biggr\}
\NN\\&&
+C_F T_F^2 \Biggl\{
\Biggl(
\frac{990283034941336}{2467763508585375}
+\frac{1255768040}{2191376187} L_\eta
+\frac{63929464}{42567525} L_\eta^2
\Biggr) \eta^3
\NN\\&&
+\Biggl(
\frac{11478584}{3361743} L_\eta
+\frac{524351089261}{97070329125}
+\frac{88972}{40425} L_\eta^2
\Biggr) \eta^2
+\Biggl(
-\frac{32427817736}{2552563125}
\NN\\&&
+\frac{64271512}{24310125} L_\eta
+\frac{376216}{231525} L_\eta^2
\Biggr) \eta
+\Biggl(
\frac{4784009}{4862025}
-\frac{55924}{15435} \left(L_2+L_1\right)
\Biggr) \zeta_2
\NN\\&&
+\frac{1786067629}{408410100} L_2
-\frac{615164}{138915} L_1^3
+\frac{223696}{19845} \zeta_3
-\frac{24797875607}{2042050500} L_1
-\frac{111848}{46305} L_1 L_2^2
\NN\\&&
+\frac{3232799}{9724050} \left(L_2^2+L_1^2\right)
+\frac{111848}{46305} L_1^2 L_2
+\frac{11119228}{4862025} L_2 L_1
-\frac{55924}{19845} L_2^3
\NN\\&&
-\frac{3161811182177}{142943535000}
\Biggr\}+O\left(\eta^4 L_\eta^3\right)~.
\end{eqnarray}

Finally, the gluonic contributions to the OMEs $A_{gq,Q}^{(3)}$ and
$A_{gg,Q}^{(3)}$ are given by
\begin{eqnarray}
\tilde{a}_{gq,Q}^{(3)}(N=2) &=&
 C_F T_F^2 \Biggl\{
 \Biggl(\frac{190500608}{281302875}
+\frac{747008}{893025} L_\eta
+\frac{1024}{2835} L_\eta^2
\Biggr) \eta^3
+\Biggl(\frac{7176352}{385875}
+\frac{64}{35} L_\eta^2
\NN\\&&
+\frac{33856}{3675} L_\eta
\Biggr) \eta^2
+\Biggl(\frac{12032}{225}
+\frac{512}{15} L_\eta
\Biggr) \eta
+\Biggl(\frac{832}{27}
+\frac{128}{3} \left(L_2+L_1\right)
\Biggr) \zeta_2
\NN\\&&
+\frac{128}{9} L_1^2 L_2
-\frac{512}{27} \zeta_3
+\frac{13600}{81} L_1
+\frac{1888}{27} L_2
+\frac{832}{27} L_2 L_1
+\frac{256}{9} L_1 L_2^2
\NN\\&&
+\frac{512}{27} L_2^3
+\frac{832}{27} \left(L_2^2+L_1^2\right)
+\frac{640}{27} L_1^3
+\frac{140128}{729}
\Biggr\}
+O\left(\eta^4 L_\eta^3\right)~,
\\
 \tilde{a}_{gq,Q}^{(3)}(N=4) &=&
 C_F T_F^2
 \Biggl\{
	 \Biggl(
	 \frac{261938336}{1406514375}
+\frac{1027136}{4465125} L_\eta
+\frac{1408}{14175} L_\eta^2
\Biggr) \eta^3
+\Biggl(
\frac{9867484}{1929375}
\NN\\&&
+\frac{88}{175} L_\eta^2
+\frac{46552}{18375} L_\eta
\Biggr) \eta^2
+\Biggl(\frac{16544}{1125}
+\frac{704}{75} L_\eta
\Biggr) \eta
+\Biggl(
\frac{2504}{675}
\NN\\&&
+\frac{176}{15} \left(L_2+L_1\right)
\Biggr) \zeta_2
+\frac{176}{45} L_1^2 L_2
-\frac{704}{135} \zeta_3
+\frac{436138}{10125} L_1
+\frac{54446}{3375} L_2
\NN\\&&
+\frac{2504}{675} L_2 L_1
+\frac{352}{45} L_1 L_2^2
+\frac{704}{135} L_2^3
+\frac{2504}{675} \left(L_2^2+L_1^2\right)
+\frac{176}{27} L_1^3
+\frac{18480197}{455625}
\Biggr\}
\NN\\&&
+O\left(\eta^4 L_\eta^3\right)~,
\\
 \tilde{a}_{gq,Q}^{(3)}(N=6) &=&
 C_F T_F^2 \Biggl\{
	 \Biggl(\frac{1047753344}{9845600625}
+\frac{4108544}{31255875} L_\eta
+\frac{5632}{99225} L_\eta^2
\Biggr) \eta^3
+\Biggl(
\frac{39469936}{13505625}
+\frac{352}{1225} L_\eta^2
\NN\\&&
+\frac{186208}{128625} L_\eta
\Biggr) \eta^2
+\Biggl(\frac{66176}{7875}
+\frac{2816}{525} L_\eta
\Biggr) \eta
+\Biggl(\frac{17632}{33075}
+\frac{704}{105} \left(L_2+L_1\right)
\Biggr) \zeta_2
\NN\\&&
+\frac{704}{315} L_1^2 L_2
-\frac{2816}{945} \zeta_3
+\frac{28089976}{1157625} L_1
+\frac{30801128}{3472875} L_2
+\frac{17632}{33075} L_2 L_1
\NN\\&&
+\frac{1408}{315} L_1 L_2^2
+\frac{2816}{945} L_2^3
+\frac{17632}{33075} \left(L_2^2+L_1^2\right)
+\frac{704}{189} L_1^3
+\frac{779635012}{40516875}
\Biggr\}
\NN\\&&
+O\left(\eta^4 L_\eta^3\right)~,
\end{eqnarray}
and
\begin{eqnarray}
 \tilde{a}_{gg,Q}^{(3)}(N=2) &=&
 C_A T_F^2 \Biggl\{
	 \Biggl(
	 \frac{19188592}{120558375}
+\frac{153892}{382725} L_\eta
+\frac{686}{1215} L_\eta^2
\Biggr) \eta^3
+\Biggl(
\frac{53824}{33075} L_\eta
+\frac{8433658}{3472875}
\NN\\&&
+\frac{296}{315} L_\eta^2
\Biggr) \eta^2
+\Biggl(
-\frac{153872}{10125}
-\frac{1412}{675} L_\eta
+\frac{14}{45} L_\eta^2
\Biggr) \eta
+\Biggl(
-\frac{556}{81}
\NN\\&&
-\frac{140}{9} \left(L_2+L_1\right)
\Biggr) \zeta_2
-\frac{214}{27} L_2
-\frac{628}{81} L_1^3
+\frac{272}{81} \zeta_3
-\frac{6682}{243} L_1
-\frac{280}{27} L_1 L_2^2
\NN\\&&
-\frac{550}{81} \left(L_2^2+L_1^2\right)
-\frac{176}{27} L_1^2 L_2
-\frac{568}{81} L_2 L_1
-\frac{524}{81} L_2^3
-\frac{71578}{2187}
\Biggr\}
\NN\\&&
+C_F T_F^2 \Biggl\{
	\Biggl(
-\frac{637103552}{843908625}
-\frac{4823552}{2679075} L_\eta
-\frac{20416}{8505} L_\eta^2
\Biggr) \eta^3
+\Biggl(
-\frac{752576}{99225} L_\eta
\NN\\&&
-\frac{116240192}{10418625}
-\frac{3904}{945} L_\eta^2
\Biggr) \eta^2
+\Biggl(\frac{702784}{30375}
-\frac{14336}{2025} L_\eta
-\frac{448}{135} L_\eta^2
\Biggr) \eta
+\Biggl(
-\frac{32}{27}
\NN\\&&
+\frac{64}{9} \left(L_2+L_1\right)
\Biggr) \zeta_2
-\frac{5024}{243} L_2
+\frac{704}{81} L_1^3
-\frac{1792}{81} \zeta_3
+\frac{736}{81} L_1
+\frac{128}{27} L_1 L_2^2
\NN\\&&
+\frac{112}{81} \left(L_2^2+L_1^2\right)
-\frac{128}{27} L_1^2 L_2
-\frac{512}{81} L_2 L_1
+\frac{448}{81} L_2^3
+\frac{4448}{243}
\Biggr\}
+O\left(\eta^4 L_\eta^3\right)~,
\NN\\&&
\\
 \tilde{a}_{gg,Q}^{(3)}(N=4) &=&
C_A T_F^2 \Biggl\{
	 \Biggl(
-\frac{311441927}{1687817250}
-\frac{1293167}{5358150} L_\eta
+\frac{1681}{34020} L_\eta^2
\Biggr) \eta^3
+\Biggl(
-\frac{205123}{99225} L_\eta
\NN\\&&
-\frac{36414571}{10418625}
-\frac{61}{189} L_\eta^2
\Biggr) \eta^2
+\Biggl(
-\frac{19131223}{303750}
-\frac{524233}{20250} L_\eta
-\frac{6943}{2700} L_\eta^2
\Biggr) \eta
\NN\\&&
+\Biggl(
-\frac{28979}{675}
-\frac{1558}{45} \left(L_2+L_1\right)
\Biggr) \zeta_2
-\frac{1057309}{13500} L_2
-\frac{1546}{81} L_1^3
+\frac{5992}{405} \zeta_3
\NN\\&&
-\frac{19874881}{121500} L_1
-\frac{3116}{135} L_1 L_2^2
-\frac{173999}{4050} \left(L_2^2+L_1^2\right)
-\frac{1588}{135} L_1^2 L_2
-\frac{86812}{2025} L_2 L_1
\NN\\&&
-\frac{6202}{405} L_2^3
-\frac{781640551}{3645000}
\Biggr\} 
\NN\\&&
+C_F T_F^2 
\Biggl\{
	\Biggl(
-\frac{22991704}{602791875}
+\frac{64016}{1913625} L_\eta
-\frac{2624}{6075} L_\eta^2
\Biggr) \eta^3
+\Biggl(
-\frac{988}{7875} L_\eta
\NN\\&&
-\frac{507478}{2480625}
-\frac{4}{9} L_\eta^2
\Biggr) \eta^2
+\Biggl(\frac{1088008}{253125}
+\frac{5456}{5625} L_\eta
-\frac{416}{1125} L_\eta^2
\Biggr) \eta
\NN\\&&
+\Biggl(\frac{14263}{10125}
+\frac{242}{225} \left(L_2+L_1\right)
\Biggr) \zeta_2
-\frac{173327}{202500} L_2
+\frac{2662}{2025} L_1^3
-\frac{6776}{2025} \zeta_3
+\frac{3676019}{607500} L_1
\NN\\&&
+\frac{484}{675} L_1 L_2^2
+\frac{31969}{20250} \left(L_2^2+L_1^2\right)
-\frac{484}{675} L_1^2 L_2
+\frac{2164}{2025} L_2 L_1
+\frac{1694}{2025} L_2^3
\NN\\&&
+\frac{314275147}{54675000}
\Biggr\}
+O\left(\eta^4 L_\eta^3\right)~,
\\
 \tilde{a}_{gg,Q}^{(3)}(N=6) &=&
C_A T_F^2 
\Biggl\{
	\Biggl(
-\frac{843352247}{4219543125}
-\frac{4137452}{13395375} L_\eta
+\frac{6431}{42525} L_\eta^2
\Biggr) \eta^3
+\Biggl(
-\frac{178615462}{52093125}
\NN\\&&
-\frac{1086976}{496125} L_\eta
-\frac{424}{1575} L_\eta^2
\Biggr) \eta^2
+\Biggl(
-\frac{435055073}{5315625}
-\frac{11624708}{354375} L_\eta
-\frac{11717}{3375} L_\eta^2
\Biggr) \eta
\NN\\&&
+\Biggl(
-\frac{5429062}{99225}
-\frac{13684}{315} \left(L_2+L_1\right)
\Biggr) \zeta_2
-\frac{2106313681}{20837250} L_2
-\frac{68252}{2835} L_1^3
+\frac{54064}{2835} \zeta_3
\NN\\&&
-\frac{1459416547}{6945750} L_1
-\frac{27368}{945} L_1 L_2^2
-\frac{54652}{2835} L_2^3
-\frac{13768}{945} L_1^2 L_2
-\frac{1808152}{33075} L_2 L_1
\NN\\&&
-\frac{3604631677201}{13127467500}
-\frac{120697}{2205} \left(L_2^2+L_1^2\right)
\Biggr\}
\NN\\&&
+C_F T_F^2 \Biggl\{
	\Biggl(
-\frac{91864096}{5907360375}
+\frac{979936}{3750705} L_\eta
-\frac{130048}{297675} L_\eta^2
\Biggr) \eta^3
+\Biggl(
-\frac{41008}{2701125}
\NN\\&&
+\frac{10816}{46305} L_\eta
-\frac{384}{1225} L_\eta^2
\Biggr) \eta^2
+\Biggl(\frac{2133088}{826875}
+\frac{3616}{3375} L_\eta
-\frac{1216}{11025} L_\eta^2
\Biggr) \eta
\NN\\&&
+\Biggl(\frac{197492}{231525}
+\frac{968}{2205} \left(L_2+L_1\right)
\Biggr) \zeta_2
+\frac{413083}{24310125} L_2
+\frac{10648}{19845} L_1^3
-\frac{3872}{2835} \zeta_3
\NN\\&&
+\frac{281801489}{72930375} L_1
+\frac{1936}{6615} L_1 L_2^2
+\frac{968}{2835} L_2^3
-\frac{1936}{6615} L_1^2 L_2
+\frac{81176}{99225} L_2 L_1
\NN\\&&
+\frac{14596284331}{5105126250}
+\frac{604598}{694575} \left(L_2^2+L_1^2\right)
\Biggr\}
+O\left(\eta^4 L_\eta^3\right).
\end{eqnarray}

In Table~1 we illustrate the ratio of the constant parts of the unrenormalized 3-loop two-mass OMEs
$\tilde{a}_{Qq(Qg,gg)}^{(3)}/\tilde{a}_{qq,Q}^{{\sf NS},(3)}$ for the fixed moments $N = 2, 4, 6$ as a function of the virtuality
$\mu^2 = 20, 100, 500$ and $1000~\GeV^2$ referring to $\eta = m_c^2/m_b^2$ and the values in 
Eqs.~(\ref{eq:mc}, \ref{eq:mb}) for the heavy quark masses. In Section~\ref{SSec-NS2MASS} we calculate 
$\tilde{a}_{qq,Q}^{{\sf NS},(3)}$ and $\tilde{a}_{gq}^{(3)}$ for general 
values of $N$. Therefore, these ratios can be used as a first estimate for these OMEs in case the two-mass contribution is only known 
for some moments.

The ratio $\tilde{a}_{Qq}^{{\sf PS},(3)}/\tilde{a}_{qq,Q}^{{\sf NS},(3)}$ is widely constant over the range $\mu^2 = 20 ... 
1000 ~\GeV^2$ and becomes smaller for larger moments. In the case of $\tilde{a}_{Qg(gg,Q)}^{(3)}/\tilde{a}_{qq,Q}^{{\sf NS},(3)}$ at low 
scales $\mu^2 \approx 20~\GeV^2$ larger ratios are obtained. They flatten out with values of $\mu^2 = 100~\GeV^2$ and larger.
Again, the ratios become smaller for larger values of $N$. This also applies to the ratio 
$\tilde{a}_{gq,Q}^{(3)}/\tilde{a}_{qq,Q}^{{\sf NS},(3)}$, starting from $\mu^2 = 100~\GeV^2$, with somewhat larger values at 
$\mu^2 = 20~\GeV^2$.

In order to obtain the results shown above, we have expanded the constant parts of the 3-loop unrenormalized OMEs for fixed even integer values of $N$.
This is a valid representation for some but not for all of the OMEs also at general values of $N$, as is shown in 
Sections~\ref{SSec-NS2MASS} and \ref{SSec-Agg2M}. In case the expansion in $\eta$ exists, one might try to reconstruct the 
$\eta$-expanded solution from the moments using guessing methods \cite{GUESS}, which have been successfully applied in other cases 
\cite{Blumlein:2009tj,Ablinger:2014yaa}. However, many more moments are needed in this case. They cannot be provided 
using {\tt Q2e}/{\tt Exp} \cite{Harlander:1997zb,Seidensticker:1999bb}, and usually require at least the analytic solution 
of part of the integrals and possibly generating function methods \cite{Ablinger:2014yaa,Brown:2008um}.\footnote{Recently,
a method has been found \cite{Blumlein:2017dxp} to generate large number 
of Mellin moments turning the integration-by-parts relations for the corresponding problem into difference 
equations. In this way it is possible to obtain $O(8000)$ moments in a massive 3-loop problem. The 
corresponding file amounts to more than 1~Gbyte.}
\renewcommand{\arraystretch}{1.3}
\begin{center}
\begin{table}[H]\centering
\begin{minipage}[c]{0.4\linewidth}
\begin{tabular}{|r||r|r|r|}
\hline
\multicolumn{4}{|c|}{$\tilde{a}_{Qq}^{{\sf PS},(3)}/\tilde{a}_{qq,Q}^{{\sf NS},(3)}$}\\
\hline
\hline
  \multicolumn{1}{|c||}{$\mu^2/\GeV^2$}
& \multicolumn{1}{c}{$N=2$}
& \multicolumn{1}{|c}{$N=4$}
& \multicolumn{1}{|c|}{$N=6$}\\
\hline
\hline
20      &  \phantom{-3}2.195  &  \phantom{-}0.197 &  \phantom{-}0.066 \\
100     &  2.110  &  0.178 &   0.058 \\
500     &  2.075  &  0.170 &   0.055 \\
1000    &  2.066  &  0.168 &   0.055 \\
\hline
\end{tabular}
\end{minipage}
\hspace*{4mm}
\begin{minipage}[c]{0.4\linewidth}
\begin{tabular}{|r||r|r|r|}
\hline
\multicolumn{4}{|c|}{$\tilde{a}_{Qg}^{(3)}/\tilde{a}_{qq,Q}^{{\sf NS},(3)}$}\\
\hline
\hline
  \multicolumn{1}{|c||}{$\mu^2/\GeV^2$}
& \multicolumn{1}{c}{$N=2$}
& \multicolumn{1}{|c}{$N=4$}
& \multicolumn{1}{|c|}{$N=6$}\\
\hline
\hline
20      &-48.563  & -5.835 &  -3.126 \\
100     & -2.351  & -1.395 &  -0.935 \\
500     & -2.254  & -1.427 &  -0.967 \\
1000    & -2.225  & -1.433 &  -0.974 \\
\hline 
\end{tabular}
\end{minipage}

\vspace*{5mm}
\begin{minipage}[c]{0.4\linewidth}
\begin{tabular}{|r||r|r|r|}
\hline
\multicolumn{4}{|c|}{$\tilde{a}_{gg,Q}^{(3)}/\tilde{a}_{qq,Q}^{{\sf NS},(3)}$}\\
\hline 
\hline 
  \multicolumn{1}{|c||}{$\mu^2/\GeV^2$}
& \multicolumn{1}{c}{$N=2$}
& \multicolumn{1}{|c}{$N=4$}
& \multicolumn{1}{|c|}{$N=6$}\\
\hline 
\hline 
20      & \phantom{-}58.777  & 29.890 &  19.795 \\
100     &  1.989  &  2.299 &   2.276 \\
500     &  2.005  &  2.467 &   2.433 \\
1000    &  2.012  &  2.505 &   2.467 \\
\hline 
\end{tabular}
\end{minipage}
\hspace*{4mm}
\begin{minipage}[c]{0.4\linewidth}
\begin{tabular}{|r||r|r|r|}
\hline
\multicolumn{4}{|c|}{$\tilde{a}_{gq,Q}^{(3)}/\tilde{a}_{qq,Q}^{{\sf NS},(3)}$}\\
\hline 
\hline 
  \multicolumn{1}{|c||}{$\mu^2/\GeV^2$}
& \multicolumn{1}{c}{$N=2$}
& \multicolumn{1}{|c}{$N=4$}
& \multicolumn{1}{|c|}{$N=6$}\\
\hline 
\hline 
20      & \phantom{4}-3.195  & -0.526 &  -0.254 \\
100     & -3.110  & -0.479 &  -0.223 \\
500     & -3.075  & -0.460 &  -0.211 \\
1000    & -3.066  & -0.456 &  -0.208 \\
\hline
\end{tabular}
\end{minipage}
\caption[]{\sf Ratios of the fixed moments for $\tilde{a}_{Qg}^{(3)}$, $\tilde{a}_{gg,Q}^{(3)}$ and $\tilde{a}_{gq,Q}^{(3)}$ to 
$\tilde{a}_{qq,Q}^{{\sf NS}(3)}$ as a function of $Q^2$ and $N$.}
\end{table}
\renewcommand{\arraystretch}{1.0}
\end{center}
\section{The Non--Singlet and \boldmath $gq$-Contributions at general Values of $N$}{\label{SSec-NS2MASS}}

\begin{figure}[H]
\[
\raisebox{-15mm}{\includegraphics[keepaspectratio = true, scale = 0.8] {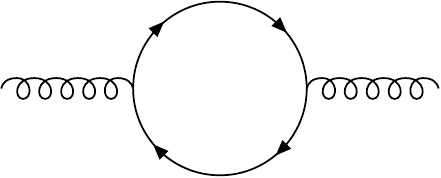}}
\qquad
\begin {aligned}
&\\
&\\
&\\
&\hat{=}~a_s T_F \frac{4}{\pi} \left(4 \pi\right)^{-\ep/2} \left(k_{\mu}k_{\nu}-k^2 g_{\mu\nu} \right)
\\&
\times
\int_{-i\,\infty}^{+i\,\infty} d \sigma
\left(\frac{m^2}{\mu^2}\right)^{\sigma} \left(-k^2\right)^{\ep/2 - \sigma}
\frac{\Gamma(\sigma-\ep/2) \Gamma^2(2-\sigma+\ep/2) \Gamma(-\sigma)}{\Gamma(4-2 \sigma+\ep)}
\end {aligned}
\]
\caption{\sf \label{fig:MassiveInsertionMB} One of the massive fermion loop insertion is effectively rendered massless via a Mellin--Barnes 
representation.}
\end{figure}

\vspace*{1mm}
\noindent
All non--singlet diagrams at 3--loop order contain two massive fermion bubbles.
One of these may be rendered effectively massless by using the Mellin--Barnes
representation\cite{MB1a,MB1b,MB2,MB3,MB4}, see Figure
{\ref{fig:MassiveInsertionMB}}. This yields similar integrals as in the case
with one massive and one massless fermionic line \cite{Ablinger:2010ty}. 

One may now introduce a Feynman parameter representation, integrate the momenta
and perform the Feynman parameter integrals in terms of Euler Beta--functions 
\begin{equation}
B(a,b) = \int_0^1 dx x^{a-1} (1-x)^{b-1} =
\frac{\Gamma(a) \Gamma(b)}{\Gamma(a+b)}.
\end{equation}

\vspace*{-2.7cm}

\noindent
The remaining contour integral is then of the general form
\begin{align}
I&\propto \Gamma \Brack{f_1(\ep,N),\ldots,f_i(\ep,N)}{f_{i+1}(\ep,N),\ldots,f_I(\ep,N)}
\int_{-i\,\infty}^{+i\,\infty} \hspace{-6mm} d \xi ~
\Gamma \Brack{g_1(\ep)+\xi,g_2(\ep)+\xi,g_3(\ep)+\xi,g_4(\ep)
-\xi,g_5(\ep)-\xi}{g_6(\ep)+\xi,g_7(\ep)-\xi}
\eta^{\xi}~, {\label{NS2MASSContInt}}
\end{align}
where the $f_j$ and the $g_j$ are linear functions. Furthermore, the notation
\begin{eqnarray}
 \Gamma \Brack{a_1,\ldots,a_i}{b_1,\ldots,b_j} &=
 {\displaystyle \frac{\Gamma(a_1)\cdots\Gamma(a_i)}{\Gamma(b_1)\cdots\Gamma(b_j)}}
\end{eqnarray}
is applied.
After closing the contour in (\ref{NS2MASSContInt}) and collecting the residues
a linear combination of generalized hypergeometric $_4F_3$--functions\cite{HYP} is obtained
\begin{eqnarray}
 I=\sum_j C_j\left(\ep,N\right){_4F_3}\Biggl[\genfrac{}{}{0pt}{}{a_1(\ep),a_2(\ep),a_3(\ep),a_4(\ep)}{b_1(\ep),b_2(\ep),b_3(\ep)} 
,\eta\Biggr]~.
\end{eqnarray}
For the flavor non-singlet (${\sf NS}$) contributions, and for $A^{(3)}_{gq}$, the arguments of the hypergeometric $_PF_Q$-function
are completely independent of the Mellin variable $N$, and each term factorizes
into contributions that describe the operator insertions and the generalized
hypergeometric functions covering the mass structure of the diagrams.  Due to
the fact that the parameters of the hypergeometric functions depend on the
dimensional regularization parameter $\ep$ only, their respective expansion may be
performed with the code {\tt HypExp 2}\cite{Huber:2007dx}. The results of these
expansions are then given in terms of the following (poly)logarithmic functions
\cite{POLYLOG1, POLYLOG2,Devoto:1983tc, Ablinger:2013jta}, 
\begin{equation}
\left\{\ln(\eta),~\ln\left(\frac{1-\eta_1}{1+\eta_1}\right),~\Li_2\left({\eta_1}\right),~
\Li_2\left(\eta\right),~\Li_3\left({\eta_1}\right)\right\}~.
\end{equation}
The pre-factor $C_j\left(\ep,N\right)$ may contain a sum stemming from the
operator insertion on the vertex, see Section~8.1~\cite{Bierenbaum:2009mv}.  This sum
is easily evaluated in terms of single harmonic sums using the summation package {\tt Sigma} \cite{SIG1,SIG2}
Applying these methods we calculate the two-mass contributions in the flavor non-singlet cases and for the OME $A_{gq}^{(3)}$. 

In the following, we denote by $\tilde{\tilde{a}}_{ij}$ the remainder constant part of the genuine 
two-mass term of the unrenormalized matrix element, omitting the terms coming from the expansion of the 
factor $(m_1 m_2/\mu^2)^{3\varepsilon/2}$ in the OMEs for brevity, cf.~(\ref{AhathatDecomp}), i.e. the 
terms $\propto L_1, L_2$, which are given in the remainder part of the OMEs $\tilde{A}_{ij}$ instead 
together
with other the terms of this kind. The expressions for $\tilde{\tilde{a}}_{ij}$ depend on $\eta$ and 
are symmetric under the interchange
\begin{equation}
\label{eq:symet}
\eta  \leftrightarrow \frac{1}{\eta}.
\end{equation}
Also the OMEs $\tilde{A}_{ij}$ are symmetric in interchanging $m_1 \leftrightarrow m_2$. One may 
furthermore check, calculating $\tilde{\tilde{a}}_{ij}(N)$ for $N = 2,4,6$ and expanding in $\eta = 
m_c^/m_b^2 < 1$ up to $O(\eta^3)$ that the values for the fixed moments in the corresponding parts of 
$\tilde{a}_{ij}(N)$ are obtained. The latter ones do not obey the symmetry interchanging the masses 
anymore, since we have chosen to expand for $\eta < 1$. To obtain the representations in 
Section~\ref{sec:x1} $L_\eta$ is given by $-\ln(\eta)$ expanding the expressions in the remainder part 
of this section or the $z$-space expressions given in Appendix~\ref{APP1}. Since the present expressions 
obey the symmetry (\ref{eq:symet}) a choice has to be made.  
 
For the single mass contributions the different OMEs receive a 3-loop correction changing from the on shell mass $m$
to the $\overline{\sf MS}$ mass expanding the OME in $a_s^{\overline{\sf MS}}$. For the two-mass contributions at 3-loop
order this is not the case for $A_{qq,Q}^{(3),\rm NS}, A_{qq,Q}^{(3),\rm NS,TR}, A_{Qq}^{(3),\rm PS}$ and $A_{qg,Q}^{(3)}$.
Terms of this kind appear in case of the genuine two-mass contributions to $A_{Qg}^{(3)}$ and $A_{gg,Q}^{(3)}$ (see also 
Eqs.~(\ref{eq:aqg2a}, \ref{eq:aqg2b}, \ref{eq:aggQ2a}, \ref{eq:aggQ2b}). They are not dealt with in the present paper.
\subsection{The flavor non-singlet contribution}

\vspace*{1mm}
\noindent
The general pole structure for the unrenormalized two-mass contribution to the
OME $A_{qq,Q}^{\sf NS}$ is given in Eq.~(\ref{AthhNS3}). The only contribution which is
not determined by the renormalization prescription is the constant part, for which we obtain
\begin{eqnarray}
\tilde{a}_{qq,Q}^{(3), \rm{NS}} &=&
 C_F T_F^2 \Biggl\{
 \left(\frac{4}{9} S_1-\frac{3 N^2+3 N+2}{9 N (N+1)}\right) \Biggl[
-24 \big(L_1^3+L_2^3+\left(L_1 L_2+2 \zeta_2+5\right) \left(L_1+L_2\right)\big)
\nonumber\\&&
+\frac{\eta+1}{\eta^{3/2}} \left(5 \eta^2+22 \eta+5\right) \biggl(
-\frac{1}{4} \ln^2(\eta) \ln\left(\frac{1+\eta_1}{1-\eta_1}\right)
+2 \ln(\eta) \Li_2\left(\eta_1\right)
-4 \Li_3\left(\eta_1\right)
\biggr)
\nonumber\\&&
+\frac{\left(\eta_1+1\right)^2}{2 \eta^{3/2}} \left(-10 \eta^{3/2}+5 \eta^2+42 \eta-10 \eta_1+5\right) \big[
\Li_3\left(\eta\right)-\ln(\eta) \Li_2\left(\eta\right)\big]
+\frac{64}{3} \zeta_3
\nonumber\\&&
+\frac{8}{3} \ln^3(\eta)
-16 \ln^2(\eta) \ln(1-\eta)
+10 \frac{\eta^2-1}{\eta} \ln(\eta)
\Biggr]
+\frac{16 \left(405 \eta^2-3238 \eta+405\right)}{729 \eta} S_1
\nonumber\\&&
+\frac{4}{3} \left(\frac{3 N^4+6 N^3+47 N^2+20 N-12}{3 N^2 (N+1)^2}-\frac{40}{3} S_1+8 S_2\right) \left[\frac{4}{3} \zeta_2+(L_1+L_2)^2\right]
\nonumber\\&&
+\frac{8}{9} \left(\frac{130 N^4+84 N^3-62 N^2-16 N+24}{3 N^3 (N+1)^3}-\frac{52}{3} S_1+\frac{80}{3} S_2-16 S_3\right) \left(L_1+L_2\right)
\nonumber\\&&
+\biggl[-\frac{R_1}{18 N^2 (N+1)^2 \eta}
+\frac{2 \left(5 \eta^2+2 \eta+5\right)}{9 \eta} S_1
+\frac{32}{9} S_2\biggr] \ln^2(\eta)
-\frac{4 R_2}{729 N^4 (N+1)^4 \eta}
\nonumber\\&&
+\frac{3712}{81} S_2
-\frac{1280}{81} S_3
+\frac{256}{27} S_4
\Biggr\}~.
 \label{eq:aqqNSQ2M}
\end{eqnarray}
Here $S_{\vec{a}} \equiv S_{\vec{a}}(N)$ denote the (nested) harmonic sums \cite{HSUM}
\begin{eqnarray}
S_{b,\vec{a}}(N) = \sum_{k=1}^N \frac{({\rm sign}(b))^k}{k^{|b|}} S_{\vec{a}}(k),~~~S_\emptyset = 1,~~~b,a_i \in \mathbb{Z} 
\backslash \{0\}~.
\end{eqnarray}
The polynomials $R_i$ read
\begin{eqnarray}
 R_1&=&15 \eta^2 N^4+78 \eta N^4+15 N^4+30 \eta^2 N^3+156 \eta N^3+30 N^3+25 \eta^2 N^2+18 \eta N^2+25 N^2
 \nonumber\\&&+10 \eta^2 N+4 \eta N+10 N+32 \eta~,
 \\
 R_2&=&1215 \eta^2 N^8-1596 \eta N^8+1215 N^8+4860 \eta^2 N^7-6384 \eta N^7+4860 N^7+8100 \eta^2 N^6
 \nonumber\\&&
 -25844 \eta N^6+8100 N^6
 +7290 \eta^2 N^5-39348 \eta N^5+7290 N^5+3645 \eta^2 N^4-20304 \eta N^4
 \nonumber\\&&
 +3645 N^4+810 \eta^2 N^3-140 \eta N^3+810 N^3+432 \eta N^2+288 \eta N+864 \eta~.
\end{eqnarray}
The two-mass part of the renormalized OME $A_{qq,Q}^{(3),\sf NS}$ is given by
\begin{eqnarray}
\tilde{A}_{qq,Q}^{(3),\sf NS} &=& 
        C_F T_F^2 \Biggl\{
                -
                \frac{4 R_3}{243 N^4 (N+1)^4}
               +\biggl[
                        -\frac{8 R_4}{81 N^3 (N+1)^3}
                        +\biggl(
                                -\frac{16 \big(
                                        3 N^2+3 N+2\big)}{3 N (N+1)}
\nonumber\\&&
                                +\frac{64}{3} S_1
                        \biggr) \zeta_2
                        +\frac{3584}{81} S_1
                        -\frac{640}{27} S_2
                        +\frac{128}{9} S_3
                \biggr] \left(L_1+L_2\right)
                +\biggl[
                         \frac{640}{27} S_1
                        -\frac{128}{9} S_2
\nonumber\\&&
                        -\frac{16 \left(3 N^4+6 N^3+47 N^2+20 N-12\right)}{27 N^2 (N+1)^2}
                \biggr] \left(L_1^2+L_2^2+\zeta_2\right)
\nonumber\\&&
                +\biggl[
                        -\frac{16 \big(
                                3 N^2+3 N+2\big)}{9 N (N+1)}
                        +\frac{64}{9} S_1
                \biggr] \left(\frac{4}{3} L_1^3+L_1^2 L_2+L_2^2 L_1+\frac{4}{3} L_2^3+\frac{4}{3} \zeta_3\right)
\nonumber\\&&
                +\frac{20992}{243} S_1
                -\frac{3584}{81} S_2
                +\frac{640}{27} S_3
                -\frac{128}{9} S_4
		\Biggr\}+\tilde{a}_{qq,Q}^{(3), \rm{NS}}.
\label{Atil:N}
\end{eqnarray}
with
\begin{eqnarray}
R_3&=&1551 N^8+6204 N^7+15338 N^6+17868 N^5+8319 N^4+944 N^3
\nonumber\\&&
+528 N^2-144 N-432~,
\\
R_4&=&219 N^6+657 N^5+1193 N^4+763 N^3-40 N^2-48 N+72~.
\end{eqnarray}
Both the constant part of the unrenormalized two-mass OME (\ref{eq:aqqNSQ2M}) and the OME (\ref{Atil:N}) vanish for $N = 1$
due to fermion number conservation for any value of the heavy quark masses. In the Appendix we present the the corresponding 
$z$-space expressions for Eqs.~(\ref{eq:aqqNSQ2M}) and (\ref{Atil:N}). The analytic continuation of the $N$-space result may also be
obtained by expressing the contributing sums in the asymptotic region $|N| \rightarrow \infty$ and using their 
recurrence relations, cf.~\cite{STRUCT}.
One may derive semi-numeric representations, cf.~\cite{ANCONT}. The inversion to $z$-space
is then done by a contour integral around the singularities of the problem.
\begin{figure}[H]
\centering
\includegraphics[width=0.8\textwidth]{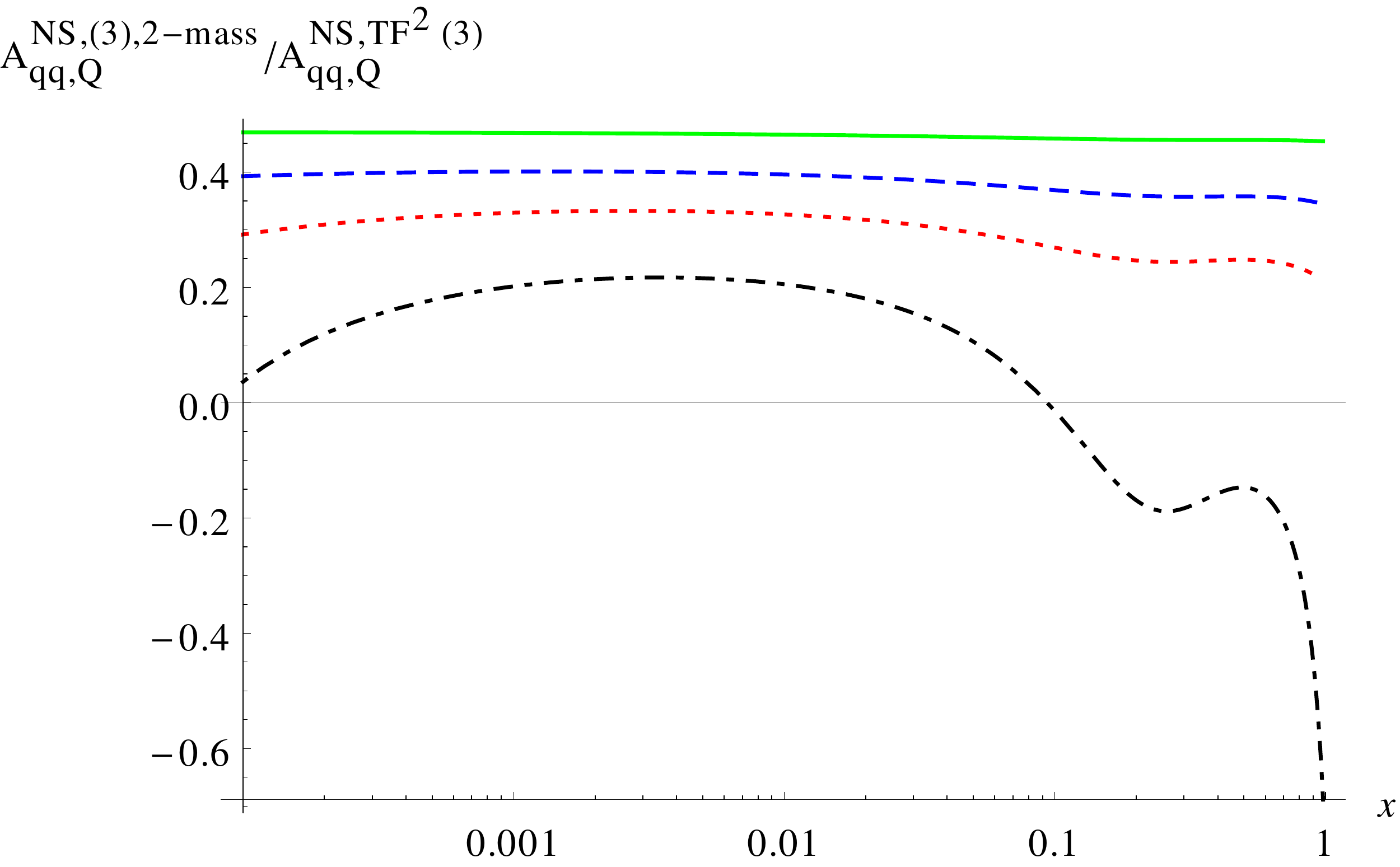}
\caption{\label{FIG:NS}
\sf \small The ratio of the genuine 2-mass contributions to $A_{qq,Q}^{(3), \rm NS}$ to the complete $T_F^2$-part of
massive 3-loop OME $A_{qq,Q}^{(3), \rm NS}$ as a function of $x$ and $Q^2$, for $m_c = 1.59~\GeV, m_b = 4.78~\GeV$ 
in the on-shell scheme.
Dash-dotted line:  $\mu^2 =   30~\GeV^2$;
Dotted line:  $\mu^2 =   50~\GeV^2$;
Dashed line:  $\mu^2 =   100~\GeV^2$;
Full line:  $\mu^2 =   1000~\GeV^2$.
The single mass contributions are given in Ref.~\cite{Ablinger:2014vwa}.}
\end{figure}
In Figure~\ref{FIG:NS} we show the ratio of the genuine 3-loop 2-mass contributions to 
$A_{qq,Q}^{(3), \rm NS}$ to the complete $T_F^2$-contribution for both masses for a range in $x$ at typical values of 
$Q^2$.
The impact of the 2-mass contribution grows with $Q^2$. At lower values of $Q^2$ it takes negative values in the
large $x$ region and at higher values of $Q^2$ it behaves almost flat. Here we illustrate the contribution to the OMEs
only. The contributions to the deep-inelastic structure functions will be given elsewhere.
\subsection{The transversity contribution}

\vspace*{1mm}
\noindent
The pole structure of the unrenormalized transversity OME corresponds to the one in Eq.~(\ref{AthhNS3}) after substituting 
the anomalous dimensions $\gamma_{qq}^{\sf NS} \rightarrow \gamma_{qq}^{\sf NS, trans}$. The constant contribution is given by
!!

\begin{eqnarray}
\tilde{a}_{qq,Q}^{(3),\rm{NS},\rm TR}&=&
C_F T_F^2 \Biggl\{
\left(4 S_1-3\right) \biggl[
-\frac{8}{3} \biggl(L_1^3+L_2^3+\biggl(L_1 L_2+2 \zeta_2+\frac{58}{9}\biggr) \left(L_1+L_2\right)\biggr)
\nonumber\\&&
+\frac{\left(\eta_1+1\right)^2}{18 \eta^{3/2}} \left(5 \eta^2-10 \eta^{3/2}+42 \eta-10 \eta_1+5\right) \big[
\Li_3(\eta)-\ln(\eta) \Li_2(\eta)\big]
\nonumber\\&&
+\frac{\eta+1} {9 \eta^{3/2}} \left(5 \eta^2+22 \eta+5\right) \biggl(
\frac{1}{4} \ln^2(\eta) \ln\left(\frac{1-\eta_1}{1+\eta_1}\right)
+2 \ln(\eta) \Li_2\left(\eta_1\right)
\nonumber\\&&
-4 \Li_3\left(\eta_1\right)
\biggr)
+\frac{8}{27} \ln^3(\eta)
-\frac{16}{9} \ln^2(\eta) \ln(1-\eta)
+10 \frac{\eta^2-1}{9 \eta} \ln(\eta)
\biggr]
\nonumber\\&&
+\frac{32}{9} \biggl(3 S_2-5 S_1+\frac{3}{8}\biggr) \left(L_1+L_2\right)^2
-\frac{64}{9} \zeta_3
+\frac{16}{9} \zeta_2
+\frac{128}{9} \left(\zeta_2+\frac{29}{9}\right) S_2
\nonumber\\&&
-\frac{1280}{81} S_3
+\frac{256}{27} S_4
+8 \biggl[-\frac{13 N^2+13 N-8}{9 N (N+1)}+\frac{80}{27} S_2-\frac{16}{9} S_3\biggr] \left(L_1+L_2\right)
\nonumber\\&&
+\frac{16}{27} \left(\frac{405 \eta^2-3238 \eta+405}{27 \eta}+16 \zeta_3-40 \zeta_2\right) S_1
-\frac{4 R_5}{243 N^2 (N+1)^2 \eta}
\nonumber\\&&
+\biggl[
-\frac{(\eta+5) (5 \eta+1)}{6 \eta}
+\frac{2 \left(5 \eta^2+2 \eta+5\right)}{9 \eta} S_1
+\frac{32}{9} S_2
\biggr] \ln^2(\eta)
\Biggr\}~,
\label{eq:aNSTR1}
\end{eqnarray}
with
\begin{eqnarray}
 R_5&=&405 \eta^2 N^4-532 \eta N^4+405 N^4+810 \eta^2 N^3-1064 \eta N^3
 +810 N^3+405 \eta^2 N^2
 \N\\&&
 -1012 \eta N^2
 +405 N^2+96 \eta N+288 \eta~.
\end{eqnarray}
The two-mass part of the renormalized OME $A_{qq,Q}^{(3),\sf NS, TR}$ reads
\begin{eqnarray}
\tilde{A}_{qq,Q}^{(3),\sf NS, TR} &=& 
C_F T_F^2 \Biggl\{
        -\frac{4 R_7}{81 N^2 (N+1)^2}
        +\biggl[
                -\frac{8 R_6}{27 N (N+1)}
                +\frac{3584}{81} S_1
                -\frac{640}{27} S_2
\nonumber\\&&
                +\frac{128}{9} S_3
                +\biggl(
                        -16
                        +\frac{64}{3} S_1
                \biggr) \zeta_2
        \biggr] \left(L_1+L_2\right)
        +\frac{20992}{243} S_1
        -\frac{3584}{81} S_2
        +\frac{640}{27} S_3
\nonumber\\&&   
        -\frac{128}{9} S_4
        +\biggl[
                -\frac{16}{3}
                +\frac{64}{9} S_1
        \biggr] \left(\frac{4}{3} L_1^3+L_1^2 L_2+L_2^2 L_1+\frac{4}{3} L_2^3+\frac{4}{3} \zeta_3\right)
\nonumber\\&&  
        +\biggl[
                -\frac{16}{9}
                +\frac{640}{27} S_1
                -\frac{128}{9} S_2   
        \biggr] \left(L_1^2+L_2^2+\zeta_2\right)
\Biggr\}+\tilde{a}_{qq,Q}^{(3),\rm{NS},\rm TR}
\label{eq:ANSTR1}
\end{eqnarray}
with
\begin{eqnarray}
R_6&=&73 N^2+73 N+24~,
\\
R_7&=&517 N^4+1034 N^3+757 N^2-48 N-144~.
\end{eqnarray}
The corresponding expressions for (\ref{eq:aqqNSQ2M},\ref{Atil:N}, \ref{eq:aNSTR1}, \ref{eq:ANSTR1}) 
in $z$-space are given in Appendix~\ref{APP1}.

As before in the equal mass case \cite{Ablinger:2014vwa} and for the $O(N_F T_F^2)$ 
contributions 
\cite{Ablinger:2010ty}, we obtain the $O(T_F^2 C_{A,F})$ terms of the 3-loop flavor non-singlet
contributions to the anomalous dimensions in the vector and transversity case from the single pole terms of the 
unrenormalized non-singlet OMEs, confirming once again the result in \cite{Moch:2004pa},
see also \cite{Velizhanin:2012nm}.
\begin{figure}[H]
\centering
\includegraphics[width=0.8\textwidth]{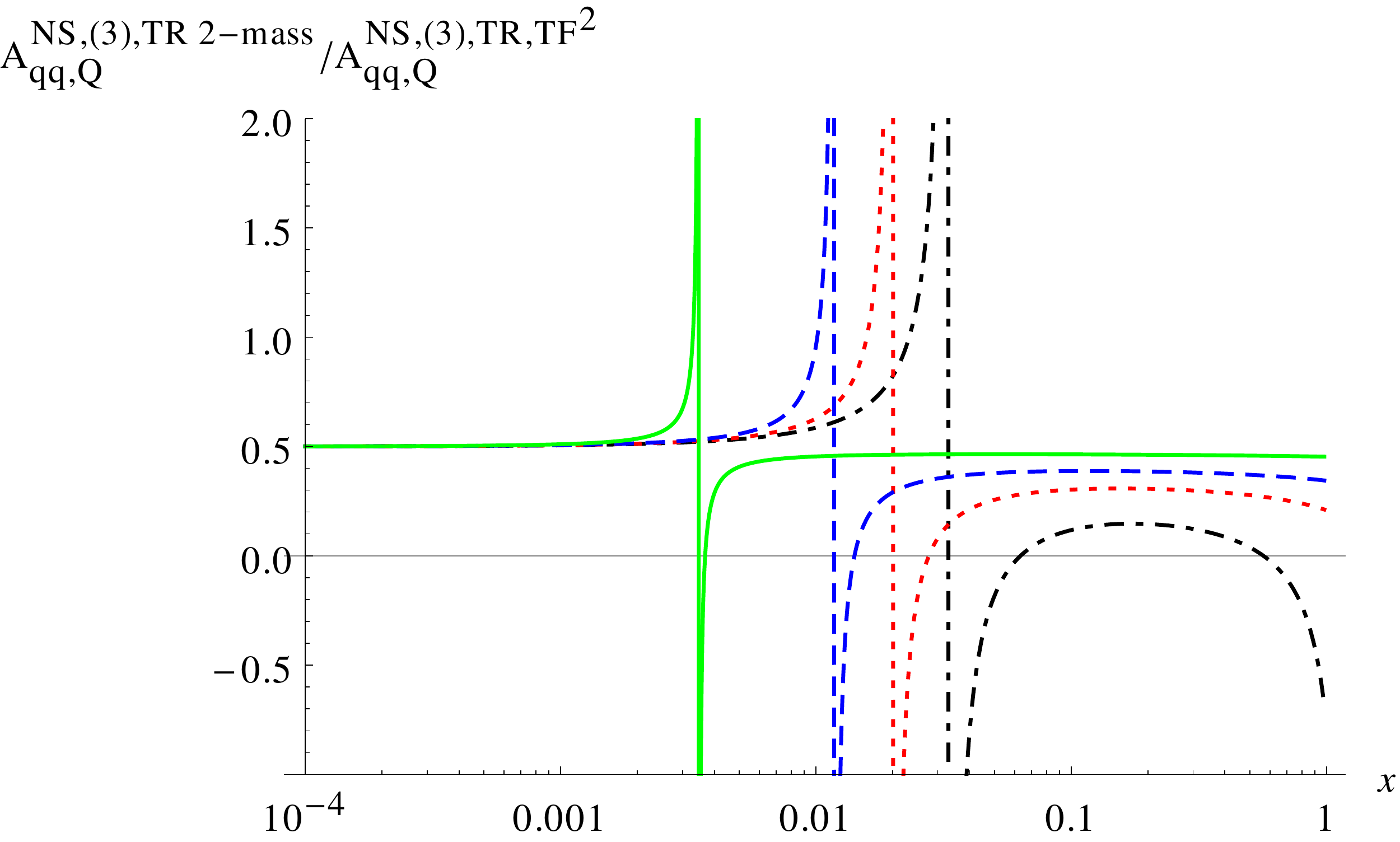}
\caption{\label{FIG:TR}
\sf \small The ratio of the genuine 2-mass contributions to $A_{qq,Q}^{(3), \rm NS, TR}$ to the complete
$T_F^2$-part of the massive 3-loop corrections to
$A_{qq,Q}^{(3), \rm NS, TR}$ as a function of $x$ and $Q^2$, for $m_c = 1.59~\GeV, m_b = 4.78~\GeV$ in
the
on-shell scheme.
Dash-dotted line:  $\mu^2 =   30~\GeV^2$;
Dotted line:  $\mu^2 =   50~\GeV^2$;   
Dashed line:  $\mu^2 =   100~\GeV^2$;
Full line:  $\mu^2 =   1000~\GeV^2$.
The single mass contributions are given in Ref.~\cite{Ablinger:2014vwa}.}
\end{figure}
In Figure~\ref{FIG:TR} we show the ratio of the genuine 2-mass contributions to the complete $T_F^2$ 3-loop term for 
transversity as a function of $x$ and $Q^2$. The spikes are due to a zero in the denominator of this ratio. Except for a
small region of $x$ around these spikes, the ratio takes values between 1.5 and -0.6. For $Q^2$ not too low, mostly values 
between 0 and 0.6 are obtained.

\subsection{The \boldmath $gq$-contribution}

\vspace*{1mm}
\noindent
The genuine two-mass contributions to the OME $A_{gq,Q}^{(3)}$ can be calculated in a similar way to $A_{qq,Q}^{{\sf 
NS},(3)}$. One obtains the constant part of the unrenormalized OME
\begin{eqnarray}
\tilde{a}_{gq,Q}^{(3)}&=&C_F T_F^2 \Biggl\{
p_{gq}^{(0)} \biggl[
16 \biggl(L_1^3+L_2^3+\left(L_1 L_2+2 \zeta_2+\frac{26}{3}\right) (L_1+L_2)\biggr)
\nonumber\\&&
-\frac{4}{3 \eta^{3/2}} \left(
\big(\sqrt{\eta }+1\big)^2 R_8  \Li_3(-\eta_1)
-\big(\sqrt{\eta }-1\big)^2 R_9 \Li_3(\eta_1)
\right)
-\frac{16}{9} \ln^3(\eta)
\nonumber\\&&
+\biggl(
\frac{2 \big(\sqrt{\eta }+1\big)^2}{3 \eta^{3/2}} R_8 \Li_2(-\eta_1)
-\frac{2 \big(\sqrt{\eta }-1\big)^2}{3 \eta^{3/2}} R_9  \Li_2(\eta_1)
-\frac{20}{3 \eta} \left(\eta^2-1\right)
\biggr) \ln(\eta)
\nonumber\\&&
+\biggl(
\frac{\big(\sqrt{\eta }+1\big)^2}{6 \eta^{3/2}} R_8 \ln(1+\eta_1)
-\frac{\big(\sqrt{\eta }-1\big)^2}{6 \eta^{3/2}} R_9  \ln(1-\eta_1)
-\frac{16}{3} S_1
\biggr) \ln^2(\eta)
\nonumber\\&&
-\frac{64}{27} S_1^3
-\frac{128}{27} S_3
-\frac{64}{3} \left(\zeta_2+\frac{1}{3} S_2\right) S_1
-\frac{128}{9} \zeta_3
\biggr]
-\frac{R_{10} \ln^2(\eta)}{3 \eta (N-1) N (N+1)^2}
\nonumber\\&&
+16 \left[-\frac{1}{(N+1)^2}+p_{gq}^{(0)} \left(\frac{8}{3}-S_1\right)\right]\left((L_1+L_2)^2-\frac{4}{3} (L_1+L_2) 
S_1\right)
\nonumber\\&&
+\left[\frac{32}{3} p_{gq}^{(0)} \left(S_2-S_1^2\right)-\frac{64 (8 N+5)}{9 (N+1)^3} \right] (L_1+L_2)
-\frac{64 R_{11} S_1}{27 (N-1) N (N+1)^3}
\nonumber\\&&
+\frac{64 \big(8 N^3+13 N^2+27 N+16\big)}{27 (N-1) N (N+1)^2} \left(
S_1^2
+S_2
+3 \zeta_2
\right)
-\frac{8 R_{12}}{243 \eta (N-1) N (N+1)^4}
\Biggr\},
\nonumber\\&&
\end{eqnarray}
with 
\begin{eqnarray}
p_{gq}^{(0)} = \frac{2+N+N^2}{(N-1)N(N+1)}
\end{eqnarray}
and the polynomials
\begin{eqnarray}
R_8&=&-10 \eta ^{3/2}+5 \eta ^2+42 \eta -10 \sqrt{\eta }+5,\\
R_9&=&10 \eta ^{3/2}+5 \eta ^2+42 \eta +10 \sqrt{\eta }+5,\\
R_{10}&=&5 \eta ^2 N^3+10 \eta ^2 N^2+15 \eta ^2 N+10 \eta ^2-14 \eta  N^3-12 \eta  N^2-58 \eta  N
-28 \eta 
+5 N^3
\nonumber\\&&
+10 N^2+15 N+10,\\
R_{11}&=&39 N^4+101 N^3+201 N^2+205 N+78,\\
R_{12}&=&405 \eta ^2 N^5+1620 \eta ^2 N^4+3240 \eta ^2 N^3+4050 \eta ^2 N^2+2835 \eta ^2 N+810 \eta ^2-5326 \eta  N^5
\nonumber\\&&
-18496 \eta  N^4-40952 \eta  N^3-55636 \eta  N^2-39370 \eta  N-10652 \eta +405 N^5+1620 N^4
\nonumber\\&&
+3240 N^3+4050 N^2+2835 N+810~.
\end{eqnarray}
The two-mass contribution to the OME is then given by
\begin{eqnarray}
\tilde{A}_{gq}^{(3)} &=& C_F T_F^2
\Biggl\{
        -\frac{64 R_{13}}{9 (N-1) N (N+1)^2} \left(L_1^2 + L_2^2\right)
        -\frac{64 R_{15}}{81 (N-1) N (N+1)^4}
\nonumber\\ &&
        -\frac{32 R_{13}}{9(N-1) N (N+1)^2} S_1^2
        -\frac{32 R_{13}}{9(N-1) N (N+1)^2} S_2
        +\frac{64 R_{14}}{27 (N-1) N (N+1)^3} S_1
\nonumber\\ &&
        +\left(L_1 + L_2\right)\Biggl[
                -\frac{64 R_{14}}{27 (N-1) N (N+1)^3}
                +\frac{64 R_{13}}{9(N-1) N (N+1)^2} S_1 \Biggr]
\nonumber\\ &&
         + p_{gq}^{(0)} \Biggl[
                -\frac{128}{9} \left(L_1^3 +L_2^3 \right)
                -\frac{32}{3} \left(L_1^2 L_2 + L_2^2 L_1\right)
                +\left(L_1 +L_2\right) \Biggl[
                        -\frac{32}{3} S_1^2
                        -\frac{32}{3} S_2
                        -32 \zeta_2
                \Biggr]
\nonumber\\ &&
                +\frac{64}{3} \left(L_1^2 + L_2^2\right) S_1
                +\Biggl[
                        \frac{32}{3} S_2
                        +\frac{64}{3} \zeta_2
                \Biggr] S_1
                +\frac{32}{9} S_1^3
                +\frac{64}{9} S_3
                -\frac{128}{9} \zeta_3
        \Biggr]
\nonumber\\ &&
        -\frac{64 R_{13}}{9 (N-1) N (N+1)^2} \zeta_2
\Biggr\} + \tilde{a}_{gq,Q}^{(3)},
\end{eqnarray}
where
\begin{eqnarray}
R_{13} &=& 8 N^3+13 N^2+27 N+16 \\ 
R_{14} &=& 43 N^4+105 N^3+224 N^2+230 N+86 \\ 
R_{15} &=& 248 N^5+863 N^4+1927 N^3+2582 N^2+1820 N+496~. 
\end{eqnarray}
The corresponding $z$-space expressions are given in Appendix~A.
Also in this case  we obtain as before in Ref.~\cite{Ablinger:2014lka} the corresponding contributions to the 3-loop anomalous 
dimension \cite{Vogt:2004mw}.

\begin{figure}[H]
\centering
\includegraphics[width=0.8\textwidth]{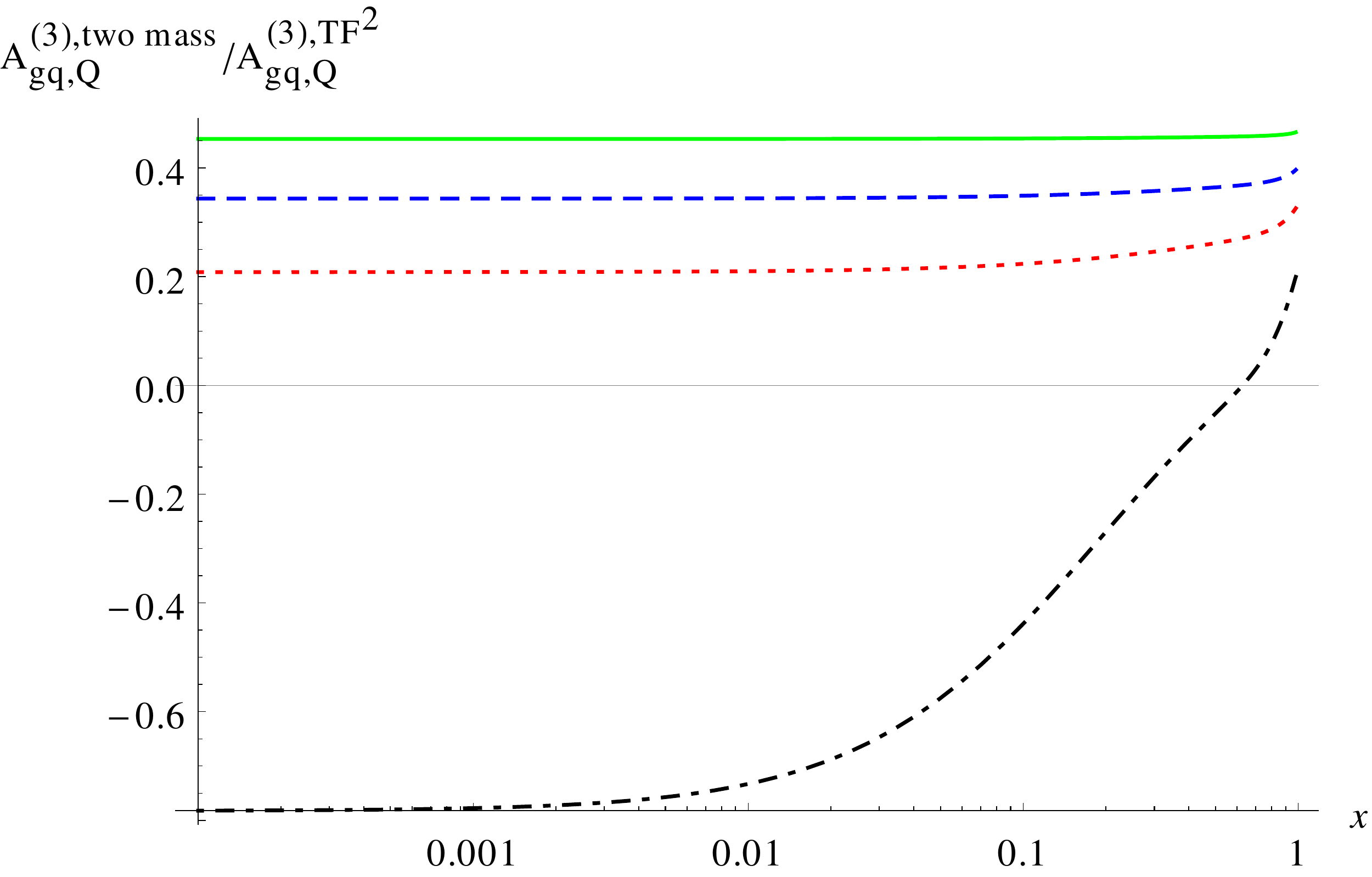}
\caption{\label{FIG:GQ}
\sf \small The ratio of the genuine 2-mass contributions to $A_{gq,Q}^{(3)}$ to the complete 
$T_F^2$-part of the massive 3-loop OME 
$A_{gq,Q}^{(3)}$ as a function of $x$ and $Q^2$, for $m_c = 1.59~\GeV, m_b = 4.78~\GeV$ in the on-shell scheme.
Dash-dotted line:  $\mu^2 =   30~\GeV^2$;
Dotted line:  $\mu^2 =   50~\GeV^2$;
Dashed line:  $\mu^2 =   100~\GeV^2$;
Full line:  $\mu^2 =   1000~\GeV^2$.
The single mass contributions are given in Ref.~\cite{Ablinger:2014lka}.}
\end{figure}
In Figure~\ref{FIG:GQ} we show the ratio of the genuine 2-mass contribution to the complete $T_F^2$
3-loop result for $A_{gq,Q}^{(3)}$ for typical values of $Q^2$ and $x$. The ratio varies between 0 and 0.5.
At higher values of $Q^2$, an almost flat behaviour is observed.
\section{Scalar \boldmath $A_{gg,Q}$ diagrams with {$m_1 \neq m_2$}}
{\label{SSec-Agg2M}}

\vspace*{1mm}
\noindent
The factorization into parts depending purely on the Mellin variable $N$ and contributions depending only on the mass ratio $\eta$, which
has been observed for the non--singlet diagrams, constitutes a very special case. Normally both variables appear in a 
more intertwined form and more advanced methods are required to perform the calculation. Since the complexity of the mathematical structures 
contributing to a Feynman diagram depends on the denominator functions and on the form of the operator insertion, we 
will first study the 
scalar topologies contributing to the OME $A_{gg,Q}$ in this paper. Due to the nesting between the Mellin variable and the mass ratio, 
novel $\eta$--dependent sums and integrals will emerge. In particular, it turns out that the expansion in $\eta$ is not possible
in general, unlike the case for fixed integer moments. Therefore, the integrals have to be calculated for general values of $\eta$.
\subsection{The Calculation Strategy}

\vspace*{1mm}
\noindent
As we expect new functions to appear in the results and since the construction of the inverse Mellin transforms for these functions turns 
out to be a non-trivial task, we opt for an approach where we derive the $z$-space representation of the respective diagrams first. The 
$N$-space representation\footnote{The steps to compute these Mellin transforms are included in the computer 
algebra package {\tt 
HarmonicSums} 
\cite{HARMONICSUMS,Ablinger:2011te,Ablinger:2013cf}.} is then obtained in a final step by using the generating function method, constructing a 
difference equation and solving it using the package {\tt Sigma}~\cite{SIG1,SIG2}. These representations can be then evaluated at fixed 
integer moments in $N$, be expanded in the parameter $\eta$ and compared to the fixed moments having been calculated using the code 
{\tt Q2e}/{\tt Exp}~\cite{Harlander:1997zb,Seidensticker:1999bb}.

First we introduce Feynman parameters and perform the momentum integration for one of the closed fermion lines. This leads to
an effective propagator, the mass of which we can detach using the Mellin-Barnes representation\cite{MB1a,MB1b,MB2,MB3,MB4}
\begin{equation}
\frac{1}{\left(A+B\right)^{\lambda}}=\frac{1}{\Gamma(\lambda)} \frac{1}{2 \pi i} \int_{-i\,\infty}^{+i\,\infty}
d \xi \frac{B^\xi}{A^{\lambda+\xi}} \Gamma(\lambda+\xi) \Gamma(-\xi)~.
\end{equation}
Then we perform the remaining momentum integrals, which leads to an expression where 
the Feynman parameter integrals are now of the generalized hypergeometric 
type~\cite{HYP} and the appropriate application of techniques used earlier in Refs.~\cite{Bierenbaum:2007qe,Bierenbaum:2008yu,Ablinger:2010ty}
allow to integrate all Feynman parameter integrals as Beta-functions, of which only one depends on both the Mellin variable $N$ and the 
Mellin-Barnes variable $\xi$ and is kept in unintegrated form. We obtain a representation of the general form 
\begin{eqnarray}
&& C(N,m_1,m_2,\ep) \frac{1}{2 \pi i} \int_{-i \infty}^{+i \infty} d\xi \int_0^1 d X \, \eta^\xi X^{\xi+N+\alpha \ep+\beta} (1-X)^{-\xi+\gamma \ep+\delta}
\NN\\&& \phantom{C(N,m_1,m_2,\ep) \frac{1}{2 \pi i} \int_{-i \infty}^{+i \infty} \int_0^1} \times
\Gamma \Brack
{a_1+b_1 \ep+c_1 \xi, \ldots, a_i+b_i \ep+c_i \xi} 
{d_1+e_1 \ep+f_1 \xi, \ldots, d_j+e_j \ep+f_j \xi}~, \label{MBintD5a}
\end{eqnarray}
where $a_k$, $d_k$, $\beta$ and $\delta$ are integers, $b_k$, $e_k$, $\alpha$ and $\gamma$ are integers or half-integers, 
$c_k \in \{-1,1\}$ and $f_k \in \{-2,-1,1,2\}$, with $\sum_{k=1}^i c_k = \sum_{k=1}^j f_k$. The dependence on $N$ of the function
$C(N,m_1,m_2,\ep)$ arises from gamma functions that depend on $N$ (and possibly on $\ep$) but not on $\xi$.

Mellin-Barnes integrals of the form
\begin{equation}
\frac{1} {2 \pi i} \int_{-i\,\infty}^{+i\,\infty} d\xi~Z^{\xi} \,
\Gamma \Brack
{a_1+b_1 \ep+c_1 \xi, \ldots, a_i+b_i \ep+c_i \xi} 
{d_1+e_1 \ep+f_1 \xi, \ldots, d_j+e_j \ep+f_j \xi} \label{MBInt}
\end{equation}
are usually solved by closing the contour either to the left or to the right and by applying
Cauchy's theorem
\begin{equation}
\oint_C f(z) d z = 2 \pi i \sum_i res_{z_i} f~.
\end{equation}
If we close the integration contour in (\ref{MBInt}) to the left(right) the
residue sum only converges for $Z>1$ ($Z<1$), respectively. In (\ref{MBintD5a})
we have 
\begin{equation}
Z = \frac{\eta X}{1-X}, 
\end{equation}
which covers both ranges for possible values of $\eta < 1$ and $\eta > 1$. In the calculations we applied 
the code {\tt MB} \cite{Czakon:2005rk}. We follow the method applied in the equal mass case in 
Ref.~\cite{Hasselhuhn:2013swa,Ablinger:2014uka}, split the integration range and remap the individual parts 
to the domain $[0,1]$:
\begin{eqnarray}
\int_{-i\,\infty}^{+i\,\infty} d\xi 
\int_0^1 d X f(\xi, X) \left(\frac{\eta X}{1-X}\right)^\xi &=& 
\int_{-i\,\infty}^{+i\,\infty} d \xi
\left(\int_0^{\frac {1} {1+\eta}} d X +\int_{\frac{1} {1+\eta}}^{1} d X \right) f(\xi, X) \left(\frac{\eta X}{1-X}\right)^\xi
\NN\\&=&
\int_{-i\,\infty}^{+i\,\infty} d\xi~\int_0^1 d T~\Biggl[
\frac{\eta}{\left(\eta+T\right)^2} 
f\left(\xi,\frac{T}{\eta+T}\right) T^{\xi}
\NN\\&&  
+ \frac{\eta}{\left(1+\eta T\right)^2} f\left(\xi,\frac{1}{1+\eta T}\right) T^{-\xi}\Biggr]
~.
\label{MBSplit}
\end{eqnarray}
A further advantage of this procedure is that the contour integration decouples the $\eta$-dependence which now only enters 
through the $T$-integration.

We follow the well known procedure of deforming the contour integral in order to separate the ascending from the descending 
poles \footnote{In some cases an additional regularization parameter was introduced in order to separate overlapping poles.} applying 
Cauchy's theorem. At this point we are left with only one integral and no overlapping singularities anymore. If necessary, we map 
$T \rightarrow 
1-T$ in order to have singularities regulated by $\ep$ only at $T=0$. They appear as $\ep$-poles after applying the following integration-by-parts 
relation:
\begin{equation}
\int_0^1 d T~T^{-a} f(T)= 
\left. \frac{1}{1-a} T^{-a+1} f(T)\right|_0^1
-\frac{1} {1-a} \int_0^1~d T~T^{-a+1} f'(T).
\label{RegTint}
\end{equation}
We may then perform the Laurent series expansion around $\ep=0$.

In the next step we rewrite the sums obtained using the package {\tt Sigma} \cite{SIG1,SIG2}\footnote{Due to the integral transformation 
(\ref{MBSplit}) these infinite sums are independent of the mass ratio $\eta$, which renders them much easier to solve.}. The sums are then 
expressed in terms of generalized harmonic sums \cite{Moch:2001zr,Ablinger:2013cf} at infinity, 
\begin{equation}
S_{b,\vec{a}}(c,\vec{d};\infty) = \sum_{k=1}^\infty \frac{c^k}{k^b} 
S_{\vec{a}}(\vec{d}; k), c, d_i \in \mathbb{R} \backslash \{0\};~~~b, a_i \in \mathbb{N} 
\backslash \{0\}, 
\end{equation}
which have to be rewritten in terms of generalized harmonic polylogarithms (GHPLs) \cite{Ablinger:2013cf} at argument $x=1$ 
using {\tt HarmonicSums}~\cite{HARMONICSUMS,Ablinger:2011te,Ablinger:2013cf}. These functions are iterated integrals over the following 
alphabet~:
\begin{equation}
\Biggl\{ 
\frac{d \tau}{\tau}, \frac{d \tau}{\tau+T}, \frac{d \tau}{1+T \tau^2}
\Biggr\}~.
\label{eq:alpabet}
\end{equation}
In order to process them, we want the remaining integration variable $T$ to only appear in the argument of the HPLs. Because of the emergence of 
letters with non-linear denominators, we cannot apply the methods used in Ref.~\cite{Ablinger:2014yaa,Brown:2008um} directly, although extensions 
of it, as it is described below, should suffice to transform these HPLs. However, due to the relatively simple structure of the 
letters in 
Eq.~(\ref{eq:alpabet}), there is a way based on applying the shuffle relations, cf.~\cite{Blumlein:2003gb}, and rescaling the internal integration 
variables, to rewrite the corresponding iterated integrals in the desired form.

Instead of computing the remaining integrals, we rather aim at
transforming them into a Mellin transform from which one can then read off the $z$-space representation.
Next, we absorb rational $N$-dependent factors into the integral, which appear both in the numerator and denominator. These factors
stem from the integration of the Feynman parameters, and are now pulled into the $T$-integration by performing a partial fraction 
decomposition and then applying the following partial integration identities repeatedly,
\begin{eqnarray}
N \int_0^1 dx g(x)^N f(x)&=&
\left.  g(x)^{N+1} \frac{f(x)}{g'(x)} \right|_0^1
-\int_0^1 d x 
\left( g(x)\right)^N 
\frac{d} {d x} \left[\frac{f(x) g(x)}{g'(x)}\right]~, 
\label{RemN1}
\\
\frac{1} {\left(N+a\right)} \int_0^1 dx g(x)^N f(x)&=&
\left.
\frac{1} {\left(N+a\right)} g(x)^{N+a} 
\left(\int_0^x d y \frac{f(y)}{g(y)^a}\right)
\right|_{x=0}^1
\NN\\&&
-\int_0^1 d x 
g(x)^{N+a-1} 
\frac{d g(x)}{d x}
\left(\int_0^x d y \frac{f(y)}{g(y)^a}\right)~.
\label{RemN2}
\end{eqnarray}
Relation (\ref{RemN1}) has to be especially handled with care, as its
application may introduce new divergences in each term. This issue is solved by
regularizing the remaining integral in (\ref{RemN1}) by a $+$-type distribution
which cancels these additional singularities, as e.g.

\begin{eqnarray}
N \int_0^1 d x~ x \HA_0\left(x\right) \left(\frac{\eta}{\eta+x^2}\right)^N
&=& \lim_{\ep \to 0} \Biggl[\left. \frac {\eta} 2 \HA_0\left(x\right)\right|_{x=\ep}^1
\NN\\&&
+\int_{\ep}^1 d x \left( \frac{\eta+x^2}{2 x} +x \HA_0\left(x\right) \right) 
\left(\frac{\eta}{\eta+x^2}\right)^N\Biggr]
\NN\\&=&
\lim_{\ep \to 0} \Biggl[
\left. \frac {\eta} 2  \HA_0\left(x\right)\right|_{x=\ep}^1
+
\int_{\ep}^1 d x~x \left(\frac 1 2+\HA_0\left(x\right)\right) 
\left(\frac{\eta}{\eta+x^2}\right)^N
\NN\\&&
+\frac \eta 2 \int_{\ep}^1 d x~\frac 1 x \left[\left(\frac{\eta}{\eta+x^2}\right)^N-1 \right]
+\left.\frac{\eta}{2} \HA_0\left(x\right)\right|_{x=\ep}\Biggr] 
\NN\\&=&
\int_0^1 d x~x \left(\frac 1 2+\HA_0\left(x\right)\right) 
\left(\frac{\eta}{\eta+x^2}\right)^N \nonumber
\NN\\&&
+\frac \eta 2 \int_0^1 d x~\frac 1 x \left[\left(\frac{\eta}{\eta+x^2}\right)^N-1 \right]~.
\end{eqnarray}
We now rewrite the remaining integral as the following Mellin transform:
\begin{align}
I&=\int_0^1 d x  \left[g(x, \eta)\right]^N f(x, \eta)
\NN\\[3mm]
&=
\begin{cases}
\qquad\qquad\int \limits_{g(0)}^{g(1)} d X  f(g^{-1}(X,\eta),\eta) X^N 
\left|\frac{d g^{-1}(X, \eta)}{d X}\right|
,&~\text{for}~g(x,\eta)>0, ~0<x<1, ~\eta>0
\NN\\[5mm]
(-1)^N \int \limits_{-g(1)}^{-g(0)} 
d X  f(-g^{-1}(X,\eta),\eta) X^N 
\left|\frac{d g^{-1}(X, \eta)}{d X}\right|
,& ~\text{for}~g(x,\eta)<0, ~0<x<1, ~\eta<0 ~.
\end{cases}
& \\
\label{ToMellin}
\end{align}
Note that the function $g$ is monotonous (cf.~Eq.~(\ref{MBSplit})) and thus the inverse function $g^{-1}$ exists.
The class of harmonic polylogarithms is not sufficient to perform this step and generalizations
are required to allow for quadratic forms in the denominator. One such generalization is given by the cyclotomic harmonic polylogarithms 
\cite{Ablinger:2011te}. We use the label $\{4,i\}$ to denote the following letter
\begin{eqnarray}
 \{4,i\} &\rightarrow& \frac{\tau^i~d\tau}{\Phi_4(\tau)}, \quad i \in \{0,1\}
 \end{eqnarray}
where $\Phi_4(\tau)=\tau^2+1$ is the fourth cyclotomic polynomial, and $d\tau$ indicates that the iteration proceeds over $\tau$. 
For example, $\HA_{0,\{4,1\}}\left(x\right)$ represents the iterated integral
\begin{eqnarray}
 \HA_{0,\{4,1\}}\left(x\right)=\int_0^x \frac{d\tau_1}{\tau_1} 
 \int_{0}^{\tau_1} 
 \frac{d\tau_2~\tau_2}{\tau_2^2+1}~.
\end{eqnarray}
More generally, we write
\begin{eqnarray}
 \{\{a,b,c\},i\} &\rightarrow& \frac{\tau^i~d\tau}{a+b \tau +c \tau^2}~, \quad i \in \{0,1\}
\end{eqnarray}
In the calculation of some of the diagrams shown in the next subsection, we thus performed simplifications
such as
\begin{align}
&\int_0^1 d x~\frac{x 
\HA_{
\{\{1,0,\eta\},1\},
\{\{1,0,\eta\},0\},
0}\left(x \right)}{\left(1+\eta x^2\right)^2} \left(\frac{\eta x^2}{1+\eta x^2}\right)^N
\NN\\=&
\frac{1}{2 \eta} \int_0^{\eta/(1+\eta)} 
d x~ \HA_{
\{\{1,0,\eta\},1\},
\{\{1,0,\eta\},0\},
0}\left(\frac{\sqrt{x}}{\sqrt{1-x} \sqrt{\eta}} \right) x^N
\NN\\=&
\frac{1}{2 \eta^{5/2}} 
\int_0^{\eta/(1+\eta)} d x
\left[
\HA_{\{4,1\},\{4,0\},0}\left(\frac{\sqrt{x}}{\sqrt{1-x}}\right)
-\frac{1}{2} \HA_0(\eta) \HA_{\{4,1\},\{4,0\}}\left(\frac{\sqrt{x}}{\sqrt{1-x}}\right)
\right] x^N~,
\end{align}
where in the last step we removed the $\eta$-dependence of the argument by
again applying a rescaling of the inner integration variables.  At this point,
it is desirable to remove the square roots in the arguments of the HPLs and to
obtain iterated integrals with the argument $x$ only. In order to obtain this
representation, we once again exploit the property that taking the derivative
reduces the transcendental weight of a hyperlogarithm and use a method
similar to the one given in \cite{Ablinger:2014yaa,Brown:2008um}. For example, 
\begin{align}
\frac{d}{d x} \HA_{\{4,1\},0}\left(\frac{\sqrt{x}}{\sqrt{1-x}}\right)
&=\frac{1}{2}~\frac{\HA_0\left(\frac{\sqrt{x}}{\sqrt{1-x}}\right)}{1-x}
\NN\\&=\frac{1}{4 (1-x)} \left[\HA_0(x)+\HA_1(x)\right]
\\
\HA_{\{4,1\},0}\left(\frac{\sqrt{x}}{\sqrt{1-x}}\right)&=\frac{1}{4} \left[\HA_{1,0}(x)+\HA_{1,1}(x)\right]~.
\end{align} 

However, not all the occurring HPLs can be expressed in terms of generalized HPLs of the previous kind and new, 
root-valued letters have to be introduced. To perform this in a systematic way, we introduce a more general class of 
iterated integrals as follows:
\begin{eqnarray}
\label{eq:GFUN}
G\left(\left\{f_1(\tau),f_2(\tau),\cdots,f_n(\tau)\right\},z\right)
&=&\int_0^z  d\tau_1~f_1(\tau_1)  
G\left(\left\{f_2(\tau),\cdots,f_n(\tau)\right\},\tau_1\right)~,
\end{eqnarray}
with the special cases
\begin{eqnarray}
 G\left(\{\},z\right)&=&1~,
\end{eqnarray}
and
\begin{eqnarray}
 G\Biggl(\Biggl\{\underbrace{\frac{1}{\tau},\frac{1}{\tau},
  \cdots,\frac{1}{\tau}}_{\text{n times}}\Biggr\},z\Biggr)
 &=&
\frac{1}{n!} \HA_0(z)^n \equiv
\frac{1}{n!} \ln^n(z)~.
\end{eqnarray}
Here $f_i(\tau)$ are real functions, with $\tau \in [0,1]$. At the moment we do not discuss matters like algebraic or
structural independence of these quantities, cf.~\cite{Blumlein:2003gb,Ablinger:2011te,Ablinger:2013cf}, 
but rather consider (\ref{eq:GFUN}) as a placeholder. Algebraic and other relations are applied later in the 
concrete cases appearing. These functions are given in explicit form in Appendix~\ref{APP2}.

Using these generalized iterated integrals we rewrite the HPLs with root-valued functions in the argument. For example one has
\begin{align}
\HA_{\{4,0\},\{\{\eta,0,1\},0\}}\left(
\frac{\sqrt{x}}{\sqrt{1-x}}\right)&=\frac{x \left(3-6 x+3\eta x +3 \eta^2 x +7 x^2 -2 \eta x^2 -5 \eta^2 x^2-3 x^3 +3 \eta^2 x^3\right)}{3 \left(\eta -1\right) \eta}
\NN\\&
-\frac{2 \left(1+\eta\right) \sqrt{1-x} \sqrt{x} \left(-1+2 x\right)}{\eta} 
G\left(\left\{\sqrt{1-\tau} \sqrt{\tau}\right\},x\right)
\NN\\&
-\frac{(\eta-1)^2 \sqrt{1-x} \sqrt{x} (-1+2 x)}{2\eta} G\left(
\left\{\frac{\sqrt{1-\tau} \sqrt{\tau}}{-\eta-\tau+\eta \tau}\right\},x\right)
\NN\\&
+\frac{8 \left(1+\eta\right)}{\eta} 
G\left(
\left\{
\sqrt{1-\tau} \sqrt{\tau}
,\sqrt{1-\tau} \sqrt{\tau}
\right\},x\right)
\NN\\&
+\frac{2 \left(\eta-1\right)^2}{\eta} G\left(
\left\{
\sqrt{1-\tau} \sqrt{\tau},\frac{\sqrt{1-\tau} \sqrt{\tau}}{\eta \tau-\eta-\tau}
\right\},x\right)
\NN\\&
-\frac{1+\eta}{2 (\eta-1)} G\left(\left\{\frac{1}{\eta \tau-\eta-\tau}\right\},x\right)~.
\end{align}
In the present computation, similar HPLs up to weight {\sf w = 3} had to be
transformed. Due to the size of the expressions and the necessity to cancel
spurious terms, all relations obeyed by these quantities have to be used.
These are
\begin{itemize}
\item{shuffle relations}
\item{integration by parts relations, such that only factors with exponents $\in ~\{1/2,-1\}$ contribute to the different letters}
\item{shuffling of single square root terms to the end and performing the integrals e.g.: 
\begin{equation}
G\left(\left\{\sqrt{\tau},\frac{1}{\tau+1}\right\},x\right)
=\frac{2}{3} \left[
-G\left(\left\{\sqrt{\tau}\right\},x\right)+x^{3/2} G\left(\left\{\frac{1}{1+\tau}\right\},x\right)
+G\left(\left\{\frac{\sqrt{\tau}}{1+\tau}\right\},x\right)
 \right]~.
\end{equation}
}
\end{itemize}
These identities have now been implemented in {\tt HarmonicSums}\cite{HARMONICSUMS,Ablinger:2011te,Ablinger:2013cf} and allow a
significant simplification of expressions with iterated integrals of this type. Finally, the integrals are merged.  After the 
mapping 
of the integration variables (\ref{ToMellin}) we are left with integrals of the form $\int_0^{f(\eta)} d x$ or 
$\int_{f(\eta)}^1 d x$  due to the splitting of the $X$-integration in Eq.~(\ref{MBSplit}). We therefore substitute
\begin{equation}
\int_{f(\eta)}^1 d x~G(x)=\int_0^1 d x ~G(x) -\int_0^{f(\eta)} d x~G(x)~.
\end{equation}
As it would have been expected, the integrals $\int_0^{f(\eta)} d x~G(x)$ completely cancel up to 
trivial integrals of the form
\begin{equation}
\int_0^{f(\eta)} d x~ x^{\alpha+N} =\left. \frac{1}{\alpha+N+1} x^{N+\alpha+1}\right|_{x=0}^{f(\eta)}~.
\end{equation}
We now use {\tt HarmonicSums}\cite{HARMONICSUMS,Ablinger:2011te,Ablinger:2013cf} to perform the inverse Mellin transform for
terms that do not contain any $x$-integration. They usually stem from integration-by-parts applied in steps (\ref{RegTint}), (\ref{RemN1}) 
or (\ref{RemN2}). We are left with a $z$-space representation for our diagram. This representation usually also includes a part proportional
to a $\delta$-distribution and one term proportional to a $+$-distribution.

As a last step, we want to generate a $N$-space representation for our result, for which the last remaining integration has to be performed.
This is done with the help of a generating function representation mapping the integral into generalized HPLs and then generating a 
recurrence relation for the $N$th coefficient of this result. This procedure is automated within the package {\tt 
HarmonicSums}~\cite{HARMONICSUMS,Ablinger:2011te,Ablinger:2013cf}. The resulting recurrences were solved using the package {\tt  
Sigma}~\cite{SIG1,SIG2}.  The result contains many generalized HPLs at argument $x = 1$, which stem from the upper integration limit. 
In case their letters are free of the mass ratio $\eta$, they can be evaluated in terms of special constants like $\pi, \ln(2)$, the Catalan 
number $\bf{C}$, $\zeta_2$ and $\zeta_3$ by using standard integration methods or applying the internal integration algorithms of computer 
algebra packages like {\tt Mathematica} or {\tt Maple}. In case these generalized HPLs are not entirely free of $\eta$, it is desirable to 
rewrite them as iterated integrals with argument $\eta$ in order to obtain algebraic independence and an easier access to series 
representations. Rewriting these generalized HPLs cannot be done by rescaling integration variables or by just applying the methods 
of \cite{Ablinger:2014yaa,Brown:2008um}, since due to the root valued letters the derivative with respect to an inner variable in general 
does not lead to a weight reduction in this case. There is, however, an extension to the ideas in \cite{Ablinger:2014yaa,Brown:2008um}:
Taking the derivative with respect to inner variables we observe, that only GHPLs of a lower weight, GHPLs independent of this 
variable and the original GHPL itself contribute, as e.g.:
\begin{align}
\frac{d}{d \eta} G\left(\left\{\sqrt{\tau}\sqrt{1-\tau},\frac{\sqrt{\tau}\sqrt{1-\tau}}{-\eta-\tau+\eta \tau}\right\},1\right)&=\frac{\left(1+\eta\right) \left(1-8 \eta +\eta^2\right)}{12 \left(\eta-1\right)^4 \eta}
\NN\\&
-\frac{\eta}{\left(\eta-1\right)^4} G\left(\left\{
\frac{1}{-\eta-\tau+\eta \tau}\right\},1\right)
\NN\\&
-\frac{2}{\left(\eta-1\right) \eta} 
G\left(
\left\{
\sqrt{\tau}\sqrt{1-\tau},
\sqrt{\tau}\sqrt{1-\tau}
\right\},1\right)
\NN\\&
-\frac{1+3 \eta}{2 \left(\eta-1\right) \eta} 
G\left(
\left\{
\sqrt{\tau}\sqrt{1-\tau},
\frac{\sqrt{\tau}\sqrt{1-\tau}}{-\eta-\tau+\eta \tau}
\right\},1\right)~.
\label{eq:GLTrans}
\end{align}
Therefore, the linear first order differential operator 
\begin{equation}
\frac{d} {d \eta}+\frac{1+ 3 \eta} {2 \left(\eta-1\right) \eta}
\end{equation}
does lead to a weight reduced expression when applied to the GHPL 
\begin{equation}
G\left(
\left\{
\sqrt{\tau}\sqrt{1-\tau},\frac{\sqrt{\tau}\sqrt{1-\tau}}{-\eta-\tau+\eta \tau}\right\},1\right).
\end{equation} 
The weight reduced expression can be rewritten with the same method and we have to undo the effect of the differential operator by using the general solution for the linear first order differential equation
\begin{align}
\frac{d} {d x} f(x)+p(x) f(x) &=q(x)\\
\rightarrow f(x)&=\exp\left(-\int_{x_0}^x p(t) d t\right) \left[f(x_0)+ \int_{x_0}^x q(t) \exp\left({\int_{x_0}^t p(u) d u}\right) d t\right]~.
\end{align} 
Applying this method to the GHPL considered above we obtain
\begin{align}
G\left(
\left\{
\sqrt{\tau}\sqrt{1-\tau},
\frac{\sqrt{\tau}\sqrt{1-\tau}}{-\eta-\tau+\eta \tau}
\right\},1\right)&=
\frac{1+4 \eta -2 \eta^2}{6 (\eta-1)^3}
-\frac{3 \left(1-4 \sqrt{\eta} +\eta\right)}{16 \left(\eta-1\right)^2} \zeta_2
\NN\\&
+\frac{\sqrt{\eta}}{8 \left(\eta-1\right)^2} 
G\left(\left\{\frac{\sqrt{\tau}}{1-\tau}, \frac{1}{\tau}\right\},\eta\right)+\frac{\left(\eta-3\right)\eta^2}{4 \left(\eta-1\right)^4} \ln(\eta)~.
\label{eq:GLTrans1}
\end{align}
For all the GHPLs considered in this section, it is always possible to construct a linear first 
order differential operator \footnote{First order linear differential operators could be used instead 
of pure differentiation in order to extend the parametric integration method. However, 
remapping parameters might be a more suitable method to integrate Feynman parameter integrals 
which are not a priori reducible. Both methods become inapplicable when non-iterative integrals 
appear, as e.g. genuine elliptic integrals and others.} which does yield a weight reduced
expression when applied to the corresponding generalized HPL, and all the GHPLs
could thus be rewritten in terms of GHPLs with argument $\eta$. 
See Appendix B for a list of relations for the GHPLs.

\subsection{The results for individual diagrams}

\vspace*{1mm}\noindent
In the following, we present the results for all scalar two-mass topologies contributing to $A_{gg,Q}$ both in $z$- and in $N$-space.
Up to a global pre-factor, all results are expressed as functions of the mass ratio $\eta$. We consider only the cases where the operator
insertion is located on a line, and not on a vertex, since the latter case can be easily derived from the former. The powers of the propagators are taken to be the highest 
ones appearing in the corresponding physical diagrams (this means that  in all of the diagrams the sum of powers of propagators equals 9).

We define the following functions which appear often in the $z$-space expressions of the diagrams,
\begin{eqnarray}
f_1(\eta,z) &=& 
\frac{1}{2} (1+\eta) \biggl[G\left(\left\{\sqrt{1-\tau} \sqrt{\tau},\frac{1}{1-\tau}\right\},z\right)
+G\left(\left\{\sqrt{1-\tau} \sqrt{\tau},\frac{1}{\tau}\right\},z\right)
\biggr]
\NN\\&&
-\frac{1}{8} (1-\eta)^2 \biggl[
\ln(\eta) G\left(\left\{\frac{\sqrt{1-\tau} \sqrt{\tau}}{1-\tau+\eta \tau}\right\},z\right)
+G\left(\left\{\frac{\sqrt{1-\tau} \sqrt{\tau}}{1-\tau+\eta \tau},\frac{1}{\tau}\right\},z\right)
\NN\\&&
+G\left(\left\{\frac{\sqrt{1-\tau} \sqrt{\tau}}{1-\tau+\eta \tau},\frac{1}{1-\tau}\right\},z\right)
\biggr]
\NN\\&&
+\left(1-\eta+\frac{1}{2} (1+\eta) \ln(\eta)\right) G\left(\left\{\sqrt{1-\tau} \sqrt{\tau}\right\},z\right)~,
\\
f_2(\eta,z) &=& 
-\frac{1}{2} (1+\eta) \biggl[G\left(\left\{\sqrt{1-\tau} \sqrt{\tau},\frac{1}{1-\tau}\right\},z\right)
+G\left(\left\{\sqrt{1-\tau} \sqrt{\tau},\frac{1}{\tau}\right\},z\right)
\biggr]
\NN\\&&
+\frac{1}{8} (1-\eta)^2 \biggl[
\ln(\eta) G\left(\left\{\frac{\sqrt{1-\tau} \sqrt{\tau}}{\eta \tau-\eta-\tau}\right\},z\right)
-G\left(\left\{\frac{\sqrt{1-\tau} \sqrt{\tau}}{\eta \tau-\eta-\tau},\frac{1}{\tau}\right\},z\right)
\NN\\&&
-G\left(\left\{\frac{\sqrt{1-\tau} \sqrt{\tau}}{\eta \tau-\eta-\tau},\frac{1}{1-\tau}\right\},z\right)
\biggr]
\NN\\&&
+\left(1-\eta+\frac{1}{2} (1+\eta) \ln(\eta)\right) G\left(\left\{\sqrt{1-\tau} \sqrt{\tau}\right\},z\right)~,
\\
f_3(\eta,z) &=& 
\frac{1}{2} (1+\eta) \biggl[
G\left(\left\{\sqrt{1-\tau} \sqrt{\tau},\sqrt{1-\tau} \sqrt{\tau},\frac{1}{\tau}\right\},z\right)
\NN\\&&
+G\left(\left\{\sqrt{1-\tau} \sqrt{\tau},\sqrt{1-\tau} \sqrt{\tau},\frac{1}{1-\tau}\right\},z\right)
\biggr]
\NN\\&&
+\left(1-\eta+\frac{1}{2} (1+\eta) \ln(\eta)\right) G\left(\left\{\sqrt{1-\tau} \sqrt{\tau},\sqrt{1-\tau} \sqrt{\tau}\right\},z\right)
\NN\\&&
-\frac{1}{8} (1-\eta)^2 \biggl[
G\left(\left\{\sqrt{1-\tau} \sqrt{\tau},\frac{\sqrt{1-\tau} \sqrt{\tau}}{1-\tau+\eta \tau},\frac{1}{\tau}\right\},z\right)
\NN\\&&
+G\left(\left\{\sqrt{1-\tau} \sqrt{\tau},\frac{\sqrt{1-\tau} \sqrt{\tau}}{1-\tau+\eta \tau},\frac{1}{1-\tau}\right\},z\right)
\NN\\&&
+\ln(\eta) G\left(\left\{\sqrt{1-\tau} \sqrt{\tau},\frac{\sqrt{1-\tau} \sqrt{\tau}}{1-\tau+\eta \tau}\right\},z\right)
\biggr]~,
\\
f_4(\eta,z) &=&
\frac{1}{2} (1+\eta) \biggl[
G\left(\left\{\sqrt{1-\tau} \sqrt{\tau},\sqrt{1-\tau} \sqrt{\tau},\frac{1}{\tau}\right\},z\right)
\NN\\&&
+G\left(\left\{\sqrt{1-\tau} \sqrt{\tau},\sqrt{1-\tau} \sqrt{\tau},\frac{1}{1-\tau}\right\},z\right)
\biggr]
\NN\\&&
-\left(1-\eta+\frac{1}{2} (1+\eta) \ln(\eta)\right) G\left(\left\{\sqrt{1-\tau} \sqrt{\tau},\sqrt{1-\tau} \sqrt{\tau}\right\},z\right)
\NN\\&&
+\frac{1}{8} (1-\eta)^2 \biggl[
G\left(\left\{\sqrt{1-\tau} \sqrt{\tau},\frac{\sqrt{1-\tau} \sqrt{\tau}}{\eta \tau-\eta-\tau},\frac{1}{\tau}\right\},z\right)
\NN\\&&                
+G\left(\left\{\sqrt{1-\tau} \sqrt{\tau},\frac{\sqrt{1-\tau} \sqrt{\tau}}{\eta \tau-\eta-\tau},\frac{1}{1-\tau}\right\},z\right)
\NN\\&&
-\ln(\eta) G\left(\left\{\sqrt{1-\tau} \sqrt{\tau},\frac{\sqrt{1-\tau} \sqrt{\tau}}{\eta \tau-\eta-\tau}\right\},z\right)
\biggr]~,
\\
f_5(\eta,z) &=&
G\left(\left\{\frac{1}{1-\tau+\eta \tau},\frac{1}{\tau}\right\},z\right)
+G\left(\left\{\frac{1}{1-\tau+\eta \tau},\frac{1}{1-\tau}\right\},z\right)
\NN\\&&
+\ln(\eta) G\left(\left\{\frac{1}{1-\tau+\eta \tau}\right\},z\right)~,
\\
f_6(\eta,z) &=&
G\left(\left\{\frac{1}{\eta \tau-\eta-\tau},\frac{1}{\tau}\right\},z\right)
+G\left(\left\{\frac{1}{\eta \tau-\eta-\tau},\frac{1}{1-\tau}\right\},z\right)
\NN\\&&
-\ln(\eta) G\left(\left\{\frac{1}{\eta \tau-\eta-\tau}\right\},z\right)~.
\end{eqnarray}

For diagram~1, each term in $z$-space factors completely into $z$- and $\eta$-dependent contributions. No iterated integrals 
involving both, $\eta$ and $z$, contribute. The $z$-space result can be  completely expressed by harmonic polylogarithms

\begin{eqnarray}
D_1(z)&=&
\left(m_1^2\right)^{\ep/2} \left(m_2^2\right)^{\ep-3} 
\Biggl\{
\frac{1+\eta^3}{105 \ep^2}
+\frac{1}{\ep}
\Biggl[
\frac{74-245\eta-245\eta^2+74\eta^3}{44100}
-\frac{\eta^3}{210} \HA_0\left(\eta\right)
\NN\\&&
-\frac{1+\eta^3}{210} \HA_1\left(z\right)
\Biggr]
+\frac{5520349-7928445\eta-7928445\eta^2+5520349\eta^3}{1185408000}
\NN\\&&
+\frac{\zeta_2}{280}\left(1+\eta^3\right)
-\frac{74-245\eta-245\eta^2+74\eta^3}{88200} \HA_1\left(z\right)
\NN\\&&
+\HA_0\left(\eta\right)
\Biggl[
\frac{-55125+37975\eta+24745\eta^2+36181\eta^3}{22579200}
+\frac{\eta^3}{420} \HA_1\left(z\right)
\Biggr]
\NN\\&&
+\frac{525-245\eta-245\eta^2+1549\eta^3}{430080} \HA_{0,0}(\eta)
+\frac{1}{420}\left(1+\eta^3\right) \HA_{1,1}\left(z\right)
\NN\\&&
-\frac{(1-\eta)^2\left(5+6\eta+5\eta^2\right)}{2048\sqrt{\eta}}
\big[\HA_{-1,0,0}\left(\sqrt{\eta}\right)+\HA_{1,0,0}\left(\sqrt{\eta}\right)\big]
\Biggr\}~.
\end{eqnarray}

\begin{figure}[H]
\begin{center}
\includegraphics[scale=1]{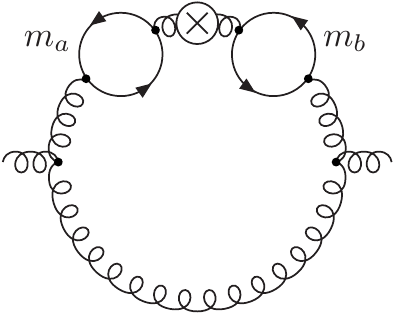}
\end{center}
\caption{\sf Diagram 1. Here both mass assignments $m_a=m_1$, $m_b=m_2$ and $m_a=m_2$, $m_b=m_1$ yield identical results.}
\end{figure}

Due to the structure in $z$-space, only harmonic sums contribute in Mellin $N$-space. 
\begin{eqnarray}
D_1(N)&=&
\left(m_1^2\right)^{\ep/2} \left(m_2^2\right)^{\ep-3} \left(\frac{1+(-1)^N}{2}\right) \frac{1}{N+1}
\Biggl\{
\frac{1+\eta^3}{105 \ep^2}
+\frac{1}{210 \ep}\Biggl[
-(1+\eta^3) S_1\left(N\right)
\NN\\&&
-\eta^3 \ln (\eta)
+\frac{(1+\eta) \big(2 \eta ^2 (37 N-68)-\eta  (319 N+109)+74 N-136\big)}{210 (N+1)}
\Biggr]
\NN\\&&
-\frac{\big(5 \eta^2+6 \eta +5\big) (1-\eta)^2}{2048 \sqrt{\eta }} 
\bigl[\HA_{-1,0,0}\left(\sqrt{\eta}\right)+\HA_{1,0,0}\left(\sqrt{\eta}\right)\bigr]
+\frac{(1+\eta^3) \zeta_2}{280}
\NN\\&&
+\ln(\eta) \biggl[
\frac{P_1}{22579200 (N+1)}
+\frac{\eta^3}{420} S_1\left(N\right)
\biggr]
+\frac{1+\eta^3}{840} \left[S_1^2(N)+S_2\left(N\right)\right]
\NN\\&&
+\frac{1549 \eta^3-245 \eta^2-245 \eta +525}{860160} \ln^2(\eta)
+\frac{(1+\eta) P_2}{1185408000 (N+1)^2}
\NN\\&&
-\frac{(1+\eta) \big(2 \eta ^2 (37 N-68)-\eta (319 N+109)+74 N-136\big)}{88200 (N+1)} S_1\left(N\right)
\Biggr\}~,
\end{eqnarray}
with the polynomials $P_i(\eta,N)$
\begin{eqnarray}
P_1&=&36181 \eta ^3 N+89941 \eta ^3+24745 \eta ^2 N+24745 \eta^2+37975 \eta  N+37975 \eta
\NN\\&&
-55125 N-55125~,
\\
P_2&=&5520349 \eta ^2 N^2+10046138 \eta ^2 N+7348189 \eta ^2-13448794 \eta  N^2
\NN\\&&
-22610228 \eta  N-11983834 \eta 
+5520349 N^2+10046138 N+7348189
~.
\end{eqnarray}
Here the factor $\frac{1}{2} (1+(-1)^N)$ comes from the operator insertion Feynman rule. This factor is removed from the $z$-space results
in all diagrams, due to the analytic continuation from the even moments.
\begin{figure}[H]
\begin{center}
\includegraphics[scale=1]{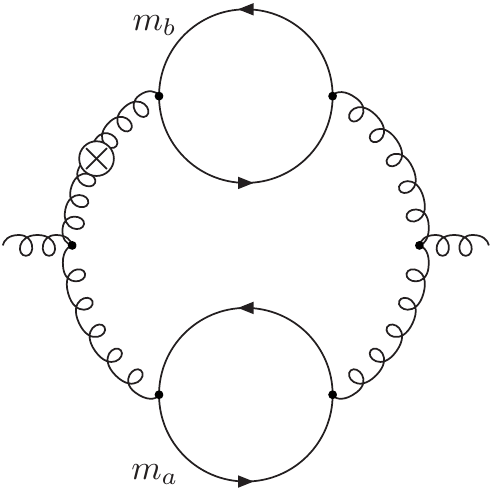}
\end{center}
\caption{\sf  Topology 2. $D_{2a}$ represents the mass assignment $m_a=m_2$, $m_b=m_1$ and $D_{2b}$ ($m_1\leftrightarrow m_2$).}
\end{figure}

Although topologically very similar to diagram $D_1$, diagrams $D_{2a}$ 
and $D_{2b}$ exhibit a much more involved mathematical structure. As we restrict 
ourselves to a representation within the class of iterated integrals of argument 
$z$, additional root-valued integration kernels had to be introduced. 
Furthermore, iterated integrals depending on both variables $\eta$ and $z$ 
contribute. 

In $z$-space diagram $D_{2a}$ consists of contribution $D_{2a}^{\sf Reg}$,
which, other than a term proportional to $\delta(1-z)$, is regular as $z\rightarrow 1$ and a contribution $D_{2a}^{+}$,
\begin{eqnarray}
D_{2a}(z)&=&D_{2a}^{\sf Reg}(z) +D_{2a}^{+}(z)~.
\end{eqnarray}
The latter term contains distributions like $\propto 1/(1-z)$ or $\propto 1/(1-z)^{3/2}$, understood as $+$-distributions.

For a distribution $f^{(+)}(x)$ of the general form
\begin{equation}
f^{(+)}(x)=\sum_{k=0}^m \left(\frac{\ln^k(1-x)}{q_k(x)}\right)_+ p_k(x)
\end{equation}
we define the Mellin transform by
\begin{equation}
\Mvec\left[f^{(+)}\right](N)=\int_0^1 dx~\sum_{k=0}^m\frac{x^N 
p_k(x)-p_k(1)}{q_k(x)}[\ln(1-x)]^k~.
\label{eq:PlusMellin}
\end{equation}
Note that in this section we use a different convention for the Mellin transform (\ref{eq:PlusMellin}), 
if compared with (\ref{eq:MT}).

This regularization is also required for the Mellin transform of the diagrams
$D_{2a}$, $D_{2b}$, $D_{8a}$ and $D_{8b}$ below. For $D_{2a}$ the $+$-part is given by 
\begin{eqnarray}
 D_{2a}^{(+)}(z)&=&
\left(m_1^2\right)^{\ep/2} \left(m_2^2\right)^{-3+\ep} \Biggl[
\frac{\eta^3}{210 (1-z)} \left(\frac{1}{\ep}+\frac{37}{210}-\frac{1}{2} \HA_1(z)-\frac{1}{2} \ln(\eta)\right)
\NN\\&&
-\frac{5 \eta^3 \sqrt{z}}{1536(1-z)^{3/2}} f_1(\eta,z)
\Biggr]~,
\end{eqnarray}
and the regular contribution to Diagram $2a$ is
\begin{eqnarray}
 D_{2a}^{\sf Reg}(z)&=&
\left(m_1^2\right)^{\ep/2} \left(m_2^2\right)^{-3+\ep} 
\Biggl\{
\delta\left(1-z\right) \frac{\eta^3}{105} \Biggl[
\frac{1}{\ep^2}
+\frac{1}{\ep} \biggl(
\frac{37}{210}-\frac{1}{2} \ln\left(\eta\right)
\biggr)
-\frac{523}{22050}
\NN\\&&
+\frac{3}{8} \zeta_2
-\frac{37}{420} \ln\left(\eta\right)
+\frac{1}{4} \HA_{0,0}\left(\eta\right)
\Biggr]
-\frac{3-6z+\left(3+7\eta^2+6\eta^3\right)z^2}{1260 z^2 \ep}
\NN\\&&
+\frac{Q_1}{11289600\eta z^2}
+\frac{Q_2}{645120\eta z} \big[\HA_1\left(z\right)+\ln(\eta)\big]
-\frac{Q_4}{645120\eta z^2} \ln\left(z\right)
\NN\\&&
+\frac{Q_3}{1536\eta\sqrt{1-z}z^{5/2}} f_1(\eta,z)
\Biggr\}~,
\end{eqnarray}
with the polynomials
\begin{eqnarray}
 Q_1&=&11025(z-1)^3z+18375\eta^5z^4+\eta^4z^2\left(-9472+25725z-62475z^2\right)
 \NN\\&&
+49\eta^3 z^2\left(-1091-900z+1350z^2\right)-\eta(z-1)^2\left(-8704-22050z+25725z^2\right)
\NN\\&&
-245\eta^2z\left(133-253z+90z^2+30z^3\right)~,
\\
Q_2&=&315(z-1)^3-525\eta^5z^3-105\eta(z-1)^2(6+z)+7\eta^3z\left(11+180z+90z^2\right)
\NN\\&&
+3\eta^4z\left(512-595z+245z^2\right)-105\eta^2\left(-1+9z-18z^2+10z^3\right)~,
\\
Q_3&=&3+(-9+4\eta)z+\left(9-8\eta-6\eta^2\right)z^2
+(\eta-1)^3(3+5\eta)z^3~,
\\
Q_4&=&
105 (\eta-1)^2 \left(17 \eta^2+22 \eta+9\right) z^3-3 \left(35 \eta^2
+302 \eta-105\right) z+\big(1715 \eta^3
\NN\\&&
+945 \eta^2-387 \eta-945\big) z^2
+105 (\eta-1)^3 (\eta+1) (5 \eta+3) z^4+768 \eta~.
\end{eqnarray}
Performing the Mellin transformation by using the regularization (\ref{eq:PlusMellin}) yields
\begin{eqnarray}
D_{2a}(N)&=&
\left(m_1^2\right)^{\ep/2} \left(m_2^2\right)^{-3+\ep} \left(\frac{1+(-1)^N}{2}\right)
\Biggl\{
\frac{\eta^3}{105 \ep^2}
+\frac{1}{\ep}
\Biggl[
-\frac{\eta^3}{210} S_1\left(N\right)
\NN\\&&
-\frac{\eta^3}{210} \ln(\eta)
+\frac{2 N \big(37 N^2-105 N+68\big) \eta ^3-245 (N-1) N \eta ^2-210}{44100 (N-1) N (N+1)}\Biggr]
\NN\\&&
+\frac{(1-\eta)^{-N-1} P_4}{6144 (2 N-3) (2 N-1) (2 N+1)} \biggl[
\frac{1}{2} \ln^2(\eta)
+\ln(\eta) S_1\left(1-\eta,N\right)
\NN\\&&
-S_2\left(1-\eta,N\right)
+S_{1,1}\left(1-\eta,1,N\right)
\biggr]
+\frac{\eta^3}{840} \big[
S_1^2(N)+S_2(N)+3 \zeta_2
\big]
\NN\\&&
+\frac{2^{-2 N-12} \binom{2 N}{N} P_6}{3 (N+1) (2 N-3) (2 N-1)} \Biggl[
-\frac{2}{\sqrt{\eta}} \big[\HA_{-1,0,0}\left(\sqrt{\eta}\right)+\HA_{1,0,0}\left(\sqrt{\eta}\right)\big]
\NN\\&&
+\frac{\ln^2(\eta)}{2 (\eta-1)}
-\frac{1}{1-\eta} \sum_{i=1}^N \frac{ 2^{2 i} (1-\eta )^{-i}}{\binom{2 i}{i} \big(1+2 i\big)} \biggl(
\frac{1}{2} \ln^2(\eta)
+\ln(\eta) S_1(1-\eta,i)
\NN\\&&
+S_{1,1}(1-\eta,1,i)
-S_2(1-\eta,i)
\biggr)
\Biggr]
+\frac{\eta^3}{420} \biggl[
\ln(\eta) S_1(N)+\frac{1}{2} \ln^2(\eta)
\biggr]
\NN\\&&
+\frac{P_3}{1185408000 (N-1)^2 N^2 (N+1)^2 (2 N-3) (2 N-1)}
\NN\\&&
+\frac{P_5}{22579200 (N+1) (2 N-3) (2 N-1)} \big[S_1\left(N\right)+\ln(\eta)\big]
\Biggr\}~,
\end{eqnarray}
with the polynomials
\begin{eqnarray}
P_3&=&14363896 \eta ^3 N^8-4 \eta ^2 (6247133 \eta +7928445) N^7
\NN\\&&
-10 \eta  \left(1788305 \eta ^2-10831254 \eta+519645\right) N^6
\NN\\&&
+\left(18840889 \eta ^3-108183915 \eta^2+18290475 \eta +24675735\right) N^5
\NN\\&&
+\left(66146587 \eta ^3+4378395 \eta ^2-17775975 \eta +1881705\right) N^4
\NN\\&&
-\left(78524357 \eta ^3-41113695 \eta ^2+8412075 \eta+86929815\right) N^3
\NN\\&&
+3 \left(7348189 \eta ^3-4635645 \eta^2+7657475 \eta +15366365\right) N^2
\NN\\&&
-40320 (245 \eta -424) N-8467200~,
\\
P_4&=&5 \eta ^3 \left(8 N^3-12 N^2-2 N+3\right)+\eta ^2 \left(-28 N^2+64 N-9\right)-3 \eta  (2 N+17)+45,
\\
P_5&=&\eta ^3 \left(71224 N^3+217316 N^2-666110 N+269823\right)+24745 \eta ^2 \left(4 N^2-8 N+3\right)
\NN\\&&
-3675 \eta  (14 N-31)-165375~,
\\
P_6&=&5 \eta ^4 \left(16 N^4-40 N^2+9\right)-12 \eta ^3 \left(8 N^3-12 N^2-2 N+3\right)-6 \eta ^2 \left(4 N^2-8 N+3\right)
\NN\\&&
+12 \eta  (2 N-3)+45~.
\end{eqnarray}
Diagram $2b$ exhibits a very similar structure and is related to diagram $2a$ 
by the interchange $m_1\leftrightarrow m_2,~\eta\rightarrow 1/\eta$. Its 
$z$-space contributions consists of a part which requires regularization via the 
$+$-distribution,
\begin{eqnarray}
D_{2b}^{(+)}(z)&=& 
\left(m_1^2\right)^{-3+\ep} \left(m_2^2\right)^{\ep/2} \Biggl[
\frac{1}{210 \eta^3 (1-z)} \left(\frac{1}{\ep}+\frac{37}{210}-\frac{1}{2} \HA_1(z)+\frac{1}{2} \ln(\eta)\right)
\NN\\&&
+\frac{5 \sqrt{z}}{1536 \eta^4 (1-z)^{3/2}} f_2(\eta,z)
\Biggr]~,
\end{eqnarray} 
and a remainder contribution
\begin{eqnarray} 
D_{2b}^{\sf Reg}(z)&=&
\left(m_1^2\right)^{-3+\ep} \left(m_2^2\right)^{\ep/2} 
\Biggl\{
\frac{\delta\left(1-z\right)}{105 \eta^3} \Biggl[
\frac{1}{\ep^2}
+\frac{1}{\ep}\biggl(
\frac{37}{210}
+\frac{1}{2} \ln(\eta)
\biggr)
-\frac{523}{22050}
\NN\\&&
+\frac{37}{420} \ln(\eta)
+\frac{1}{8} \ln^2(\eta)
+\frac{3}{8}\zeta_2
\Biggr]
-\frac{3 \eta^3 (z-1)^2+6 z^2+7 \eta z^2}{1260 \eta^3 z^2 \ep}
\NN\\&&
+\frac{Q_7}{11289600 \eta^4 z^2}
+\frac{Q_6}{645120 \eta^4 z^2} \ln(z)
+\frac{Q_5}{645120 \eta^4 z} \big[\HA_1\left(z\right)-\ln(\eta)\big]
\NN\\&&
+\frac{Q_8}{1536 \eta^4 \sqrt{1-z} z^{5/2}} f_2(\eta,z)
\Biggr\}~,
\end{eqnarray}
with
\begin{eqnarray}
Q_5&=&315 \eta^5 (z-1)^3-105 \eta^4 (z-1)^2 (z+6)
-105 \eta^3 \big(10 z^3-18 z^2+9 z-1\big)
\NN\\&&
+7 \eta^2 z \big(90 z^2+180 z+11\big)
+3 \eta  z \big(245 z^2-595 z+512\big)-525 z^3~,
\\
Q_6&=&315 \eta^5 (z-1)^3 z
-3 \eta^4 (z-1)^2 \big(35 z^2+210 z+256\big)
-105 \eta^3 z \big(10 z^3-18 z^2
\NN\\&&
+9 z-1\big)+35 \eta^2 z^2 \big(18 z^2+36 z-49\big)+105 \eta  z^3 (7 z-17)-525 z^4~,
\\
Q_7&=&11025 \eta^5 (z-1)^3 z
-\eta^4 (z-1)^2 \big(25725 z^2-22050 z-8704\big)
-245 \eta^3 z \big(30 z^3
\NN\\&&
+90 z^2-253 z+133\big)+49 \eta^2 z^2 \big(1350 z^2-900 z-1091\big)+\eta z^2 \big(-62475 z^2
\NN\\&&
+25725 z-9472\big)+18375 z^4~,
\\
Q_8&=&
        3 \eta ^4 (z-1)^3
        -4 \eta ^3 (z-1)^2 z
        -6 \eta ^2 (z-1) z^2
        -5 z^3
        +12 \eta  z^3~.
\end{eqnarray}
In Mellin $N$-space one obtains
\begin{eqnarray}
D_{2b}(N)&=&
\left(m_1^2\right)^{\ep/2} \left(m_2^2\right)^{-3+\ep} \left(\frac{1+(-1)^N}{2}\right)
\Biggl\{
\frac{1}{105 \eta^3 \ep^2}
+\frac{1}{210 \eta^3 \ep}
\Biggl[
-S_1\left(N\right)
+\ln(\eta)
\NN\\&&
-\frac{210 \eta^3+245 (N-1) N \eta-2 N \big(37 N^2-105 N+68\big)}{210 (N-1) N (N+1)}\Biggr]
\NN\\&&
+\frac{(\eta-1)^{-N-1} \eta^{N-2} P_8}{6144 (2 N-3) (2 N-1) (2 N+1)} \biggl[
\frac{1}{2} \ln^2(\eta)
-\ln(\eta) S_1\left(\frac{\eta-1}{\eta},N\right)
\NN\\&&
-S_2\left(\frac{\eta-1}{\eta},N\right)
+S_{1,1}\left(\frac{\eta-1}{\eta},1,N\right)
\biggr]
+\frac{1}{840 \eta^3} \big[
S_1^2(N)+S_2(N)+3 \zeta_2
\big]
\NN\\&&
+\frac{2^{-2 N-12} \binom{2 N}{N} P_9}{3 \eta^3 (N+1) (2 N-3) (2 N-1)} \Biggl[
-\frac{2}{\sqrt{\eta}} \big[\HA_{-1,0,0}\left(\sqrt{\eta}\right)+\HA_{1,0,0}\left(\sqrt{\eta}\right)\big]
\NN\\&&
+\frac{\ln^2(\eta)}{2 (1-\eta)}
-\frac{1}{\eta-1} \sum_{i=1}^N \frac{ 2^{2 i} (\eta-1)^{-i} \eta^i}{\binom{2 i}{i} \big(1+2 i\big)} \biggl[
\frac{1}{2} \ln^2(\eta)
-\ln(\eta) S_1\left(\frac{\eta-1}{\eta},i\right)
\NN\\&&
+S_{1,1}\left(\frac{\eta-1}{\eta},1,i\right)
-S_2\left(\frac{\eta-1}{\eta},i\right)
\biggr]
\Biggr]
+\frac{1}{420 \eta^3} \biggl[
\frac{1}{2} \ln^2(\eta)-\ln(\eta) S_1(N)
\biggr]
\NN\\&&
+\frac{P_{10}}{1185408000 \eta^3 (N-1)^2 N^2 (N+1)^2 (2 N-3) (2 N-1)}
\NN\\&&
+\frac{P_7}{22579200 \eta^3 (N+1) (2 N-3) (2 N-1)} \big[\ln(\eta)-S_1\left(N\right)\big]
\Biggr\}~,
\end{eqnarray}
with the polynomials
\begin{eqnarray}
P_7&=&165375 \eta ^3+3675 \eta ^2 (14 N-31)-24745 \eta \left(4 N^2-8 N+3\right)-71224 N^3
\NN\\&&
-217316 N^2
+666110 N-269823~, \\ 
P_8&=&45 \eta ^3-3 \eta ^2 (2 N+17)+\eta  \left(-28 N^2+64 N-9\right)+5 \left(8 N^3-12 N^2-2 N+3\right),
\\
P_9&=&45 \eta ^4+12 \eta ^3 (2 N-3)-6 \eta ^2 \left(4 N^2-8 N+3\right)-12 \eta  \left(8 N^3-12 N^2-2 N+3\right)
\NN\\&&
+5 \left(16 N^4-40 N^2+9\right)~,
\\
P_{10}&=&105 \eta ^3 \left(235007 N^5+17921 N^4-827903 N^3+439039 N^2+162816 N-80640\right)
\NN\\&&
-25725 \eta ^2 (N-1)^2 N \left(202 N^3-307 N^2-125 N+384\right)
\NN\\&&
-5145 \eta  (N-1)^2 N^2 \left(6164 N^3-8724 N^2-2585 N+2703\right)
\NN\\&&
+(N-1)^2 N^2 \bigl(14363896 N^4+3739260 N^3-24768426 N^2
\NN\\&&
-34435223 N+22044567\bigr)~.
\end{eqnarray}
\begin{figure}[H]
\begin{center}
\includegraphics[scale=1]{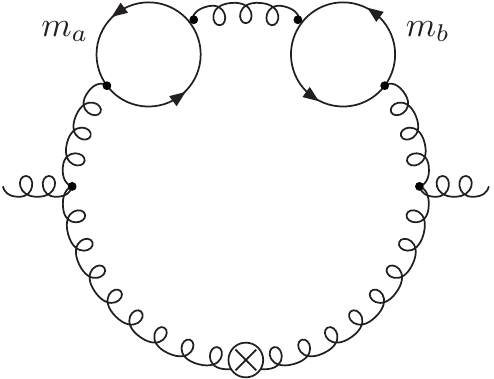}
\end{center}
\caption{\sf Diagram 3. Due to the symmetry of the diagram both mass assignments $m_a=m_1$, $m_b=m_2$ and $m_a=m_2$, $m_b=m_1$ yield identical results.}
\end{figure}
{\noindent
Diagram $3$ displays a particularly simple structure and does only depend on the 
logarithms $\HA_0(\eta)=\ln(\eta)$ and $\HA_0(z)=\ln(z)$ in $z$-space,}
\begin{eqnarray}
D_3(z)&=&
\left(m_1^2\right)^{\ep/2} \left(m_2^2\right)^{\ep-3} \frac{1-z}{210 z}
 \Biggl[
(1+\eta^3) \left(\frac{1}{\ep}+\frac{37}{210}+\frac{1}{2} \ln\left(z\right)\right)
\NN\\&&
-\frac{7}{12} \eta (1+\eta)
-\frac{1}{2} \eta^3 \ln\left(\eta\right)
\Biggr]~.
\end{eqnarray}
In $N$-space this corresponds to an expression in terms of rational functions and $\ln(\eta)$ only. It is given by
\begin{eqnarray}
D_{3}(N)&=&
\left(m_1^2\right)^{\ep/2} \left(m_2^2\right)^{\ep-3} 
\left(\frac{1+(-1)^N}{420 N (N+1)}\right)
\Biggl[
\frac{1+\eta^3}{\ep}
+\frac{(1+\eta) P_{11}}{420 N (N+1)}
-\frac{\eta^3}{2} \ln(\eta )\Biggr],
\end{eqnarray}
with the polynomial
\begin{eqnarray}
P_{11}&=&\eta ^2 \left(74 N^2-346 N-210\right)+\eta 
 \left(-319 N^2+101 N+210\right)
 +74 N^2-346 N-210~.
 \NN\\&&
\end{eqnarray}
\begin{figure}[H]
\begin{center}
\includegraphics[scale=1]{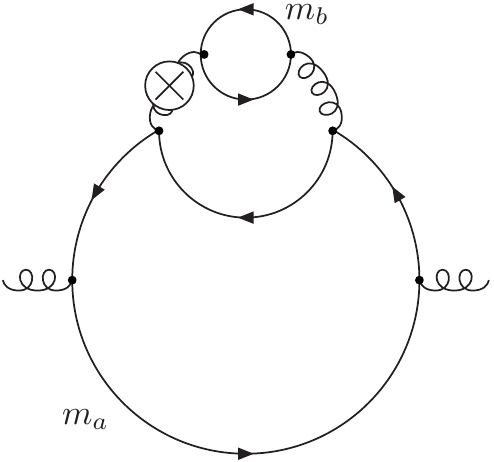}
\end{center}
\caption{\sf  Topology 4. $D_{4a}$ is given by assigning $m_a=m_2$,$m_b=m_1$ and  $D_{4b}$ by assigning $m_a=m_1$, $m_b=m_2$ respectively.}
\end{figure}
{\noindent
The $z$-space expressions for Diagrams $D_{4a}$ and $D_{4b}$ are completely 
regular as $z \rightarrow 1$. For $D_{4a}(z)$ one obtains}
\begin{eqnarray}
D_{4a}(z)&=&
\left(m_1^2\right)^{\ep/2} \left(m_2^2\right)^{-3+\ep} 
\Biggl[
\frac{1}{\ep} 
\left(\frac{(1-z)\left(-1+5z+2z^2\right)}{60z^2}
+\frac{1}{10} \ln\left(z\right)
\right)
\NN\\&&
+\frac{Q_9}{57600\eta z^2}
-\frac{\eta^2\zeta_2}{60}
+\frac{(1-z) Q_{10}}{1920\eta z} \HA_1\left(z\right)
+\frac{Q_{13}}{28800\eta z^2} \ln\left(z\right)
\NN\\&&
+\frac{(1-z)^{3/2} Q_{11}}{240 \eta z^{5/2}} f_1(\eta,z)
-\frac{\eta(5+3 \eta)(7-5 \eta)}{7680} \ln^2(\eta)
\NN\\&&
+\frac{\eta}{480} (35+3 \eta-8 \eta^2) f_5(\eta,z)
+\frac{(1-\eta) (7+2 \eta-\eta^2)}{8 \eta} f_3(\eta,z)
\NN\\&&
+\frac{3+\eta^2}{120} \ln^2(z)
+\frac{Q_{12}}{1920 \eta z} \ln(\eta)
+\frac{\eta^2}{60} \HA_{0,1}(z)
+\frac{\eta^2}{60} \ln(z) \ln(\eta)
\NN\\&&
+\frac{7-5\eta-3\eta^2+\eta^3}{512\sqrt{\eta}} G\left(\left\{\frac{\sqrt{\tau}}{1-\tau},\frac{1}{\tau},\frac{1}{\tau}\right\},\eta\right)
\Biggr]~,
\end{eqnarray}
with the polynomials
\begin{eqnarray}
 Q_9&=&-150\eta^3z^3\left(10+41z-62z^2+24z^3\right)+75\eta^4z^2\left(12-22z+43z^2-38z^3+12z^4\right)
 \NN\\&&
-20\eta^2z\left(82-57z+65z^2+135z^3-240z^4+90z^5\right)
\NN\\&&
-15z\left(24-152z+662z^2+51z^3-830z^4+420z^5\right)
\NN\\&&
+2\eta\left(104+696z-2568z^2+7798z^3+3195z^4-11850z^5+5400z^6\right)~,
\\
 Q_{10}&=&
-6+67z-81z^2-85z^3+105z^4+5\eta^4(z-1)^2z(3 z-1)
\NN\\&&
-10\eta^3(z-1)^2z(1+3z)
+\eta^2\left(2-50z+20z^2+160z^3-120z^4\right)
\NN\\&&
+2\eta\left(-6+99z+7z^2-45z^3+15z^4\right)~,
\\
Q_{11}&=&3+(-13+5\eta)z-15\left(-3+2\eta+\eta^2\right)z^2
+15\left(7-5\eta-3\eta^2+\eta^3\right)z^3~,
\\
Q_{12}&=&
-6+38z-148z^2-4z^3+190z^4-105z^5-5\eta^4z\left(3-6z+12z^2
-10z^3+3z^4\right)
\NN\\&&
+10\eta^3z\left(3+6z^2-8z^3+3z^4\right)
-2\eta\left(6-30z
+92z^2+52z^3-60z^4+15z^5\right)
\NN\\&&
+\eta^2\left(2-17z+70z^2+140z^3
-280z^4+120z^5\right)~,
\\
Q_{13}&=&
150\eta^3z^4\left(6-8z+3z^2\right)
-75\eta^4z^3\left(-6+12z-10z^2+3z^3\right)
-15z\left(6-38z
\right.
\NN\\&&
\left.
+148z^2+4z^3-190z^4+105z^5\right)
+10\eta^2z\left(3-203z+105z^2+210z^3-420z^4
\right.
\NN\\&&
\left.
+180z^5\right)
-6\eta\left(40-210z-86z^2+460z^3+260z^4-300z^5+75z^6\right)~.
\end{eqnarray}

Performing the Mellin transform yields
\begin{eqnarray}
D_{4a}(N)&=&
\left(m_1^2\right)^{\ep/2} \left(m_2^2\right)^{-3+\ep} \left(\frac{1+(-1)^N}{2 (N+1)^2 (N+2)}\right)
\Biggl\{
-\frac{1}{5 (N-1) N \ep}
\NN\\&&
+(1-\eta )^{-N} \eta \frac{\big(4 N^2-4 N-3\big) \eta ^2+(4 N-6) \eta -35}{512 (2 N-3)} \biggl[
\frac{1}{2} \ln^2(\eta)
\NN\\&&
+\ln(\eta) S_1(1-\eta,N)
+S_{1,1}(1-\eta,1,N)
-S_2(1-\eta,N)
\biggr]
\NN\\&&
-\eta \frac{3 \big(4 N^2-4 N-3\big) \eta ^2+12 (2 N-3) \eta -35}{768 (2 N-3)} \big[\ln(\eta)+S_1(N)\big]
\NN\\&&
+\frac{2^{-2 N-7} \binom{2 N-2}{N-1} P_{13}}{N (2 N-3)} \Biggl[
\frac{1}{\sqrt{\eta}} \big[\HA_{-1,0,0}(\sqrt{\eta})+\HA_{1,0,0}(\sqrt{\eta})\big]
\NN\\&&
-\frac{1}{4} \ln^2(\eta)+\ln(\eta)-2
+\frac{\eta}{8} \sum_{i=1}^N \frac{2^{2 i}}{\binom{2 i-2}{i-1}} \biggl(-\frac{1}{i^2}+\frac{\ln(\eta)}{i}+\frac{S_1(i)}{i}\biggr)
\NN\\&&
+\frac{\eta}{8} \sum_{i=1}^N \frac{2^{2 i} (1-\eta)^{-i}}{\binom{2 i-2}{i-1}} \biggl(
-\ln(\eta) S_1(1-\eta,i)
-S_{1,1}(1-\eta,1,i)
\NN\\&&
+S_2(1-\eta,i)
-\frac{1}{2} \ln^2(\eta)
\biggr)
\Biggr]
+\frac{P_{12}}{28800 (N-1)^2 N^2 (N+1) (2 N-3)}
\Biggr\}~,
\end{eqnarray}
with the polynomials
\begin{eqnarray}
P_{12}&=&900 \eta ^3 N^7-900 \eta ^2 (2 \eta -1) N^6-25 \eta  \left(27 \eta ^2+90 \eta +163\right) N^5+\bigl(2475 \eta^3
\NN\\&&
+450 \eta ^2+8875 \eta
 +7296\bigr) N^4+\left(-225 \eta ^3+2250 \eta ^2-725 \eta +6336\right) N^3
 \NN\\&& 
 -\bigl(675 \eta^3+1350 \eta ^2+8875 \eta
 +33216\bigr) N^2+192 (25 \eta +27) N+8640~,
\\
P_{13}&=&\eta ^3 \left(8 N^3-12 N^2-2 N+3\right)+3 \eta ^2 \left(4 N^2-8 N+3\right)+\eta  (45-30 N)-105
~.
\end{eqnarray}
Interchanging the masses $m_1 \leftrightarrow m_2$ one obtains
\begin{eqnarray}
 D_{4b}(z)&=&
\left(m_1^2\right)^{-3+\ep} \left(m_2^2\right)^{\ep/2} 
\Biggl[
\frac{1}{\ep}
\left(
-\frac{(z-1) \big(2 z^2+5 z-1\big)}{60 z^2}
+\frac{1}{10} \HA_0\left(z\right)
\right) 
\NN\\&&     
+\frac{Q_{17}}{57600 \eta^3 z^2}
-\frac{\zeta_2}{60 \eta^2}
+\frac{(1-z) Q_{14}}{1920 \eta^3 z} \HA_1(z)
+\frac{Q_{16}}{28800 \eta^3 z^2} \ln(z)
\NN\\&&     
-\frac{(1-z)^{3/2} Q_{18}}{240 \eta^3 z^{5/2}} f_2(\eta,z)
+\frac{105 \eta^3-180 \eta^2-5 \eta+64}{7680 \eta^2} \ln^2(\eta) 
\NN\\&&
-\frac{35 \eta^2+3 \eta-8}{480 \eta^2} f_6(\eta,z)
-\frac{(1-\eta) (7 \eta^2+2 \eta-1)}{8 \eta^3} f_4(\eta,z)
\NN\\&&
+\frac{3 \eta^2+1}{120 \eta^2} \ln^2(z)
+\frac{Q_{15}}{1920 \eta^3 z} \ln(\eta)
+\frac{1}{60 \eta^2} \big[\HA_{0,1}(z)-\ln(\eta) \ln(z)\big]
\NN\\&&
+\frac{7 \eta^3-5 \eta^2-3 \eta+1}{512 \eta ^{5/2}} G\left(\left\{\frac{\sqrt{\tau}}{1-\tau},\frac{1}{\tau},\frac{1}{\tau}\right\},\eta\right)
\Biggr]~,
\end{eqnarray}
where
\begin{eqnarray}
Q_{14}&=&\eta ^4 \big(105 z^4-85 z^3-81 z^2+67 z-6\big)
+2 \eta ^3 \big(15 z^4-45 z^3+7 z^2+99 z
\NN\\&&
-6\big)
+\eta ^2 \big(-120 z^4+160 z^3+20 z^2-50 z+2\big)
-10 \eta  (z-1)^2 z (3 z+1)
\NN\\&&
+5 (z-1)^2 z (3 z-1)~,
\\
Q_{15}&=&\eta ^4 \big(105 z^5-190 z^4+4 z^3+148 z^2-143 z+6\big)
+2 \eta ^3 \big(15 z^5-60 z^4
\NN\\&&
+52 z^3+92 z^2-45 z+6\big)
-\eta ^2 \big(120 z^5-280 z^4+140 z^3+70 z^2
\NN\\&&
-137 z+2\big)
-10 \eta  z^3 \big(3 z^2-8 z+6\big)
+5 z^2 \big(3 z^3-10 z^2+12 z-6\big)~,
\\
Q_{16}&=&-15 \eta ^4 z \big(105 z^5-190 z^4+4 z^3+148 z^2-38 z+6\big)
-6 \eta ^3 \big(75 z^6-300 z^5
\NN\\&&
+260 z^4+460 z^3-86 z^2-210 z+40\big)
+10 \eta ^2 z \big(180 z^5-420 z^4+210 z^3
\NN\\&&
+105 z^2-203 z+3\big)
+150 \eta  z^4 \big(3 z^2-8 z+6\big)
-75 z^3 \big(3 z^3-10 z^2+12 z-6\big),
\nonumber\\
\\
Q_{17}&=&-15 \eta ^4 z \big(420 z^5-830 z^4+51 z^3+662 z^2-572 z+24\big)
+2 \eta ^3 \big(5400 z^6
\NN\\&&
-11850 z^5+3195 z^4+7798 z^3-7968 z^2+696 z+104\big)
-20 \eta ^2 z \big(90 z^5
\NN\\&&
-240 z^4+135 z^3+65 z^2-147 z+82\big)
-150 \eta  z^2 \big(24 z^4-62 z^3+41 z^2+10 z
\NN\\&&
-24\big)+75 z^3 \big(12 z^3-38 z^2+43 z-22\big)~,
\\
Q_{18}&=&
        15 z^3
        -15 \eta  z^2 (3 z+1)
        -5 \eta ^2 z \big(
                15 z^2+6 z-1\big)
        +\eta ^3 \big(
                105 z^3+45 z^2-13 z+3\big)~.
\nonumber\\ 
\end{eqnarray}

In Mellin space $D_{4b}$ takes the form
\begin{eqnarray}
D_{4b}(N)&=&
\left(m_1^2\right)^{-3+\ep} \left(m_2^2\right)^{\ep/2} 
\left(\frac{1+(-1)^N}{2 (N+1)^2 (N+2)}\right)
\Biggl\{
- \frac{1}{\ep} \frac{1}{5 (N-1) N}
\NN\\&&
+\frac{35 \eta^2+(6-4 N) \eta-4 N^2+4 N+3}{512 \eta^3 (2 N-3)} \left(\frac{\eta}{\eta-1}\right)^N \biggl[
-\frac{1}{2} \ln^2(\eta)
\NN\\&&
+\ln(\eta) S_1\left(\frac{\eta -1}{\eta },N\right)
+S_2\left(\frac{\eta -1}{\eta },N\right)
-S_{1,1}\left(\frac{\eta -1}{\eta },1,N\right)
\biggr]
\NN\\&&
+\frac{35 \eta ^2+(36-24 N) \eta -12 N^2+12 N+9}{768 \eta^3 (2 N-3)} \big[S_1(N)-\ln(\eta)\big]
\NN\\&&
+\frac{P_{15}}{28800 \eta^3 (N-1)^2 N^2 (N+1) (2 N-3)}
\NN\\&&
+\frac{2^{-2 N-7} \binom{2 N-2}{N-1} P_{14}}{\eta^3 N (2 N-3)} \Biggl[
-\sqrt{\eta} \big[\HA_{-1,0,0}(\sqrt{\eta})+\HA_{1,0,0}(\sqrt{\eta})\big]
\NN\\&&
+\frac{1}{4} \ln^2(\eta)+\ln(\eta)+2
+\frac{1}{8 \eta} \sum_{i=1}^N\frac{2^{2 i}}{\binom{2 i-2}{i-1}} \biggl(\frac{1}{i^2}+\frac{\ln(\eta)}{i}-\frac{S_1(i)}{i}\biggr)
\NN\\&&
+\frac{1}{8 \eta} \sum_{i=1}^N\frac{2^{2 i} (\eta-1)^{-i} \eta ^{i}}{\binom{2 i-2}{i-1}} \biggl[
-\ln(\eta) S_1\left(\frac{\eta-1}{\eta },i\right)
\NN\\&&
+S_{1,1}\left(\frac{\eta-1}{\eta },1,i\right)
-S_2\left(\frac{\eta-1}{\eta},i\right)
+\frac{1}{2} \ln^2(\eta)
\biggr]
\Biggr]
\Biggr\}~,
\end{eqnarray}
with the polynomials
\begin{eqnarray}
P_{14}&=&105 \eta ^3+15 \eta ^2 (2 N-3)-3 \eta  \left(4 N^2-8 N+3\right)-8 N^3+12 N^2+2 N-3,
\\
P_{15}&=&192 \eta ^3 \left(38 N^4+33 N^3-173 N^2+27 N+45\right)-25 \eta ^2 (N-1)^2 N \bigl(163 N^2
\NN\\&&
-29 N-192\bigr)
+450 \eta  (N-1)^2 N^2 \left(2 N^2-N-3\right)
\NN\\&&
+225 (N-1)^2 N^2 \left(4 N^3-7 N-3\right)~.
\end{eqnarray}
\begin{figure}[!htb]
\begin{center}
\includegraphics[scale=1]{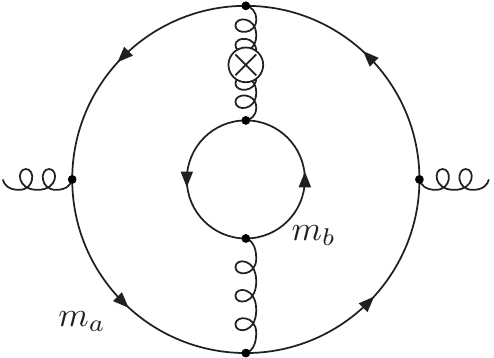}
\end{center}
\caption{\sf This diagram depicts $D_{5a}$ with $m_a=m_2$, $m_b=m_1$ and $D_{5b}$ with $m_a=m_1$, $m_b=m_2$ respectively.}
\end{figure}
Diagram $D_{5a}$ is given in $z$-space by
\begin{eqnarray}
D_{5a}(z)&=&
\left(m_1^2\right)^{\ep/2} \left(m_2^2\right)^{-3+\ep}
\Biggl[
\frac{1}{\ep}
\left(
-\frac{(1-z)\left(1-5 z-2 z^2\right)}{45 z^2}
+\frac{2}{15}\ln\left(z\right)
\right)
\NN\\&&
-\frac{Q_{19}}{201600 \eta z^2}
-\frac{\zeta_2}{105} \eta^3 z
-\frac{(1-z) Q_{20}}{6720 \eta z} \HA_1\left(z\right)
+\frac{Q_{21}}{100800\eta z^2} \ln(z)
\NN\\&&
-\frac{(1-z)^{3/2} Q_{22}}{840 \eta z^{5/2}} f_1(\eta,z)
+\frac{\eta^3 z}{105} \big[\HA_{0,1}(z)+\HA_{0,0}(\eta)\big]
\NN\\&&
+\frac{\eta \left(175+35 \eta+16 \eta^2z-16 \eta^3 z\right)}{1680} f_5(\eta,z)
+\frac{Q_{23}}{6720 \eta z} \ln(\eta)
\NN\\&&
+\frac{(1-\eta) (5+2\eta+\eta^2)}{4 \eta} f_3(\eta,z)
+\frac{7+\eta^3 z}{210} \ln^2(z)
+\frac{\eta^3 z}{105} \ln(\eta) \ln(z)
\NN\\&&
+\frac{5-3\eta-\eta^2-\eta^3}{256\sqrt{\eta}} G\left(\left\{\frac{\sqrt{\tau}}{1-\tau},\frac{1}{\tau},\frac{1}{\tau}\right\},\eta\right)
-\frac{\eta}{768} \left(5+3 \eta^2\right) \ln^2(\eta)
\Biggr]~,
\end{eqnarray}
with the polynomials
\begin{eqnarray}
 Q_{19}&=&30\eta^3z^2\left(216-146z-105z^2+70z^3\right)+525\eta^4z^2\left(12-22z+43z^2
 \right.
 \NN\\&&
 \left.
 -38z^3+12z^4\right)
+75z\left(24-152z+662z^2+51z^3-830z^4+420z^5\right)
\NN\\&&
+20\eta^2z\bigl(290-237z+535z^2
+756z^3-1554z^4+630z^5\bigr)
-2\eta\bigl(672
\NN\\&&
+2248z-11984z^2+37234z^3+15345z^4-55590z^5
+25200z^6\bigr)~,
\\
Q_{20}&=&
35\eta^4(z-1)^2z(3 z-1)+2\eta^3z\left(3+35z-175z^2+105z^3\right)+5\bigl(6
\NN\\&&
-67z+81z^2
+85z^3-105z^4\bigr)
  -2\eta\left(-30+509z+17z^2-251z^3
  \right.
  \NN\\&&
  \left.
  +105z^4\right)+2\eta^2\bigl(-5+47z-28z^2
  -294z^3+210z^4\bigr)~,
\\
Q_{21}&=&
525\eta^4z^3\left(-6+12z-10z^2+3z^3\right)
+30\eta^3z^2\left(-108-32z+210z^2
\right.
\NN\\&&
\left.
-280z^3+105z^4\right)
-75z\left(6-38z+148z^2+4z^3-190z^4+105z^5\right)
\NN\\&&
+10\eta^2z\bigl(15-751z+225z^2
+798z^3-1512z^4+630z^5\bigr)
-2\eta\bigl(560
\NN\\&&
-2910z-854z^2+7380z^3+4020z^4
-5340z^5+1575z^6\bigr)~,
\\
Q_{22}&=&
-15+(65-21\eta)z+\left(-225+126\eta+35\eta^2\right)z^2
+105\left(-5+3\eta
\right.
\NN\\&&
\left.
+\eta^2+\eta^3\right)z^3~,
\\
Q_{23}&=&
35\eta^4z\left(3-6z+12z^2-10z^3+3z^4\right)+2\eta^3z\left(-3-32z+210z^2
\right.
\NN\\&&
\left.
-280z^3+105z^4\right)
-2\eta\left(30-154z+492z^2+268z^3-356z^4+105z^5\right)
\NN\\&&
-5\bigl(6-38z+148z^2
+4z^3-190z^4+105z^5\bigr)
+\eta^2\left(10-69z+150z^2
\right.
\NN\\&&
\left.
+532z^3-1008z^4+420z^5\right)~.
\end{eqnarray}
In Mellin-space one obtains
\begin{eqnarray}
D_{5a}(N)&=&
\left(m_1^2\right)^{\ep/2} \left(m_2^2\right)^{-3+\ep} 
\left(\frac{1+(-1)^N}{2 (N+1) (N+2)}\right)
\Biggl\{
-\frac{4}{15 (N-1) N (N+1) \ep}
\NN\\&&
-(1-\eta )^{-N-1} \eta \frac{\big(4 N^2-8 N+3\big) \eta^2-4 (N+1) \eta +25}{128 (2 N-3) (2 N-1)} \biggl[
\frac{1}{2} \ln^2(\eta)
\NN\\&&
+\ln(\eta) S_1(1-\eta,N)
+S_{1,1}(1-\eta,1,N)
-S_2(1-\eta,N)
\biggr]
\NN\\&&
+\eta \frac{3 \big(4 N^2-4 N-3\big) \eta^2+25}{384 (N+1) (2 N-3)} \big[S_1\left(N\right)+\ln(\eta)\big]
\NN\\&&
+\frac{2^{-2 N-7} \binom{2 N}{N} P_{17}}{(N+1) (2 N-3) (2 N-1)} \Biggl[
-\frac{1}{\sqrt{\eta}} \big[\HA_{-1,0,0}(\sqrt{\eta})+\HA_{1,0,0}(\sqrt{\eta})\big]
\NN\\&&
+\frac{\eta}{2 (1-\eta)} \sum_{i=1}^N \frac{2^{2 i} (1-\eta)^{-i}}{\binom{2 i}{i}} \biggl(
\ln(\eta) S_1(1-\eta,i)
+S_{1,1}(1-\eta,1,i)
\NN\\&&
-S_2(1-\eta,i)
+\frac{1}{2} \ln^2(\eta)
\biggr)
+\frac{\ln^2(\eta)}{4 (1-\eta)}-\ln(\eta)+2
\Biggr]
\NN\\&&
-\frac{P_{16}}{14400 (N-1)^2 N^2 (N+1)^2 (2 N-3)}
\Biggr\}~,
\end{eqnarray}
where the polynomials read
\begin{eqnarray}
P_{16}&=&900 \eta ^3 N^7-900 \eta ^2 (2 \eta +1) N^6-25 \eta  \left(27 \eta ^2-90 \eta -89\right) N^5
+\bigl(2475 \eta^3
\NN\\&&
-450 \eta ^2-4625 \eta
 -5504\bigr) N^4-\left(225 \eta ^3+2250 \eta ^2-175 \eta +3264\right) N^3
\NN\\&&
+\bigl(-675 \eta ^3+1350 \eta ^2+4625 \eta 
+22784\bigr) N^2-96 (25 \eta +46) N-5760~,
\\
P_{17}&=&\eta ^3 \left(8 N^3-12 N^2-2 N+3\right)+\eta ^2 \left(-4 N^2+8 N-3\right)+9 \eta  (2 N-3)+75~.
\end{eqnarray}
The mass-reversed diagram $D_{5b}$ obeys the $z$-space representation
\begin{eqnarray}
D_{5b}(z)&=&
\left(m_1^2\right)^{-3+\ep} \left(m_2^2\right)^{\ep/2} 
\Biggl\{
\frac{1}{\ep} \Biggl[
-\frac{(z-1) \big(2 z^2+5 z-1\big)}{45 z^2}
+\frac{2}{15} \HA_0\left(z\right)
\Biggr]
\NN\\&&
-\frac{Q_{27}}{201600 \eta^3 z^2}
-\left(\frac{1}{16}-\frac{1}{384 \eta}-\frac{5 \eta}{128}-\frac{z}{105 \eta^3}\right) \HA_{0,0}\left(\eta\right)
\NN\\&&
-\frac{(1-z)^{3/2} Q_{28}}{840 \eta^3 z^{5/2}} f_2(\eta,z)
+\frac{1}{105} \left(7+\frac{z}{\eta^3}\right) \HA_{0,0}(z)
\NN\\&&
-\frac{35 \eta^2+175 \eta^3-16 z+16 \eta z}{1680 \eta^3} f_6(\eta,z)
-\frac{\zeta_2 z}{105 \eta^3}
\NN\\&&
-\frac{(1-\eta) (5 \eta^2+2 \eta+1)}{4 \eta^3} f_4(\eta,z)
+\frac{Q_{25}}{6720 \eta^3 z} \ln(\eta)
\NN\\&&
+\frac{(1-z) Q_{24}}{6720 \eta^3 z} \HA_1(z)
+\frac{5 \eta^3-3 \eta^2-\eta-1}{256 \eta^{5/2}} G\left(\left\{\frac{\sqrt{\tau}}{1-\tau},\frac{1}{\tau},\frac{1}{\tau}\right\},\eta\right)
\NN\\&&
+\frac{Q_{26}}{100800 \eta^3 z^2} \ln(z)
+\frac{z}{105 \eta^3} \big[\HA_{0,1}(z)-\ln(\eta) \ln(z)\big]
\Biggr\}~,
\end{eqnarray}
where
\begin{eqnarray}
Q_{24}&=&5 \eta ^4 \big(105 z^4-85 z^3-81 z^2+67 z-6\big)
+2 \eta ^3 \big(105 z^4-251 z^3+17 z^2
\NN\\&&
+509 z-30\big)
+\eta ^2 \big(-420 z^4+588 z^3+56 z^2-94 z+10\big)
-2 \eta  z \big(105 z^3
\NN\\&&
-175 z^2+35 z+3\big)-35 (z-1)^2 z (3 z-1)~,
\\
Q_{25}&=&5 \eta ^4 \big(105 z^5-190 z^4+4 z^3+148 z^2-143 z+6\big)
+\eta ^3 \big(210 z^5-712 z^4
\NN\\&&
+536 z^3+984 z^2-518 z+60\big)
-\eta ^2 \big(420 z^5-1008 z^4+532 z^3+150 z^2
\NN\\&&
-489 z+10\big)
+2 \eta  z \big(-105 z^4+280 z^3-210 z^2+32 z+108\big)
\NN\\&&
-35 z^2 \big(3 z^3-10 z^2+12 z-6\big)~,
\\
Q_{26}&=&-75 \eta ^4 z \big(105 z^5-190 z^4+4 z^3+148 z^2-38 z+6\big)
-2 \eta ^3 \big(1575 z^6
\NN\\&&
-5340 z^5+4020 z^4+7380 z^3-854 z^2-2910 z+560\big)
+10 \eta ^2 z \big(630 z^5
\NN\\&&
-1512 z^4+798 z^3+225 z^2-751 z+15\big)
+30 \eta  z^2 \big(105 z^4-280 z^3
\NN\\&&
+210 z^2-32 z-108\big)
+525 z^3 \big(3 z^3-10 z^2+12 z-6\big)~,
\\
Q_{27}&=&75 \eta ^4 z \big(420 z^5-830 z^4+51 z^3+662 z^2-572 z+24\big)
-2 \eta ^3 \big(25200 z^6
\NN\\&&
-55590 z^5+15345 z^4+37234 z^3-37184 z^2
+2248 z+672\big)
\NN\\&&
+20 \eta ^2 z \big(630 z^5-1554 z^4+756 z^3+535 z^2
-867 z+290\big)
\NN\\&&
+30 \eta  z^2 \big(70 z^3-105 z^2-146 z+216\big)
\NN\\&&
+525 z^3 \big(12 z^3-38 z^2+43 z-22\big)~,
\\
Q_{28}&=&-105 z^3
        -35 \eta  z^2 (3 z+1)
        -21 \eta ^2 z \big(
                15 z^2+6 z-1\big)
\NN\\&&                
        +5 \eta ^3 \big(
                105 z^3+45 z^2-13 z+3\big)~.
\end{eqnarray}

In Mellin space one obtains
\begin{eqnarray}
D_{5b}(N)&=&
\left(m_1^2\right)^{-3+\ep} \left(m_2^2\right)^{\ep/2} 
\left(\frac{1+(-1)^N}{2 (N+1) (N+2)}\right)
\Biggl\{
-\frac{4}{15 (N-1) N (N+1) \ep}
\NN\\&&
+\frac{25 \eta ^2-4 (N+1) \eta +4 N^2-8 N+3}{128 (\eta-1) \eta^2 (2 N-3) (2 N-1)} \left(\frac{\eta}{\eta-1}\right)^N \biggl[
-\frac{1}{2} \ln^2(\eta)
\NN\\&&
+\ln(\eta) S_1\left(\frac{\eta-1}{\eta },N\right)
+S_2\left(\frac{\eta -1}{\eta},N\right)
-S_{1,1}\left(\frac{\eta -1}{\eta},1,N\right)
\biggr]
\NN\\&&
+\frac{25 \eta ^2+12 N^2-12 N-9}{384 \eta^3 (N+1) (2 N-3)} \big[S_1(N)-\ln(\eta)\big]
\NN\\&&
+\frac{P_{19}}{14400 \eta ^3 (N-1)^2 N^2 (N+1)^2 (2 N-3)}
\NN\\&&
+\frac{2^{-2 N-7} \binom{2 N}{N} P_{18}}{\eta^2 (N+1) (2 N-3) (2 N-1)} \Biggl[
-\frac{1}{\sqrt{\eta}} \big[\HA_{-1,0,0}(\sqrt{\eta})+\HA_{1,0,0}(\sqrt{\eta})\big]
\NN\\&&
+\frac{1}{\eta} \big[2+\ln(\eta)\big]
+\frac{1}{2 (\eta-1) \eta} \sum_{i=1}^N \frac{2^{2 i} (\eta-1)^{-i} \eta^{i}}{\binom{2 i}{i}} \biggl[
-\ln(\eta) S_1\left(\frac{\eta-1}{\eta},i\right)
\NN\\&&
+S_{1,1}\left(\frac{\eta-1}{\eta},1,i\right)
-S_2\left(\frac{\eta-1}{\eta},i\right)
+\frac{1}{2} \ln^2(\eta)
\biggr]
+\frac{\ln^2(\eta)}{4 (\eta-1)}
\Biggr]
\Biggr\}~,
\end{eqnarray}
where we abbreviated the polynomials
\begin{eqnarray}
P_{18}&=&75 \eta ^3+9 \eta ^2 (2 N-3)+\eta  \left(-4 N^2+8 N-3\right)+8 N^3-12 N^2-2 N+3,
\\
P_{19}&=&64 \eta ^3 \left(86 N^4+51 N^3-356 N^2+69 
N+90\right)-25 \eta ^2 (N-1)^2 N \bigl(89 N^2
\NN\\&&
-7 N-96\bigr)
+450 \eta  (N-1)^2 N^2 \left(2 N^2-N-3\right)
\NN\\&&
-225 (N-1)^2 N^2 
\left(4 N^3-7 N-3\right)~.
\end{eqnarray}
\begin{figure}[!htb]
\begin{center}
\includegraphics[scale=1]{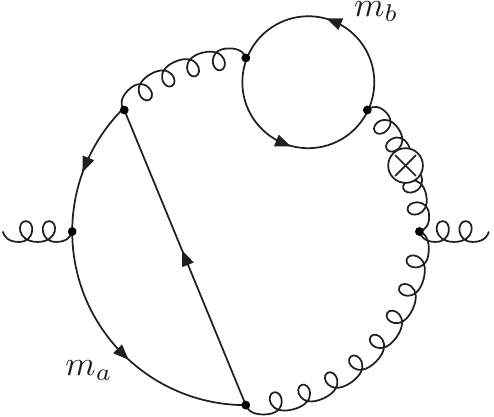}
\end{center}
\caption{\sf $D_{6a}$ with $m_a=m_1$, $m_b=m_2$ and  $D_{6b}$ with $m_a=m_2$, $m_b=m_1$ respectively.}
\label{fig:D6ab}
\end{figure}

\begin{figure}[!htb]
\begin{center}
\includegraphics[scale=1]{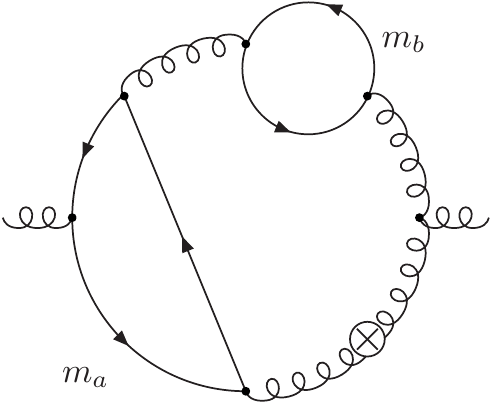}
\end{center}
\caption{\sf $D_{8a}$ with $m_a=m_2$, $m_b=m_1$ and  $D_{8b}$ with $m_a=m_1$, $m_b=m_2$, respectively.}
\label{fig:D8a}
\end{figure}

\noindent
The diagrams $D_{6a,b}$ and $D_{8a,b}$, see Figures \ref{fig:D6ab} and \ref{fig:D8a}, 
respectively, consist of one
fermionic triangle and one fermion-bubble. For $D_{6a}(z)$ one obtains
\begin{eqnarray}
D_{6a}(z)&=&
\left(m_1^2\right)^{-3+\ep} \left(m_2^2\right)^{\ep/2} 
\Biggl\{
-\frac{1}{45 \ep^2}
+\frac{1}{90 \ep}
\biggl[
\frac{11}{10}
+\frac{3}{2 \eta^2}
+\frac{5}{4 \eta }
+\frac{1}{2 z^2}
-\frac{2}{z}
\NN\\&&
+\HA_1\left(z\right)
-\HA_0\left(z\right)
\biggr]
-\frac{1}{180} \big[\HA_{1,1}\left(z\right)+\HA_{0,0}\left(z\right)\big]
-\left(\frac{1}{120}-\frac{\HA_1\left(z\right)}{420 \eta^3}\right) \zeta_2
\NN\\&&
+\frac{1}{201600 (\eta-1) \eta^3 z^2} \big[z Q_{30} \HA_1(z)+Q_{31} \HA_0(z)\big]
+\frac{Q_{32}}{18144000 \eta^3 z^2}
\NN\\&&
-\frac{\sqrt{1-z} Q_{33}}{3360 \eta^3 z^{5/2}} f_2(\eta,z)
+\frac{1}{420 \eta^3} \big[\HA_{1,0}(z) \HA_0(\eta)-\HA_1(z) \HA_{0,0}(\eta)\big]
\NN\\&& 
-\frac{75 \eta^3-63 \eta^2-35 \eta-105}{840 \eta^3} f_4(\eta,z)
+\frac{25 \eta^3-26 \eta^2-23 \eta-8}{3360 (\eta-1) \eta^2} f_6(\eta,z)
\NN\\&&
+\frac{Q_{29}}{40320 (\eta-1) \eta^3 z} \ln(\eta)
+\frac{\eta-1}{420 \eta ^3} \Biggl[
G\left(\left\{\frac{1}{1-\tau},\frac{1}{\eta \tau-\eta-\tau},\frac{1}{1-\tau}\right\},z\right)
\NN\\&&
+G\left(\left\{\frac{1}{1-\tau},\frac{1}{\eta \tau-\eta-\tau},\frac{1}{\tau}\right\},z\right)
-\ln(\eta) G\left(\left\{\frac{1}{1-\tau},\frac{1}{\eta \tau-\eta-\tau}\right\},z\right)      
\Biggr] 
\NN\\&&
-\frac{1}{420 \eta^3} \big[\HA_{1,0,1}\left(z\right)+\HA_{1,0,0}\left(z\right)\big]
\Biggr\}~,
\end{eqnarray}
with
\begin{eqnarray}
Q_{29}&=&-15 \eta ^5 \big(15 z^5-25 z^4-8 z^3+34 z^2-14 z+3\big)
+3 \eta ^4 \big(63 z^5-55 z^4-96 z^3
\NN\\&&
-10 z^2+34 z-15\big)
+\eta ^3 \big(330 z^5-826 z^4+294 z^3+987 z^2-478 z+105\big)
\NN\\&&
+\eta ^2 \big(126 z^5-154 z^4+294 z^3-273 z^2+178 z-15\big)
+\eta  z \big(-105 z^4
\NN\\&&
+35 z^3+210 z^2-105 z-12\big)
-105 z^2 \big(3 z^3-7 z^2+6 z-3\big)~,
\\
Q_{30}&=&75 \eta ^5 (z-1)^2 \big(15 z^3+5 z^2-13 z+3\big)
+\eta ^4 \big(-945 z^5+825 z^4+1440 z^3
\NN\\&&
+150 z^2-1247 z+225\big)
-\eta ^3 \big(1650 z^5-4130 z^4+1470 z^3+4935 z^2
\NN\\&&
-2602 z+525\big)
-5 \eta ^2 \big(126 z^5-154 z^4+294 z^3-273 z^2+78 z-15\big)
\NN\\&&
+35 \eta  z \big(15 z^4-5 z^3-30 z^2+15 z+53\big)
+525 (z-1)^2 z \big(3 z^2-z+1\big)~,
\\
Q_{31}&=&75 \eta ^5 z \big(15 z^5-25 z^4-8 z^3+34 z^2-14 z+3\big)
+\eta ^4 \big(-945 z^6+825 z^5
\NN\\&&
+1440 z^4+150 z^3+722 z^2-2015 z+560\big)
+\eta ^3 \big(-1650 z^6+4130 z^5
\NN\\&&
-1470 z^4-4935 z^3+2558 z^2+1715 z-560\big)
-5 \eta ^2 z \big(126 z^5-154 z^4
\NN\\&&
+294 z^3-273 z^2+122 z-15\big)
+5 \eta  z^2 \big(105 z^4-35 z^3-210 z^2+105 z
\NN\\&&
-324\big)
+525 z^3 \big(3 z^3-7 z^2+6 z-3\big)~,
\\
Q_{32}&=&3375 \eta ^4 z \big(60 z^5-110 z^4-27 z^3+146 z^2-56 z+12\big)
-8 \eta ^3 \big(46575 z^6
\NN\\&&
-93825 z^5-4050 z^4+114075 z^3-51319 z^2+7605 z+3780\big)
\NN\\&&
+450 \eta ^2 z \big(168 z^5-434 z^4+77 z^3+574 z^2-232 z+290\big)
-360 \eta  z^2 \big(525 z^4
\NN\\&&
-875 z^3+700 z^2-175 z-853\big)
+23625 z^3 \big(12 z^3-26 z^2+19 z-10\big)~,
\\
Q_{33}&=&
        105 (1-2 z) z^3
        +35 \eta  z^2 \big(
                -2 z^2+z+1\big)
        -21 \eta ^2 z \big(
                6 z^3-3 z^2-4 z+1\big)
\NN\\&&                
        +5 \eta ^3 \big(
                30 z^4-15 z^3-23 z^2+11 z-3\big)~.
\end{eqnarray}
Performing the Mellin transformation using {\tt HarmonicSums}  \cite{HARMONICSUMS,Ablinger:2011te,Ablinger:2013cf} one obtains
\begin{eqnarray}
D_{6a}(N)&=&
\left(m_1^2\right)^{-3+\ep} \left(m_2^2\right)^{\ep/2} 
\left(\frac{1+(-1)^N}{2 (N+1)}\right)
\Biggl\{
-\frac{1}{45 \ep^2}
+\frac{1} {\ep} \Biggl[
\frac{1}{90} S_1\left(N\right)
\NN\\&&
+\frac{P_{24}}{1800 \eta ^2 (N-1) N (N+1)} \Biggr]
-\frac{75 \eta^2-38 \eta-41}{53760 \eta ^2} \ln^2(\eta)
\NN\\&&
+\frac{P_{21}}{26880 \eta^3 (N+1) (2 N-3) (2 N-1)} \left(\frac{\eta}{\eta-1}\right)^{N+1} \biggl[
\frac{1}{2} \ln^2(\eta)
\NN\\&&
-\ln(\eta) S_1\left(\frac{\eta-1}{\eta},N\right)
+S_{1,1}\left(\frac{\eta-1}{\eta},1,N\right)
-S_2\left(\frac{\eta-1}{\eta},N\right)
\biggr]
\NN\\&&
+\frac{2^{-2 N-8} \binom{2 N}{N} P_{25}}{105 \eta^3 (N+1) (2 N-3) (2 N-1)} \Biggl[
\frac{\eta}{4 (1-\eta)} \ln^2(\eta)-\ln(\eta)-2
\NN\\&&
+\frac{1}{2 (\eta-1)} \sum_{i=1}^N \frac{2^{2 i} (\eta-1)^{-i} \eta^i}{\binom{2 i}{i}} \biggl[
\ln(\eta) S_1\left(\frac{\eta-1}{\eta},i\right)
+S_2\left(\frac{\eta-1}{\eta},i\right)
\NN\\&&
-S_{1,1}\left(\frac{\eta-1}{\eta},1,i\right)
-\frac{1}{2} \ln^2(\eta)
\biggr]
+\sqrt{\eta} \big[\HA_{-1,0,0}(\sqrt{\eta})+\HA_{1,0,0}(\sqrt{\eta})\big]
\Biggr]
\NN\\&&
+\frac{1}{420 \eta ^3} \Biggl[
- S_3(N)
- S_1\left(\frac{\eta }{\eta-1},N\right) S_{1,1}\left(\frac{\eta -1}{\eta },1,N\right)
\NN\\&&
-S_{1,2}\left(\frac{\eta -1}{\eta },\frac{\eta }{\eta-1},N\right)
+ S_{1,2}\left(\frac{\eta }{\eta -1},\frac{\eta-1}{\eta },N\right)
\NN\\&&
+S_{1,1,1}\left(\frac{\eta -1}{\eta},1,\frac{\eta }{\eta -1},N\right)
+S_{1,1,1}\left(\frac{\eta -1}{\eta },\frac{\eta }{\eta -1},1,N\right)
\NN\\&&
+\ln(\eta) S_{1,1}\left(\frac{\eta }{\eta -1},\frac{\eta -1}{\eta },N\right)
-\ln(\eta) S_2(N)
\Biggr]
\NN\\&&
+\frac{75 \eta^3-63 \eta^2-35 \eta-105}{13440 \eta^{5/2}} \big[\HA_{-1,0,0}\left(\sqrt{\eta}\right)+\HA_{1,0,0}\left(\sqrt{\eta}\right)\big]
\NN\\&&
-\frac{1}{403200 \eta^3 (N+1) (2 N-3)} \big[P_{23} S_1\left(N\right)+5 P_{22} \ln(\eta)\big]
\NN\\&&
-\frac{\ln^2(\eta)}{840 \eta^3} S_1\left(\frac{\eta }{\eta-1},N\right)
-\frac{1}{360} \big[S_1^2(N)+S_2\left(N\right)\big]
-\frac{\zeta_2}{120}
\NN\\&&
-\frac{P_{20}}{9072000 \eta ^3 (N-1)^2 N^2 (N+1)^2 (2 N-3)}
\Biggr\}~.
\end{eqnarray}
Here we abbreviated the polynomials
\begin{eqnarray}
P_{20}&=& \eta^3 \big(196228 N^7-334662 N^6-190856 N^5+437484 N^4+770788 N^3-1514022 N^2
\NN\\&&
+131040 N+302400\big)+\eta^2 \big(-309600 N^7+590400 N^6+150075 N^5-522675 N^4
\NN\\&&
-29475 N^3-67725 N^2+189000 N\big)+\eta \big(-401580 N^7+899370 N^6+60660 N^5
\NN\\&&
-1056240 N^4+340920 N^3+156870 N^2\big)-283500 N^7+567000 N^6+212625 N^5
\NN\\&&
-779625 N^4+70875 N^3+212625 N^2~,
\\
P_{21}&=& \left(375 \eta ^2-138 \eta -13\right) N+3 \left(125 \eta ^2-38 \eta +41\right)-4 (6 \eta +169) N^2+420 N^3~,
\\
P_{22}&=& 4 \left(225 \eta ^3-164 \eta ^2-117 \eta -315\right) N^2-2 \left(225 \eta ^3-164 \eta ^2-99 \eta-630\right) N
\NN\\&&
-3 \left(450 \eta ^3-203 \eta ^2-252 \eta -315\right)~,
\\
P_{23}&=& -28 \left(64 \eta ^3-200 \eta ^2-240 \eta -225\right) N^2+4 \left(1344 \eta ^3-700 \eta ^2-795 \eta-1575\right) N
\NN\\&&
-9 \left(448 \eta ^3+725 \eta ^2+1150 \eta +525\right)~,
\\
P_{24}&=& \left(-8 \eta^2+25 \eta+30\right) N^3+20 \eta^2 N^2+\left(-12 \eta^2-25 \eta-30\right) N+40 \eta^2~,
\\
P_{25}&=& 14 \left(27 \eta ^2+20 \eta -15\right) N+3 \left(375 \eta ^3-189 \eta ^2-35 \eta +105\right)-140 (\eta +9) N^2
\NN\\&&
+840 N^3~.
\end{eqnarray}
Assigning the masses for this diagram the other way yields diagram $D_{6b}$. 
In $z$-space it reads
\begin{eqnarray}
 D_{6b}(z)&=&
\left(m_1^2\right)^{\ep/2} \left(m_2^2\right)^{-3+\ep} 
\Biggl\{
-\frac{1}{45\ep^2}
+\frac{1}{\ep}
\biggl[
\frac{10-40 z+\left(22+25\eta+30\eta^2\right)z^2}{1800z^2}
\NN\\&&
+\frac{1}{90} \big[\HA_1\left(z\right)-\ln\left(z\right)\big]
\biggr]
+\frac{Q_{35} \ln(z)+z Q_{37} \HA_1(z)-5 z Q_{38} \ln(\eta)}{201600 (\eta-1) \eta z^2} 
\NN\\&&
-\frac{\sqrt{1-z} Q_{36}}{3360 \eta z^{5/2}} f_1(\eta,z)
-\frac{\eta \left(25-26 \eta-23 \eta^2-8 \eta^3\right)}{3360 (1-\eta)} f_5(\eta,z)
-\frac{\zeta_2}{120}
\NN\\&&
-\frac{75-63 \eta-35 \eta^2-105 \eta^3}{840 \eta} f_3(\eta,z)
+\frac{Q_{34}}{18144000 \eta z^2}
-\frac{\eta^3}{420} \HA_1(z) \HA_{0,0}(\eta)
\NN\\&&
-\frac{\eta^3}{420} \HA_{1,0}(z) \ln(\eta)
+\frac{\eta^3 (\eta-1)}{420} \Biggl[
G\left(\left\{\frac{1}{1-\tau},\frac{1}{1-\tau+\eta \tau},\frac{1}{1-\tau}\right\},z\right)
\NN\\&&
+G\left(\left\{\frac{1}{1-\tau},\frac{1}{1-\tau+\eta \tau},\frac{1}{\tau}\right\},z\right)
+\ln(\eta) G\left(\left\{\frac{1}{1-\tau},\frac{1}{1-\tau+\eta \tau}\right\},z\right)
\Biggr]
\NN\\&&
-\frac{\eta^3}{420} \big[
\HA_{1,0,1}(z)
+\HA_{1,0,0}(z)
-\zeta_2 \HA_1(z)
\big]
-\frac{1}{180} \big[\HA_{1,1}(z)+\HA_{0,0}(z)\big]
\Biggr\}~,
\end{eqnarray}
where
\begin{eqnarray}
 Q_{34}&=&
 23625\eta^4z^3\left(-10+19z-26z^2+12z^3\right)-360\eta^3z^2\bigl(-853-175z
 \NN\\&&
 +700z^2-875z^3+525z^4\bigr)+3375z\left(12-56z+146z^2-27z^3-110z^4+60z^5\right)
 \NN\\&&
 +450\eta^2z\left(290-232z+574z^2+77z^3-434z^4+168z^5\right)-8\eta\bigl(3780+
 \NN\\&&
 7605z-51319z^2+114075z^3-4050z^4-93825z^5+46575z^6\bigr)~,
\\
Q_{35}&=&
-525\eta^5z^3\left(-3+6z-7z^2+3z^3\right)-5\eta^4z^2\left(-324+105z-210z^2-35z^3+105z^4\right)
\NN\\&&
-75z\left(3-14z+34z^2-8z^3-25z^4+15z^5\right)+5\eta^3z\bigl(-15+122z-273z^2+294z^3
\NN\\&&
-154z^4+126z^5\bigr)+\eta\left(-560+2015z-722z^2-150z^3-1440z^4-825z^5+945z^6\right)
\NN\\&&
+\eta^2\left(560-1715z-2558z^2+4935z^3+1470z^4-4130z^5+1650z^6\right)~,
\\
Q_{36}&=&
15+(-55+21\eta)z+\left(115-84\eta-35\eta^2\right)z^2
-\left(-75+63\eta+35\eta^2+105\eta^3\right)z^3
\NN\\&&
+2\left(-75+63\eta+35\eta^2+105\eta^3\right)z^4~,
\\
Q_{37}&=&
-525\eta^5(z-1)^2z\left(1-z+3z^2\right)-75(z-1)^2\left(3-13z+5z^2+15z^3\right)
\NN\\&&
-35\eta^4z\left(53+15z-30z^2-5z^3+15z^4\right)+5\eta^3\bigl(-15+78z-273z^2+294z^3
\NN\\&&
-154z^4+126z^5\bigr)+\eta\left(-225+1247z-150z^2-1440z^3-825z^4+945z^5\right)
\NN\\&&
+\eta^2\left(525-2602z+4935z^2+1470z^3-4130z^4+1650z^5\right)~,
\\
Q_{38}&=&
105 \eta^5 z^2 \left(3 z^3-7 z^2+6 z-3\right)
+\eta^4 z \left(105 z^4-35 z^3-210 z^2+105 z+12\right)
\NN\\&&
+\eta^3 \left(-126 z^5+154 z^4-294 z^3+273 z^2-178 z+15\right)
-\eta^2 \big(330 z^5-826 z^4
\NN\\&&
+294 z^3+987 z^2-478 z+105\big)
-3 \eta \left(63 z^5-55 z^4-96 z^3-10 z^2+34 z-15\right)
\NN\\&&
+15 \left(15 z^5-25 z^4-8 z^3+34 z^2-14 z+3\right)~.
\end{eqnarray}
In Mellin space this corresponds to
\begin{eqnarray}
D_{6b}(N)&=&
\left(m_1^2\right)^{\ep/2} \left(m_2^2\right)^{-3+\ep} 
\left(\frac{1+(-1)^N}{2 (N+1)}\right)
\Biggl\{
-\frac{1}{45 \ep^2}
+\frac{1}{\ep} \Biggl[
\frac{1}{90} S_1\left(N\right)
\NN\\&&
+\frac{P_{26}}{1800 (N-1) N (N+1)}
\Biggr]
+\frac{41 \eta ^2+38 \eta -75}{53760} \ln^2(\eta)
-\frac{\zeta_2}{120}
\NN\\&&
+\frac{\eta (1-\eta)^{-N-1} P_{31}}{26880 (N+1) (2 N-3) (2 N-1)} \biggl[
\frac{1}{2} \ln^2(\eta)
-S_2\left(1-\eta ,N\right)
\NN\\&&
+S_{1,1}\left(1-\eta,1,N\right)
+\ln(\eta) S_1(1-\eta,N)
\biggr]
-\frac{1}{360} \big[S_1^2(N)+S_2(N)\big]
\NN\\&&
+\frac{2^{-2 N-8} \binom{2 N}{N} P_{30}}{105 (N+1) (2 N-3) (2 N-1)} \Biggl[
\frac{1}{\sqrt{\eta}} \big[\HA_{-1,0,0}(\sqrt{\eta})+\HA_{1,0,0}(\sqrt{\eta})\big]
\NN\\&&
+\frac{\eta}{2 (\eta-1)} \sum_{i=1}^N \frac{2^{2 i} (1-\eta)^{-i}}{\binom{2 i}{i}} \biggl(
S_{1,1}\left(1-\eta ,1,i\right)
-S_2\left(1-\eta,i\right)
\NN\\&&
+\frac{1}{2} \ln^2(\eta)
+\ln(\eta) S_1(1-\eta,i)
\biggr)
+\frac{\ln^2(\eta)}{4 (\eta-1)}+\ln(\eta)-2
\Biggr]
\NN\\&&
+\frac{\eta^3}{420}
\Biggl[
\ln(\eta) S_2(N)
-S_3(N)
-S_1\left(\frac{1}{1-\eta },N\right) S_{1,1}\left(1-\eta ,1,N\right)
\NN\\&&
+S_{1,2}\left(\frac{1}{1-\eta },1-\eta ,N\right)
-S_{1,2}\left(1-\eta ,\frac{1}{1-\eta },N\right)
\NN\\&&
+S_{1,1,1}\left(1-\eta ,1,\frac{1}{1-\eta},N\right)
+S_{1,1,1}\left(1-\eta ,\frac{1}{1-\eta},1,N\right)
\NN\\&&
-\ln(\eta) S_{1,1}\left(\frac{1}{1-\eta},1-\eta,N\right)
-\frac{1}{2} \ln^2(\eta) S_1\left(\frac{1}{1-\eta},N\right)
\Biggr]
\NN\\&&
-\frac{105 \eta^3+35 \eta^2+63 \eta-75}{13440 \sqrt{\eta}} \big[\HA_{-1,0,0}\left(\sqrt{\eta}\right)+\HA_{1,0,0}\left(\sqrt{\eta}\right)\big]
\NN\\&&
-\frac{4 P_{27}-(N-1)^2 N^2 (N+1) \left(450 P_{28} \ln(\eta)-90 P_{29} S_1(N)\right)}{36288000 (N-1)^2 N^2 (N+1)^2 (2 N-3)}
\Biggr\}~,
\end{eqnarray}
where
\begin{eqnarray}
P_{26}&=&\left(30 \eta ^2+25 \eta -8\right) N^3-\left(30 \eta ^2+25 \eta +12\right) N+20 N^2+40~,
\\
P_{27}&=&-4 \left(70875 \eta ^3+100395 \eta ^2+77400 \eta -49057\right) N^7
+6 \bigl(94500 \eta ^3+149895 \eta ^2
\NN\\&&
+98400 \eta -55777\bigr) 
N^6+\left(212625 \eta ^3+60660 \eta ^2+150075 \eta -190856\right)
N^5
\NN\\&&
-3 \left(259875 \eta ^3+352080 \eta ^2+174225 \eta -145828\right)
N^4+\bigl(70875 \eta ^3+340920 \eta ^2
\NN\\&&
-29475 \eta +770788\bigr)
N^3+3 \left(70875 \eta ^3+52290 \eta ^2-22575 \eta -504674\right)
N^2+
\NN\\&&
2520 (75 \eta +52) N+302400~,
\\
P_{28}&=&-315 \eta ^3 \left(4 N^2-4 N-3\right)+\eta ^2 \left(-468 N^2+198 
N+756\right)+\eta  \bigl(-656 N^2+328 N
\NN\\&&
+609\bigr)+450 \left(2 N^2-N-3\right)~,
\\
P_{29}&=&1575 \eta^3 \left(4 N^2-4 N-3\right)+30 \eta ^2 \left(224 N^2-106 N-345\right)+25 \eta  \bigl(224 N^2-112 N
\NN\\&&
-261\bigr)
-448 (3-2 N)^2~,
\\
P_{30}&=&105 \eta ^3 \left(8 N^3-12 N^2-2 N+3\right)-35 \eta ^2 \left(4 N^2-8 N+3\right)+189 \eta  (2 N-3)
\NN\\&&
+1125~,
\\
P_{31}&=&\eta ^2 \left(420 N^3-676 N^2-13 N+123\right)-6 \eta  \left(4 N^2+23 N+19\right)+375 (N+1)~.
\end{eqnarray}
\begin{figure}[H]
\begin{center}
\includegraphics[scale=1.5]{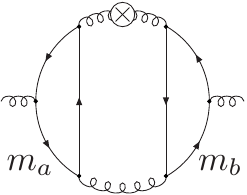}
\end{center}
\caption{\sf $D_{7}$, both mass assignments $m_a=m_2$,$m_b=m_1$ and $m_a=m_1$, 
$m_b=m_2$ yield the same result due to symmetry reasons.}
\end{figure}
The ladder-type diagram $D_{7}$ is symmetric under $m_1 \rightarrow m_2$ and 
only one mass assignment has to be considered. It evaluates to
\begin{eqnarray}
 D_{7}(z)&=&
\left(m_1^2\right)^{-3+3/2 \ep} 
\Biggl\{
-\frac{\eta+1}{24 \eta^2 \ep}
+\frac{Q_{40}}{8640 \eta ^3 z}
+\frac{\left(\eta ^3+1\right) \left(z^2-1\right)}{180 \eta ^3 z} 
\bigl[
\HA_{1,1}\left(z\right)
+\HA_{1,0}\left(z\right)
\bigr]
\NN\\&&
+\biggl(\frac{16 \eta-\eta^2-8 z-\eta^3 (8 z-27)}{1440 \eta^3}+\frac{\left(\eta^3+1\right)}{180 \eta^3} \HA_1\left(z\right)\biggr) \HA_{0,0}\left(\eta\right)
\NN\\&&
+\frac{\sqrt{1-z}}{360 \eta^3 \sqrt{z}} \left(27+2 (5 \eta+27) z+\left(27 \eta^2-10 \eta-81\right) z^2\right) f_1(\eta,z)
\NN\\&&
+\frac{\sqrt{1-z}}{360 \eta^3 \sqrt{z}} \left(10 \eta (z-1) z-27 z^2+27 \eta ^2 \left(3 z^2-2 z-1\right)\right) f_2(\eta,z)
\NN\\&&
-\frac{1}{90} \HA_{0,0,0}\left(\eta\right)
-\frac{27 \eta^2-10 \eta-81}{180 \eta^3} f_3(\eta,z)
+\frac{81 \eta^2+10 \eta-27}{180 \eta^3} f_4(\eta,z)
\NN\\&&
-\frac{\eta^3+1}{90 \eta ^3} \HA_{1,0,0}\left(\eta\right)
+\frac{Q_{41}}{720 (\eta-1) \eta^3 z} f_5(\eta,z)
-\frac{Q_{43}}{720 (\eta-1) \eta^3 z} f_6(\eta,z)
\NN\\&&
+\frac{27 \eta ^2+10 \eta +27}{2880 \eta ^{5/2}} G\left(\left\{\frac{\sqrt{\tau}}{1-\tau},\frac{1}{\tau},\frac{1}{\tau}\right\},\eta\right)
-\frac{\eta^3+1}{90 \eta^3}\bigl[\HA_{0,1,1}\left(z\right)+\HA_{0,1,0}\left(z\right)\bigr]
\NN\\&&
+\frac{\eta^3+1}{180 \eta^3} \bigl[\HA_{1,0,1}\left(z\right)+\HA_{1,0,0}\left(z\right)\bigr]
-\frac{\big(\eta^3+1\big) z}{180 \eta^3} \bigl[\HA_{0,1}\left(z\right)+\HA_{0,0}\left(z\right)\bigr]
\NN\\&&
+\frac{\eta^3-1}{180 \eta^3} \big[\HA_{1,0}\left(z\right)+2 \HA_{0,1}\left(z\right)\big] \ln(\eta)
-\frac{\eta+1}{1440 \eta^3} \big[Q_{42} \HA_1\left(z\right)+Q_{44} \HA_0\left(z\right)\big]
\NN\\&&
+\frac{z \left(1-\eta^3\right)}{180 \eta ^3} \HA_0\left(z\right) \ln(\eta)
+\frac{1-\eta}{180} \Biggl[ 
G\left(\left\{\frac{1}{1-\tau},\frac{1}{1-\tau+\eta \tau},\frac{1}{1-\tau}\right\},z\right)
\NN\\&&
+G\left(\left\{\frac{1}{1-\tau},\frac{1}{1-\tau+\eta \tau },\frac{1}{\tau}\right\},z\right)
+\ln(\eta) G\left(\left\{\frac{1}{1-\tau},\frac{1}{1-\tau+\eta \tau}\right\},z\right)
\Biggr]
\NN\\&&
-\frac{\big(\eta^3-1\big)\big(z^2-1\big)}{180 \eta^3 z} \HA_1\left(z\right) \ln(\eta)
+\frac{1-\eta}{90} \Biggl[
G\left(\left\{\frac{1}{\tau},\frac{1}{\eta \tau-\eta-\tau},\frac{1}{1-\tau}\right\},z\right)
\NN\\&&
+G\left(\left\{\frac{1}{\tau},\frac{1}{\eta \tau-\eta-\tau},\frac{1}{\tau}\right\},z\right)
-\ln(\eta) G\left(\left\{\frac{1}{\tau},\frac{1}{\eta \tau-\eta-\tau}\right\},z\right)
\Biggr]
\NN\\&&
+\frac{\eta^3+1}{180 \eta^3} \zeta_2 \bigl[z-\HA_1\left(z\right)\bigr]
+\frac{(1-\eta)}{180 \eta^3} \Biggl[
G\left(\left\{\frac{1}{1-\tau},\frac{1}{\eta \tau-\eta-\tau},\frac{1}{1-\tau}\right\},z\right)
\NN\\&&
+G\left(\left\{\frac{1}{1-\tau},\frac{1}{\eta \tau-\eta-\tau},\frac{1}{\tau}\right\},z\right)
-\ln(\eta) G\left(\left\{\frac{1}{1-\tau},\frac{1}{\eta \tau-\eta-\tau}\right\},z\right)
\Biggr]
\NN\\&&
+\frac{Q_{39}}{4320 (\eta-1) \eta^3} \ln(\eta)
+\frac{1-\eta}{90 \eta^3} \Biggl[
G\left(\left\{\frac{1}{\tau},\frac{1}{1-\tau+\eta \tau},\frac{1}{1-\tau}\right\},z\right)
\NN\\&&
+G\left(\left\{\frac{1}{\tau},\frac{1}{1-\tau+\eta \tau},\frac{1}{\tau}\right\},z\right)
+\ln(\eta) G\left(\left\{\frac{1}{\tau},\frac{1}{1-\tau+\eta \tau}\right\},z\right)
\Biggr]
\Biggr\}~,
\end{eqnarray}
with
\begin{eqnarray}
Q_{39}&=& 27 \left(\eta^4+1\right) z \left(6 z^3-11 z^2+3 z+3\right)-2 \left(\eta^2+1\right) \eta \left(22 z^3-66 z^2-15 z
\right.
\NN\\&&
\left.
+63\right)
-\eta^2 \left(324 z^4-682 z^3+426 z^2+30 z-84\right)-162 \eta^4+330 \eta
\\
Q_{40}&=& -27 \left(\eta^3+1\right) z^2 \left(24 z^3-46 z^2+11 z+16\right)+648 \eta^3 z+408 \eta z
\NN\\&&
+(\eta+1) \eta \left(648 z^5-1370 z^4+873 z^3-12 z^2-390 z-180\right)~,
\\
Q_{41}&=&
        4 \eta ^4 (1-2 z) z
        +4 \eta ^5 z^2
        +\eta ^3 z (4 z+3)
        +4 \big(
                z^2-1\big)
        -8 \eta  \big(
                z^2+z-1\big)
\NN\\&&                
        +\eta ^2 \big(
                4 z^2-15 z-4\big)~,
\\
Q_{42}&=&
9 (1+\eta^2) (z-1)^3 (3 z+2)
        +2 \eta  \big(
                5 z^4-33 z^3+42 z^2
                +14 z-43\big)~,
\\
Q_{43}&=&
                -4 z^2
                +4 \eta  z (2 z-1)
                -\eta ^2 z (4 z+3)
                +\eta ^3 \big(
                        -4 z^2+15 z+4\big)
\NN\\&&                        
                -4 \eta ^5 \big(
                        z^2-1\big)
                +8 \eta ^4 \big(
                        z^2+z-1\big)~,
\\                        
Q_{44}&=&
9 (1+\eta^2) z \big(3 z^3-7 z^2+3 z+3\big)
+2 \eta  \big(5 z^4
-33 z^3+42 z^2+14 z+11\big)~.
\end{eqnarray}
The Mellin space-expression for this diagram reads
\begin{eqnarray}
D_7(N)&=&
 \left(m_1^2\right)^{-3+3/2 \ep} 
\left(\frac{1+(-1)^N}{2}\right)
\Biggl\{
-\frac{\eta+1}{24 \ep \eta^2 (N+1)}
-\frac{(\eta^3-1) \ln(\eta)}{180 \eta^3 (N+1)} S_2(N)
\NN\\&&
+\frac{(\eta^3+1) S_3(N)}{180 \eta^3 (N+1)}
-\frac{32+32 \eta ^3+11 \eta (1+\eta) N \big(N^2+3 N+2\big)}{5760 \eta ^3 N (N+1)^2 (N+2)} \ln^2(\eta)
\NN\\&&
+\frac{P_{35}}{5760 \eta^3 N (N+1)^2 (N+2)} \left(\frac{\eta}{\eta-1}\right)^{N+1} \biggl[
-\ln(\eta) S_1\left(\frac{\eta-1}{\eta },N\right)
\NN\\&&
+S_{1,1}\left(\frac{\eta -1}{\eta},1,N\right)
-S_2\left(\frac{\eta -1}{\eta},N\right)
+\frac{1}{2} \ln^2(\eta)
\biggr]
\NN\\&&
+\frac{(1-\eta )^{-N-1} P_{36}}{5760 \eta^3 N (N+1)^2 (N+2)} \biggl[
-\ln(\eta) S_1\left(1-\eta ,N\right)
+S_2\left(1-\eta ,N\right)
\NN\\&&
-S_{1,1}\left(1-\eta ,1,N\right)
-\frac{1}{2} \ln^2(\eta)
\biggr]
+\frac{(\eta^3-1) \ln(\eta)}{90 \eta^3 N (N+1)^2 (N+2)} S_1(N)
\NN\\&&
-3 N \frac{(\eta+1) 2^{-2 N-7} \binom{2 N}{N}}{5 \eta^2 (N+1)^2} \ln^2(\eta)
-\frac{(\eta-1) 2^{-2 N-7} \binom{2 N}{N} P_{40}}{45 \eta^3 (N+1)^2 (N+2)} \ln(\eta)
\NN\\&&
+\frac{2^{-2 N-8} \binom{2 N}{N} P_{33}}{45 (\eta-1) \eta^3 (N+1)^2 (N+2)} \sum_{i=1}^N \frac{2^{2 i} (\eta-1)^{-i} \eta^i}{\binom{2 i}{i}} \biggl[
S_{1,1}\left(\frac{\eta-1}{\eta},1,i\right)
\NN\\&&
-S_2\left(\frac{\eta-1}{\eta},i\right)
-\ln(\eta) S_1\left(\frac{\eta-1}{\eta},i\right)
+\frac{1}{2} \ln^2(\eta)
\biggr]
\NN\\&&
+\frac{2^{-2 N-8} \binom{2 N}{N} P_{39}}{45 (\eta-1) \eta (N+1)^2 (N+2)} \sum_{i=1}^N \frac{2^{2 i} (1-\eta)^{-i}}{\binom{2 i}{i}} \biggl[
-\ln(\eta) S_1\left(1-\eta,i\right)
\NN\\&&
+S_2\left(1-\eta ,i\right)
-S_{1,1}\left(1-\eta,1,i\right)
-\frac{1}{2} \ln^2(\eta)
\biggr]
+\frac{2^{-2 N-6} \binom{2 N}{N} P_{41}}{45 \eta^3 (N+1)^2 (N+2)}
\NN\\&&
+\frac{1}{\eta^{5/2}}\biggl(
\frac{27 \eta^2+10 \eta+27}{1440 (N+1)} 
-\frac{2^{-2 N-6} \binom{2 N}{N} P_{32}}{45 (N+1)^2 (N+2)} \biggr) \bigl[\HA_{-1,0,0}\left(\sqrt{\eta}\right)
\NN\\&&
+\HA_{1,0,0}\left(\sqrt{\eta}\right)\bigr]
+\frac{\eta^3+1}{180 \eta^3 N (N+1)^2 (N+2)} \bigl[S_2\left(N\right)-S_1^2(N)\bigr]
\NN\\&&
+\frac{1}{2880 \eta^3 (N+1)^2 (N+2)} \left(\frac{\eta+1}{2} P_{34} S_1(N)-\frac{\eta+1}{N} P_{38}-\frac{P_{37}}{2} \ln(\eta)\right)
\NN\\&&
+\frac{1}{180 (N+1)} \biggl[
S_{1,2}\left(1-\eta ,\frac{1}{1-\eta},N\right)
-S_{1,2}\left(\frac{1}{1-\eta},1-\eta,N\right)
\NN\\&&
+S_1\left(\frac{1}{1-\eta},N\right) S_{1,1}\left(1-\eta ,1,N\right)
-S_{1,1,1}\left(1-\eta ,1,\frac{1}{1-\eta},N\right)
\NN\\&&
-S_{1,1,1}\left(1-\eta ,\frac{1}{1-\eta},1,N\right)
+\ln(\eta) S_{1,1}\left(\frac{1}{1-\eta},1-\eta,N\right)
\biggr]
\NN\\&&
+\frac{1}{180 \eta^3 (N+1)} \biggl[
S_{1,2}\left(\frac{\eta-1}{\eta },\frac{\eta}{\eta-1},N\right)
-S_{1,2}\left(\frac{\eta}{\eta-1},\frac{\eta-1}{\eta },N\right)
\NN\\&&
+S_1\left(\frac{\eta }{\eta-1},N\right) S_{1,1}\left(\frac{\eta -1}{\eta},1,N\right)
-S_{1,1,1}\left(\frac{\eta-1}{\eta },1,\frac{\eta }{\eta-1},N\right)
\NN\\&&
-S_{1,1,1}\left(\frac{\eta-1}{\eta},\frac{\eta }{\eta-1},1,N\right)
-\ln(\eta) S_{1,1}\left(\frac{\eta}{\eta-1},\frac{\eta-1}{\eta },N\right)
\biggr]
\NN\\&&
+\frac{\ln^2(\eta)}{360 (N+1)} \left[S_1\left(\frac{1}{1-\eta},N\right)+\frac{1}{\eta^3} S_1\left(\frac{\eta}{\eta-1},N\right)\right]
\Biggr\}~.
\end{eqnarray}
This expression contains the polynomials
\begin{eqnarray}
P_{32}&=&27 \left(1+\eta^2\right) \left(2 N^2+4 N+3\right)-10 \eta  (2 N+1)~,\\
P_{33}&=&81 \eta^2-10 \eta  (2 N+1)+27 \left(4 N^2+8 N+3\right)~,\\
P_{34}&=&27 \left(1+\eta^2\right) \left(4 N^2+8 N+3\right)+2 \eta  \left(6 N^2+73 N+115\right)~,\\
P_{35}&=&64 \eta ^3-64 \eta ^2 (N+1)+5 \eta  N (N+1)-N \left(54 N^2+103 N+17\right)~,
\\
P_{36}&=&\eta ^3 N \left(54 N^2+103 N+17\right)-5 \eta ^2 N (N+1)+64 \eta (N+1)-64~,
\\
P_{37}&=&
27 \left(1-\eta^3\right) \left(4 N^2+8 N+3\right)+\eta^2 \left(196 N^2+586 N+449\right)
\NN\\&&
+\eta \bigl(164 N^2+494 N+271\bigr)~,
\\
P_{38}&=&
27 \left(1+\eta^2\right) N \left(4 N^2+8 N+3\right)+2 \eta  \left(27 N^3+76 N^2+78 N+60\right)~,
\\
P_{39}&=&27 \big(4 N^2+8 N+3\big) \eta ^2-10 (2 N+1) \eta +81~,\\
P_{40}&=& 27 \left(1+\eta^2\right) \left(4 N^2+8 N+3\right)+2 \eta \left(54 N^2+98 N-5\right)~,\\
P_{41}&=& 
27 \left(1+\eta^3\right) \left(4 N^2+8 N+3\right)+\eta \left(1+\eta\right) (71-20 N)~.
\end{eqnarray}

\noindent
Finally, we turn to the diagrams $D_{8a,b}$. In $z$-space they contain 
contributions which have to be regularized as in (\ref{eq:PlusMellin}).
For $D_{8a}$ this contribution is given by
\begin{eqnarray}
D_{8a}^{(+)}(z)&=&
\left(m_1^2\right)^{\ep/2} \left(m_2^2\right)^{-3+\ep} 
\Biggl[
\frac{1}{90 (1-z)} \left(-\frac{1}{\ep}-\frac{1}{5}+\frac{1}{2} \HA_1\left(z\right)\right)
\NN\\&&
-\frac{25+(63 \eta-100) (1-z)}{3360 \eta (1-z)^{3/2}} \sqrt{z} f_2(\eta,z)
\Biggr]~.
 \end{eqnarray} 
The regular contribution to $D_{8a}(z)$ reads
\begin{eqnarray}
D_{8a}^{\sf Reg}(z) &=&
\left(m_1^2\right)^{\ep/2} \left(m_2^2\right)^{-3+\ep} 
\Biggl\{
\left(
-\frac{1}{45\ep^2}
-\frac{1}{225\ep}
+\frac{7}{20250}
-\frac{\zeta_2}{120}\right)\delta\left(1-z\right)
\NN\\&&
-\frac{1}{45\ep^2}
+\frac{1}{\ep}
\Biggl[
\frac{\eta}{72}+\frac{1}{450}\left(8-\frac{5}{z}\right)
+\frac{1}{90} \HA_1\left(z\right)
-\frac{1}{90} \ln\left(z\right)
\Biggr]
\NN\\&&
+\HA_1\left(z\right)
\left(\frac{Q_{46}}{201600(\eta-1)\eta}
+\frac{\eta^3 (z-1)}{420z} \ln\left(\eta\right)\right)
-\frac{\zeta_2}{120}
\NN\\&&
-\frac{\sqrt{1-z} Q_{48}}{3360 \eta \sqrt{z}} f_2(\eta,z)
-\frac{3 \eta^3 (z-1)+7 z}{1260z} \HA_{1,1}\left(z\right)
-\frac{\eta^3(z-1)}{420z}\HA_{1,0}\left(z\right)
\NN\\&&
+\frac{\eta Q_{50}}{3360 (\eta-1) z} f_6(\eta,z)
-\frac{\eta\left(25-6\eta+105\eta^2\right)}{26880}\HA_{0,0}\left(\eta\right)
-\frac{1}{180}\HA_{0,0}\left(z\right)
\NN\\&&
-\frac{Q_{47}}{201600(1-\eta)\eta z}\ln\left(z\right)
-\left(
\frac{Q_{49}}{40320 (1-\eta) \eta}
+\frac{\eta^3}{420} \HA_{0,1}\left(z\right)
\right)
\ln\left(\eta\right)
\NN\\&&
+\frac{Q_{45}}{18144000\eta z}
-\frac{\eta^3}{420}(1-\eta) \Biggl[
G\left(\left\{\frac{1}{\tau},\frac{1}{\eta \tau-\eta-\tau},\frac{1}{1-\tau}\right\},z\right)
\NN\\&&
+G\left(\left\{\frac{1}{\tau},\frac{1}{\eta \tau-\eta-\tau},\frac{1}{\tau}\right\},z\right)
-\ln(\eta) G\left(\left\{\frac{1}{\tau},\frac{1}{\eta \tau-\eta-\tau}\right\},z\right)
\Biggr]
\NN\\&&
+\frac{75-63 \eta-35 \eta^2-105 \eta^3}{840\eta} \biggl[
f_4(\eta,z)
+\frac{\sqrt{\eta}}{64} G\left(\left\{\frac{\sqrt{\tau}}{1-\tau},\frac{1}{\tau},\frac{1}{\tau}\right\},\eta\right)
\biggr]
\NN\\&&
+\frac{\eta^3}{420}
\bigl[
\HA_{1,0,0}\left(\eta\right)
+\HA_{0,1,1}\left(z\right)
+\HA_{0,1,0}\left(z\right)
+\HA_{0,0,0}\left(\eta\right)
\bigr]
\Biggr\}~,
\end{eqnarray}
with
\begin{eqnarray}
 Q_{45}&=&16875z^2\left(2+3z-26z^2+12z^3\right)+23625\eta^4z\left(-12+14z+z^2-22z^3+12z^4\right)
 \NN\\&&
-12600\eta^3\left(-12-3z+25z^2+5z^3-35z^4+15z^5\right)+450\eta^2\left(280+30z\right.
\NN\\&&
\left.+238z^2-301z^3-238z^4+168z^5\right)
-8\eta\left(5040-10864z-675z^2+6075z^3\right.
\NN\\&&
\left.
-92475z^4+46575z^5\right)~,
\\
Q_{46}&=&525\eta^5(z-1)^2\left(-1+z+3z^2\right)+\eta^2\left(-1332+1845z+270z^2+2470z^3-1650z^4\right)
\NN\\&&
-175\eta^4\left(13-9z-12z^2+11z^3-3z^4\right)
-5\eta^3\left(4+315z+210z^2-350z^3+126z^4\right)
\NN\\&&
+3\eta\left(619-500z-690z^2+985z^3-315z^4\right)
+375\left(1+2z^2-7z^3+3z^4\right)~,
\\
Q_{47}&=&
375z^3\left(2-7z+3z^2\right)+525\eta^5z^2\left(3-5z^2+3z^3\right)
+5\eta^4z\left(-96+315z+420z^2
\right.
\NN\\&&
\left.
-385z^3+105z^4\right)-5\eta^3z\left(-376+315z+210z^2-350z^3+126z^4\right)
+\eta^2\left(-1120
\right.
\NN\\&&
\left.
-728z+1845z^2+270z^3+2470z^4-1650z^5\right)-\eta\left(-1120+672z+1500z^2
\right.
\NN\\&&
\left.
+2070z^3-2955z^4+945z^5\right)~,
\\
Q_{48}&=&75(1-2z)z+63\eta z(-1+2z)+35\eta^2z(-1+2z)+105\eta^3\left(-1-z+2z^2\right)~,
\\
Q_{49}&=&
-75z^2\left(2-7z+3z^2\right)+3\eta z\left(100+138z-197z^2+63z^3\right)-35\eta^4\left(-3+9z
\right.
\NN\\&&
\left.
+12z^2-11z^3+3z^4\right)
-105\eta^5\left(-3+3z-5z^3+3z^4\right)+\eta^3\left(-395+315z
\right.
\NN\\&&
\left.
+210z^2-350z^3+126z^4\right)
+\eta^2\left(-25-369z-54z^2-494z^3+330z^4\right)~,
\\
Q_{50}&=&
-8\eta^4(z-1)-25z+26\eta z+8\eta^3(-2+3z)+\eta^2(8+15z)~.
\end{eqnarray}
In Mellin space one obtains
\begin{eqnarray}
D_{8a}(N)&=&
\left(m_1^2\right)^{\ep/2} \left(m_2^2\right)^{-3+\ep} 
\left(\frac{1+(-1)^N}{2 (N+1)}\right)
\Biggl\{
-\frac{N+2}{45 \ep^2}
\NN\\&&
+\frac{1}{\ep}\Biggl[
\frac{1}{90} (N+2) S_1\left(N\right)
-\frac{8 N^3+(4-25 \eta ) N^2-(25 \eta +24) N+20}
{1800 N (N+1)}\Biggr]
\NN\\&&
-\frac{7 N \big(N^2+3 N+2\big)-3 \eta^3}{2520 N (N+1)} S_1^2(N)
+\frac{10 P_{44} \ln(\eta)-2 P_{45} S_1\left(N\right)}{806400 \eta  (N+1)}
\NN\\&&
-\frac{7 N \big(N^2+3 N+2\big)+3 \eta^3}{2520 N (N+1)} S_2(N)
+\eta^3 \frac{\ln^2(\eta)-2 \ln(\eta) S_1(N)}{840 N (N+1)}
\NN\\&&
+\frac{(\eta-1)^{-N-1} \eta^N}{26880 N (N+1)} P_{42} \biggl[
\ln(\eta) S_1\left(\frac{\eta-1}{\eta},N\right)
+S_2\left(\frac{\eta-1}{\eta},N\right)
\NN\\&&
-S_{1,1}\left(\frac{\eta-1}{\eta},1,N\right)
-\frac{1}{2} \ln^2(\eta)
\biggr]
+\frac{P_{46}}{9072000 \eta N^2 (N+1)^2}
\NN\\&&
-\frac{\zeta_2}{120} (N+2)
+\frac{2^{-2 N-8} \binom{2 N}{N} P_{43}}{105 \eta (N+1)} \Biggl[
\sqrt{\eta} \big[\HA_{-1,0,0}(\sqrt{\eta})+\HA_{1,0,0}(\sqrt{\eta})\big]
\NN\\&&
+\frac{1}{2 (\eta-1)} \sum_{i=1}^N \frac{2^{2 i} (\eta-1)^{-i} \eta^i}{\binom{2 i}{i}} \biggl(
S_2\left(\frac{\eta-1}{\eta},i\right)
-S_{1,1}\left(\frac{\eta-1}{\eta},1,i\right)
\NN\\&&
+\ln(\eta) S_1\left(\frac{\eta-1}{\eta},i\right)
-\frac{1}{2} \ln^2(\eta)
\biggr)
-\frac{\eta}{4 (\eta-1)} \ln^2(\eta)-\ln(\eta)-2
\Biggr]
\Biggr\}~,
\end{eqnarray}
with the polynomials
\begin{eqnarray} 
P_{42}&=&64 \eta ^4-64 \eta ^3 (N+1)-3 \eta ^2 N (N+1)+2 \eta  N \left(88 N^2+245 N+157\right)
\NN\\&&
-25 N \left(4 N^3+20 N^2+31 N+15\right)~,
\\
P_{43}&=&105 \eta ^3-35 \eta ^2 (2 N+1)-63 \eta  \left(4 N^2+8 N+3\right)+25 \bigl(8 N^3+36 N^2
\NN\\&&
+46 N+15\bigr)~,
\\
P_{44}&=&192 \eta ^3-\eta ^2 (538 N+547)-6 \eta  \left(76 N^2+52 N-93\right)+75 \bigl(8 N^3+36 N^2
\NN\\&&
+46 N+15\bigr)~,
\\
P_{45}&=&960 \eta ^3+5 \eta ^2 (22 N+13)+\eta  \left(-3176 N^2-2008 N+5478\right)
\NN\\&&
+375 \left(8 N^3+36 N^2+46 N+15\right)~,
\\
P_{46}&=&10800 \eta ^3 N \left(9 N^2+16 N+7\right)+225 \eta ^2 N \left(478 N^3+945 N^2+747 N+280\right)
\NN\\&&
-4 \eta  \left(58616 N^5+203774 N^4+241285 N^3+101167 N^2-32760 N-12600\right)
\NN\\&&
+16875 N^2 \left(8 N^4+44 N^3+82 N^2+61 N+15\right)~.
\end{eqnarray}

For diagram $D_{8b}$ the part that requires regularization reads
\begin{eqnarray}
 D_{8b}^{(+)}(z) &=&
\left(m_1^2\right)^{-3+\ep} \left(m_2^2\right)^{\ep/2} 
\Biggl\{
\frac{1}{90 (1-z)} \left(-\frac{1}{\ep}-\frac{1}{5}+\frac{1}{2} \HA_1\left(z\right)\right)
\NN\\&&
+\frac{63 (1-z)+25 \eta (4 z-3)}{3360 \eta (1-z)^{3/2}} \sqrt{z} f_1(\eta,z)
\Biggr\}~,
\end{eqnarray}
and the regular contribution is given by
\begin{eqnarray}
 D_{8b}^{\sf Reg}(z)&=&
\left(m_1^2\right)^{-3+\ep} \left(m_2^2\right)^{\ep/2} 
\Biggl\{
\left(
        -\frac{1}{45 \ep^2}
        -\frac{1}{225 \ep}
        +\frac{7}{20250}
        -\frac{\zeta_2}{120}
\right) \delta\left(1-z\right)
\NN\\&&
-\frac{1}{45 \ep^2}
+\frac{1}{\ep}
\left[
\frac{1}{72 \eta }
+\frac{1}{450} \left(
        8-\frac{5}{z}\right)
+\frac{1}{90} \HA_1\left(z\right)
-\frac{1}{90} \ln\left(z\right)
\right]
\NN\\&&
+\HA_1\left(z\right) \left(\frac{Q_{52}}{201600 (\eta-1) \eta^3}
+\frac{1-z}{420 \eta^3 z} \ln\left(\eta\right)\right)
+\frac{Q_{54}}{18144000 \eta^3 z}
\NN\\&& 
-\frac{\sqrt{1-z} Q_{55}}{3360 \eta^3 \sqrt{z}} f_1(\eta,z)
+\frac{Q_{53}}{201600 (\eta-1) \eta^3 z} \ln\left(z\right)  
\NN\\&&
+\frac{8 (\eta-1)^2-\left(25 \eta^4-26 \eta^3-15 \eta^2-24 \eta+8\right) z}{3360 (\eta-1) \eta^3 z} f_5(\eta,z)
\NN\\&&
+\frac{75 \eta^3-138 \eta^2+3 \eta-64}{26880 \eta^2} \HA_{0,0}\left(\eta\right)
+\frac{3-3 z-7 \eta^3 z}{1260 \eta^3 z } \HA_{1,1}\left(z\right)
\NN\\&&
+\left(
\frac{Q_{51}}{40320 (\eta-1) \eta^3}     
+\frac{1}{420 \eta^3} \HA_{0,1}\left(z\right)
\right) \ln\left(\eta\right)
+\frac{1-z}{420 \eta ^3 z} \HA_{1,0}\left(z\right)  
\NN\\&&
-\frac{1}{180} \HA_{0,0}\left(z\right)
-\frac{\zeta_2}{120}
+\frac{\eta-1}{420 \eta^3} \Biggl[
G\left(\left\{\frac{1}{\tau},\frac{1}{1-\tau+\eta \tau },\frac{1}{1-\tau}\right\},z\right)
\NN\\&&
+G\left(\left\{\frac{1}{\tau},\frac{1}{1-\tau+\eta \tau },\frac{1}{\tau}\right\},z\right)
+\ln(\eta) G\left(\left\{\frac{1}{\tau},\frac{1}{1-\tau+\eta \tau }\right\},z\right)
\Biggr]
\NN\\&&
+\frac{75 \eta^3-63 \eta^2-35 \eta-105}{840 \eta^3} \biggl[
f_3(\eta,z)
+\frac{\sqrt{\eta}}{64} G\left(\left\{\frac{\sqrt{\tau}}{1-\tau},\frac{1}{\tau},\frac{1}{\tau}\right\},\eta\right)
\biggr]
\NN\\&&
      +\frac{1}{420 \eta^3} \bigl[
        \HA_{0,1,1}\left(z\right)
        +\HA_{0,1,0}\left(z\right)
        +\HA_{1,0,0}\left(\eta\right)
\bigr]      
\Biggr\}~,
\end{eqnarray}
with the polynomials
\begin{eqnarray}
Q_{51}&=&-75 \eta ^5 \big(3 z^4-7 z^3+2 z^2+3\big)
+3 \eta ^4 \big(63 z^4-197 z^3+138 z^2+100 z+63\big)
\NN\\&&
+\eta ^3 \big(330 z^4-494 z^3-54 z^2-369 z+305\big)
+\eta ^2 \big(126 z^4-350 z^3+210 z^2
\NN\\&&
+315 z-269\big)
-35 \eta  z \big(3 z^3-11 z^2+12 z+9\big)
-105 z \big(3 z^3-5 z^2+3\big)~,
\\
Q_{52}&=&-375 \eta ^5 \big(3 z^4-7 z^3+2 z^2+1\big)
+3 \eta ^4 \big(315 z^4-985 z^3+690 z^2+500 z-619\big)
\NN\\&&
+\eta ^3 \big(1650 z^4-2470 z^3-270 z^2-1845 z+1332\big)
+5 \eta ^2 \big(126 z^4-350 z^3
\NN\\&&
+210 z^2+315 z+4\big)
-175 \eta  \big(3 z^4-11 z^3+12 z^2+9 z-13\big)
\NN\\&&
-525 (z-1)^2 \big(3 z^2+z-1\big)~,
\\
Q_{53}&=&-375 \eta ^5 z^3 \big(3 z^2-7 z+2\big)
+\eta ^4 \big(945 z^5-2955 z^4+2070 z^3+1500 z^2+672 z
\NN\\&&
-1120\big)
+\eta ^3 \big(1650 z^5-2470 z^4-270 z^3-1845 z^2+728 z+1120\big)
\NN\\&&
+5 \eta ^2 z \big(126 z^4-350 z^3+210 z^2+315 z-376\big)
-5 \eta  z \big(105 z^4-385 z^3+420 z^2
\NN\\&&
+315 z-96\big)
-525 z^2 \big(3 z^3-5 z^2+3\big)~,
\\
Q_{54}&=&16875 \eta ^4 z \big(12 z^4-26 z^3+3 z^2+2 z+12\big)
-8 \eta ^3 \big(46575 z^5-92475 z^4+6075 z^3
\NN\\&&
-675 z^2+35711 z+5040\big)
+450 \eta ^2 \big(168 z^5-238 z^4-301 z^3+238 z^2
\NN\\&&
+198 z+280\big)
-12600 \eta  \big(15 z^5-35 z^4+5 z^3+25 z^2+12 z-12\big)
\NN\\&&
+23625 z^2 \big(12 z^3-22 z^2+z+14\big)~,
\\
Q_{55}&=&
        105
        +\big(
                -75 \eta ^3+63 \eta ^2+35 \eta +105\big) z
        +2 \big(
                75 \eta ^3-63 \eta ^2-35 \eta -105\big) z^2~.
\end{eqnarray}

Finally, one obtains the $N$-space representation
\begin{eqnarray}
D_{8b}(N)&=&
\left(m_1^2\right)^{-3+\ep} \left(m_2^2\right)^{\ep/2} 
\left(\frac{1+(-1)^N}{2 (N+1)}\right)
\Biggl\{
-\frac{N+2}{45 \ep^2}
\NN\\&&
+\frac{1}{\ep} \Biggl[
\frac{1}{90} (N+2) S_1\left(N\right)
-\frac{4 \eta \big(2 N^3+N^2-6 N+5\big)-25 N (N+1)}{1800 \eta  N (N+1)}\Biggr]
\NN\\&&
-\frac{7 \eta ^3 N \big(N^2+3 N+2\big)-3}{2520 \eta^3 N (N+1)} S_1^2(N)
-\frac{10 P_{49} \ln(\eta)+2 P_{50} S_1\left(N\right)}{806400 \eta^2 (N+1)}
\NN\\&&
-\frac{7 \eta ^3 N \big(N^2+3 N+2\big)+3}{2520 \eta^3 N (N+1)} S_2(N)
+\frac{\ln^2(\eta)+2 \ln(\eta) S_1(N)}{840 \eta^3 N (N+1)}
\NN\\&&
+\frac{(1-\eta)^{-N-1} P_{47}}{26880 \eta^3 N (N+1)} \biggl[
\ln(\eta) S_1\left(1-\eta,N\right)
-S_2\left(1-\eta,N\right)
\NN\\&&
+S_{1,1}\left(1-\eta,1,N\right)
+\frac{1}{2} \ln^2(\eta)
\biggr]
+\frac{P_{51}}{9072000 \eta^2 N^2 (N+1)^2}
\NN\\&&
-\frac{\zeta_2}{120} (N+2)
+\frac{2^{-2 N-8} \binom{2 N}{N} P_{48}}{105 \eta^2 (N+1)} \Biggl[
\frac{1}{\sqrt{\eta}} \big[\HA_{-1,0,0}(\sqrt{\eta})+\HA_{1,0,0}(\sqrt{\eta})\big]
\NN\\&&
+\frac{\eta}{2 (\eta-1)} \sum_{i=1}^N \frac{2^{2 i} (1-\eta)^{-i}}{\binom{2 i}{i}} \biggl(
S_{1,1}\left(1-\eta,1,i\right)
-S_2\left(1-\eta,i\right)
\NN\\&&
+\ln(\eta) S_1\left(1-\eta,i\right)
+\frac{1}{2} \ln^2(\eta)
\biggr)
+\frac{\ln^2(\eta)}{4 (\eta-1)}+\ln(\eta)-2
\Biggr]
\Biggr\}~.
\end{eqnarray}
Here the polynomials read
\begin{eqnarray} 
P_{47}&=&25 \eta ^4 N \left(4 N^3+20 N^2+31 N+15\right)-2 \eta ^3 N \left(88 N^2+245 N+157\right)
\NN\\&&
+3 \eta ^2 N (N+1)
+64 \eta  (N+1)-64~,
\\
P_{48}&=&25 \eta ^3 \left(8 N^3+36 N^2+46 N+15\right)-63 \eta ^2 \left(4 N^2+8 N+3\right)-35 \eta  (2 N+1)
\NN\\&&
+105~,
\\
P_{49}&=&75 \eta ^3 \left(8 N^3+36 N^2+46 N+15\right)-6 \eta ^2 \left(76 N^2+52 N-93\right)-\eta  (538 N+547)
\NN\\&&
+192~,
\\
P_{50}&=&375 \eta ^3 \left(8 N^3+36 N^2+46 N+15\right)+\eta ^2 \left(-3176 N^2-2008 N+5478\right)
\NN\\&&
+5 \eta  (22 N+13)
+960~,
\\
 P_{51}&=&16875 \eta ^3 N^2 \left(8 N^4+44 N^3+82 N^2+61 N+15\right)-4 \eta ^2 \bigl(58616 N^5+203774 N^4
\NN\\&& 
 +241285 N^3+101167 N^2-32760 N-12600\bigr)+225 \eta  N \bigl(478 N^3+945 N^2
\NN\\&&  
 +747 N+280\bigr)
 +10800 N \left(9 N^2+16 N+7\right)~.
\end{eqnarray}
With the exception of $D_1$ and $D_3$, in $z$-space the scalar $A_{gg,Q}$ diagrams cannot 
be expressed
within the class of the usual harmonic polylogarithms\cite{Remiddi:1999ew}, but generalizations 
thereof occur. These are given in
terms of iterated integrals over the following letters
\begin{eqnarray}
 \left\{
 \frac{d\tau}{1-\tau},\frac{d\tau}{\tau},\frac{\sqrt{\tau} \, d\tau}{1-\tau},\sqrt{1-\tau} \sqrt{\tau} \, d\tau,
 \frac{d\tau}{\eta \tau-\tau+1},\frac{\sqrt{1-\tau} \sqrt{\tau} \, d\tau}{\eta \tau-\tau+1},
 \frac{d\tau}{\eta \tau-\eta-\tau},\frac{\sqrt{1-\tau} \sqrt{\tau} \, d\tau}{\eta \tau-\eta-\tau}
 \right\}~.&& \NN\\
\end{eqnarray}

In Mellin-space all scalar $A_{gg,Q}$-diagrams can be expressed in terms of
$\ln(\eta)$, the harmonic polylogarithms
$\HA_{-1,0,0}(\sqrt{\eta})$, $\HA_{1,0,0}(\sqrt{\eta})$, alternating harmonic sums,
$\eta$-dependant generalized harmonic sums and $\eta$-dependent finite binomial sums.
For fixed values of the Mellin variable $N$, these $\eta$-dependent sums turn
into rational functions in $\eta$. Thus for fixed Mellin moments, all diagrams
are given in terms of the $\ln(\eta)$ and the combination
$\HA_{-1,0,0}(\sqrt{\eta})+\HA_{1,0,0}(\sqrt{\eta})$ with rational coefficients in
$\eta$.

The summands of many of these sums diverge for $\eta \rightarrow 1$ due to
factors as $(1-\eta )^{-j}$, where $j$ is a summation index which assumes
positive integer values. Furthermore, also contributions $\propto(1-\eta)^{-N}$
emerge. Physically the limit $\eta \rightarrow 1$ represents the equal mass
case $m_1=m_2$ \cite{Ablinger:2014uka} and thus the diagrams are expected to be convergent 
in this limit. Due to the many individually divergent terms this is highly non-trivial
to prove for general values of $N$. However, evaluating a series of Mellin
moments $N=2 \ldots 30$, yields convergent results for $\eta=1$, which agree
with the results given in Ref. \cite{Ablinger:2014uka} previously. This
indicates that these apparent divergences are just a relic of this specific
representation which has been applied. By induction one may prove that the result is valid
at general values of $N$. 
The diagrams ($D_{2a}$, $D_{2b}$),
($D_{4a}$, $D_{4b}$), ($D_{5a}$, $D_{5b}$), ($D_{6a}$, $D_{6b}$) and ($D_{8a}$,
$D_{8b}$) have all been computed independently. One notes that as expected the
respective $z$- and Mellin-space results can be translated into each other by
interchanging the masses $m_1 \leftrightarrow m_2,~ \eta\rightarrow 1/\eta$.
Furthermore, the results for the mass-symmetric diagrams $D_{1}$, $D_{3}$ and
$D_{7}$ turn out to be invariant under this interchange, which constitutes a further check
of these results.

For all scalar $A_{gg,Q}^{(3)}$-topologies, series expansions up to $O(\eta^3
\ln^3(\eta))$ for a series of fixed Mellin moments ($N=2,4,6$) have been
computed using the code {\tt Q2e}/{\tt Exp}~\cite{Harlander:1997zb,
Seidensticker:1999bb}. All the general $N$ and general-$\eta$ results agree
with these expansions. 

\section{Conclusions \label{sec:conc}}

\vspace{1mm}
\noindent
Genuine two-mass contributions to the Wilson coefficients and the transition matrix elements in the VFNS
occur at 3-loop order in QCD. We derived the renormalization of these contributions, which extends the 
single mass case considered earlier in Ref.~\cite{Bierenbaum:2009mv}. Although the new contributions 
manifest themselves as two-mass contributions in single diagrams carrying local operators, it is possible 
to assign a diagram to either of the heavy flavor distributions in the VFNS by the quark species carrying 
the operator. The diagrams arise from separating off the massless Wilson coefficient in the light-cone 
expansion. Through this, one knows the charge assignment for the corresponding diagram. In this way an asymmetric 
separation of the otherwise symmetric OMEs under $m_1 \leftrightarrow m_2$ occurs. This only applies to the 
OMEs $A_{Qq}^{(3), \sf PS}$ and $A_{Qg}^{(3)}$. All other OMEs enter the VFNS in a mass-symmetric way.

In a first step we have calculated a series of moments $(N = 2,4,6)$ for all contributing massive OMEs and 
presented the constant part of the unrenormalized genuine two-mass OME. With current technologies 
\cite{Harlander:1997zb,Seidensticker:1999bb}, the 6th moment required one CPU year of computational time. For 
a series of OMEs, the solution for general values of the mass ratio $\eta$, and at general values of the 
Mellin variable $N$, could be derived along with its $z$-space representation. This is the case for the OMEs 
$A_{qq,Q}^{{\sf NS}, (3)}, A_{qq,Q}^{{\sf NS,TR}, (3)}$ and $A_{qg,Q}^{(3)}$. The corresponding expressions 
depend on harmonic sums, weighted with a (poly)logarithmic dependence on the mass ratio. In these cases we 
presented also numerical results studying their relative contribution to the complete $O(T_F^2)$-term of 
the OMEs $A_{ij}^{(3)}$ in a wide range of $x$ and $Q^2$, in order to illustrate the two mass effects compared to 
the single mass contributions. In all cases these ratios vary between 0 and $\sim 0.5$ in 
part of the kinematic region, exhibiting scaling violations.

We have also calculated all the scalar topologies appearing in the more involved case of the
OME $A_{gg,Q}^{(3)}$. Here, more advanced computation methods were required. The corresponding integrals do 
not allow an expansion in the mass ratio at general values of $N$, so we calculated these integrals exactly.
In $z$-space the corresponding integrals could be represented in terms of iterated two-variate and partly 
root-valued integrals, the $G$-functions, see also Appendix~\ref{APP2}. Associated to it, one obtains in 
Mellin-$N$ space, sum representations containing functions of $\eta$ in denominators, with a formally 
divergent behaviour as $\eta \rightarrow 1$. However, since $N \in \mathbb{N}$, one obtains convergent 
representations for {\it each} individual integer $N$ in this limit. Also because of this behaviour, the 
inverse Mellin transform to $z$-space requires a series of special steps, which we have outlined. It is 
expected that the corresponding representation in the case of the two-mass contributions to the OME 
$A_{Qg}^{(3)}$ is even more involved, since already in the equal mass case elliptic integrals and iteration
of other letters over them contribute.

\appendix 
\section{Massive Operator Matrix Elements in \boldmath $z$-Space \label{APP1}} 

\vspace{1mm}
\noindent
In the following, we present a series of genuine two-mass contributions in $z$-space. These are 
distribution-valued and consist of the three parts $A_{ij}^\delta, A_{ij}^+(z)$ and $A_{ij}^{\sf reg}(z)$.
The Mellin convolution of the OMEs with a function $f(z)$ is defined by, cf.~e.g.~\cite{Blumlein:1989gk},
\begin{eqnarray}
A(z) \otimes f(z) &=&  A^\delta f(1) 
                   + \int_z^1 dy A^+(y) \left[\frac{1}{y} f\left(\frac{z}{y}\right)-f(z)\right]
                   - f(z) \int_0^z dy A^+(y) 
\nonumber\\ &&     
                   + \int_z^1 \frac{dy}{y} A^{\sf reg}(y) f\left(\frac{z}{y}\right)~. 
\end{eqnarray}

In the flavor non-singlet case, the parts of the OME are given by

\allowdisplaybreaks[4]
\begin{eqnarray}
\tilde{A}_{qq,Q}^{{\sf NS},(3), \delta}&=&
        C_F T_F^2 \Biggl\{
                -\frac{2068}{81}
                -\biggl[
                        \frac{584}{27}
                        +16 \zeta_2
                \biggr] \left(L_1+L_2\right)
                -\frac{16}{9}\left(L_1^2+L_2^2\right)
\nonumber\\&&
                -\frac{16}{3}\left(L_1^2 L_2+L_2^2 L_1\right)
                -\frac{64}{9}\left(L_1^3+L_2^3\right)
                -\frac{16}{9} \zeta_2
                -\frac{64}{9} \zeta_3
        \Biggr\} + \tilde{a}_{qq.Q}^{\sf NS, (3), \delta},
\\
\tilde{A}_{qq,Q}^{{\sf NS},(3), +}&=&
\frac{C_F T_F^2}{z-1} \Biggl\{
        \biggl[
                \frac{3584}{81}
                +\frac{640}{27} \ln (z)
                +\frac{64}{9} \ln ^2(z)
                +\frac{64}{3} \zeta_2
        \biggr] \left(L_1+L_2\right)
\nonumber\\&&
        +\biggl[
                \frac{640}{27}
                +\frac{128}{9} \ln (z)
        \biggr] \left(L_1^2+L_2^2\right)
        +\frac{64 }{9}\left(L_1^2 L_2+L_2^2 L_1\right)
\nonumber\\&&
        +\frac{256 }{27}\left(L_1^3+L_2^3\right)
        +\biggl(
                \frac{3584}{81}
                +\frac{128}{9} \zeta_2
        \biggr) \ln (z)
        +\frac{320}{27} \ln ^2(z)
\nonumber\\&&
        +\frac{64}{27} \ln^3(z)
        +\frac{640}{27} \zeta_2
        +\frac{256}{27} \zeta_3
        +\frac{20992}{243}
\Biggr\} + \tilde{a}_{qq,Q}^{{\sf NS},(3),+}~,
\\
\tilde{A}_{qq,Q}^{{\sf NS},(3), \sf Reg}&=&
C_F T_F^2 \Biggl\{
        \frac{64}{243} (377 z-49)
        +\biggl[
                \frac{64}{81} (67 z-11)
                +\frac{64}{27} (11 z-1) \ln (z)
\nonumber\\&&
                +\frac{32}{9} (z+1) \left(\ln ^2(z)+3 \zeta_2\right)
        \biggr] \left(L_1+L_2\right)
        +\biggl[
                \frac{64}{27} (11 z-1)
\nonumber\\&&
                +\frac{64}{9} (z+1) \ln (z)
        \biggr] \left(L_1^2+L_2^2\right)
        +\frac{32}{9} (z+1) \left(L_1^2 L_2+L_2^2 L_1\right)
\nonumber\\&&
        +\frac{128}{27} (z+1) \left(L_1^3+L_2^3\right)
        +\biggl[
                \frac{64}{81} (67 z-11)
                +\frac{64}{9} (z+1) \zeta_2
        \biggr] \ln (z)
\nonumber\\&&
        +\frac{32}{27} (11 z-1) \left(\ln^2(z)+2 \zeta_2\right)
        +\frac{32}{27} (z+1) \left(\ln^3(z)+4 \zeta_3\right)
\Biggr\} + \tilde{a}_{qq,Q}^{\sf NS, (3), Reg}~.
\end{eqnarray}
The contributions to the constant two-mass term of the unrenormalized non-singlet OME are
\begin{eqnarray}
\tilde{a}_{qq,Q}^{\sf NS, (3),\delta}&=&
C_F T_F^2 \Biggl\{
8 \big(L_1^3+L_2^3+(L_1 L_2+2 \zeta_2+5) (L_1+L_2)\big)+\frac{4}{3}(L_1+L_2)^2
\nonumber\\&&
+\frac{4}{243} \biggl(-405 \eta +532-\frac{405}{\eta }\biggr)
-\frac{10 \big(\eta ^2-1\big)}{3 \eta} \ln(\eta)
        -\frac{8}{9} \ln^3(\eta)
        +\frac{16 \zeta_2}{9}
        -\frac{64 \zeta_3}{9}
\nonumber\\&&
+\frac{\eta+1}{\eta^{3/2}} \big(5 \eta ^2+22 \eta +5\big) \biggl[
\frac{1}{12} \ln^2(\eta) \ln\left(\frac{1+\eta_1}{1-\eta_1}\right)
-\frac{2}{3} \ln(\eta) \Li_2(\eta_1)
+\frac{4}{3} \Li_3(\eta_1)
\biggr]
\nonumber\\&&
+\frac{T_1}{6 \eta^{3/2}} \big[\ln(\eta) \Li_2(\eta)-\Li_3(\eta)\big]
        -\biggl[
                 \frac{(\eta +5) (5 \eta +1)}{6 \eta}
                -\frac{16}{3} \ln (1-\eta)
        \biggr] \ln^2(\eta)
\Biggr\}~, \nonumber\\
\end{eqnarray}
where
\begin{equation}
T_1 = 5 \eta ^3+27 \eta ^2+64 \eta ^{3/2}+27 \eta +5,
\end{equation}
\begin{eqnarray}
\tilde{a}_{qq,Q}^{\sf NS, (3) +}&=&
\frac{C_F T_F^2}{z-1} \Biggl\{
-\frac{32}{3} \biggl[L_1^3+L_2^3+\biggl(L_1 L_2+2 \zeta_2+\frac{58}{9}\biggr) (L_1+L_2)\biggr]
        +\frac{32}{27} \ln^3(\eta)
\nonumber\\&&
-\biggl(\frac{160}{9}+\frac{32}{3} \ln(z)\biggr) (L_1+L_2)^2
-\frac{64}{9} \biggl(\ln^2(z)+\frac{10}{3} \ln(z)\biggr) (L_1+L_2)
\nonumber\\&&
        +\frac{16 \big(405 \eta ^2-3238 \eta +405\big)}{729 \eta}
                +\frac{40 \big(\eta^2-1\big)}{9 \eta} \ln (\eta)
        -\frac{640}{27} \zeta_2
        +\frac{256}{27} \zeta_3
\nonumber\\&&
        -\frac{2 T_1}{9 \eta^{3/2}} \big[\ln(\eta) \Li_2(\eta)-\Li_3(\eta)\big]
        -\biggl(
                \frac{3712}{81}
                +\frac{128}{9} \zeta_2
        \biggr) \ln (z)
        -\frac{128}{81} \ln^3(z)
\nonumber\\&&
-\frac{5 \eta^3+27 \eta^2+27 \eta+5}{9 \eta^{3/2}} \biggl[ 
\ln^2(\eta) \ln\left(\frac{1+\eta_1}{1-\eta_1}\right)
-8 \ln (\eta) \Li_2(\eta_1)
+16 \Li_3(\eta_1)
\biggr]
\nonumber\\&&
        +\biggl[
                \frac{2 \big(5 \eta^2+2 \eta+5\big)}{9 \eta}
                -\frac{64}{9} \ln(1-\eta)
                -\frac{32}{9} \ln(z)
        \biggr] \ln^2(\eta)
        -\frac{640}{81} \ln^2(z)
\Biggr\}~,
\\
\tilde{a}_{qq,Q}^{\sf NS, (3), Reg}&=&
C_F T_F^2 \Biggl\{
-\frac{16}{3} (z+1) \biggl[L_1^3+L_2^3+\biggl(L_1 L_2+\frac{2}{3} \ln^2(z)+2 \zeta_2+\frac{2}{3}\biggr) (L_1+L_2)\biggr]
\nonumber\\&&
+\frac{16}{9} \big(1-11 z-3 (z+1) \ln(z)\big) (L_1+L_2)^2
+\frac{20 \left(\eta^2-1\right) (z+1)}{9 \eta} \ln(\eta)
\nonumber\\&&
-\frac{32}{27} \big(52 z+(22z-2) \ln(z)\big) (L_1+L_2)
        +\frac{128}{27} (z+1) \biggl[\frac{1}{8} \ln^3(\eta)-\frac{1}{6} \ln^3(z)
\nonumber\\&&
+\zeta_3\biggr]
+\frac{8 T_2}{729 \eta}
+\frac{T_3}{9 \eta^{3/2}} \left(\sqrt{\eta}+1\right)^2 (z+1) \big[\Li_3(\eta)-\ln(\eta) \Li_2(\eta)\big]
\nonumber\\&&
%
-\frac{z+1}{18 \eta^{3/2}} \left(5 \eta^3+27 \eta^2+27 \eta+5\right) \biggl[
\ln^2(\eta) \ln\left(\frac{1+\eta_1}{1-\eta_1}\right)
-8 \ln(\eta) \Li_2(\eta_1)
\nonumber\\&&
+16 \Li_3(\eta_1)
\biggr]
        +\biggl[
                \frac{T_4}{9 \eta}
                -\frac{16}{9} (z+1) \left[2 \ln (1-\eta)+\ln(z)\right]
       \biggr] \ln^2(\eta)
\nonumber\\&&
        -\frac{64}{81} \big[
                9 (z+1) \zeta_2
                +55 z+3
        \big] \ln (z)
%
        -\frac{64}{27} (11 z-1) \left[\frac{1}{3} \ln^2(z)+\zeta_2\right]
\Biggr\},
\end{eqnarray}
with
\begin{eqnarray}
T_2&=&
                \eta  (1658-8134 z)
                +405 (z+1)
                +405 \eta ^2 (z+1)~,
\\
T_3&=&5 \eta ^2-10 \eta ^{3/2}+42 \eta -10 \sqrt{\eta }+5~,
\\
T_4&=&
                        \eta  (34-30 z)
                        +5 (z+1)
                        +5 \eta ^2 (z+1)~.
\end{eqnarray}
For transversity one obtains
\begin{eqnarray}
\tilde{A}_{qq,Q}^{{\sf NS,TR, (3), \delta}}&=&
C_F T_F^2 \Biggl\{
                -\frac{2068}{81}
                -\left(
                        \frac{584}{27}
                        +16 \zeta_2
                \right) \left(L_1+L_2\right)
                -\frac{16 }{9}\left(L_1^2+L_2^2\right)
\nonumber\\&&
                -\frac{16 }{3}\left(L_1^2 L_2+L_2^2 L_1\right)
                -\frac{64 }{9}\left(L_1^3+L_2^3\right)
                -\frac{16}{9} \zeta_2
                -\frac{64}{9} \zeta_3
\Biggr\} + \tilde{a}_{qq,Q}^{\sf NS, TR, (3), \delta},
\\
\tilde{A}_{qq,Q}^{{\sf NS,TR, (3), +}}&=&
\frac{C_F T_F^2}{z-1} \Biggl\{
        \frac{20992}{243 }
        +\biggl[
                \frac{3584}{81}
                +\frac{640}{27} \ln (z)
                +\frac{64}{9} \ln^2(z)
                +\frac{64}{3} \zeta_2
        \biggr] \left(L_1+L_2\right)
\nonumber\\&&
        +\biggl[
                \frac{640}{27}
                +\frac{128}{9} \ln(z)
        \biggr] \left(L_1^2+L_2^2\right)
        +\frac{64}{9}\left(L_1^2 L_2+L_2^2 L_1\right)
        +\frac{256}{27}\left(L_1^3+L_2^3\right)
\nonumber\\&&
        +\left(
                \frac{3584}{81}
                +\frac{128}{9} \zeta_2
        \right) \ln (z)
        +\frac{320}{27} \ln^2(z)
        +\frac{64}{27} \ln^3(z)
        +\frac{640}{27} \zeta_2
\nonumber\\&&
        +\frac{256}{27} \zeta_3
\Biggr\} + \tilde{a}_{qq,Q}^{\sf NS, TR, (3), +},
\\
\tilde{A}_{qq,Q}^{{\sf NS,TR, (3), Reg}}&=&
C_F T_F^2 \Biggl\{
        \frac{64}{243} (45 z+283)
        +\biggl[
                \frac{64}{81} (9 z+47)
                +\frac{640}{27} \ln (z)
                +\frac{64}{9} \ln^2(z)
\nonumber\\&&
                +\frac{64}{3} \zeta_2
        \biggr] \left(L_1+L_2\right)
        +\frac{64 }{9}\left(L_1^2 L_2+L_2^2 L_1\right)
        +\frac{256 }{27}\left(L_1^3+L_2^3\right)
\nonumber\\&&
        +\left(
                \frac{64}{81} (9 z+47)
                +\frac{128}{9} \zeta_2
        \right) \ln (z)
        +\frac{320}{27} \ln^2(z)
        +\frac{64}{27} \ln^3(z)
\nonumber\\&&
        +\left[\frac{128}{9} \ln (z)+\frac{640}{27}\right] \left(L_1^2+L_2^2\right) 
        +\frac{640}{27} \zeta_2
        +\frac{256}{27} \zeta_3
\Biggr\} + \tilde{a}_{qq,Q}^{\sf NS, TR, (3), Reg},
\end{eqnarray}
with
\begin{eqnarray}
\tilde{a}_{qq,Q}^{\sf NS,TR, (3), \delta}&=&
C_F T_F^2 \Biggl\{
8 \big(L_1^3+L_2^3+(L_1 L_2+2 \zeta_2+5) (L_1+L_2)\big)+\frac{4}{3} (L_1+L_2)^2
\nonumber\\&&
        +\frac{4}{243} \left(-405 \eta +532-\frac{405}{\eta}\right)
        +\frac{10 \left(1-\eta^2\right)}{3 \eta} \ln(\eta)
        -\frac{8}{9} \ln^3(\eta)
\nonumber\\&&
        +\biggl[
                -\frac{(\eta +5) (5 \eta +1)}{6 \eta }
                +\frac{16}{3} \ln (1-\eta)
        \biggr] \ln^2(\eta)
        +\frac{16}{9} \zeta_2
        -\frac{64}{9} \zeta_3
\nonumber\\&&
                +\frac{2 (\eta+1)}{3 \eta^{3/2}} (5 \eta ^2+22 \eta +5) \biggl[ 
                             2 \Li_3(\eta_1)
                             -\ln(\eta) \Li_2(\eta_1)
\nonumber\\&&
                             +\frac{1}{8} \ln^2(\eta) \ln\left(\frac{1+\eta_1}{1-\eta_1}\right)
                 \biggr]
        +\frac{(1+\sqrt{\eta})^2}{6 \eta^{3/2}} T_3 \big[\ln(\eta) \Li_2(\eta)-\Li_3(\eta)\big]
\Biggr\}~,
\\
\tilde{a}_{qq,Q}^{\sf NS,TR, (3), + }&=&
\frac{C_F T_F^2}{z-1} \Biggl\{
-\frac{32}{3} L_1^3-\frac{32}{3} L_2^3-\frac{32}{3} \biggl(L_1 L_2+2 \zeta_2+\frac{58}{9}\biggr) (L_1+L_2)
\nonumber\\&&
-\biggl(\frac{160}{9}+\frac{32}{3} \ln(z)\biggr) (L_1+L_2)^2
-\frac{64}{9} \biggl(\ln^2(z)+\frac{10}{3} \ln(z)\biggr) (L_1+L_2)
\nonumber\\&&
        +\frac{16 \big(405 \eta^2-3238 \eta +405\big)}{729 \eta}
        -\frac{40 \big(1-\eta^2\big)}{9 \eta} \ln(\eta)
        +\frac{32}{27} \ln^3(\eta)
\nonumber\\&&
        +\biggl[
                \frac{2 \big(5 \eta ^2+2 \eta +5\big)}{9 \eta}
                -\frac{64}{9} \ln(1-\eta)
                -\frac{32}{9} \ln(z)
        \biggr] \ln^2(\eta)
\nonumber\\&&
        -\biggl(
                \frac{3712}{81}
                +\frac{128}{9} \zeta_2
        \biggr) \ln(z)
               -\frac{8 (\eta+1)}{9 \eta^{3/2}} (5 \eta ^2+22 \eta +5) \biggl[ 
                             2 \Li_3(\eta_1)
\nonumber\\&&
                             -\ln(\eta) \Li_2(\eta_1)
                             +\frac{1}{8} \ln^2(\eta) \ln\left(\frac{1+\eta_1}{1-\eta_1}\right)
                 \biggr]
        -\frac{640}{27} \zeta_2
        +\frac{256}{27} \zeta_3
\nonumber\\&&
        -\frac{2 (1+\sqrt{\eta})^2}{9 \eta^{3/2}} T_3 \big[\ln(\eta) \Li_2(\eta)-\Li_3(\eta)\big]
        -\frac{640}{81} \ln^2(z)
        -\frac{128}{81} \ln^3(z)
\Biggr\},
\\
\tilde{a}_{qq,Q}^{{\sf NS,TR}, (3), \sf Reg}&=&
C_F T_F^2 \Biggl\{
-\frac{32}{3}L_1^3-\frac{32}{3} L_2^3-\frac{32}{3} \biggl(L_1 L_2+\frac{20}{9} \ln(z)+\frac{2}{3} \ln^2(z)
\nonumber\\&&
+\frac{2}{3} z+2 \zeta_2+\frac{52}{9}\biggr) (L_1+L_2)
-\biggl(\frac{160}{9}+\frac{32}{3} \ln(z)\biggr) (L_1+L_2)^2
\nonumber\\&&
+\frac{16 \left(405 \eta^2-2878 \eta-360 \eta z+405\right)}{729 \eta}
-\frac{40 \left(1-\eta^2\right)}{9 \eta} \ln (\eta)
+\frac{256}{27} \zeta_3
\nonumber\\&&
-\zeta_2 \left(\frac{128}{9} \ln(z)+\frac{640}{27}\right)
-\frac{128}{81} \ln^3(z)
-\frac{640}{81} \ln^2(z)
-\frac{128}{81} (3 z+26) \ln(z)
\nonumber\\&&
+\biggl[\frac{2 \left(5 \eta ^2+2 \eta +5\right)}{9 \eta}
        -\frac{64}{9} \ln(1-\eta)
        -\frac{32}{9} \ln(z)
\biggr] \ln^2(\eta)
\nonumber\\&&
+\frac{32}{27} \ln^3(\eta)
               -\frac{8 (\eta+1)}{9 \eta^{3/2}} (5 \eta ^2+22 \eta +5) \biggl[ 
                             2 \Li_3(\eta_1)
                             -\ln(\eta) \Li_2(\eta_1)
\nonumber\\&&
                             +\frac{1}{8} \ln^2(\eta) \ln\left(\frac{1+\eta_1}{1-\eta_1}\right)
                 \biggr]
        -\frac{2 (1+\sqrt{\eta})^2}{9 \eta^{3/2}} T_3 \big[\ln(\eta) \Li_2(\eta)-\Li_3(\eta)\big]
\Biggr\}~.
\end{eqnarray}
Using the shorthand
\begin{eqnarray}
\tilde{p}_{gq}^{(0)} = \frac{1+(1-z)^2}{z}
\end{eqnarray}
the OME $\tilde{A}_{gq,Q}^{(3)}$ is given by
\begin{eqnarray}
\tilde{A}_{gq,Q}^{(3)}&=&
        C_F T_F^2 \Biggl\{
                -\frac{512 \left(31 z^2-41 z+41\right)}{81 z}
                -\frac{128}{9} p_{gq}^{(0)} \zeta_3
                -\biggl[
                        \frac{64}{27 z} \left(43 z^2-56 z+56\right)
\nonumber\\&&
                        +\frac{128}{9 z} \left(4 z^2-5 z+5\right) \ln(1-z)
                        +\frac{32}{3} p_{gq}^{(0)} \ln^2(1-z)
                        + 32 p_{gq}^{(0)} \zeta_2
                \biggr] \left(L_1+L_2\right)
\nonumber\\&&
                -\biggl[
                        \frac{128}{9 z} \left(4 z^2-5 z+5\right)
                        + p_{gq}^{(0)} \frac{64}{3} \ln(1-z)
                \biggr] \left(L_1^2+L_2^2\right)
                -\frac{32}{9} 
p_{gq}^{(0)} \ln^3(1-z)
\nonumber\\&&
                -\biggl(
                        \frac{64}{27 z} \left(43 z^2-56 z+56\right)
                        +\frac{64}{3} p_{gq}^{(0)} \zeta_2
                \biggr) \ln (1-z)
                -\frac{32}{3} p_{gq}^{(0)} \left(L_1^2 L_2+L_2^2 L_1\right)
\nonumber\\&&
                -\frac{128}{9} 
p_{gq}^{(0)} \left(L_1^3+L_2^3\right)
                -\frac{64}{9 z} \left(4 z^2-5 z+5\right) \big[\ln^2(1-z)+2 \zeta_2\big]
        \Biggr\}
+ \tilde{a}_{gq,Q}^{(3)},
\end{eqnarray}
with
\begin{eqnarray}
\tilde{a}_{gq,Q}^{(3)}&=&C_F T_F^2 \Biggl\{
16 p_{gq}^{(0)} \biggl[L_1^3+L_2^3+\biggl(L_1 L_2+\frac{20}{9} \ln(1-z)+\frac{2}{3} \ln^2(1-z)
\nonumber\\&&
+2 \zeta_2+\frac{58}{9}\biggr) (L_1+L_2)\biggr]
+16 \biggl[z+p_{gq}^{(0)} \biggl(\frac{5}{3}+\ln(1-z)\biggr)\biggr] (L_1+L_2)^2
\nonumber\\&&
+z \biggl(\frac{320}{9}+\frac{64}{3} \ln(1-z)\biggr) (L_1+L_2)
-\frac{20 (\eta^2-1)}{3 \eta} p_{gq}^{(0)} \ln(\eta)
\nonumber\\&&
        -\frac{8 T_8}{243 \eta  z}
        +\frac{64}{3} \left(
                \frac{39 z^2-58 z+58}{9 z}
                +\zeta_2 p_{gq}^{(0)}
        \right) \ln(1-z)
        -\frac{128}{9} \zeta_3 p_{gq}^{(0)}
\nonumber\\&&
+\frac{\left(\sqrt{\eta}+1\right)^2}{\eta^{3/2}} p_{gq}^{(0)} T_5 \biggl[
\frac{2}{3} \ln(\eta) \Li_2(-\eta_1)
+\frac{1}{6} \ln^2(\eta) \ln (1+\eta_1)
-\frac{4}{3} \Li_3(-\eta_1)
\biggr]
\nonumber\\&&
+\frac{\left(\sqrt{\eta}-1\right)^2}{\eta^{3/2}} p_{gq}^{(0)} T_6 \biggl[
-\frac{2}{3} \ln(\eta) \Li_2(\eta_1)
-\frac{1}{6} \ln^2(\eta) \ln(1-\eta_1)
+\frac{4}{3} \Li_3(\eta_1)
\biggr]
\nonumber\\&&
+\frac{64}{27} p_{gq}^{(0)} \ln^3(1-z)
        -\biggl[
                \frac{T_7}{3 \eta  z}
                - \frac{16}{3} p_{gq}^{(0)} \ln(1-z)
        \biggr] \ln^2(\eta)
-\frac{16}{9} p_{gq}^{(0)} \ln^3(\eta)
\nonumber\\&&
+\frac{128}{9 z} \left(4 z^2-5 z+5\right) \biggl[\frac{1}{3} \ln^2(1-z)+\zeta_2\biggr]
\Biggr\}~,
\end{eqnarray}
and the polynomials
\begin{eqnarray}
T_5&=&-10 \eta ^{3/2}+5 \eta ^2+42 \eta -10 \sqrt{\eta }+5~,\\
T_6&=&10 \eta ^{3/2}+5 \eta ^2+42 \eta +10 \sqrt{\eta }+5~,\\
T_7&=&5 \eta ^2 z^2-10 \eta ^2 z+10 \eta ^2-14 \eta  z^2-4 \eta  z+4 \eta +5 z^2-10 z+10~,\\
T_8&=&405 \eta ^2 z^2-810 \eta ^2 z+810 \eta ^2-5326 \eta  z^2+6476 \eta  z-6476 \eta\nonumber\\&&
+405 z^2-810 z+810~.
\end{eqnarray}

\section{Formulae \label{APP2}}

\vspace{1mm}
\noindent
In the following, we list a series of useful relations between the iterated $G$-integrals and some of their
special values. In the case where the letters in the alphabet are restricted to ${\frac{1}{x},\frac{1}{1-x},\frac{1}{1+x}}$, 
the $G$-integrals correspond, of course, to the standard harmonic polylogarithms.
\begin{equation}
G\left(\left\{{\sf w}_{a_1},...,{\sf w}_{a_k}\right\},z\right) 
\equiv 
\HA_{a_1,...,a_k}(z), \quad {\sf w}_{a_i}=\frac{a_i^2+a_i-1}{a_i-x}, \quad a_i \in \{0,1,-1\}~.
\end{equation}

For $G$-functions of weight one, we have the following identities:
\begin{eqnarray}
G\left(\left\{\sqrt{x(1-x)}\right\},z\right) &=& \frac{1}{4}\left[
-\sqrt{z(1-z)} (1-2z) + \arcsin(\sqrt{z})\right]~, 
\\
G\left(\left\{\frac{\sqrt{x(1-x)}}{1-x(1-\eta)}\right\},z\right) &=& 
\frac{1}{(1-\eta)^2}\biggl[(1+\eta) \arcsin(\sqrt{z})
-(1-\eta) \sqrt{z(1-z)}
\nonumber \\ &&
-2 \sqrt{\eta} \arctan\left(\sqrt{\frac{\eta z}{1-z}}\right)\biggr]~,
\\
G\left(
        \left\{\frac{\sqrt{x}}{1-x}\right\},z\right)
&=&
 -2 \sqrt{z}
+\HA_{-1}\big(
        \sqrt{z}\big)
+\HA_1\big(
        \sqrt{z}\big)~,
\\
G\left(\left\{\frac{1}{1-x(1-\eta)}\right\},1\right) 
&=& 
- \frac{\HA_0(\eta)}{1-\eta}~,
\\
G\left(\left\{\frac{\sqrt{x(1-x)}}{1-x(1-\eta)}\right\},1\right) 
&=& 
\frac{\pi}{2(1+\sqrt{\eta})^2}~,
\\
G\left(\left\{\frac{1}{1+\sqrt{x}}\right\},1\right)
&=&
2 - 2 \ln(2)~,
\\
G\left(\left\{\frac{\sqrt{x}}{\sqrt{1-x}}\right\},1\right) 
&=& 
\frac{\pi}{2}~.
\end{eqnarray}

For $G$-functions of weight two, we have,
\begin{eqnarray}
G\left(\left\{\frac{\sqrt{x}}{1-x},\frac{1}{x}\right\},z\right)
&=&
2 \HA_{-1}\left(\sqrt{z}\right) \HA_0\left(\sqrt{z}\right)
-2 \HA_{0,-1}\left(\sqrt{z}\right)
\nonumber \\ &&
+2 \HA_0\left(\sqrt{z}\right) \HA_1\left(\sqrt{z}\right)
-2 \HA_{0,1}\left(\sqrt{z}\right)
\nonumber \\ &&
-4 \sqrt{z} \left[-1+\HA_0\left(\sqrt{z}\right)\right]~,
\\
G\left(\left\{\frac{\sqrt{x}}{1-x},\frac{1}{x}\right\},\frac{1}{z}\right)
&=&
\frac{4 (z+1)}{\sqrt{z }}
-6 \zeta_2
-\frac{2 \left(z -1\right)}{\sqrt{z}} \HA_0\left(z\right)
\nonumber \\ &&
-G\left(\left\{\frac{\sqrt{x}}{1-x},\frac{1}{x}\right\},z\right)~,
\\
G\left(\left\{\frac{1}{1-x(1-\eta)},\frac{1}{1-x}\right\},1\right) 
&=&
\frac{1}{1-\eta}\biggl[
\zeta_2 + \frac{\ln^2(\eta)}{2}
+\ln(\eta) \HA_1(\eta) 
-\HA_{0,1}(\eta)\biggr],
\\
G\left(\left\{\frac{\sqrt{x (1-x)}}{x\eta -\eta-x},\frac{1}{x}\right\},1\right)
&=&
\frac{\pi}{(1-\eta)^2} \Biggl\{
(\eta+1) \ln(2)-\frac{5 \eta-1}{2}
+\sqrt{\eta} \biggl[
\HA_0(\eta) 
\nonumber \\ &&
+\HA_1(\eta) 
-G\left(\left\{\frac{\sqrt{x}}{1-x}\right\},\eta\right)
\biggr]
\Biggr\}~,
\\
G\left(\left\{\frac{\sqrt{x (1-x)}}{x \eta -\eta-x},\frac{1}{1-x}\right\},1\right)
&=&
\frac{\pi}{(1-\eta)^2} \Biggl\{
\frac{1}{2} (3 \eta+1)-(\eta+1) \ln(2)
\nonumber \\ &&
+\sqrt{\eta} \biggl[
-\HA_1(\eta)+G\left(\left\{\frac{\sqrt{x}}{1-x}\right\},\eta\right)
\biggr]
\Biggr\}~,
\\
G\left(\left\{\frac{1}{\eta x-x+1},\frac{1}{x}\right\},1\right)
&=&
\frac{\text{Li}_2(1-\eta)}{\eta-1}~,
\\
G\left(\left\{\sqrt{x (1-x)},\frac{\sqrt{x (1-x)}}{x \eta -\eta-x}\right\},1\right)
&=&
\frac{(\eta -3) \eta^2 \HA_0(\eta)}{4 (\eta-1)^4}
+\frac{\sqrt{\eta} G\left(\left\{\frac{\sqrt{x}}{1-x},\frac{1}{x}\right\},\eta \right)}{8 \left(\eta-1\right)^2}
\nonumber \\ &&
-\frac{2 \eta ^2-4\eta -1}{6 (\eta -1)^3}-\frac{3\left(\eta -4 \sqrt{\eta}+1\right) \zeta_2}{16 (\eta-1)^2}~,
\\
G\left(\left\{\frac{1}{x},\frac{1}{\eta x-x+1}\right\},1\right)
&=&
   -G\left(\left\{\frac{1}
   {\eta 
   x-x+1},\frac{1}{x}\right\},1\right)~,
\\
G\left(\left\{\frac{1}{x \eta -\eta-x},\frac{1}{x}\right\},1\right)
&=&
-\frac{1}{\eta-1}\bigl[\zeta_2
+\HA_{1,0}\left(\eta\right)
+\HA_{0,0}\left(\eta \right)
\bigr]~,
\\
G\left(\left\{\frac{1}{x \eta-\eta-x},\frac{1}{1-x}\right\},1\right)
&=&
\frac{1}{\eta-1} \bigl[\zeta_2
+\HA_{1,0}\left(\eta \right)
\bigr]~,
\\
G\left(\left\{\frac{1}{x},\frac{1}{x \eta-\eta -x}\right\},1\right)
&=&
\frac{1}{\eta-1} \bigl[\zeta_2
+\HA_{1,0}\left(\eta\right)
+\HA_{0,0}\left(\eta\right)
\bigr]~,
\\
G\left(\left\{\frac{\sqrt{x (1-x)}}{\eta x-x+1},\frac{1}{x}\right\},1\right) 
&=& 
-\frac{\pi}{2 (\eta-1)^2}  \bigl[2 (1+\eta) \ln(2)+\eta-1 
\nonumber \\ &&
-4 \sqrt{\eta} \ln\left(1+\sqrt{\eta}\right)
\bigr]~,
\\
G\left(\left\{\frac{\sqrt{x (1-x)}}{\eta x-x+1},\frac{1}{1-x}\right\},1\right) 
&=& 
\frac{\pi}{2 (\eta-1)^2}  \bigl[2 (1+\eta) \ln(2)-\eta+1 
\nonumber \\ &&
+2 \sqrt{\eta} \left(\ln(\eta)-2 \ln\left(1+\sqrt{\eta}\right)\right)
\bigr]~,
\\
G\left(\left\{\frac{1}{\frac{1}{1-\eta}-x},\frac{1}{x+1}\right\},1\right)
&=&
- 2 \ln(2) \ln\left(\frac{\eta}{2-\eta}\right) + \HA_{0,1}\left(\frac{1-\eta}{2-\eta}\right) 
\nonumber\\ &&
-  \HA_{0,1}\left(\frac{2(1-\eta)}{2-\eta}\right)
\nonumber\\  &&
%
%
\\
G\left(\left\{\frac{1}{\eta 
   x-x+1},\frac{1}{x-1}\right\},1\right)
&=&
   -\frac{\HA_{0,1}\left(\frac{\eta
   -1}{\eta }\right)}{\eta
   -1}~,
\\
G\left(\left\{\sqrt{x (1-x)},\frac{\sqrt{x (1-x)}}{\eta x-x+1}\right\},1\right)
&=&
\frac{(\eta -3) \eta^2 \ln(\eta )}{4 (\eta-1)^4}
-\frac{2 \eta ^2-4\eta -1}{6 (\eta -1)^3}
\nonumber \\ &&
+\frac{\sqrt{\eta } G\left(\left\{\frac{\sqrt{x}}{1-x},\frac{1}{x}\right\},\eta \right)}{8 \left(\eta-1\right)^2}
+\frac{3 (\eta +1) \zeta_2}{16 (\eta-1)^2}~,
\\
G\left(\left\{\sqrt{x (1-x)},\sqrt{x (1-x)}\right\},1\right) 
&=& 
\frac{\pi^2}{128}~,
\\
G\left(\left\{\sqrt{x (1-x)},\frac{1}{1-x}\right\},1\right) 
&=&  
-\frac{\pi}{16} (1-4 \ln(2))~,
\\
G\left(\left\{\sqrt{x (1-x)},\frac{1}{x}\right\},1\right) 
&=&  
\frac{\pi}{16} (1-4 \ln(2))~,
\\
G\left(\left\{\frac{1}{x},\frac{\sqrt{x}}{\sqrt{1-x}}\right\},1\right) 
&=& 
   \pi \left(-\frac{1}{2}+\ln(2)\right)~,
\\
G\left(\left\{\frac{1}{x+1},\frac{\sqrt{x}}{\sqrt{1-x}}\right\},1\right)
&=&
\frac{\pi}{2} \left[-3+2 \sqrt{2}-2 \ln\left(1+\sqrt{2}\right)+3 \ln(2)\right]~,
\\
G\left(\left\{\frac{\sqrt{x}}{\sqrt{1-x}},\frac{1}{x+1}\right\},1\right)
&=&
\pi \left[ - \ln(2)+  \ln\left(1+\sqrt{2}\right)-\sqrt{2}  +\frac{3 }{2} \right]~,
\\
G\left(\left\{\frac{\sqrt{x}}{\sqrt{1-x}},\frac{\sqrt{x (1-x)}}{x+1}\right\},1\right) 
&=&
-\sqrt{2} \text{Li}_2\left(3-2 \sqrt{2}\right)
+\sqrt{2} \text{Li}_2\left(-3+2 \sqrt{2}\right)
\nonumber \\ &&
+2
+\frac{\pi^2}{4} \left(\frac{3}{2}-\sqrt{2}\right)
-2\ln(2)~,
\\
G\left(\left\{\frac{\sqrt{1-x}}{\sqrt{x}-1},\frac{\sqrt{1-x}}{\sqrt{x}-1}\right\},1\right)
&=&
\frac{1}{8}(4+\pi)^2~,
\\
G\left(\left\{\frac{\sqrt{1-x}}{\sqrt{x}-1},\frac{\sqrt{1-x}}{\sqrt{x}+1}\right\},1\right)
&=&
\frac{2}{3}-\pi+\frac{\pi^2}{8}~,
\\
G\left(\left\{\frac{\sqrt{1-x}}{\sqrt{x}-1},\frac{\sqrt{1-x}}{\sqrt{x}}\right\},1\right)
&=&
-\frac{11}{3}-\frac{\pi^2}{8}~,
\\
G\left(\left\{\frac{\sqrt{1-x}}{\sqrt{x}+1},\frac{\sqrt{1-x}}{\sqrt{x}-1}\right\},1\right)
&=& 
-\frac{14}{3}+\pi +\frac{\pi^2}{8}~,
\\
G\left(\left\{\frac{\sqrt{1-x}}{\sqrt{x}+1},\frac{\sqrt{1-x}}{\sqrt{x}+1}\right\},1\right)
&=& \frac{1}{8} (-4+\pi)^2~,
\\
G\left(\left\{\frac{\sqrt{1-x}}{\sqrt{x}+1},\frac{\sqrt{1-x}}{\sqrt{x}}\right\},1\right) 
&=&
\frac{5}{3}-\frac{\pi^2}{8}~,
\\
G\left(\left\{\frac{\sqrt{x}}{\sqrt{1-x}},\frac{1}{x}\right\},1\right)
&=&
\frac{\pi }{2}-\pi  \ln(2)~,
\\
G\left(\left\{\frac{\sqrt{1-x}}{\sqrt{x}+1},\frac{1}{x}\right\},1\right)
&=& -4+\pi  \left(-\frac{1}{2}+\ln(2)\right)+4 \ln(2)~,
\\
G\left(\left\{\frac{\sqrt{1-x}}{\sqrt{x}-1},\frac{1}{x-1}\right\},1\right)
&=&
4+\frac{\pi}{2} (1+2 \ln(2))~,
\\
G\left(\left\{\frac{\sqrt{1-x}}{\sqrt{x}-1},\frac{1}{x}\right\},1\right)
&=&
4+\pi  \left(-\frac{1}{2}+\ln(2)\right)-4 \ln(2)~,
\\
G\left(\left\{\frac{\sqrt{1-x}}{\sqrt{x}+1},\frac{1}{x-1}\right\},1\right)
&=&
-4+\frac{\pi}{2} (1+2\ln(2))~,
\\
G\left(\left\{\frac{\sqrt{x}}{\sqrt{1-x}},\frac{1}{x-1}\right\},1\right)
&=&
-\frac{\pi}{2} (1+2 \ln(2)) ~,
\\
G\left(\left\{\frac{\sqrt{1-x}}{\sqrt{x}},\frac{1}{x}\right\},1\right)
&=&
-\frac{\pi}{2} (1+ 2\ln(2))~,
\\
G\left(\left\{\frac{\sqrt{1-x}}{\sqrt{x}},\frac{1}{x-1}\right\},1\right)
&=&
\frac{\pi }{2}-\pi \ln(2)~.
\end{eqnarray}

The following identities hold for weight three functions:
\begin{eqnarray}
G\left(\left\{\frac{\sqrt{x}}{1-x},\frac{1}{x},\frac{1}{x}\right\},z\right)
&=&
\left[8 \sqrt{z}
-4 \HA_{0,-1}\left(\sqrt{z}\right)
-4 \HA_{0,1}\left(\sqrt{z}\right)
\right] \HA_0\left(\sqrt{z}
\right)
\nonumber \\ &&
+\left[-4 \sqrt{z}
+2 \HA_{-1}\left(\sqrt{z}\right)
+2 \HA_1\left(\sqrt{z}\right)
\right] \HA_0\left(\sqrt{z}\right)^2
\nonumber \\ &&
+4 \HA_{0,0,-1}\left(\sqrt{z}\right)
+4 \HA_{0,0,1}\left(\sqrt{z}\right)
-8 \sqrt{z}~,
\\
G\left(\left\{\frac{1}{x},\frac{\sqrt{x}}{1-x},\frac{1}{x}\right\},z\right)
&=&
4 \left[-2 \sqrt{z}
+\HA_{0,-1}\left(\sqrt{z}\right)
+\HA_{0,1}\left(\sqrt{z}\right)
\right] \HA_0\left(\sqrt{z}\right)
\nonumber \\ &&
-8 \HA_{0,0,-1}\left(\sqrt{z}\right)
-8 \HA_{0,0,1}\left(\sqrt{z}\right)
+16 \sqrt{z}~,
\\
G\left(\left\{\frac{1}{1-x},\frac{\sqrt{x}}{1-x},\frac{1}{x}\right\},z\right)
&=&
-4 H_{-1,0}\left(\sqrt{z}\right)
-4 H_{1,0}\left(\sqrt{z}\right)
-2 H_{-1,-1,0}\left(\sqrt{z}\right)
\nonumber \\ &&
-2 H_{-1,1,0}\left(\sqrt{z}\right)
+2 H_{1,-1,0}\left(\sqrt{z}\right)
+2 H_{1,1,0}\left(\sqrt{z}\right)
\nonumber \\ &&
+4 H_{-1}\left(\sqrt{z}\right)
+4 H_1\left(\sqrt{z}\right)
+8 \sqrt{z} H_0\left(\sqrt{z}\right)
\nonumber \\ &&
-16 \sqrt{z}~,
\\
G\left(\left\{\frac{\sqrt{x}}{1-x},\frac{1}{1-x},\frac{1}{x}\right\}, z\right)
&=& 
\Bigl[
-8 \sqrt{z}
+\left(4+4 \sqrt{z} - 2 \HA_1\left(\sqrt{z}\right)\right) \HA_{-1}\left(\sqrt{z}\right)
\nonumber \\ &&
-\HA_{-1}(\sqrt{z})^2 
+4 \left(1-\sqrt{z}\right) \HA_1(\sqrt{z})
+\HA_1(\sqrt{z})^2
\nonumber \\ &&
+4 \HA_{-1,1}(\sqrt{z})
\Bigr] \HA_0(\sqrt{z})
+8 \sqrt{z}
-2 \HA_{0,-1,-1}(\sqrt{z})
\nonumber \\ &&
-2 \Bigl[2+2 \sqrt{z} 
-\HA_{-1}(\sqrt{z})
-\HA_1(\sqrt{z})
\Bigr] \HA_{0,-1}(\sqrt{z})
\nonumber \\ &&
-2 \Bigl[
2-2 \sqrt{z} +\HA_{-1}(\sqrt{z})+\HA_1(\sqrt{z})
\Bigr] \HA_{0,1}(\sqrt{z})
\nonumber \\ &&
-2 \HA_{0,-1,1}(\sqrt{z}) 
+2 \HA_{0,1,-1}(\sqrt{z})
+2 \HA_{0,1,1}(\sqrt{z})~,
\\
G\left(\left\{\frac{1}{1-x},\frac{1}{x},\frac{1}{x}\right\},\frac{1}{z}\right)
&=&
\HA_{1,0,0}\left(z\right)
+\HA_{0,0,0}\left(z\right)~,
\\
G\left(\left\{\frac{\sqrt{x}}{1-x},\frac{1}{x},\frac{1}{x}\right\},\frac{1}{z}\right)
&=&
\frac{2 (z -1) \HA_{0,0}\left(z\right)}{\sqrt{z}}
-\frac{4(z +1) \HA_0\left(z\right)}{\sqrt{z}}
\nonumber \\ &&
+\frac{8 (z-1)}{\sqrt{z }}
+G\left(\left\{\frac{\sqrt{x}}{1-x},\frac{1}{x},\frac{1}{x}\right\},z\right)~,
\end{eqnarray}

\begin{eqnarray}
G\left(\left\{\sqrt{x (1-x)},\frac{\sqrt{x (1-x)}}{x \eta -\eta-x},\frac{1}{1-x}\right\},1\right)
&=& 
-\frac{(\eta - 3) \eta^2}{4 (\eta - 1)^4} \HA_{1, 0}(\eta) -
 \frac{3 \zeta_2 \sqrt{\eta}}{4 (\eta - 1)^2} \HA_1(\eta)
\nonumber \\ &&
+\frac{\sqrt{\eta}}{8 (\eta - 1)^2}
  \biggl[(2 \zeta_2 + 4) G\left(\textstyle{\left\{\frac{\sqrt{x}}{1 - x}\right\}}, \eta\right) 
\nonumber \\ &&
-G\left(\textstyle{\left\{\frac{1}{1 - x}, \frac{\sqrt{x}}{1 - x}, \frac{1}{x}\right\}}, \eta\right)
-G\left(\textstyle{\left\{\frac{\sqrt{x}}{1 - x}, \frac{1}{1 - x}, \frac{1}{x}\right\}}, \eta\right)\biggr]
\nonumber \\ &&
+\frac{53 \eta ^2-16 \eta +35}{144 (\eta -1)^3}
+\frac{5 \eta ^3-3 \eta ^2+3 \eta +3}{16 (\eta -1)^4} \zeta_2
\nonumber \\ &&
+\frac{(\eta +1)}{(1-\eta )^2} \left(\frac{7 \zeta_3}{32}-\frac{3}{8} \zeta_2 \ln(2)\right)~,
\\
G\left(\left\{\sqrt{x (1-x)},\frac{\sqrt{x (1-x)}}{x \eta -\eta-x},\frac{1}{x}\right\},1\right)
&=&
\frac{(\eta-3) \eta^2}{4 (\eta -1)^4} \bigl[ 
 \HA_{0,0}\left(\eta\right)
+\HA_{1,0}\left(\eta\right)
\bigr]
\nonumber \\ &&
+\frac{\sqrt{\eta}}{8 (\eta-1)^2} \biggl[
6 \zeta_2 \HA_1(\eta)
+6 \zeta_2 \HA_0(\eta)
\nonumber \\ &&
+G\left({\textstyle\left\{\frac{1}{1-x},\frac{\sqrt{x}}{1-x},\frac{1}{x}\right\}},\eta \right)
+G\left({\textstyle\left\{\frac{1}{x},\frac{\sqrt{x}}{1-x},\frac{1}{x}\right\}},\eta \right)
\nonumber \\ &&
+G\left({\textstyle\left\{\frac{\sqrt{x}}{1-x},\frac{1}{1-x},\frac{1}{x}\right\}},\eta \right)
+G\left({\textstyle\left\{\frac{\sqrt{x}}{1-x},\frac{1}{x},\frac{1}{x}\right\}},\eta \right)
\biggr]
\nonumber \\ &&
-\frac{179 \eta^2-160 \eta +53}{144 (\eta-1)^3}
+\frac{7 (\eta +1)\zeta_3}{32 (\eta-1)^2}
\nonumber \\ &&
-\frac{(\zeta_2+2) \sqrt{\eta}}{4 (\eta-1)^2} G\left({\textstyle\left\{\frac{\sqrt{x}}{1-x}\right\}},\eta\right)
+\frac{3(\eta +1) \ln (2)\zeta_2}{8 (\eta-1)^2}
\nonumber \\ &&
-\frac{11 \eta ^3-21 \eta^2+21 \eta -3}{16 (\eta-1)^4} \zeta_2~,
\\
G\left(\left\{\frac{1}{x},\frac{1}{x \eta-\eta-x},\frac{1}{1-x}\right\},1\right)
&=&
\frac{1}{\eta-1} \bigl[\zeta_3
-2\zeta_2 \HA_1\left(\eta\right)
-2 \HA_{1,1,0}\left(\eta\right)
\nonumber \\ &&
-\HA_{1,0,0}\left(\eta\right)
-\HA_{0,1,0}\left(\eta\right)
\bigr]~,
\\
G\left(\left\{\frac{1}{x},\frac{1}{x \eta -\eta-x},\frac{1}{x}\right\},1\right)
&=&
\frac{2}{\eta-1} \bigl[\zeta_2 \HA_1\left(\eta\right)
+\zeta_2 \HA_0\left(\eta\right)
+\HA_{1,1,0}\left(\eta\right)
\nonumber \\ &&
+\HA_{1,0,0}\left(\eta\right)
+\HA_{0,1,0}\left(\eta\right)
+\HA_{0,0,0}\left(\eta\right)
\bigr]~,
\\
G\left(\left\{\sqrt{x (1-x)},\frac{\sqrt{x (1-x)}}{\eta x-x+1},\frac{1}{1-x}\right\},1\right)
&=&
\frac{3 \ln(2) \zeta_2 (\eta +1)}{8 (\eta-1)^2}
-\frac{(\eta -3) \eta^2}{8 (\eta-1)^4} \left[\ln^2(\eta )+2 \HA_{1,0}\left(\eta\right)\right]
\nonumber \\ &&
+\frac{\sqrt{\eta}}{8 (\eta-1)^2} \biggl[ 
4 (1-\zeta_2)G\left({\textstyle\left\{\frac{\sqrt{x}}{1-x}\right\}},\eta\right)
\nonumber \\ &&
-G\left({\textstyle\left\{\frac{1}{1-x},\frac{\sqrt{x}}{1-x},\frac{1}{x}\right\}},\eta \right)
-G\left({\textstyle\left\{\frac{1}{x},\frac{\sqrt{x}}{1-x},\frac{1}{x}\right\}},\eta\right)
\nonumber \\ &&
-G\left({\textstyle\left\{\frac{\sqrt{x}}{1-x},\frac{1}{1-x},\frac{1}{x}\right\}},\eta \right)
-G\left({\textstyle\left\{\frac{\sqrt{x}}{1-x},\frac{1}{x},\frac{1}{x}\right\}},\eta\right)
\biggr]
\nonumber \\ &&
+\frac{1}{(\eta-1)^4} \biggl[
\left(-\frac{19}{16} \zeta_2-\frac{7}{32} \zeta_3+\frac{179}{144}\right) \eta^3
\nonumber \\ &&
+\left(\frac{45}{16} \zeta_2+\frac{7}{32} \zeta_3-\frac{113}{48}\right) \eta^2
+\frac{3}{16} \zeta_2-\frac{7}{32} \zeta_3
\nonumber \\ &&
+\left(-\frac{21}{16} \zeta_2+\frac{7}{32} \zeta_3+\frac{71}{48}\right) \eta 
-\frac{53}{144}
\biggr]~,
\\
G\left(\left\{\sqrt{x (1-x)},\frac{\sqrt{x (1-x)}}{\eta x-x+1},\frac{1}{x}\right\},1\right)
&=&
\frac{(\eta -3) \eta ^2}{4 (\eta-1)^4} \HA_{1,0}\left(\eta\right)
-\frac{3 \ln(2) \zeta_2 (\eta +1)}{8 (\eta-1)^2}
\nonumber \\ &&
+\frac{\sqrt{\eta}}{8 (\eta-1)^2} \biggl[
4 \left(\zeta_2-1\right) G\left({\textstyle\left\{\frac{\sqrt{x}}{1-x}\right\}},\eta\right)
\nonumber \\ &&
+G\left({\textstyle\left\{\frac{1}{1-x},\frac{\sqrt{x}}{1-x},\frac{1}{x}\right\}},\eta \right)
+G\left({\textstyle\left\{\frac{\sqrt{x}}{1-x},\frac{1}{1-x},\frac{1}{x}\right\}},\eta \right)
\biggr]
\nonumber \\ &&
+\frac{1}{(\eta-1)^4} \biggl[
\left(\frac{13}{16} \zeta_2-\frac{7}{32} \zeta_3-\frac{53}{144}\right) \eta^3
\nonumber \\ &&
+\left(-\frac{27}{16} \zeta_2+\frac{7}{32} \zeta_3+\frac{23}{48}\right) \eta^2
+\frac{3}{16} \zeta_2
+\frac{35}{144}
\nonumber \\ &&
+\left(\frac{3}{16} \zeta_2+\frac{7}{32} \zeta_3-\frac{17}{48}\right) \eta
-\frac{7}{32} \zeta_3
\biggr]~,
\\
G\left(\left\{\sqrt{x (1-x)},\sqrt{x (1-x)},\frac{1}{x}\right\},1\right) 
&=&  
-\frac{7 \zeta_3}{128}
+\frac{7}{192}
+\frac{\pi^2}{256} [1-4 \ln(2)]~,
\\
G\left(\left\{\sqrt{x (1-x)},\sqrt{x (1-x)},\frac{1}{1-x}\right\},1\right)
&=&
-\frac{7 \zeta_3}{128}
+\frac{7}{192}
-\frac{\pi^2}{256} [1-4 \ln(2)]~,
\\
G\left(\left\{\frac{1}{x},\frac{\sqrt{x}}{\sqrt{1-x}},\frac{\sqrt{x}}{\sqrt{1-x}}\right\},1\right)
&=&
\frac{3}{4}-\frac{\pi^2}{8} [1-2 \ln(2)]-\frac{7}{8} \zeta_3~,
\\
G\left(\left\{\frac{1}{x+1},\frac{\sqrt{x}}{\sqrt{1-x}},\frac{\sqrt{x}}{\sqrt{1-x}}\right\},1\right)
&=&
   \frac{5}{4}-\ln(2)+\frac{\pi^2}{8} \left[2\sqrt{2}-3+\ln \left(24-16
   \sqrt{2}\right)\right]
\nonumber \\ &&
-\sqrt{2} \text{Li}_2\left(3-2 \sqrt{2}\right)
+\sqrt{2} \text{Li}_2\left(-3+2 \sqrt{2}\right)
\nonumber \\ &&
-\frac{1}{2} \text{Li}_3\left(3-2 \sqrt{2}\right)
+\frac{1}{2} \text{Li}_3\left(-3+2 \sqrt{2}\right)~,
\\
G\left(\left\{\frac{\sqrt{x}}{\sqrt{1-x}},\frac{\sqrt{x}}{\sqrt{1-x}},\frac{1}{x}\right\},1\right)
&=&
-\frac{3 \zeta_2 \ln(2)}{2}
+\frac{3 \zeta_2}{4}
-\frac{7 \zeta_3}{8}
+\frac{3}{4}~,
\\
   {G}\left(\left\{\frac{\sqrt{x}
   }{\sqrt{1-x}},\frac{\sqrt{x}}{\sqrt{1-x}},\frac{1}{x+1}\right\},1\right) 
&=&  
-\frac{15 \zeta_2 \ln(2)}{4}
-\ln(2)-\frac{3 \zeta_2}{\sqrt{2}}
+\frac{9 \zeta_2}{4}
+\frac{5}{4} 
\nonumber \\ &&
+\frac{1}{2} \text{Li}_3\left(-3+2 \sqrt{2}\right)
-\frac{1}{2} \text{Li}_3\left(3-2 \sqrt{2}\right)
\nonumber \\ &&
+\sqrt{2} \text{Li}_2\left(-3+2 \sqrt{2}\right)
-\sqrt{2} \text{Li}_2\left(3-2 \sqrt{2}\right)
\nonumber \\ &&
+\frac{3}{4} \zeta_2 \ln\left(24-16 \sqrt{2}\right)
+3 \zeta_2 \ln\left(1+\sqrt{2}\right)~.
\end{eqnarray}

\noindent
{\bf Acknowledgment.}~
We would like to thank M.~Steinhauser for providing the codes {\tt MATAD 3.0} and {\tt Q2E}/{\tt Exp}. Discussions
with A.~Behring and A.~von Manteuffel are gratefully acknowledged. The graphs have been drawn using 
{\tt Axodraw}~\cite{Vermaseren:1994je}. This work was supported in part by Studienstiftung des Deutschen Volkes, the 
Austrian Science Fund (FWF) grant SFB F50 (F5009-N15), the European Commission through PITN-GA-2012-316704 
({HIGGSTOOLS}), and by FP7 ERC Starting Grant  257638 PAGAP.

\newpage


\begin{thebibliography}{99}
%
\bibitem{Bierenbaum:2009mv}
  I.~Bierenbaum, J.~Bl\"umlein and S.~Klein,
  Nucl.\ Phys.\ B {\bf 820} (2009) 417
  [arXiv:0904.3563 [hep-ph]].
%
\bibitem{HLO}
  E.~Witten,
  Nucl.\ Phys.\  B {\bf 104} (1976) 445;\\
  J.~Babcock, D.W.~Sivers and S.~Wolfram,
  Phys.\ Rev.\  D {\bf 18} (1978) 162;\\
  M.A.~Shifman, A.I.~Vainshtein and V.I.~Zakharov,
  Nucl.\ Phys.\  B {\bf 136} (1978) 157
  [Yad.\ Fiz.\  {\bf 27} (1978) 455];\\
  J.P.~Leveille and T.J.~Weiler,
  Nucl.\ Phys.\  B {\bf 147} (1979) 147;\\
  M.~Gl\"uck, E.~Hoffmann and E.~Reya,
  Z.\ Phys.\  C {\bf 13} (1982) 119.
%
\bibitem{HNLO}
  E.~Laenen, S.~Riemersma, J.~Smith and W.L.~van Neerven,
  Nucl.\ Phys.\  B {\bf 392} (1993) 162;
  229;\\
  S.~Riemersma, J.~Smith and W.L.~van Neerven,
  Phys.\ Lett.\  B {\bf 347} (1995) 143
  [arXiv:hep-ph/9411431].
%
\bibitem{Alekhin:2003ev}
  S.I.~Alekhin and J.~Bl\"umlein,
  Phys.\ Lett.\  B {\bf 594} (2004) 299
  [arXiv:hep-ph/0404034].
%
\bibitem{Bethke:2011tr}
  S.~Bethke et al., 
  {\sf Proceedings of the 2011 Workshop on Precision Measurements of $\alpha_s$},
  arXiv:1110.0016 [hep-ph];\\
  S.~Moch, S.~Weinzierl et al., 
  arXiv:1405.4781 [hep-ph].  \\    
  D.~d'Enterria and P.~Z.~Skands,
  arXiv:1512.05194 [hep-ph];\\
  S.~Alekhin, J.~Bl\"umlein and S.~O.~Moch,
  Mod.\ Phys.\ Lett.\ A {\bf 31} (2016) no.25,  1630023;\\
  J.~Bl\"umlein, H.~B\"ottcher and A.~Guffanti,
  Nucl.\ Phys.\  B {\bf 774} (2007) 182
  [arXiv:hep-ph/0607200];\\
  J.~Bl\"umlein,
  Prog.\ Part.\ Nucl.\ Phys.\  {\bf 69} (2013) 28
  [arXiv:1208.6087 [hep-ph]].
%
\bibitem{Blumlein:2006be}
  J.~Bl\"umlein, H.~B\"ottcher and A.~Guffanti,
  Nucl.\ Phys.\ B {\bf 774} (2007) 182
  [hep-ph/0607200].
%
\bibitem{Alekhin:2017kpj}
  S.~Alekhin, J.~Bl\"umlein, S.~Moch and R.~Placakyte,
  {\it Parton Distribution Functions, $\alpha_s$ and Heavy-Quark Masses for LHC Run II},
  arXiv:1701.05838 [hep-ph].
%
\bibitem{LHC}
  R.~Hamberg, W.L.~van Neerven and T.~Matsuura,
  Nucl.\ Phys.\ B {\bf 359} (1991) 343; Erratum ibid. B {\bf 644} (2002) 403];\\
  J.~Currie, A.~Gehrmann-De Ridder, E.W.N.~Glover and J.~Pires,
  JHEP {\bf 1401} (2014) 110
  [arXiv:1310.3993 [hep-ph]];\\
  M.~Czakon, P.~Fiedler and A.~Mitov,
  Phys.\ Rev.\ Lett.\  {\bf 110} (2013) 252004
  [arXiv:1303.6254 [hep-ph]];\\
  C.~Anastasiou, C.~Duhr, F.~Dulat, F.~Herzog and B.~Mistlberger,
  Phys.\ Rev.\ Lett.\  {\bf 114} (2015) 21,  212001
  [arXiv:1503.06056 [hep-ph]];
  C.~Anastasiou, C.~Duhr, F.~Dulat, E.~Furlan, F.~Herzog and B.~Mistlberger,
  JHEP {\bf 1508} (2015) 051
  [arXiv:1505.04110 [hep-ph]];\\
  C.~Anastasiou, C.~Duhr, F.~Dulat, E.~Furlan, T.~Gehrmann, F.~Herzog, A.~Lazopoulos and B.~Mistlberger,
  arXiv:1602.00695 [hep-ph].
%
\bibitem{Boer:2011fh}
  D.~Boer, M.~Diehl, R.~Milner, R.~Venugopalan, W.~Vogelsang, D.~Kaplan, H.~Montgomery and S.~Vigdor {\it et al.},
  arXiv:1108.1713 [nucl-th].
%
\bibitem{AbelleiraFernandez:2012cc}
  J.L.~Abelleira Fernandez {\it et al.}  [LHeC Study Group Collaboration],
  J.\ Phys.\ G {\bf 39} (2012) 075001
  [arXiv:1206.2913 [physics.acc-ph]].
%
\bibitem{Buza:1995ie}
  M.~Buza, Y.~Matiounine, J.~Smith, R.~Migneron and W.L.~van Neerven,
  Nucl.\ Phys.\  B {\bf 472} (1996) 611
  [arXiv:hep-ph/9601302].
%
\bibitem{Bierenbaum:2007qe}
  I.~Bierenbaum, J.~Bl\"umlein and S.~Klein,
  Nucl.\ Phys.\  B {\bf 780} (2007) 40
  [arXiv:hep-ph/0703285].
%
\bibitem{Buza:1996wv}
  M.~Buza, Y.~Matiounine, J.~Smith and W.L.~van Neerven,
  Eur.\ Phys.\ J.\  C {\bf 1} (1998) 301
  [arXiv:hep-ph/9612398].
%
\bibitem{WIL2}
  W.L.~van Neerven and E.B.~Zijlstra,
  Phys.\ Lett.\  B {\bf 272} (1991) 127;\\
  E.B.~Zijlstra and W.L.~van Neerven,
  Phys.\ Lett.\  B {\bf 273} (1991) 476;
  Nucl.\ Phys.\  B {\bf 383} (1992) 525;\\
  S.A.~Larin and J.A.M.~Vermaseren,
  Z.\ Phys.\  C {\bf 57} (1993) 93;\\
  S.~Moch and J.A.M.~Vermaseren,
  Nucl.\ Phys.\  B {\bf 573} (2000) 853
  [arXiv:hep-ph/9912355].
%
  \bibitem{Moch:2004pa}
  S.~Moch, J.A.M.~Vermaseren and A.~Vogt,
  Nucl.\ Phys.\ B {\bf 688} (2004) 101
  [hep-ph/0403192].
%
\bibitem{Vogt:2004mw}
  A.~Vogt, S.~Moch and J.A.M.~Vermaseren,
  Nucl.\ Phys.\ B {\bf 691} (2004) 129
  [hep-ph/0404111].
%
\bibitem{Blumlein:2006mh}
  J.~Bl\"umlein, A.~De Freitas, W.L.~van Neerven and S.~Klein,
  Nucl.\ Phys.\  B {\bf 755} (2006) 272
  [arXiv:hep-ph/0608024].
%
\bibitem{Behring:2014eya}
  A.~Behring, I.~Bierenbaum, J.~Bl\"umlein, A.~De Freitas, S.~Klein and F.~Wi\ss{}brock,
  Eur.\ Phys.\ J.\ C {\bf 74} (2014) 9,  3033
  [arXiv:1403.6356 [hep-ph]].
%
\bibitem{Ablinger:2010ty}
  J.~Ablinger, J.~Bl\"umlein, S.~Klein, C.~Schneider and F.~Wi\ss{}brock,
  Nucl.\ Phys.\ B {\bf 844} (2011) 26
  [arXiv:1008.3347 [hep-ph]].
%
\bibitem{Blumlein:2012vq}
  J.~Bl\"umlein, A.~Hasselhuhn, S.~Klein and C.~Schneider,
  Nucl.\ Phys.\ B {\bf 866} (2013) 196
  [arXiv:1205.4184 [hep-ph]].
%
\bibitem{Ablinger:2014vwa}
  J.~Ablinger, A.~Behring, J.~Bl\"umlein, A.~De Freitas, A.~Hasselhuhn, A.~von Manteuffel, 
  M.~Round, C.~Schneider, and F.~Wi\ss{}brock,
  Nucl.\ Phys.\ B {\bf 886} (2014) 733
  [arXiv:1406.4654 [hep-ph]];\\
  A.~Behring, J.~Bl\"umlein, A.~De Freitas, A.~von Manteuffel and C.~Schneider,
  Nucl.\ Phys.\ B {\bf 897} (2015) 612
  [arXiv:1504.08217 [hep-ph]].
%
\bibitem{Ablinger:2014nga}
  J.~Ablinger, A.~Behring, J.~Bl\"umlein, A.~De Freitas, A.~von Manteuffel and C.~Schneider,
  Nucl.\ Phys.\ B {\bf 890} (2014) 48
  [arXiv:1409.1135 [hep-ph]].
%
\bibitem{Ablinger:2014lka}
  J.~Ablinger, J.~Bl\"umlein, A.~De Freitas, A.~Hasselhuhn, A.~von Manteuffel, M.~Round, C.~Schneider and 
  F.~Wi\ss{}brock,
  Nucl.\ Phys.\ B {\bf 882} (2014) 263
  [arXiv:1402.0359 [hep-ph]].
%
\bibitem{Ablinger:2014uka}
  J.~Ablinger, J.~Bl\"umlein, A.~De Freitas, A.~Hasselhuhn, A.~von Manteuffel, M.~Round and C.~Schneider,
  Nucl.\ Phys.\ B {\bf 885} (2014) 280
  [arXiv:1405.4259 [hep-ph]].
%
\bibitem{AGG}
  J.~Ablinger et al., DESY 15--112~.
%
\bibitem{Blumlein:2014zxa}
  J.~Bl\"umlein, A.~De Freitas and C.~Schneider,
  Nucl.\ Part.\ Phys.\ Proc.\  {\bf 261-262} 185
  [arXiv:1411.5669 [hep-ph]];\\
  J.~Ablinger, A.~Behring, J.~Bl\"umlein, A.~De~Freitas, A. Hasselhuhn, A.~von~Manteuffel, C.G.~Raab,
  M.~Round, C.~Schneider, and F.~Wi\ss{}brock,
  PoS (EPS-HEP2015) 504 [arXiv:1602.00583 [hep-ph]];\\
J.~Ablinger, A.~Behring, J.~Bl\"umlein, A.~De Freitas, A.~Hasselhuhn, A.~von Manteuffel, M.~Round, 
C.~Schneider and F.~Wi\ss{}brock, 
  PoS (QCDEV2016) 052
  [arXiv:1611.01104 [hep-ph]].
%
\bibitem{Ablinger:2012qm}
  J.~Ablinger, J.~Bl\"umlein, A.~Hasselhuhn, S.~Klein, C.~Schneider and F.~Wi\ss{}brock,
  Nucl.\ Phys.\ B {\bf 864} (2012) 52
  [arXiv:1206.2252 [hep-ph]].
%
\bibitem{Ablinger:2014yaa}   
  J.~Ablinger, J.~Bl\"umlein, C.~Raab, C.~Schneider and F.~Wi\ss{}brock,
  Nucl.\ Phys.\ B {\bf 885} (2014) 409
  [arXiv:1403.1137 [hep-ph]].
%
\bibitem{Ablinger:2015tua}
  J.~Ablinger, A.~Behring, J.~Bl\"umlein, A.~De Freitas, A.~von Manteuffel and C.~Schneider,
  Comput.\ Phys.\ Commun.\  {\bf 202} (2016) 33
  [arXiv:1509.08324 [hep-ph]].
%
\bibitem{CC}
  T.~Gottschalk,
  Phys.\ Rev.\ D {\bf 23} (1981) 56;\\
  M.~Gl\"uck, S.~Kretzer and E.~Reya,
  Phys.\ Lett.\ B {\bf 398} (1997) 381
   [Phys.\ Lett.\ B {\bf 405} (1997) 392]
  [hep-ph/9701364];\\
  M.~Buza and W.L.~van Neerven,
  Nucl.\ Phys.\ B {\bf 500} (1997) 301
  [hep-ph/9702242];\\
  J.~Bl\"umlein, A.~Hasselhuhn, P.~Kovacikova and S.~Moch,
  Phys.\ Lett.\ B {\bf 700} (2011) 294
  [arXiv:1104.3449 [hep-ph]];\\
  J.~Bl\"umlein, A.~Hasselhuhn and T.~Pfoh,
  Nucl.\ Phys.\ B {\bf 881} (2014) 1
  [arXiv:1401.4352 [hep-ph]];\\
  A.~Behring, J.~Bl\"umlein, A.~De Freitas, A.~Hasselhuhn, A.~von Manteuffel and C.~Schneider,
  Phys.\ Rev.\ D {\bf 92} (2015) 11,  114005
  [arXiv:1508.01449 [hep-ph]];\\
  A.~Behring, J.~Bl\"umlein, G.~Falcioni, A.~De Freitas, A.~von Manteuffel and C.~Schneider,
  Phys.\ Rev.\ D {\bf 94} (2016) no.11,  114006
  [arXiv:1609.06255 [hep-ph]];\\
  J.~Bl\"umlein, G.~Falcioni and A.~De Freitas,
  Nucl.\ Phys.\ B {\bf 910} (2016) 568
  [arXiv:1605.05541 [hep-ph]].
%
\bibitem{Ablinger:2011pb}
  J.~Ablinger, J.~Bl\"umlein, S.~Klein, C.~Schneider and F.~Wi\ss{}brock,
  arXiv:1106.5937 [hep-ph].
%
\bibitem{Ablinger:2012qj}
  J.~Ablinger, J.~Bl\"umlein, A.~Hasselhuhn, S.~Klein, C.~Schneider and F.~Wi\ss{}brock,
  PoS {(RADCOR2011)} 031
  [arXiv:1202.2700 [hep-ph]].
%
\bibitem{Wissbrock:2015faa}
F.~Wi\ss{}brock,
{\sf $O(\alpha_s^3)$ contributions to the heavy flavor Wilson coefficients of the structure function $F_2(x, Q^2)$ 
at $Q^2 \gg m^2$}, DESY-THESIS-2015-040.
%
\bibitem{Alekhin:2012vu}
  S.I.~Alekhin, J.~Bl\"umlein, K.~Daum, K.~Lipka and S.~Moch,
  Phys.\ Lett.\ B {\bf 720} (2013) 172
  [arXiv:1212.2355 [hep-ph]].
%
\bibitem{PDG}
K.A. Olive et al. (Particle Data Group), Chin. Phys. C {\bf 38} (2014) 090001.
%
\bibitem{WIL1}
  W.~Furmanski and R.~Petronzio,
  Z.\ Phys.\  C {\bf 11} (1982) 293 and references therein.
%
\bibitem{Vermaseren:2005qc}
  J.A.M.~Vermaseren, A.~Vogt and S.~Moch,
  Nucl.\ Phys.\  B {\bf 724} (2005) 3
  [arXiv:hep-ph/0504242].
%
\bibitem{LCE}
  K.G.~Wilson,
  Phys.\ Rev.\  {\bf 179} (1969) 1499;\\
R.A.~Brandt and G.~Preparata, Fortschr. Phys. {\bf 18} (1970)
249;\\
W.~Zimmermann, {\sf Lect. on Elementary Particle Physics and
Quantum
Field Theory}, Brandeis Summer Inst., Vol.~1,
(MIT Press, Cambridge, 1970),~p. 395;\\
  Y.~Frishman,
  Annals Phys.\  {\bf 66} (1971) 373;\\
  B.~Geyer, D.~Robaschik and E.~Wieczorek,
  Fortsch.\ Phys.\  {\bf 27} (1979) 75;\\
  E.~Reya,
  Phys.\ Rept.\  {\bf 69} (1981) 195.
%
\bibitem{Moch:2001zr}
  S.~Moch, P.~Uwer and S.~Weinzierl,
  J.\ Math.\ Phys.\  {\bf 43} (2002) 3363
  [arXiv:hep-ph/0110083].
%
\bibitem{Ablinger:2013cf}
  J.~Ablinger, J.~Bl\"umlein and C.~Schneider,
  J.\ Math.\ Phys.\  {\bf 54} (2013) 082301
  [arXiv:1302.0378 [math-ph]].
%
\bibitem{Ablinger:2014bra}
  J.~Ablinger, J.~Bl\"umlein, C.G.~Raab and C.~Schneider,
  J.\ Math.\ Phys.\  {\bf 55} (2014) 112301
  [arXiv:1407.1822 [hep-th]].
%
\bibitem{Ablinger:2013eba}
  J.~Ablinger, J.~Bl\"umlein and C.~Schneider,
  J.\ Phys.\ Conf.\ Ser.\  {\bf 523} (2014) 012060
  [arXiv:1310.5645 [math-ph]].
%
\bibitem{Ablinger:2013jta}
  J.~Ablinger and J.~Bl\"umlein,
  arXiv:1304.7071 [math-ph], In~: {\sf Integration, Summation and Special Functions in Quantum Field 
  Theory}, 
  eds.~J.~Bl\"umlein and C.~Schneider, (Springer, Wien, 2013) 1. 
%
\bibitem{'tHooft:1972fi}
  G.~'t Hooft and M.J.G.~Veltman,
  Nucl.\ Phys.\ B {\bf 44} (1972) 189.
%
\bibitem{Bierenbaum:2007dm}
  I.~Bierenbaum, J.~Bl\"umlein and S.~Klein,
  Phys.\ Lett.\  B {\bf 648} (2007) 195
  [arXiv:hep-ph/0702265].
%
\bibitem{Bierenbaum:2008yu}
  I.~Bierenbaum, J.~Bl\"umlein, S.~Klein and C.~Schneider,
  Nucl.\ Phys.\  B {\bf 803} (2008) 1
  [arXiv:0803.0273 [hep-ph]].
%
\bibitem{Bierenbaum:2009zt}
  I.~Bierenbaum, J.~Bl\"umlein and S.~Klein,
  Phys.\ Lett.\  B {\bf 672} (2009) 401
  [arXiv:0901.0669 [hep-ph]].
%
\bibitem{Gray:1990yh}
  N.~Gray, D.J.~Broadhurst, W.~Grafe and K.~Schilcher,
  Z.\ Phys.\  C {\bf 48},  (1990) 673.
%
\bibitem{Melnikov:2000zc}
  K.~Melnikov and T.~van Ritbergen,
  Nucl.\ Phys.\ B {\bf 591} (2000) 515
  [hep-ph/0005131].
%
\bibitem{Bekavac:2007tk}
  S.~Bekavac, A.~Grozin, D.~Seidel and M.~Steinhauser,
  JHEP {\bf 0710} (2007) 006
  [arXiv:0708.1729 [hep-ph]].
%
\bibitem{Marquard:2016dcn}
  P.~Marquard, A.V.~Smirnov, V.A.~Smirnov, M.~Steinhauser and D.~Wellmann,
  Phys.\ Rev.\ D {\bf 94} (2016) no.7,  074025
  [arXiv:1606.06754 [hep-ph]].
%
\bibitem{Tarrach:1980up}
  R.~Tarrach,
  Nucl.\ Phys.\  B {\bf 183} (1981) 384.
%
\bibitem{Nachtmann:1981zg}
  O.~Nachtmann and W.~Wetzel,
  Nucl.\ Phys.\  B {\bf 187} (1981) 333.
%
\bibitem{Broadhurst:1991fy}
  D.J.~Broadhurst, N.~Gray and K.~Schilcher,
  Z.\ Phys.\  C {\bf 52} (1991) 111.
%
\bibitem{Fleischer:1998dw}
  J.~Fleischer, F.~Jegerlehner, O.V.~Tarasov and O.L.~Veretin,
  Nucl.\ Phys.\  B {\bf 539} (1999) 671
  [Erratum-ibid.\  B {\bf 571} (2000) 511]
  [arXiv:hep-ph/9803493].
%
\bibitem{Remiddi:1999ew}
  E.~Remiddi and J.A.M.~Vermaseren,
  Int.\ J.\ Mod.\ Phys.\ A {\bf 15} (2000) 725
  [hep-ph/9905237].
%
\bibitem{Blumlein:2009cf}
  D.~J.~Broadhurst, J.~A.~Gracey and D.~Kreimer,
  Z.\ Phys.\ C {\bf 75} (1997) 559
  [hep-th/9607174];\\
  J.~Bl\"umlein, D.J.~Broadhurst and J.A.M.~Vermaseren,
  Comput.\ Phys.\ Commun.\  {\bf 181} (2010) 582
  [arXiv:0907.2557 [math-ph]].
%
\bibitem{Abbott:1980hw}
  L.F.~Abbott,
  Nucl.\ Phys.\  B {\bf 185} (1981) 189.
%
\bibitem{Rebhan:1985yf}
  A.~Rebhan,
  Z.\ Phys.\  C {\bf 30} (1986) 309.
%
\bibitem{Jegerlehner:1998zg}
  F.~Jegerlehner and O.V.~Tarasov,
  Nucl.\ Phys.\  B {\bf 549} (1999) 481
  [arXiv:hep-ph/9809485].
%
\bibitem{Khriplovich:1969aa}
  I.B.~Khriplovich,
  Yad.\ Fiz.\  {\bf 10} (1969) 409.
%
\bibitem{tHooft:unpub}
G. t'Hooft, unpublished.
%
\bibitem{Politzer:1973fx}
  H.D.~Politzer,
  Phys.\ Rev.\ Lett.\  {\bf 30} (1973) 1346.
%
\bibitem{Gross:1973id}
  D.J.~Gross and F.~Wilczek,
  Phys.\ Rev.\ Lett.\  {\bf 30} (1973) 1343.
%
\bibitem{Caswell:1974gg}
  W.E.~Caswell,
  Phys.\ Rev.\ Lett.\  {\bf 33} (1974) 244.
%
\bibitem{Jones:1974mm}
  D.R.T.~Jones,
  Nucl.\ Phys.\  B {\bf 75}  (1974) 531.
%
\bibitem{DeWitt:1967ub}
  B.S.~DeWitt,
  Phys.\ Rev.\  {\bf 162} (1967) 1195.
%
\bibitem{YND}
F.J.~Yndurain, {\sf The Theory of Quark and Gluon Interactions}, (Springer, Berlin, 2006, 4th Edition).
%
\bibitem{Matiounine:1998ky}
  Y.~Matiounine, J.~Smith and W.L.~van Neerven,
  Phys.\ Rev.\  D {\bf 57} (1998) 6701
  [arXiv:hep-ph/9801224].
%
\bibitem{CS1}
  K.G.~Chetyrkin and M.~Steinhauser,
  Nucl.\ Phys.\  B {\bf 573}, 617 (2000)
  [arXiv:hep-ph/9911434];
  Phys.\ Rev.\ Lett.\  {\bf 83}, 4001 (1999)
  [arXiv:hep-ph/9907509].
%
\bibitem{Chetyrkin:2008jk}
  K.G.~Chetyrkin, B.A.~Kniehl and M.~Steinhauser,
  Nucl.\ Phys.\  B {\bf 814} (2009) 231
  [arXiv:0812.1337 [hep-ph]].
%
\bibitem{B4cite}
  D.J.~Broadhurst,
  Z.\ Phys.\  C {\bf 54} (1992) 599;\\
  L.~Avdeev, J.~Fleischer, S.~Mikhailov and O.~Tarasov,
  Phys.\ Lett.\  B {\bf 336} (1994) 560
  [Erratum-ibid.\  B {\bf 349} (1995) 597]
  [arXiv:hep-ph/9406363];\\
  S.~Laporta and E.~Remiddi,
  Phys.\ Lett.\  B {\bf 379} (1996) 283
  [arXiv:hep-ph/9602417].
  D.J.~Broadhurst,
  Eur.\ Phys.\ J.\  C {\bf 8} (1999) 311
  [arXiv:hep-th/9803091];\\
  R.~Boughezal, J.B.~Tausk and J.J.~van der Bij,
  Nucl.\ Phys.\  B {\bf 713} (2005) 278
  [arXiv:hep-ph/0410216].
%
\bibitem{Harlander:1997zb}
  R.~Harlander, T.~Seidensticker and M.~Steinhauser,
  Phys.\ Lett.\ B {\bf 426} (1998) 125
  [hep-ph/9712228].
%
\bibitem{Seidensticker:1999bb}
  T.~Seidensticker,
  hep-ph/9905298.
%
\bibitem{Klein:2009ig}
  S.W.G.~Klein,
  {\sf Mellin Moments of Heavy Flavor Contributions to $F_2(x,Q^2)$ at NNLO}, PhD Thesis,
  TU Dortmund (2009) arXiv:0910.3101 [hep-ph].
%
\bibitem{Steinhauser:2000ry}
    M.~Steinhauser,
    Comput.\ Phys.\ Commun.\  {\bf 134} (2001) 335,
    [arXiv:hep-ph/0009029]; code~{\sf MATAD 3.0}.
%
\bibitem{Nogueira:1991ex}
  P.~Nogueira,
  J.\ Comput.\ Phys.\  {\bf 105} (1993) 279.
%
\bibitem{vanRitbergen:1998pn}
  T.~van Ritbergen, A.N.~Schellekens and J.A.M.~Vermaseren,
  Int.\ J.\ Mod.\ Phys.\  A {\bf 14}  (1999) 41
  [arXiv:hep-ph/9802376].
%
\bibitem{Blumlein:2003gb}
  J.~Bl\"umlein,
  Comput.\ Phys.\ Commun.\  {\bf 159} (2004) 19
  [arXiv:hep-ph/0311046].
%
\bibitem{Chetyrkin:1988zz}
  K.G.~Chetyrkin,
  Theor.\ Math.\ Phys.\  {\bf 75} (1988) 346
   [Teor.\ Mat.\ Fiz.\  {\bf 75} (1988) 26].
%
\bibitem{Chetyrkin:1988cu}
  K.G.~Chetyrkin,
  Theor.\ Math.\ Phys.\  {\bf 76} (1988) 809
   [Teor.\ Mat.\ Fiz.\  {\bf 76} (1988) 207].
%
\bibitem{Tkachov:1997gz}
  F.V.~Tkachov,
  Sov.\ J.\ Part.\ Nucl.\  {\bf 25} (1994) 649
  [hep-ph/9701272].
%
\bibitem{Gorishnii:1986gn}
  S.G.~Gorishnii and S.A.~Larin,
  Nucl.\ Phys.\ B {\bf 283} (1987) 452.
%
\bibitem{FORM}
  J.A.M.~Vermaseren,
  math-ph/0010025;\\
  M.~Tentyukov and J.A.M.~Vermaseren,
  Comput.\ Phys.\ Commun.\  {\bf 181} (2010) 1419
  [hep-ph/0702279].
%
\bibitem{ANDI1}
  S.A.~Larin, T.~van Ritbergen and J.A.M.~Vermaseren,
  Nucl.\ Phys.\  B {\bf 427} (1994) 41;\\
  S.A.~Larin, P.~Nogueira, T.~van Ritbergen and J.A.M.~Vermaseren,
  Nucl.\ Phys.\  B {\bf 492} (1997) 338
  [arXiv:hep-ph/9605317];\\
  A.~Retey and J.A.M.~Vermaseren,
  Nucl.\ Phys.\  B {\bf 604} (2001) 281
  [arXiv:hep-ph/0007294];\\
  J.~Bl\"umlein and J.A.M.~Vermaseren,
  Phys.\ Lett.\  B {\bf 606} (2005) 130
  [arXiv:hep-ph/0411111].
%
\bibitem{GUESS}
M. Kauers, {\sf Guessing Handbook}, Technical Report RISC 09-07, J.Kepler University, Linz. 
%
\bibitem{Blumlein:2009tj}
  J.~Bl\"umlein, M.~Kauers, S.~Klein and C.~Schneider,
  Comput.\ Phys.\ Commun.\  {\bf 180} (2009) 2143
  [arXiv:0902.4091 [hep-ph]].
%
\bibitem{Brown:2008um}
  F.~Brown,
  Commun.\ Math.\ Phys.\  {\bf 287} (2009) 925
  [arXiv:0804.1660 [math.AG]].
%
\bibitem{Blumlein:2017dxp}
  J.~Bl\"umlein and C.~Schneider,
  arXiv:1701.04614 [hep-ph]. 
%
\bibitem{MB1a}
E.W.~Barnes, 
Proc. Lond. Math. Soc. (2) {\bf 6} (1908) 141.
%
\bibitem{MB1b}
E.W.~Barnes,
Quarterly Journal of Mathematics {\bf 41} (1910) 136.
%
\bibitem{MB2}
H.~Mellin,
Math. Ann. {\bf 68}, no. 3 (1910) 305.
%
\bibitem{MB3}
E.T.~Whittaker and G.N.~Watson, {\sf A Course of Modern Analysis}, (Cambridge University Press, Cambridge, 1927;
                   reprinted 1996) 616~p.
%
\bibitem{MB4}
E.C.~Titchmarsh,
{\sf Introduction to the Theory of Fourier Integrals},
(Calendron Press, Oxford, 1937; 2nd Edition 1948).
%
\bibitem{HYP}
W.N.~Bailey, {\sf Generalized Hypergeometric Series}, (Cambridge University
Press,  Cambridge, 1935);\\
L.J.~Slater, {\sf Generalized Hypergeometric Functions}, (Cambridge University
Press, Cambridge, 1966);\\
P.~Appell and J.~Kamp\'{e} de F\'{e}riet, {\sf Fonctions
Hyperg\'{e}om\'{e}triques et Hypersp\'{e}riques, Polynomes D' Hermite},
(Gauthier-Villars, Paris, 1926);\\
P.~Appell, {\sf Les Fonctions Hyperg\"{e}om\'{e}triques de Plusieur
Variables}, (Gauthier-Villars, Paris, 1925);\\
J.~Kamp\'{e} de F\'{e}riet, {\sf La fonction
hyperg\"{e}om\'{e}trique},(Gauthier-Villars, Paris, 1937);\\
H.~Exton, {\sf Multiple Hypergeometric Functions and Applications},
(Ellis Horwood, Chichester, 1976);\\
H.~Exton, {\sf Handbook of Hypergeometric Integrals},
(Ellis Horwood, Chichester, 1978);\\
H.M.~Srivastava and P.W. Karlsson, {\sf Multiple Gaussian Hypergeometric
Series}, (Ellis Horwood, Chicester, 1985).
%
\bibitem{Huber:2007dx}
  T.~Huber and D.~Maitre,
  Comput.\ Phys.\ Commun.\  {\bf 178} (2008) 755
  [arXiv:0708.2443 [hep-ph]].
%
\bibitem{POLYLOG1}
L.~Lewin, {\sf Dilogarithms and associated functions}, (Macdonald, London, 1958).
%
\bibitem{POLYLOG2}
L.~Lewin, {\sf Polylogarithms and associated functions},
(North Holland, New York, 1981).
%
\bibitem{Devoto:1983tc}
  A.~Devoto and D.W.~Duke,
  Riv.\ Nuovo Cim.\  {\bf 7N6} (1984) 1.
%
\bibitem{SIG1}
C.~Schneider, {S\'em.~Lothar. Combin.\/} {\bf 56} (2007) 1, 
 article B56b.
%
\bibitem{SIG2}
C.~Schneider, {{\sf Computer Algebra in Quantum Field Theory: Integration,
  Summation and Special Functions}\/} Texts and Monographs in Symbolic
  Computation eds. C.~Schneider and J.~Bl\"umlein  (Springer, Wien, 2013) 325, 
  arXiv:1304.4134 [cs.SC].
%
\bibitem{HSUM}
  J.A.M.~Vermaseren,
  Int.\ J.\ Mod.\ Phys.\ A {\bf 14} (1999) 2037
  [hep-ph/9806280].
  J.~Bl\"umlein and S.~Kurth,
  Phys.\ Rev.\  D {\bf 60} (1999) 014018
  [arXiv:hep-ph/9810241].
%
\bibitem{STRUCT}
  J.~Bl\"umlein,
  arXiv:0901.3106 [hep-ph];
  arXiv:0901.0837 [math-ph].
%
\bibitem{ANCONT}
  J.~Bl\"umlein,
  Comput.\ Phys.\ Commun.\  {\bf 133} (2000) 76
  [arXiv:hep-ph/0003100];\\
  J.~Bl\"umlein and S.O.~Moch,
  Phys.\ Lett.\  B {\bf 614} (2005) 53
  [arXiv:hep-ph/0503188].
%
\bibitem{Velizhanin:2012nm}
  V.N.~Velizhanin,
  Nucl.\ Phys.\ B {\bf 864} (2012) 113
  [arXiv:1203.1022 [hep-ph]].
%
\bibitem{Ablinger:2011te}
  J.~Ablinger, J.~Bl\"umlein and C.~Schneider,
  J.\ Math.\ Phys.\  {\bf 52} (2011) 102301
  [arXiv:1105.6063 [math-ph]].
%
\bibitem{HARMONICSUMS}
  J.~Ablinger,
  PoS {(LL2014)} 019;
  {\sf Computer Algebra Algorithms for Special Functions in Particle Physics}, Ph.D. Thesis, J. Kepler University 
Linz, 2012,
  arXiv:1305.0687 [math-ph];\\
  {\sf A Computer Algebra Toolbox for Harmonic Sums Related to Particle Physics}, Diploma Thesis, J. Kepler University Linz, 2009,
  arXiv:1011.1176 [math-ph].
%
\bibitem{Czakon:2005rk}
  M.~Czakon,
  Comput.\ Phys.\ Commun.\  {\bf 175} (2006) 559
  [hep-ph/0511200];\\
  A.V.~Smirnov and V.A.~Smirnov,
  Eur.\ Phys.\ J.\ C {\bf 62} (2009) 445
  [arXiv:0901.0386 [hep-ph]].
%
\bibitem{Hasselhuhn:2013swa}
  A.~Hasselhuhn,
  {\sf 3-Loop Contributions to Heavy Flavor Wilson Coefficients of Neutral and Charged Current 
  Deep-Inelastic Scattering},
  DESY-THESIS-2013-050.
%
\bibitem{Blumlein:1989gk}
  J.~Bl\"umlein,
  Z.\ Phys.\ C {\bf 47} (1990) 89.
%
\bibitem{Vermaseren:1994je}
  J.A.M.~Vermaseren,
  Comput.\ Phys.\ Commun.\  {\bf 83} (1994) 45.
\end{thebibliography}
\end{document}